\documentclass[12pt]{article}
\usepackage[margin=1in]{geometry}

\usepackage{cancel}
\usepackage{amsmath,amssymb,bm,url,mathrsfs}
\usepackage{algorithmic}
\usepackage[ruled,linesnumbered]{algorithm2e}
\usepackage{amsthm}
\usepackage{hyperref}
\usepackage{makecell}
\usepackage{tablefootnote}
\usepackage{comment}
\usepackage{soul}
\usepackage{graphics,graphicx}
\usepackage{subcaption,caption,cleveref}
\hypersetup{
    colorlinks=true,
    linkcolor=blue,
    citecolor=blue,
    filecolor=magenta,      
    urlcolor=cyan,
}
\usepackage{booktabs}
\usepackage{multirow}
\usepackage{multicol}
\usepackage{natbib}
\usepackage{xcolor}
\usepackage{dsfont}
\usepackage{bbm}
\usepackage{enumitem}
\usepackage{blkarray}
\usepackage{amsfonts}
\usepackage{caption}
\usepackage{xr}
\makeatletter
\newcommand*{\addFileDependency}[1]{
  \typeout{(#1)}
  \@addtofilelist{#1}
  \IfFileExists{#1}{}{\typeout{No file #1.}}
}
\makeatother

\DeclareMathOperator{\diag}{diag}

\usepackage{footmisc}
\def\bar{\overline}

\newcommand{\mb}{\mathbf }

\def\calA{{\mathcal A}}
\def\calB{{\mathcal B}}
\def\calC{{\mathcal C}}
\def\calD{{\mathcal D}}
\def\calE{{\mathcal E}}

\def\calG{{\mathcal G}}
\def\calH{{\mathcal H}}
\def\calI{{\mathcal I}}
\def\calJ{{\mathcal J}}

\def\calN{{\mathcal N}}

\def\calP{{\mathcal P}}

\def\calS{{\mathcal S}}
\def\calT{{\mathcal T}}

\def\calV{{\mathcal V}}

\def\calZ{{\mathcal Z}}

\def\bcalI{{\boldsymbol{\mathcal I}}}

\def\EE{{\mathbb E}}

\def\II{{\mathbb I}}

\def\NN{{\mathbb N}}
\def\OO{{\mathbb O}}
\def\PP{{\mathbb P}}

\def\RR{{\mathbb R}}
\def\SS{{\mathbb S}}

\DeclareMathOperator*{\argmax}{arg\,max}

\def\bs{\boldsymbol{s}}

\def\calA{{\cal  A}} 
\def\calB{{\cal  B}} 
\def\calC{{\cal  C}} \def\scrC{{\mathscr  C}}
\def\calD{{\cal  D}} 
\def\calE{{\cal  E}} 
 
\def\calG{{\cal  G}} 
\def\calH{{\cal  H}} 
\def\calI{{\cal  I}} 
\def\calJ{{\cal  J}}

\def\calN{{\cal  N}} 
 
\def\calP{{\cal  P}}

\def\calS{{\cal  S}} 
\def\calT{{\cal  T}} 
 
\def\calV{{\cal  V}}

\def\calZ{{\cal  Z}}

\newcommand{\bfm}[1]{\ensuremath{\mathbf{#1}}}

   \def\bA{\bfm A}  
   \def\bB{\bfm B}  
     
   \def\bD{\bfm D}  
\def\be{\bfm e}   \def\bE{\bfm E}  \def\EE{\mathbb{E}}

   \def\bH{\bfm H}  
   \def\bI{\bfm I}  \def\II{\mathbb{I}}

   \def\bM{\bfm M}  
     \def\NN{\mathbb{N}}
   \def\bO{\bfm O}  \def\OO{\mathbb{O}}
   \def\bP{\bfm P}  \def\PP{\mathbb{P}}
     
   \def\bR{\bfm R}  \def\RR{\mathbb{R}}
\def\bs{\bfm s}     \def\SS{\mathbb{S}}
     
   \def\bU{\bfm U}  
   \def\bV{\bfm V}  
   \def\bW{\bfm W}

   \def\bZ{\bfm Z}  
                   
\def\btheta{\bfm \theta}  \def\bTheta{\bfm \Theta}
 
  \def\bOmega{\bfm \Omega}
\def\bSigma{\bfm \Sigma}

\def\bDelta{\bfm \Delta}
\def\bPi{\bfm \Pi}

\def\mPsi{\boldsymbol{\Psi}}

\def\hat{\widehat}
\def\wt{\widetilde}

\newcommand\inp[2]{\left\langle #1, #2 \right\rangle}
\newcommand\fro[1]{\left\| #1 \right\|_{\rm F}}
\newcommand\op[1]{\left\|#1\right\|}
\newcommand\infn[1]{\left\|#1\right\|_{2,\infty}}
\newcommand\brac[1]{\left(#1\right)}
\newcommand\ebrac[1]{\left\{#1\right\}}
\newcommand\sqbrac[1]{\left[#1\right]}
\newcommand\ab[1]{\left|#1\right|}

\newtheorem{theorem}{Theorem}
\newtheorem{definition}{Definition}
\newtheorem{proposition}{Proposition}
\newtheorem{remark}{Remark}

\newtheorem{assumption}{Assumption}
\newtheorem{lemma}{Lemma}

\theoremstyle{remark}

\newcommand{\E}{\mathbb{E}}
\newcommand{\Prob}{\mathbb{P}}

\newcommand\blfootnote[1]{%
  \begingroup
  \renewcommand\thefootnote{}\footnote{#1}%
  \addtocounter{footnote}{-1}%
  \endgroup
}

\newcommand{\blind}{1}

\def\spacingset#1{\renewcommand{\baselinestretch}%
{#1}\small\normalsize}
\spacingset{1}

\begin{document}

\if1\blind
{
\title{Degree-heterogeneous Latent Class Analysis for High-dimensional Discrete Data}

\author{Zhongyuan Lyu$^{*\dagger}$ \and 
Ling Chen$^*$
\and Yuqi Gu$^{*\dagger}$}

\date{$^*$Department of Statistics \& $^\dagger$Data Science Institute, Columbia University}

\maketitle
} \fi

\begin{abstract}
The latent class model is a widely used mixture model for multivariate discrete data. Besides the existence of qualitatively heterogeneous latent classes, real data often exhibit additional quantitative heterogeneity nested within each latent class. The modern latent class analysis also faces extra challenges, including the high-dimensionality, sparsity, and heteroskedastic noise inherent in discrete data. Motivated by these phenomena, we introduce the Degree-heterogeneous Latent Class Model and propose an easy-to-implement \emph{HeteroClustering} algorithm for it. HeteroClustering uses heteroskedastic PCA with $\ell_2$ normalization to remove degree effects and perform clustering in the top singular subspace of the data matrix. We establish the result of exact clustering under minimal signal-to-noise conditions. We further investigate the estimation and inference of the high-dimensional continuous item parameters in the model, which are crucial to interpreting and finding useful markers for latent classes. We provide comprehensive procedures for global testing and multiple testing of these parameters with valid error controls. The superior performance of our methods is demonstrated through extensive simulations and applications to three diverse real-world datasets from political voting records, genetic variations, and single-cell sequencing.
\end{abstract}


\blfootnote{Correspondence to Yuqi Gu, Department of Statistics, Columbia University, New York, NY 10027. Email: \texttt{yuqi.gu@columbia.edu}. Research is partially supported by NSF grant DMS-2210796.}

\section{Introduction}\label{sec:intro}
\spacingset{1}

Uncovering interpretable hidden patterns in high-dimensional data is of great interest in scientific applications and statistical learning \citep{bishop2006pattern}.
One fundamental and basic hidden pattern is the mixture structure, where the data form several latent classes. For example, advances in next-generation sequencing technologies 
made it possible to access high-dimensional single-cell sequencing data, where the mixture is given by a plethora of distinct cell types \citep{kiselev2019challenges}. Latent social groups uncovered from extensive survey response data also widely exist, where the groups are defined by distinct behavioral patterns, political attitudes, or demographic traits \citep{nylund2018ten}. In these applications, it is often meaningful to perform  Latent Class Analysis  \citep[LCA,][]{hagenaars2002applied}. The Latent Class Model (LCM) is a mixture model for multivariate discrete data that assumes a subject's multivariate observed responses are conditionally independent given its latent class membership \citep{goodman1974exploratory}. 
LCM was first proposed by \cite{lazarsfeld1950logical} to uncover latent subgroups from people's binary responses to questionnaire items, and it has attracted great interest from both statisticians and practitioners ever since.
However, there are at least two challenges when adapting LCA for modern data analyses. The first is the ultra-large dimensions of data, as such data become increasingly common in genetics and genomics \citep{kiselev2019challenges},
health sciences \citep{zhang2012latent}, and  social sciences \citep{chen2020structured, gu2023joint}. 
The second challenge stems from LCA's nature as a model for multivariate discrete, often binary, data instead of continuous data. Binary data are ubiquitous in various fields including presence/absence of symptoms in electronic health records, yes/no responses in social science surveys, and correct/wrong answers in educational tests.
However, binary data are challenging to deal with since they are heteroskedastic, often less informative, and potentially very sparse in real-world applications.

\begin{figure}[h!]
\centering
     \begin{subfigure}[b]{0.3\textwidth}
         \centering
         \includegraphics[width=\textwidth]{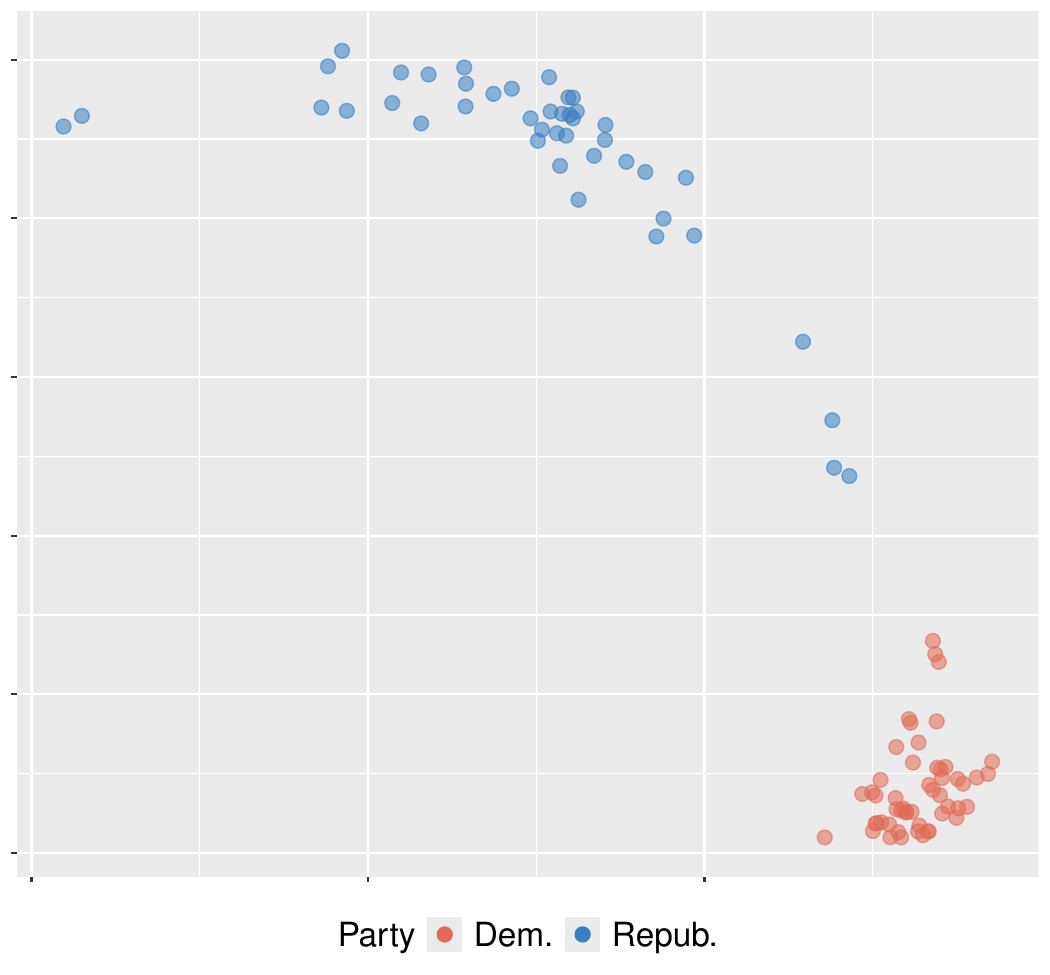}
     \end{subfigure}
     \hfill
     \begin{subfigure}[b]{0.3\textwidth}
         \centering
         \includegraphics[width=\textwidth]{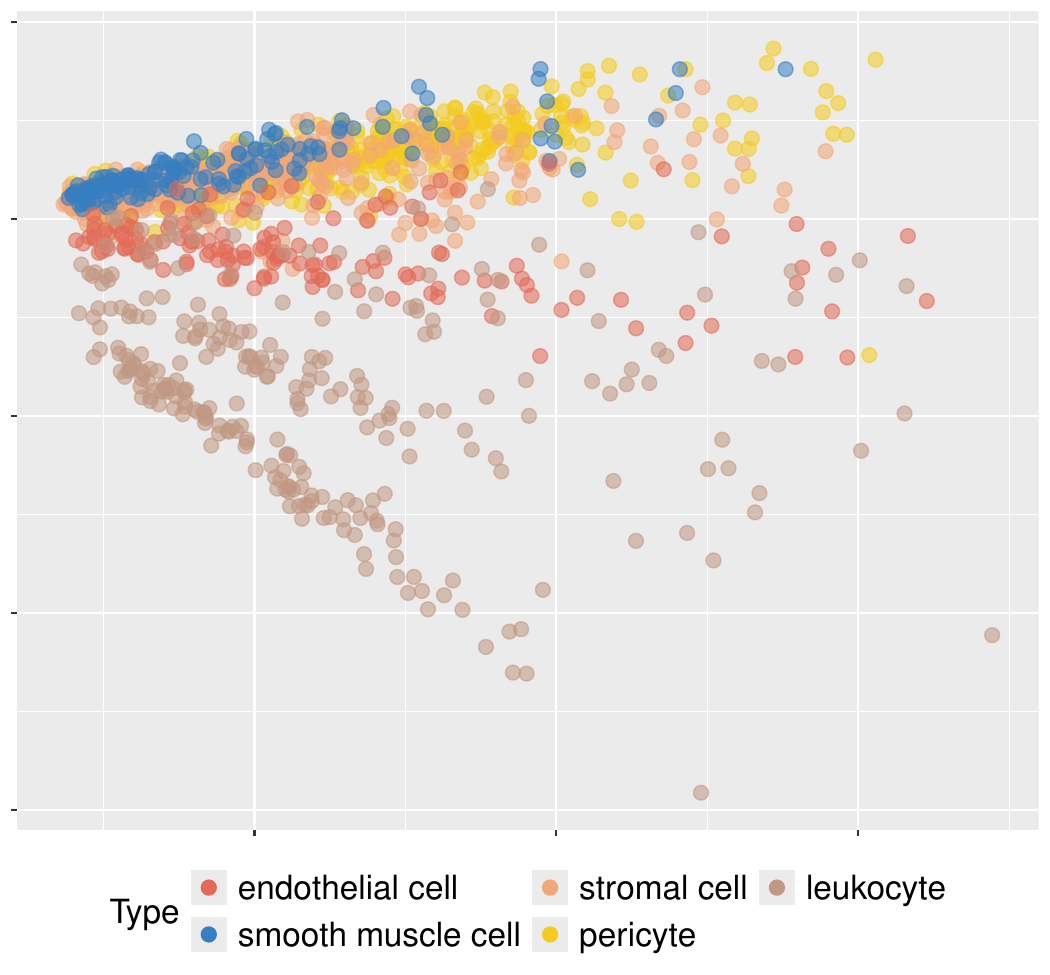}
     \end{subfigure}
     \hfill
     \begin{subfigure}[b]{0.3\textwidth}
         \centering
         \includegraphics[width=\textwidth]{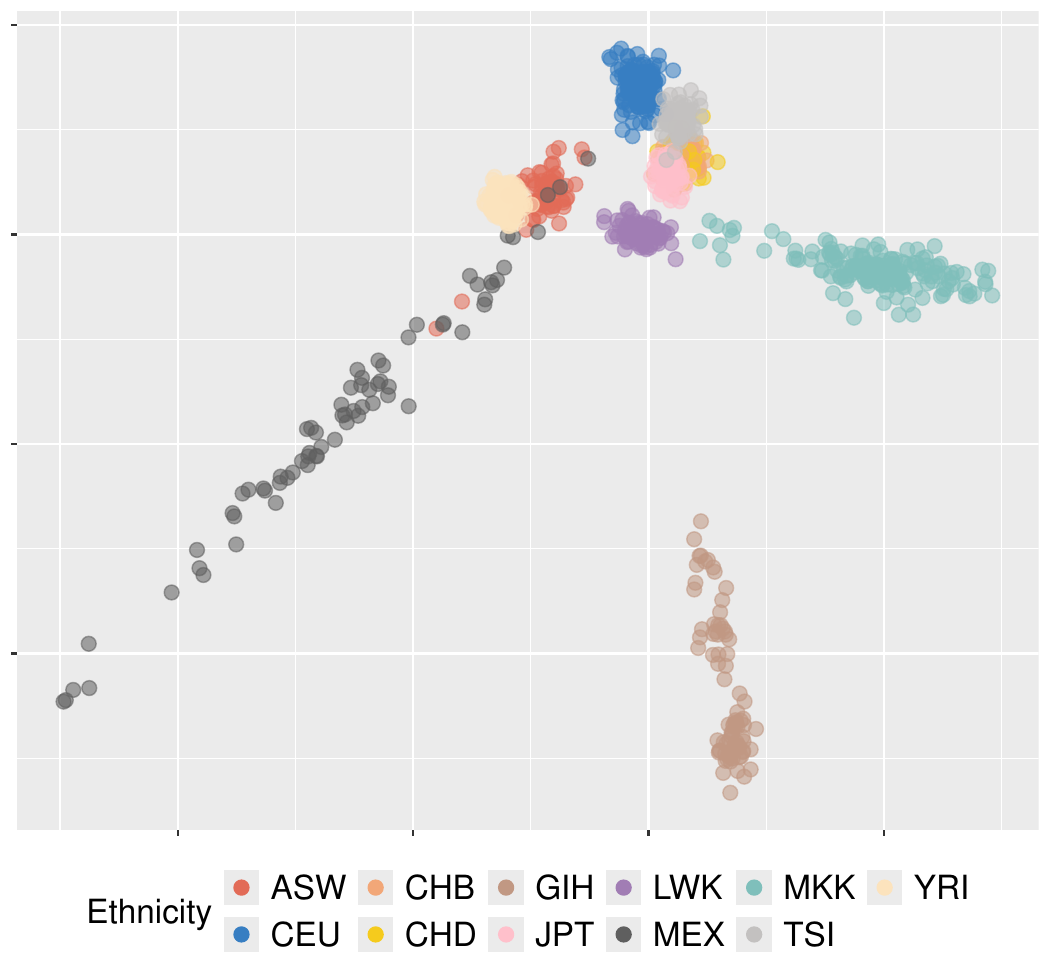}
     \end{subfigure}
\caption{Scatter plots of singular subspace embeddings given by HeteroPCA. Left: first and second singular vectors of the Senate voting data. Middle: first and second singular vectors of the single-cell data. Right: fourth and sixth singular vectors of the genetic variation data. 
}
\label{fig:data}
\end{figure}

LCA itself is a method to capture heterogeneity across different latent classes. More precisely, it identifies \emph{qualitative heterogeneity} via a discrete variable indicating each subject's latent class membership. However, a notable number of applications reveal an additional layer of heterogeneity within these latent classes, which is substantial and cannot be overlooked. We consider three motivating examples of high-dimensional discrete data: a U.S. Senate roll call votes dataset \citep{chen2021unfolding}, a single-cell sequencing dataset \citep{lengyel2022molecular}, and a genetic variation dataset of single nucleotide polymorphisms (SNPs) \citep{international2010integrating}.  Figure \ref{fig:data} shows that in each dataset, the (singular subspace) embeddings of persons/cells reveal clear ``radial streaks" \citep{rohe2023varimax} corresponding to meaningful clusters; see Section \ref{sec:real-data} for details. This pattern underscores a significant shortcoming of most existing LCA methods: their inability to account for \emph{quantitative heterogeneity} at the individual level within each latent class.
Our observations in these diverse datasets validate the wide presence of such heterogeneity, and highlight the necessity for statistical approaches that can capture such complexities.

Motivated by the phenomena of quantitative heterogeneity nested within qualitative heterogeneity, we introduce the Degree-heterogeneous Latent Class Model  (DhLCM).  
We make DhLCM versatile and accommodate a range of discrete distributions, including sparse Bernoulli, Binomial, and Poisson distributions; see Section \ref{sec:model} (and also Section \ref{sec:gen-type} in the Supplementary Material). In the context of binary data where $N$ subjects' $J$ features are observed, the data can be collected in a $N\times J$ matrix $\mb R = (R_{i,j})\in\{0,1\}^{N\times J}$. In a DhLCM, we endow each subject $i$ with a  continuous degree parameter $\omega_i>0$ and assume that given $\omega_i$ and the latent class membership $s_i=k\in[K]$, the $J$ observed variables $R_{i,1},\ldots,R_{i,J}$ are conditionally independent with 
$$
\Prob\brac{R_{i,j}=1\big|s_i=k}=\omega_i\theta_{j,k},$$
where $\theta_{j,k}$ represents the conditional probability of observing a positive response to feature $j$ by a ``typical'' subject from latent class $k$ with degree one $\omega_i=1$.
The DhLCM presents two fundamental and challenging questions that form the crux of this paper. First, with the additional degree effect $\omega_i$'s, what is the appropriate method for clustering? Second, how to  leverage the clustering outcomes for downstream tasks including estimation and statistical inference of continuous parameters? Addressing these two questions is crucial to fully realizing the potential of DhLCM in complex data analysis scenarios.

In terms of methodology for clustering in LCA, the existing literature mainly focuses on likelihood-based approaches \citep{scrucca2023model, fienberg2007maximum,zeng2023tensor}, which is usually computationally inefficient and algorithmically unstable under the high-dimensional regime. Introducing additional degree effects further complicates the optimization problem, rendering it more difficult to solve. On the other hand,  our clustering task is  related to the community detection task in degree-corrected network models, for which spectral methods have been widely adopted and thoroughly investigated \citep{qin2013regularized, jin2015fast, sarkar2015role, gao2018community, zhang2020detecting, deng2023strong}.
Nonetheless, it is still unclear how to appropriately gear spectral methods towards LCA and its variants with rigorous theoretical guarantees.

To fill this gap, 
we propose an easy-to-implement spectral clustering method for DhLCM.
One key observation is that the large $N\times J$ data matrix $\bR$ under a DhLCM can be written as a sum of a low-rank ``signal'' matrix and a ``noise" matrix, where the essential information 
is encoded in the signal matrix. This observation motivates us to exploit the top singular vectors of the data matrix for statistical analyses.
To accurately estimate the top singular subspace, we apply the recently proposed heteroskedastic PCA \citep[HeteroPCA,][]{zhang2022heteroskedastic} due to the high-dimensionality and heteroskedastic noise caused by discrete data. In light of the streak structure  shown in Figure \ref{fig:data}, we further propose to $\ell_2$-normalize the singular subspace embeddings before clustering.
The simultaneous consideration of HeteroPCA and $\ell_2$ normalization not only distinguishes our approach from existing methods, but also poses technical challenges that call for in-depth theoretical analyses.  We manage to show that HeteroClustering in DhLCMs leads to 
exact clustering under the minimal signal-to-noise condition. Notably, it remains unclear whether likelihood-based methods for LCA enjoys comparable theoretical and algorithmic guarantees.

In addition to clustering, another important task in LCA is estimating high-dimensional continuous parameters in $\bTheta=\brac{\theta_{j,k}}\in\RR^{J\times K}$. They characterize how the $J$ observed variables are linked to the $K$ latent classes and give the basis for interpreting each latent class. Estimation and inference of $\bTheta$ are meaningful in many  applications. For example, in single-cell data \citep{stein2018enter}, 
$\bTheta$ is equivalent to the ``amplitude matrix", which has genes as rows and latent factors as columns to reflect gene co-regulation patterns. Distributional results on $\bTheta$ is crucial to determining the differentially expressed genes or biomarkers for clusters \citep{squair2021confronting, lengyel2022molecular}. Another example is in educational assessments, where $\bTheta$ are called ``item parameters" of the assessment questions. They encode the questions' fundamental properties, such as their difficulty levels and discriminatory power across different latent subgroups of students. Accurate estimation of  $\bTheta$ is crucial to identifying questions with high discriminatory power and facilitates their calibration and administration in future tests.
Motivated by these applications, we propose a comprehensive procedure for estimation and inference of $\bTheta$. Blessed with the nice property of our clustering algorithm, we define a simple estimator of $\bTheta$ by averaging the samples belonging to the same estimated latent class, after correcting the degree effects. We establish the sharp estimation error bound and rich distributional results for our estimator $\hat \bTheta$.
These results facilitate methods for global testing and multiple testing with valid error controls.

\medskip

\medskip
\noindent\textbf{Organization.}
Section \ref{sec:model} formally defines the model and reviews related literature. Section \ref{sec:clust} presents the  HeteroClustering algorithm and provides theoretical results of exact clustering. Section \ref{sec:est-theta} proposes procedures for estimation and inference of $\bTheta$ with theoretical guarantees. Section \ref{sec:sim} includes extensive simulation studies. 
Real data applications are presented in Section \ref{sec:real-data}. Section \ref{sec-discussion} gives concluding remarks. The Supplementary Material contains extensions to the Binomial and Poisson models, discussions on key technical quantities,  additional numerical results, and all technical proofs of the theoretical results. 
The \texttt{R} codes implementing the algorithms and reproducing the experiments are available on Github at \href{https://github.com/lscientific/DhLCM}{https://github.com/lscientific/DhLCM}.

\section{Degree-heterogeneous Latent Class Model}\label{sec:model}
In this section, we formally introduce our model DhLCM. First define some notations.
We use  bold capital letters such as $\bA,\bB,\cdots$, to denote matrices. 
For any positive integer $m$, denote $[m]:=\{1,\cdots,m\}$. For any matrix $\bA\in\RR^{N\times J}$ and any $i\in[N]$ and $j\in[J]$, we use $A_{i,j}$ to denote its entry on the $i$-th row and $j$-th column, and use $\bA_{i,:}$ (or $\bA_{:,j}$) to denote its $i$-th row (or $j$-th column) vector. Let $\sigma_k\brac{\bA}$ denote the $k$-th largest singular value of $\bA$ for $k=1,\cdots,\min\ebrac{N,J}$. For any square matrix $\bA$, we define $\calH(\bA)$ to be the diagonal-deleted version of $\bA$ such that $[\calH(\bA)]_{ij}=\bA{_{ij}}$ if $i\ne j$ and $[\calH(\bA)]_{ii}=0$, and $\calD(\bA):=\bA-\calH(\bA)$. Let $\op{\cdot}$ denote the spectral norm (operator norm) for matrices and $\ell_2$ norm for vectors, and $\op{\cdot}_{\sf max}$, $\infn{\cdot},\fro{\cdot}$ denote the max norm, two-to-infinity norm and Frobenius norm for matrices respectively. For two sequences $\{a_N\},\{b_N\}$, we use $a_N\lesssim b_N$ (or $a_N\gtrsim b_N$) if and only if there exists some constant $C>0$ independent of $N$ such that $a_N\le Cb_N$ (or $b_N\le Ca_N$), and use $a_N\asymp b_N$ if and only if $a_N\lesssim b_N$ and $b_N\lesssim a_N$ hold simultaneously.

Consider a dataset with $N$ subjects, each responding to $J$ items/features with binary responses\footnote{For generalizations to Binomial or unbounded count data, see Section \ref{sec:gen-type} in the Supplementary Material.}. The data can be structured as a $N\times J$ binary matrix $\bR=(R_{i,j})\in \{0,1\}^{N\times J}$. Assume there are $K$ latent classes, with the latent class memberships summarized by a vector $s = (s_1,\ldots,s_N)\in[K]^{N}$.  For each $k\in[K]$, there are $J$ item-specific Bernoulli parameters $\{\theta_{j,k}: j=1,\cdots J\}$. We collect $\theta_{j,k}$'s in an item parameter matrix $\bTheta=\brac{\theta_{j,k}}\in\RR^{J\times K}$. In addition, we introduce a vector of degree effects $\omega=\brac{\omega_1,\cdots,\omega_N}$  to capture the quantitative heterogeneity on the subject level. For the $i$-th subject belonging to the $k$-th latent class, the probability of providing a positive response to the $j$th item is 
\begin{align}\label{eq:model-original}
\PP\brac{R_{i,j}=1|\omega,\bTheta} =  \omega_i\theta_{j,k}.
\end{align}
In other words, the probability of observing a certain response is based on latent-class-specific item parameters, modulated by individual-specific degree parameters.

A fundamental question in LCA is to infer the latent class membership of each subject, i.e., recover the vector $s$ up to a permutation of the $K$ latent class labels. One of the most popular ways to achieve this goal is to perform  model-based clustering based on finite mixture modeling \citep{fraley2002model, scrucca2023model}. 
Many model-based clustering methods boil down to specifying a generative model and then estimating parameters by maximizing likelihood. A traditional method is to maximize the {\it marginal likelihood function}, whose form  for a DhLCM is
\begin{align}\label{eq:margin-mle}
L\brac{\pi,\omega,\bTheta|\bR}=\prod_{i=1}^N\sum_{k=1}^K\pi_k\prod_{j=1}^J{\brac{\omega_i\theta_{j,k}}^{\II\brac{R_{i,j}=1}}\brac{1-\omega_i\theta_{j,k}}^{\II\brac{R_{i,j}=0}}},
\end{align}
where $\pi_k:=\PP\brac{s_i=k}$ denotes the mixture proportion of $k$-th latent class satisfying $\pi_k>0$ and $\sum_{k=1}^K\pi_k=1$. Eq.~\eqref{eq:margin-mle} is called the marginal likelihood function because it marginalizes out the latent variables $s$.
The above perspective treats class labels $\{s_i\}_{i=1}^N$ as random variables following a categorical distribution with parameters $\pi=\brac{\pi_1,\cdots,\pi_K}^\top$. 
Another perspective in model-based clustering treats  $\{s_i\}_{i=1}^N$ as fixed unknown parameters and maximizes  the \emph{joint likelihood function} with respect to both $s$ and other parameters:
\begin{align}\label{eq:joint-mle}
L\brac{s,\omega,\bTheta|\bR}=\prod_{i=1}^N\prod_{j=1}^J\sum_{k=1}^K\sqbrac{\brac{\omega_i\theta_{j,k}}^{\II\brac{R_{i,j}=1}}\brac{1-\omega_i\theta_{j,k}}^{\II\brac{R_{i,j}=0}}}^{\II\brac{s_i=k}}.
\end{align}
Unfortunately, both likelihood functions \eqref{eq:margin-mle} and \eqref{eq:joint-mle} suffer from non-convexity, which poses significant challenges to efficient computation. This drawback becomes more acute in the setting of our interest -- that is, when the dimension of the data $J$ is substantially large and even much larger than the sample size $N$.  For instance, even in the degree-homogeneous case when $\omega=1_N^\top $, the task of directly maximizing the function in \eqref{eq:joint-mle} with respect to $\brac{s,\bTheta}$ is hard. People often use the iterative expectation-maximization (EM) algorithms to maximize likelihood in standard LCMs \citep[e.g.,][]{zeng2023tensor}.

In summary, there are three drawbacks in pursuing likelihood-based clustering methods with iterative algorithms. First, the  distribution of data needs to be specified a priori, and the algorithm has to be modified substantially when dealing with different types of data. Second, there is generally no theoretical guarantee for the global convergence of EM or similar algorithms, with only a few exceptions for relatively simple mixture models such as the Gaussian mixture \citep[e.g.,][]{balakrishnan2017em}.
Third, especially in the context of DhLCMs, one has to spare additional effort on estimating the degrees $\ebrac{\omega_i,i\in[N]}$, although our primary goal is often to recover the latent class labels $\ebrac{s_i,i\in[N]}$. Trying to estimate $\omega$ not only brings extra computational load, but also potentially leads to an increased inaccuracy in clustering due to the statistical error of estimating $\omega_i$'s.

To remedy these issues, we propose a spectral approach to DhLCMs by taking a different perspective on the model defined in \eqref{eq:model-original}. Our motivation  hinges on the observation that the $(i,j)$-th entry of  $N\times J$ data matrix $\bR$ can be written as $R_{i,j}=\omega_i\theta_{j,s_i}+ E_{i,j}$ with a mean zero noise $E_{i,j}=R_{i,j}-\omega_i\theta_{j,s_i}$. This observation leads to the  following  decomposition:
\begin{align}\label{eq:model-decomp}
\bR=\underbrace{\bOmega\bZ\bTheta^\top}_{\text{low-rank ``signal"}}+\underbrace{\bE}_{\text{``noise"}},
\end{align}
where $\bOmega=\textsf{diag}(\omega)\in \RR^{N\times N}$, $\bZ=(Z_1^\top,\cdots,Z_N^\top)^\top\in\{0,1\}^{N\times K}$ with $Z_i=e_k$ if and only if   $s_i=k$, and $\bE$ consists of independent Bernoulli noise with $\mathbb E [\bE] = \mathbf 0_{N\times J}$. The  data matrix $\bR$ in \eqref{eq:model-decomp} is written as a sum of a ``signal''  matrix $\bOmega\bZ\bTheta^\top$  and a ``noise'' matrix $\bE$. The signal matrix has rank at most $K$, which is typically low-rank as $K\ll \min\ebrac{N,J}$ usually holds. Our spectral method (Algorithm \ref{alg:HeteroClustering}) can efficiently extract the information in the signal part by leveraging the low-rankness.  Built upon the estimated  latent class labels  $\ebrac{\hat s_i,i\in[N]}$, we further perform downstream tasks including estimation and inference of $\bTheta$.

The form of \eqref{eq:model-decomp} is  related to degree-corrected stochastic block models (DCSBM) for network data \citep{qin2013regularized, gao2018community, zhang2020detecting, hu2022multiway, deng2023strong}. 
An undirected network with $N$ nodes can be represented by a symmetric binary adjacency matrix $\mathbf A\in\{0,1\}^{N\times N}$. The DCSBM assumes $\EE[\bA]=\bOmega\bZ\bB\bZ^\top\bOmega$, where  $\bB = (B_{k,\ell}) \in\mathbb R^{K\times K}$ has entries $B_{k,\ell}$ denoting the probability of two nodes in communities $k$ and  $\ell$ being adjacent.
We remark that there are two major differences between DhLCMs and DCSBMs.
\emph{First}, we focus on the high-dimensional regime in DhLCMs with $J\gtrsim N$ and even $J\gg N$, and the data matrix $\bR$ is highly asymmetric. 
In contrast, most network literature  focus on $N\times N$ adjacency matrices, except the bipartite SBM  
\citep{cai2021subspace}. 
For an $N\times N$ matrix $\mathbf A$,  standard PCA would produce a sufficiently accurate eigenspace estimator, which can be further used for clustering. 
However, when our data matrix $\bR$ has size $N\times J$ with $J\gg N$, PCA of $\bR\bR^\top$ or SVD of $\bR$ becomes suboptimal, which motivates us to use HeteroPCA \citep{zhang2022heteroskedastic} to estimate the top eigenspace of $\mathbf R\mathbf R^\top$  due to the high-dimensionality and discrete nature of data.  
\emph{Second}, another major difference lies in the downstream task after clustering or community detection. We are particularly interested in estimation and inference of high-dimensional parameters in $\bTheta_{J\times K}$.  The counterpart of $\bTheta$ in network modeling is the low-dimensional connection probabilities in $\bB_{K\times K}$. It often holds less significance for estimation or inference of $\bB$, because it describes the relationships between latent communities only.
On the contrary, our $\bTheta$ characterizes how the observed features are linked to the latent classes and gives the very basis for interpreting each latent class.
In fact, $\bTheta$ plays a crucial role in various applications as discussed in Section \ref{sec:intro}.

\section{HeteroClustering Algorithm and Its Analysis}
Let $s$ denote the ground truth latent class labels and $s'\in[K]^N$ denote an estimator of the labels.
We define the following mis-clustering error to measure the performance of clustering: 
\begin{align*}
	h\brac{s,s^\prime}:=\min_{\pi\in \Pi_K}\frac{1}{N}\sum_{i=1}^N\II\brac{s_i\ne \pi\brac{s^\prime_i}},
\end{align*}
where $\Pi_K$ collects all permutations on $[K]$.

\label{sec:clust}
\subsection{HeteroClustering Algorithm}

We briefly explain the high level idea of our clustering method. First, in the oracle degree-homogeneous and noiseless setting of DhLCM where $\bOmega=\bI_N$ and $\bE=\mathbf 0_{N\times J}$ in \eqref{eq:model-decomp}, $\bR=\bZ \bTheta^\top = \bU\bSigma\bV^\top$ and the $N\times K$ left singular matrix $\bU$ of $\bR$ spans the same column space of the cluster indicator matrix $\bZ$. 
In this case, the $N$ row vectors of $\bU$ become $K$ points in $\mathbb R^K$ corresponding to $K$ latent classes, so cluster labels can be perfectly read off from $\bU$. Second, in the degree-heterogeneous and noiseless setting with $\bOmega\ne \bI$ and $\bE = \mathbf 0_{N\times J}$, we have $\bR = \bOmega \bZ \bTheta^\top = \bU\bSigma\bV^\top$. The $N$ rows of the left singular matrix $\bU$ exhibit a pattern of $K$ rays in $\mathbb R^K$, each corresponding to a latent class.
In this case, we need to perform certain normalizations to remove the degree effects and ``collapse'' the $K$ rays into $K$ points.
Third, in the realistic setting with $\bOmega\ne \bI$ and $\bE \ne \mathbf 0_{N\times J}$, we use a spectral method to extract the top singular subspace to approximate the noiseless case.
The above insights lead to the following \textit{HeteroClustering} for DhLCMs.
We first use HeteroPCA (Algorithm \ref{alg:HeteroPCA}) to obtain the top $K$ left singular vectors of $\bR$ collected in a $N\times K$ matrix $\wt\bU$, 
and then apply an $\ell_2$ normalization on its rows to obtain $\hat\bU$. Finally, we apply $K$-means clustering  on rows of  $\hat\bU$. This procedure is summarized in Algorithm \ref{alg:HeteroClustering}.

To understand the validity of $\ell_2$ normalization, we first take a careful look at the oracle noiseless setting with $\bR = \bOmega \bZ \bTheta^\top = \bU\bSigma\bV^\top$ where $\bU$ has size $N\times K$. Let  $\calC_k:=\{i\in[N]:s_i=k\}$ be the collection of indices in the $k$-th latent class, and define 
$$\bW:=\textsf{diag}\left(\sqrt{\sum_{i\in\calC_1}\omega_i^2},\cdots,\sqrt{\sum_{i\in\calC_K}\omega_i^2}\right)\in\RR^{K\times K},\quad \bar\bU:=\textsf{diag}\left(\op{\bU_{1,:}}^{-1},\cdots,\op{\bU_{N,:}}^{-1}\right)\bU.$$
where $\bar\bU$ serves as the population counterpart of $\wt\bU$.
In addition, denote $\wt\bTheta^\top :=\bW\bTheta^\top$ and let $\wt\bTheta^\top=\bU^\dag\bSigma^\dag\bV^{\dag\top }$ be its compact SVD with $(\bU^\dag)^{\top}\bU^\dag = (\bV^\dag)^{\top}\bV^\dag = \bI_K$.

\begin{lemma}\label{lem:U-prop}
	The row vectors of $\bU$ in the noiseless SVD $\bOmega \bZ \bTheta^\top = \bU\bSigma\bV^\top$ can be written as 
\begin{align}\label{eq:Ui-Udagi}
	\bU_{i,:}=\frac{\omega_i}{\sqrt{\sum_{j\in\calC_k}\omega_j^2}}\bU^\dag_{k,:}, \quad \forall i\in\calC_k.
\end{align}
The $N\times K$ matrix $\bar\bU$ has $K$ distinct row vectors, i.e., $\bar\bU_{i,:}=\bU^\dag_{k,:}$ for $k\in[K]$ and $i\in\calC_k$. Moreover, $\op{\bar\bU_{i,:}-\bar\bU_{j,:}}=\sqrt{2}$ for any  $i\in\calC_k$, $j\in\calC_l$ and $k\ne l\in[K]$.
\end{lemma}
Lemma \ref{lem:U-prop} indicates that we can read off the true latent class labels from the rows of the $\ell_2$-normalized $\bar\bU$. In the noisy setting, $\bar\bU$ would be well-approximated by $\wt \bU$ if the noise matrix has a relatively small spectral norm $\op{\bE}$, which is implied by  classical matrix perturbation theory such as Wedin’s theorem. However, standard SVD on $\bR$ (or PCA on $\bR\bR^\top$) turns out to be suboptimal when $\op{\bE}$ (or equivalently, $\op{\bE\bE^\top }$) becomes substantially large  in the high-dimensional and heteroskedastic setting. In these scenarios, HeteroPCA proposed by \cite{zhang2022heteroskedastic} emerges as a robust alternative. Roughly speaking, HeteroPCA iteratively imputes the diagonal entries of $\bR\bR^\top $ using the off-diagonals of its low-rank approximation.  This iterative procedure alleviates the bias introduced by the diagonals in $\bE\bE^\top$.
Recently,
\cite{yan2021inference} obtained the precise characterization for the perturbation of singular vectors in HeteroPCA, which serve as the building block of our analysis in this section.

Our normalization approach carries a similar spirit as the SCORE normalization\footnote{The supplement to \cite{jin2015fast} conceptually introduces $\text{SCORE}_{q}$ as an extension to SCORE, which is equivalent to our $\ell_2$ normalization when $q=2$. But we still differentiate our method from SCORE, as the term ``SCORE'' in the literature typically refers to the original version of the algorithm, not its extensions.
} for DCSBM \citep{jin2015fast, ke2019community, jin2021improvements, ke2023special}. 
Consider a $N\times K$ top singular matrix $\hat\bU$. The SCORE-normalized and $\ell_2$-normalized row vectors are:
\begin{align*}
    \hat\bU^{\text{SCORE}}_{i,:} =  \hat\bU_{i,2:K} / \hat U_{i,1},\quad 
    \hat\bU^{\ell_2}_{i,:} = \hat\bU_{i,:} / \|\hat\bU_{i,:}\|_2,\quad \forall i\in[N].
\end{align*}
The above vectors $\{\hat\bU^{\text{SCORE}}_{i,:}: i\in[N]\}$ and $\{\hat\bU^{\ell_2}_{i,:}: i\in[N]\}$ are subsequently used for $K$-means clustering, respectively.
Since $\|\hat\bU_{i,:}\|_2 \geq \hat U_{i,1}$ always holds, $\ell_2$ normalization can lead to more robust clustering performance. 
In practice, we observe the $\ell_2$ normalization outperforms SCORE in simulations and real-data applications; see Section \ref{sec:sim} and \ref{sec:real-data} for details.

\begin{algorithm}[!ht]
\footnotesize
\caption{HeteroClustering}\label{alg:HeteroClustering}
	\KwData{Matrix $\bR\in\RR^{N\times J}$, rank $K$, number of iterations $T_0$ 
 }
	\KwResult{$\{\hat\bs_i,i\in[N]\}$}
	\textbf{HeteroPCA}:
	$\hat\bU=\text{HeteroPCA}(\bR,K,T_0)$ \quad (\text{see Algorithm 2})
	\\\textbf{$\ell_2$ normalization}: $\wt\bU=\textsf{diag}\left(\op{\hat\bU_{1,:}}_2^{-1},~\cdots,~\op{\hat\bU_{N,:}}_2^{-1}\right)\hat\bU$
	\\\textbf{K-means}: $\hat\bs=\text{$K$-means~clustering~on~rows~of~} \wt\bU$
\end{algorithm}

\begin{algorithm}[!ht]
\footnotesize
	\caption{HeteroPCA \citep{zhang2022heteroskedastic}}\label{alg:HeteroPCA}
	\KwData{Matrix $\bR\in\RR^{N\times J}$, rank $K$, number of iterations $T_0$}
	\KwResult{$\hat\bU\in\RR^{N\times K}$}
	\textbf{Initialization}:
	$\bM^{(0)}=\calH\brac{\bR\bR^\top} $
	
	\For{$t=0,1,\cdots,T_0-1$}{
		$\bar\bM^{(t)}=\text{best~rank~} K\text{~approximation~of~} \bM^{(t)}$
		\\$\bM^{(t+1)}=\calH\brac{\bM^{(t)}}+\calD\brac{\bar\bM^{(t)}}$
	}
$\hat\bU=\text{leading~}K \text{~eigenvectors~of~}\hat \bM:=\bM^{(T_0)}$
\end{algorithm}

\subsection{Technical Assumptions and Key Quantities}\label{subsec:assump-key}
We introduce the following two technical assumptions.
\begin{assumption}[Balanced cluster sizes]\label{cond:balanced} 
	There exists an absolute constant $\beta\in\brac{0,1}$ such that $\min_{k\in[K]}|\calC_k|\ge  \beta\max_{k\in[K]}|\calC_k|$.
\end{assumption}
\begin{assumption}[Constant degrees]\label{cond:const-degee} 
There exists absolute constants $c_0,C_0>0$ such that $c_0<\min_{i\in[N]}\omega_i\le \max_{i\in[N]}\omega_i< C_0$.
\end{assumption}
Assumption \ref{cond:balanced} requires  the sizes of clusters to be balanced, which is  standard  in clustering analysis, latent class models, and network models \citep{loffler2019optimality,jing2022community, zeng2023tensor}. 
Assumption \ref{cond:const-degee} essentially requires that (a) all $\omega_i$'s are of the same order; and (b) the order is of constant order.  Conditions in terms of (a) are not uncommon and can be found in \cite{fan2022simplea} and \cite{bhattacharya2023inferences}. We emphasize that even when (a) fails, our results still hold under a weaker balanced cluster degrees assumption:
\begin{align}\label{cond:balanced-cluster} 
\max_{k\in[K]}\sqrt{\sum_{i\in\calC_k}\omega_i^2}\lesssim \min_{k\in[K]}\sqrt{\sum_{i\in\calC_k}\omega_i^2}.
\end{align}
Our purpose of assuming (a) is for clarity of presentation in the main text. All general conclusions by relaxing (a) to \eqref{cond:balanced-cluster} can be found in Section \ref{sec:notations_assumptions} in the Supplementary Material. In terms of (b), we note that $R_{i.j}$ has an expectation of $\omega_i\theta_{j,s_i}$ and hence it is impossible to identify $\omega_i$ and $\theta_{j,s_i}$ separately without any constraint in a DhLCM.  We thereby cast (b) in Assumption \ref{cond:const-degee}  such that the sparsity of the data matrix $\bR$, defined as $\max_{i,j}\EE R_{i,j}$,  is purely characterized by $\bTheta$. As a result, assuming (b) that the order of the degrees $\omega_i$' is a constant  is a major difference from the typical assumption in the degree-corrected network models, e.g., \cite{jin2015fast} and \cite{jin2021improvements}, where the sparsity is purely determined by $\omega_i$'s. 

To facilitate our theoretical analysis, the following key quantities are needed:
\begin{align*}
    \Delta:=\min_{a\ne b\in[K]}\op{\bTheta_{:,a}-\bTheta_{:,b}},\quad \sigma_\star:=\sigma_K\brac{\bTheta},\quad \kappa:=\op{\bTheta}/\sigma_\star, \quad \mu_{\bTheta}:={J\infn{\bTheta}^2}/{\fro{\bTheta}^2}.
\end{align*}
See Section \ref{sec:supp-discussion} in the Supplementary Material for a comprehensive discussion on these quantities.   Hereafter, our subsequent theorems focus on the scenario when $\kappa=O(1)$, merely for clarity of presentation.  All general theorems  with explicit dependency on $\kappa$ and relaxed assumption \eqref{cond:balanced-cluster} are included the Supplementary Material.

\subsection{Exact Clustering}

The following theorem establishes the exact clustering result for Algorithm \ref{alg:HeteroClustering}.
\begin{theorem}\label{thm:exp-clustering-error-simple}
Suppose Assumptions \ref{cond:balanced}-\ref{cond:const-degee} hold, $J\gtrsim N$ and $\kappa=O(1)$. In addition, assume  $N\gtrsim \log^2J$, $J\gtrsim \log^4J$, the number of iterations $T_0$ satisfies 
    $T_0\gtrsim \log\left(\frac{N\sigma_\star^2}{\theta_{\sf max}\sqrt{NJ}K\log J+\sigma_\star N\sqrt{\theta_{\sf max}K\log J}}\right)$,
then exact clustering occurs with high probability:
$$
\mathbb P(h(\hat s,s)=0) \geq 1-\brac{N+J}^{-20},
$$
provided that the following condition holds for some absolute constant $C_{\sf exact}>0$:
\begin{align}\label{eq:delta-cond-exact}
	\frac{\Delta^2}{\theta_{\sf max}}\ge  C_{\sf exact}\mu_{\bTheta}K^2\sqrt{\frac{J}{N}}\log J.
\end{align}
\end{theorem}

A key message of  Theorem \ref{thm:exp-clustering-error-simple} is that our spectral method, the HeteroClustering algorithm, itself leads to the exact clustering of all $N$ data points with high probability. This differentiates our method from those in the existing literature that primarily treats the spectral method as a warm initialization \citep{gao2018community, hu2022multiway, lyu2022optimal, lyu2023optimal}. Our finding aligns with the fact that spectral clustering alone leads to exact clustering under minimal conditions in recent literature; see, e.g.,
\cite{ndaoud2022sharp, han2022exact}. In practice, we observe that Algorithm \ref{alg:HeteroClustering}'s performance is not sensitive to the choice of $T_0$ in HeteroPCA, and we set $T_0=20$ in all simulations and real-data applications. 

\begin{remark}
    We emphasize that in addition to exact clustering, an exponential error rate can be achieved under a relaxed degree assumption Assumption \ref{cond:general-degree} (See  Theorem \ref{thm:exp-clustering-error} in the Supplementary Material and an example that follows).
\end{remark}

\paragraph{Comparison to the degree-corrected hypergraph stochastic block model in \cite{deng2023strong}.}
For degree-corrected SBMs, the only sharp analysis for $\ell_2$-normalized spectral clustering that gives exact clustering is \cite{deng2023strong}.
A key distinction between our work and \cite{deng2023strong} lies in the focus of the latter on modeling symmetric network/hypergraph data and directly using PCA. In contrast, our work is centered on  mixture models where the data matrix exhibits significant asymmetry and high-dimensionality with $J\gg N$, for which we employ HeteroPCA.
For illustration, consider a DhLCM with $\theta_{\sf max}\asymp \min_{j,k}\theta_{j,k}$ and $\mu_{\bTheta},K=O(1)$. In this case, our condition \eqref{eq:delta-cond-exact} can be equivalently written as $(NJ)^{1/2}\theta_{\sf max}\gtrsim\log J$, which can be interpreted as a sparsity condition on $\bTheta$. For a direct comparison, consider a SBM, which is a special case of the model in \cite{deng2023strong}, and the sparsity condition therein reads as $N\theta_{\sf max}\gtrsim \log N$. So our condition aligns with theirs when $N\asymp J$, which means $(NJ)^{1/2} \asymp N$.  Notably, a direct application of PCA in DhLCMs would lead to the suboptimal sparsity condition of $N\theta_{\sf max}\gtrsim \log J$, which is more stringent than our current condition  $(NJ)^{1/2}\theta_{\sf max}\gtrsim\log J$ in the regime with $J\gg N$.

\paragraph{Comparison to the Gaussian Mixture Model.}
Another related line of research is clustering in mixture models.  In the context of (degree-homogeneous) isotropic Gaussian mixture model, we can write $\bR_{i,:}=\bTheta_{:,s_i}+\bE_{i,:}$ where $\bE_{i,:}\overset{i.i.d.}{\sim}N\brac{0,\sigma^2 \bI_J}$ by adopting our notations. To simplify the narrative, we assume $\mu_{\bTheta},K=O(1)$ and $J\gtrsim N\log N$. \cite{chen2021cutoff} derives  the threshold for exact clustering of the general $K$-component Gaussian mixture model in the form of $ {\Delta^2}/{\sigma^2}\gtrsim \sqrt{{J}/{N}}\log^{1/2} N$.
\textcolor{black}{
In comparison, our signal-to-noise condition \eqref{eq:delta-cond-exact} admits a similar form, whose optimality under the Bernoulli setting remains unexplored. To this end, we derive a  lower bound regarding the condition for exact clustering by introducing  the following parameter space  for $\mathbb E[\bR]$:
\begin{align*}
 \calP_K(s,\bOmega,\bTheta):=\ebrac{\wt\bR:s\in[K]^N,\wt R_{i,j}=\omega_{i}\theta_{j,s_i}\in[0,1], \omega_i\ge 0,\theta_{j,s_i}\ge 0,  \forall i\in[N], j\in[J]}. 
\end{align*} 
It suffices to consider the parameter space for two-component LCMs $\calP_2(s,\bI_N,\bTheta)\subset \calP_K(s,\bOmega,\bTheta)$:
\begin{theorem}\label{thm:exact-lower-bound}
Let $\epsilon\in(1/2,1)$ be a constant. Assume $J\gg N$ and $\theta_{j,k}\asymp N^{-(1-\epsilon)}$  for $j\in[J]$, $k\in[2]$. There exists some absolute constants $c_0,c_1>0$ such that if $\frac{\Delta^2}{\theta_{\sf max}}\le c_0\sqrt{\frac{J}{N}\log J}$, then 
        $\inf_{\hat s}\sup_{\calP_2(s,\bI_N,\bTheta)}\EE h\brac{\hat s,s} \ge c_1$.
\end{theorem}
Theorem \ref{thm:exact-lower-bound} justifies the near-optimality (up to a logarithmic factor) of condition \eqref{eq:delta-cond-exact} for exact clustering under DhLCMs. Due to technical reasons, we need the sparsity level to be larger than $N^{-1/2}$. To the best of our knowledge, this is first lower bound for exact clustering  regarding the mixture model under the high-dimensional Bernoulli setting.
}

\section{Estimation and Statistical Inference of $\bTheta$}\label{sec:est-theta}
In this section, we study the identifiability, estimation, and inference of item parameters $\bTheta$.
\subsection{Identifiability of $\bTheta$}
\begin{definition}[$\brac{ \bOmega, \bTheta}$-identifiable]
	The degree-heterogeneous LCM with parameter set $\brac{ \bOmega,\bZ, \bTheta}$ is said to be $\brac{ \bOmega, \bTheta}$-identifiable, if for any other valid parameter set $\brac{\wt \bOmega,\bZ,\wt \bTheta}$, $\bOmega\bZ \bTheta=\wt \bOmega\bZ\wt \bTheta$ holds if and only if  $\brac{ \bOmega,\bZ, \bTheta}$ and $\brac{\wt \bOmega,\bZ,\wt \bTheta}$ are identical.
\end{definition}
Consider the DhLCM with parameter $\brac{\bOmega,\bZ,\bTheta}$ without additional assumptions on $\bOmega$ or $\bTheta$. We can construct another parameter pair $(\wt\bTheta,\wt\bOmega)$ satisfying $\wt\bTheta=\bTheta\bD$ and $\wt\bOmega=\bOmega\wt\bD$, where $\bD=\textsf{diag}(D_1,\cdots, D_K)\in\RR^{K\times K}$
for arbitrary $D_1,\cdots,D_K>0$, and $\wt\bD=\textsf{diag}(D^{-1}_{s_1},\cdots,D^{-1}_{s_n})\in\RR^{N\times N}$.
It is not hard to verify that  $\wt\bOmega\bZ\wt\bTheta^\top=\bOmega\bZ\bTheta^\top$ and the model is not $\brac{\bOmega,\bTheta}$-identifiable.
To tackle this issue, we cast the following identifiability condition. 

\begin{assumption}[Identifiability of $\brac{\bOmega,\bTheta}$]\label{cond:iden-theta} $\sum_{i\in \calC_{k}}\omega_i^2=|\calC_k|$ for all $k\in[K]$.
\end{assumption}
\begin{proposition}\label{prop:identifiability}
	Under Assumption \ref{cond:iden-theta}, the DhLCM with $\brac{ \bOmega,\bZ, \bTheta}$ is $\brac{ \bOmega, \bTheta}$-identifiable.
\end{proposition}
Assumption \ref{cond:iden-theta} is a sufficient condition for identifiability. Other identification conditions are also possible, e.g., $\sum_{i\in\calC_{k}}\omega_i=|\calC_k|$ or $\max_{i\in\calC_{k}}\omega_i=\omega_0$ for given $\omega_0>0$. Here we adopt Assumption \ref{cond:iden-theta} as it is most consistent with our general degree condition \eqref{cond:balanced-cluster}, see also the general assumption in Section \ref{sec:notations_assumptions} in the Supplementary Material.

\subsection{Estimation of $\bTheta$}
\label{subsec:theta}

Lemma \ref{lem:U-prop} implies that $\op{\bU_{i,:}}=\omega_i/\sqrt{\sum_{j\in\calC_k}\omega_j^2}\op{\bU_{k,:}^\dagger} =\omega_i/\sqrt{\sum_{j\in\calC_k}\omega_j^2}$ for any $i\in \calC_k$. Under Assumption \ref{cond:iden-theta}, we estimate the degrees by $\hat\bOmega:=\textsf{diag}\brac{{\ab{\calC_{\hat s_1}}^{1/2}}\op{\hat \bU_{1,:}},\cdots,{\ab{\calC_{\hat s_N}}^{1/2}}\op{\hat \bU_{N,:}}}$ and define the following simple estimator for $\bTheta^\top$ based on the HeteroClustering result $\hat \bZ$:
\begin{align}\label{eq:theta-est}
	\hat\bTheta^\top =\brac{\hat\bZ^\top \hat\bZ}^{-1} \hat\bZ^\top \hat\bOmega^{-1}\bR.
\end{align}

 \begin{theorem}\label{thm:Thetaerr-ave-simple}
	Suppose the conditions of Theorem \ref{thm:exp-clustering-error-simple} and Assumption \ref{cond:iden-theta} hold.  
Then we have with probability exceeding $1-\brac{N+J}^{-20}$ that 
	\begin{align}\label{eq:thetaer-event}
		\min_{\bPi\in \SS_{K}}\op{\hat\bTheta-\bTheta\bPi}_{\sf max}&\le C\mu^{1/2}_{\bTheta}K^{3/2}\sqrt{\frac{\theta_{\sf max}\log J}{N}}\color{black}{+\frac{K\log J}{N}},
	\end{align}
	for some large constant $C>0$, where $\SS_K$ stands for the set of $K\times K$ permutation matrices.
\end{theorem}

{\textcolor{black}{
The minimax optimality (up to logarithmic factors) of Theorem \ref{thm:Thetaerr-ave-simple} can be justified by the following lower bound.
\begin{theorem}\label{thm:est-theta-lb}
Assume $\theta_{\sf max}\lesssim\min_{j}\theta_{j,k}=o(1)$, then there exists a constant $c_0>0$ such that 
	\begin{align*}
	\inf_{\hat \bTheta} \sup_{\calP_2\brac{s,\bOmega,\bTheta}}\EE\min_{\bPi\in\SS_2}\op{\hat \bTheta-\bTheta\bPi}_{\sf max}\ge c_0\brac{\sqrt{\frac{\theta_{\sf max}}{N}}+\frac{1}{N}}.
\end{align*}
\end{theorem}}
}
\subsection{Distributional Results for $\bTheta$}
Let $\calT_+:=\left\{\brac{j,k}\in[J]\times [K]: \theta_{j,k}> 0\right \}$ and $\calJ_+:=\left\{j\in[J]:\brac{j,k}\in \calT_+,\forall k\in[K]\right \}$. For inference on $\bTheta$, we restrict our interest in the rows in $\calJ_+$ and entries in $\calT_+$. Denote $\theta^*_{\sf min}:=\min_{(j,k)\in \calT_+}\theta_{j,k}$. Although equipped with the exact clustering guarantee in Theorem \ref{thm:exp-clustering-error-simple},  inference on $\bTheta$ is still non-trivial due to the presence of the degree parameters $\bOmega$.

Consider an arbitrary fixed index set  $\calJ_0:=\left\{j_1,\cdots,j_{M}\right \}\subseteq \calJ_+$ with cardinality $M=|\calJ_0|\le \ab{\calJ_+}$, where $M$ is a constant.
Let $\bTheta_{\calJ_0,:}\in\RR^{M\times K}$ be a sub-matrix of $\bTheta$ by restricting rows in $\calJ_0$, and define $\hat\bTheta_{\calJ_0,:}$ similarly. For $\brac{j,k}\in \calT_+$ we define 
\begin{align}\label{eq:theta-variance}
	 \sigma^2_{j,k}= \frac{\theta_{j,k}}{\ab{ \calC_k}^2}\sum_{i\in\calC_k}\frac{1-\omega_i \theta_{j,k}}{\omega_i},\quad \hat \sigma^2_{j,k}=\frac{\hat \theta_{j,k}}{\ab{\hat \calC_k}^2}\sum_{i\in\hat\calC_k}\frac{1-\hat\omega_i\hat \theta_{j,k}}{\hat\omega_i}.
\end{align}
where the former is the asymptotic variance of $\btheta_{j,k}$ and the latter serves as a plug-in estimator for it based on the clustering results and the estimated parameters. 
We start with presenting a general distributional result on $\bTheta$.
\begin{theorem}\label{thm:Thetaerr-gen-inf-simple}
Suppose the conditions of Theorem \ref{thm:exp-clustering-error-simple} and Assumption \ref{cond:iden-theta} hold. Assume that $MK=O(1)$,  $J\gtrsim \mu^3_{\bTheta}\brac{{\theta_{\sf max}}/{\theta^*_{\sf min}}}\log^2(N+J)$ and  
\begin{align}\label{eq:theta-inf-cond}
	\frac{\Delta^2}{\theta_{\sf max}}\ge C_{\sf inf}\mu^2_{\bTheta}K^3\brac{\frac{\theta_{\sf max}}{\theta^*_{\sf min}}}\frac{J}{N}\log^3J,
\end{align}
for some absolute constant $C_{\sf inf}>0$, then we have
\begin{align*}
    \sup_{\calC\in\scrC^{MK}}\ab{\PP\brac{ \bSigma^{-1/2}_{\bTheta,\calJ_0}\textsf{vec}\brac{\hat \bTheta^\top_{\calJ_0,:}-\bTheta^\top _{\calJ_0,:}}\in \calC}-   \PP\brac{\calN\brac{0,\bI_{MK}}\in \calC}}\lesssim \frac{\brac{MK}^{5/4}}{\sqrt{\log\brac{N+J}}},
\end{align*}
where $\scrC^{MK}$ is the set of all convex sets in $\RR^{MK}$ and $\bSigma_{\bTheta,\calJ_0}$ is a diagonal matrix defined as $\bSigma_{\bTheta,\calJ_0}:=\textsf{diag}\brac{\bSigma_{\bTheta,j_1},\cdots,\bSigma_{\bTheta,j_{M}}}\in \RR^{MK\times MK}$
with $\bSigma_{\bTheta,j_m}:=\textsf{diag}\brac{\left\{\sigma^2_{j_m,k}\right\}_{k=1,\cdots,K}}$ for $m\in[M]$. Moreover, the conclusion continues to hold if we replace $\bSigma_{\bTheta,\calJ_0}$ by its plug-in estimator $\hat \bSigma_{\bTheta,\calJ_0}:=\textsf{diag}\brac{\left\{\hat \sigma^2_{j_m,k}\right\}_{m\in[M],k\in[K]}}$.
\end{theorem}
\begin{remark}
    Suppose $\mu_\bTheta,K=O(1)$,  condition \eqref{eq:theta-inf-cond}  is more stringent than \eqref{eq:delta-cond-exact} required for exact clustering  in Theorem \ref{thm:exp-clustering-error-simple} when $J\gg N$.  Intuitively, characterizing the distribution of entries of $\hat\bTheta$ boils down to consistently estimating the right singular space $\mathbf V$ of  $\EE[\bR]$, for which  the minimal signal-to-noise condition \eqref{eq:delta-cond-exact} for solely estimating $\bU$  would be inadequate. It turns out that the condition for consistently estimating $\bV$ is equivalent to \eqref{eq:theta-inf-cond}, implying that \eqref{eq:theta-inf-cond} is likely a necessary condition for inference  on $\bTheta$ using our method.
\end{remark}  

It is worth noting that the asymptotic covariance matrix $\bSigma_{\bTheta,\calJ_0}$ in Theorem \ref{thm:Thetaerr-gen-inf-simple} is a diagonal matrix.
Therefore, Theorem \ref{thm:Thetaerr-gen-inf-simple} implies that any finite subset of entries in our high-dimensional estimator $\hat\bTheta$ are not only asymptotically normal, but also asymptotically independent. 
Theorem \ref{thm:Thetaerr-gen-inf-simple} enables one to  construct confidence intervals for all $\theta_{j,k}$'s and  perform hypothesis testing of the form $H_0:\bTheta_{\calS}=\bTheta^*_{\calS}$ against $ H_a:\bTheta_{\calS}\ne \bTheta^*_{\calS}$, 
for any $\calS\subset \calT_+$ with $\ab{\calS}=O\brac{1}$ and $\bTheta^*$ is  some pre-specified matrix.

\subsection{Hypothesis Testing of $\bTheta$}
\subsubsection{Global Testing for a Subset of Items in $\bTheta$}\label{subsec:test}
In this section, we focus on testing whether a pre-specified group of features are useful and relevant for clustering.
Note that $\theta_{j,k}$ represents the conditional probability of providing a positive response to feature $j$ given a typical subject from latent class $k$ with degree one $\omega_i=1$. So, we say a feature $j\in[J]$ is useful for clustering if the $\theta_{j,k}$'s across the $K$ latent classes are not identical.
We consider the following global hypothesis testing problem:
\begin{align}\label{eq:null}&H_0:\theta_{j,1}=\theta_{j,2}=\cdots=\theta_{j,K},\quad \forall j\in \calJ_0\notag\\
&H_a:\ab{\theta_{j,k_1}-\theta_{j,k_2}}\ge d_N \text{ for some } j\in\calJ_0\text{~and~}k_1\ne k_2\in[K],
\end{align}
for some $d_N=o(1)$, corresponding to testing against a local alternative. 
Define test statistic
\begin{align}\label{eq:test-statistic}
T:=\max_{j\in\calJ_0}\max_{k_1<k_2\in[K]}T_j(k_1,k_2),\quad \text{where }T_j(k_1,k_2):={\brac{\hat\theta_{j,k_1}-\hat\theta_{j,k_2}}^2}/\brac{\hat\sigma^2_{j,k_1}+\hat\sigma^2_{j,k_2}}.
\end{align}

\begin{theorem}\label{thm:test-theta-gen-simple}
Suppose the conditions of Theorem \ref{thm:exp-clustering-error-simple} and Assumption \ref{cond:iden-theta} hold.  In addition, assume that  $J\gtrsim \mu^3_{\bTheta}\brac{\frac{\theta_{\sf max}}{\theta^*_{\sf min}}}\log^2 J$ and there exists some absolute constant $C_{\sf inf}>0$ such that 
\begin{align}\label{eq:test-simple-cond}
	\frac{\Delta^2}{\theta_{\sf max}}\ge C_{\sf inf}\mu^2_{\bTheta}K^3\brac{\frac{\theta_{\sf max}}{\theta^*_{\sf min}}}\frac{J}{N}\log^3J,
\end{align}
then the following conclusions hold:
\begin{enumerate}
	\item[(a)] Under the null hypothesis $H_0$, 
\begin{itemize}
\item[(i)] If $MK^2=O(1)$ , we have 
 $\sup_{t\in \RR}\ab{\Prob\brac{T\le t}-\sqbrac{\Prob\brac{\chi_1^2\le t}}^{M{K\choose 2}}}=o(1)$;
\item[(ii)]  If $MK^2\rightarrow \infty$ and $MK^2=o\brac{\log^{1/2}\brac{N+J}}$, we have  
\begin{align*}
	\sup_{t\in\RR}\ab{\Prob\brac{\frac{T-c_{M,K}}{2}\le t}-\calG(t)}=o(1),
\end{align*}
where $\calG(x):=\exp(-e^{-x})$ is the Gumbel distribution and 
	$c_{M,K}:=2\brac{\log M+\log {K\choose 2}}-\log\brac{\log M+\log {K\choose 2}}-\log \pi$.
 \end{itemize}
	\item[(b)] \sloppy Under the local alternative hypothesis $H_a$ such that $d_N\gg \sqrt{{\theta_{\sf max}(\log \log J)}/{N}}$,
    we have $\Prob\brac{T>C}=1-o(1)$ for any constant $C>0$.
\end{enumerate}
\end{theorem}

In Theorem \ref{thm:test-theta-gen-simple}, part (a) states that under appropriate conditions, our test statistic $T$ in \eqref{eq:test-statistic} converges to the $\chi^2$ distribution (when $MK^2$ is finite) or the Gumbel distribution (when $MK^2$ goes to infinity) under the null hypothesis; furthermore, part (b) states that our test procedure enjoys full power asymptotically against local alternatives.
We remark that Theorem \ref{thm:test-theta-gen-simple} is not a trivial corollary of the asymptotic normality result in Theorem  \ref{thm:Thetaerr-gen-inf-simple}, mainly because Theorem  \ref{thm:Thetaerr-gen-inf-simple} requires $MK^2$ to be finite. In contrast, Theorem \ref{thm:test-theta-gen-simple} allows $MK^2$ to grow slowly as $\log^{1/2}\brac{N+J}$, and the proof of it requires a careful investigation of the asymptotic expression of the test statistic $T$ defined in \eqref{eq:test-statistic}. As a consequence, Theorem \ref{thm:test-theta-gen-simple} indicates  the following practical testing procedure for a given significance level $\alpha_0$:
\begin{enumerate}
    \item[(a)] For small  $MK^2$, we will reject $H_0$ if $T>\chi^2_{1,\beta_{M}\brac{\alpha_0}}$ with $\beta_{M}\brac{\alpha}:=\brac{1-\alpha}^{\frac{1}{M{K\choose 2}}}$, where $\chi^2_{1,\alpha}$ is the $\alpha$ upper quantile of the $\chi_1^2$ distribution;
    \item[(b)] For large $MK^2$, we will reject $H_0$ if $T>2g_{1-\alpha_0}+c_{M,K}$, where $g_{\alpha}$ is the $\alpha $ upper quantile of the Gumbel distribution.
\end{enumerate}
\subsubsection{Multiple Testing across Many Rows of $\bTheta$}
We briefly explore the effectiveness of our method in a multiple testing context. In practice, it may be interesting to test whether each observed feature is relevant for clustering to discover useful markers to differentiate latent classes. Consider a family of hypothesis $\calH_{\calJ_0}:=\ebrac{\brac{H_{0,j},H_{a,j}}:j\in\calJ_0}$ with the following null and alternative hypotheses:
\begin{align}\label{eq:multi-null}
H_{0,j}:\theta_{j,1}=\theta_{j,2}=\cdots=\theta_{j,K}, \quad \text{versus} \quad H_{a,j}:\ab{\theta_{j,k_1}-\theta_{j,k_2}}\ge d_N, ~\exists k_1\ne k_2\in[K].
\end{align}
Let $\calN_{\calJ_0}:=\ebrac{j\in\calJ_0: \theta_{j,1}=\theta_{j,2}=\cdots=\theta_{j,K}}$ be the index set of all true null hypotheses. For each $j\in\calJ_0$, our test statistic for the single hypothesis $H_{0,j}\text{~versus~}H_{a,j}$ is defined as
\begin{align}\label{eq:multi-test-statistic}
    T_j:=\max_{k_1<k_2\in[K]}T_j\brac{k_1,k_2},
\end{align}
where $T_j(k_1,k_2)$ is defined in \eqref{eq:test-statistic}. We reject $H_{0,j}$ if $T_j>\chi^2_{1,\beta_1\brac{\alpha_0}}$
for some pre-specified level $\alpha_0\in(0,1)$.
The following theorem demonstrates that our test statistic combined with the celebrated Benjamini-Hochberg (BH) procedure \citep{benjamini1995controlling}, effectively controls the false discovery rate ($\textsf{FDR}$) in multiple testing.

\begin{theorem}\label{thm:fdr-control}
    Suppose the conditions of Theorem \ref{thm:test-theta-gen-simple} hold. Assume $K=o\brac{\log^{1/4}J}$, then applying the BH procedure for a given $\alpha_0\in(0,1)$ gives the following as $J\rightarrow\infty$, 
    \begin{align*}
&\textsf{FDR}:=\EE\brac{\frac{\sum_{j\in\calN_{\calJ_0}}\II\brac{H_{0,j}\text{~is~rejected~}}}{1\vee \sum_{j\in\calJ_0}\II\brac{H_{0,j}\text{~is~rejected~}}}}=\frac{\ab{\calN_{\calJ_0}}}{M}\cdot \alpha_0\brac{1+o\brac{1}}.
\end{align*}
\end{theorem}

Theorem \ref{thm:fdr-control} justifies the validity of simply using the BH procedure to control FDR in our setting. Intuitively, the reason why BH can succeed is that our estimators for individual entries in $\bTheta$ are asymptotically independent.
On a related note, the influential features PCA \citep[IF-PCA,][]{jin2016ifpca} is a popular method for screening useful features for clusters before (instead of after) clustering. However, IF-PCA relies on the Gaussian assumption of the noise to establish theoretical guarantees, whereas we do not make this assumption.

\section{Simulation Studies}\label{sec:sim}

We perform extensive simulation studies in the high-dimensional settings with $J \ge N$ to (a) evaluate the clustering accuracy of Algorithm \ref{alg:HeteroClustering}; and (b) validate the statistical inference results. 
We only consider the Bernoulli and the Poisson models in the simulation studies, as the estimation and inference procedures for the Binomial model closely parallel those of the Bernoulli model. The true latent class labels $s_i$ are uniformly sampled from $[K]$, and the degree parameters $\omega_i$ are independently sampled from $\text{Uniform}[0.1, 1.5]$. The parameters $\theta_{j,k}$ are independently sampled from $(2/3) \cdot
\text{Beta}(0.1, 1)$ for the Bernoulli model and \text{Gamma}(0.5, 1) for the Poisson model. We further scale the degree parameters for the identifiability Assumption \ref{cond:iden-theta} to hold. 
We generate 500 independent replicates in each simulation setting.

\paragraph{Simulation Study 1: Clustering.}
We compare the clustering accuracy of  different clustering approaches {for the Bernoulli model}.
For estimating the top left singular subspace of $\bR$, we consider both HeteroPCA and SVD. For normalizing the singular subspace embedding, consider $\ell_2$ normalization, SCORE normalization, and no normalization. 
We consider a challenging clustering scenario with $N=200, J=1000$, and $K=10$ latent classes. 

\begin{figure}[h!]
    \centering
    \includegraphics[width=0.7\textwidth]{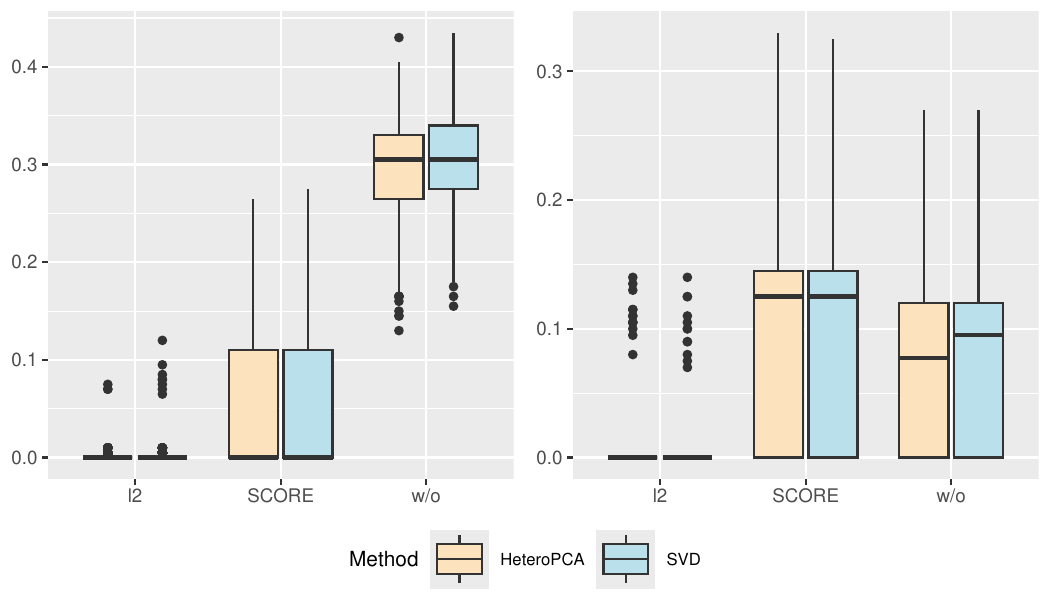}
    \caption{Clustering error boxplots for Bernoulli model with (left) and without (right) degree heterogeneity, with $N=200, J=1000, K=10$ and 500 replications. From left to right in each figure are $\ell_2$ normalization, SCORE normalization, and no normalization. }
    \label{fig:clustering_err_strong}
\end{figure}

Figure \ref{fig:clustering_err_strong} presents the classification error boxplots for the six spectral clustering methods from $\{\text{SVD,~ HeteroPCA}\}\bigotimes\{\ell_2,~ \text{SCORE,~ no normalization}\}$ across 500 independent simulation replicates. 
The left panel of Figure \ref{fig:clustering_err_strong} presents results where the true model is a DhLCM, whereas the right panel presents results where the true model is a traditional LCM with $\omega_i=1$ for all $i\in[N]$.
The $\ell_2$ normalization yields higher clustering accuracy compared to SCORE normalization and no normalization, regardless of whether there exists degree heterogeneity. When there does exist degree heterogeneity, although both $\ell_2$ and SCORE contribute to a reduction in the clustering error compared to no normalization, the $\ell_2$ normalization demonstrates a better and more stable clustering performance.  However, in the absence of degree heterogeneity, SCORE normalization leads to suboptimal results even compared to no normalization. 
In contrast, even if the true model does not have degree heterogeneity, $\ell_2$ normalization will not degrade and can even improve the clustering performance. We also observe that HeteroPCA yields better results compared to SVD without normalization. In other scenarios, HeteroPCA and SVD give similar results. In real data analyses in Section \ref{sec:real-data}, HeteroPCA enjoys slight advantages over SVD in two large datasets.
In Section \ref{sec:tables_figures} in the Supplementary Material, we also compare HeteroClustering with two likelihood-based clustering methods for LCA.
In summary, HeteroClustering has superior statistical and computational performance across various settings. 

{\color{black}
To provide more empirical insights into the differences between HeteroPCA and SVD, we conduct additional simulation studies. 
Consider $J=3000, N=500, K=2, \bTheta_{:, 1}=(0.3\cdot \mathbf{1}_{J/2}^\top, 0.5\cdot 
\mathbf{1}_{J/2}^\top)^\top, \bTheta_{:, 2}=(0.1\cdot \mathbf{1}_{J/2}^\top, 0.06\cdot 
\mathbf{1}_{J/2}^\top)^\top$ for both the Bernoulli and the Poisson models, and generate $100$ independent replications. Figure \ref{fig:svd_hetero_compare} displays the boxplots of the Frobenius norm error of $\hat\bU$ (defined as $\|\bU\bU^\top - \hat\bU \hat\bU^\top\|_F$) on the left, clustering error in the middle, and the maximum absolute error of $\hat\bTheta$ on the right. The upper and lower rows correspond to the Bernoulli model and the Poisson model, respectively. 
These results show that HeteroPCA typically yields higher accuracy in terms of singular space estimation, clustering, as well as item parameter estimation compared to SVD under both models.

\begin{figure}[h!]
    \centering
    \includegraphics[width=0.85\linewidth]{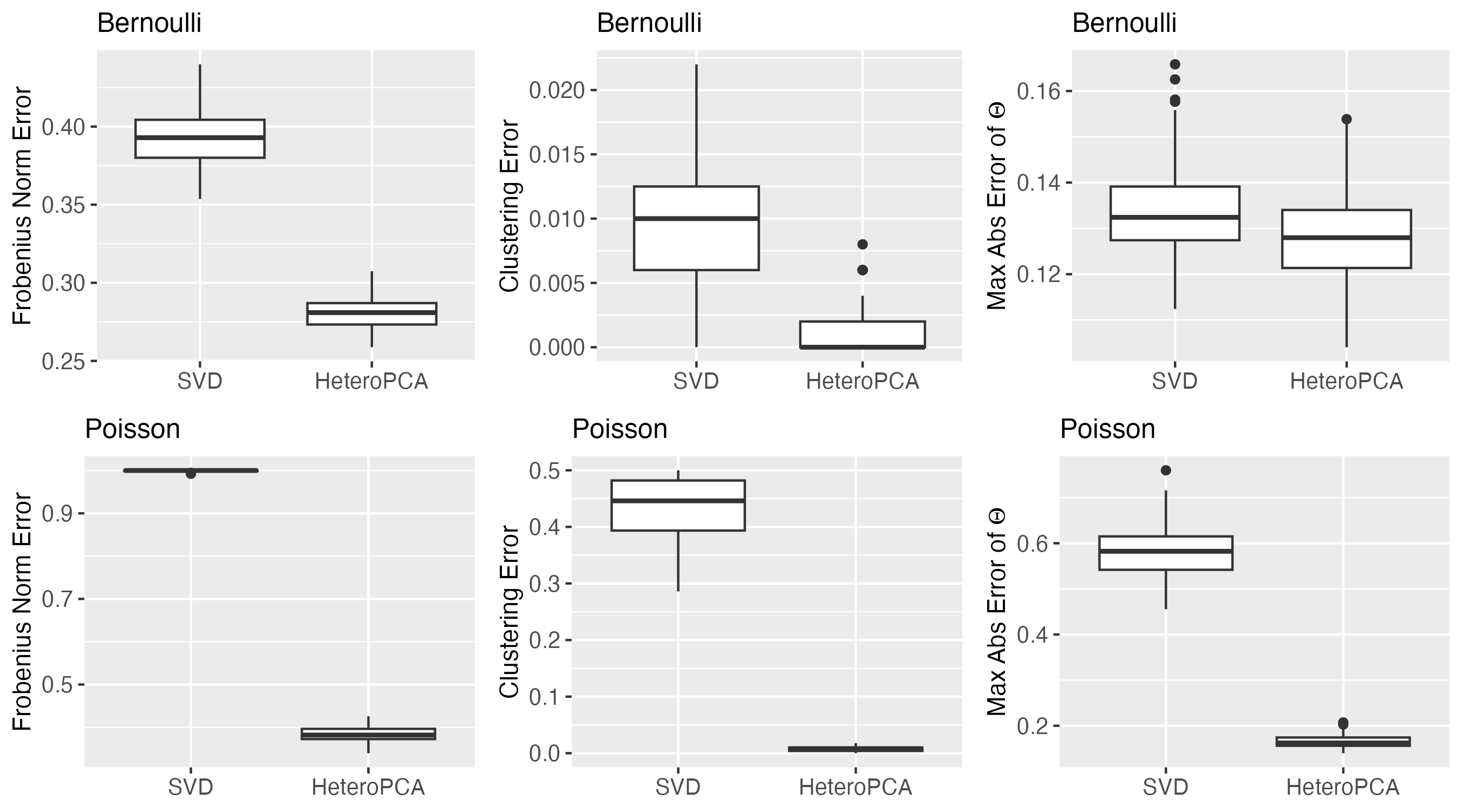}
    \caption{Boxplots comparing SVD and HeteroPCA in terms of the Frobenius norm error of $\hat \bU$ (left), clustering error (middle), and the maximum absolute error of $\hat\bTheta$ (right) for the Bernoulli model (upper row) and the Poisson model (lower row). 
    }
    \label{fig:svd_hetero_compare}
\end{figure}
}

\begin{figure}[h!]
\centering
     \begin{subfigure}[b]{0.24\textwidth}
         \centering
         \includegraphics[width=0.95\textwidth]{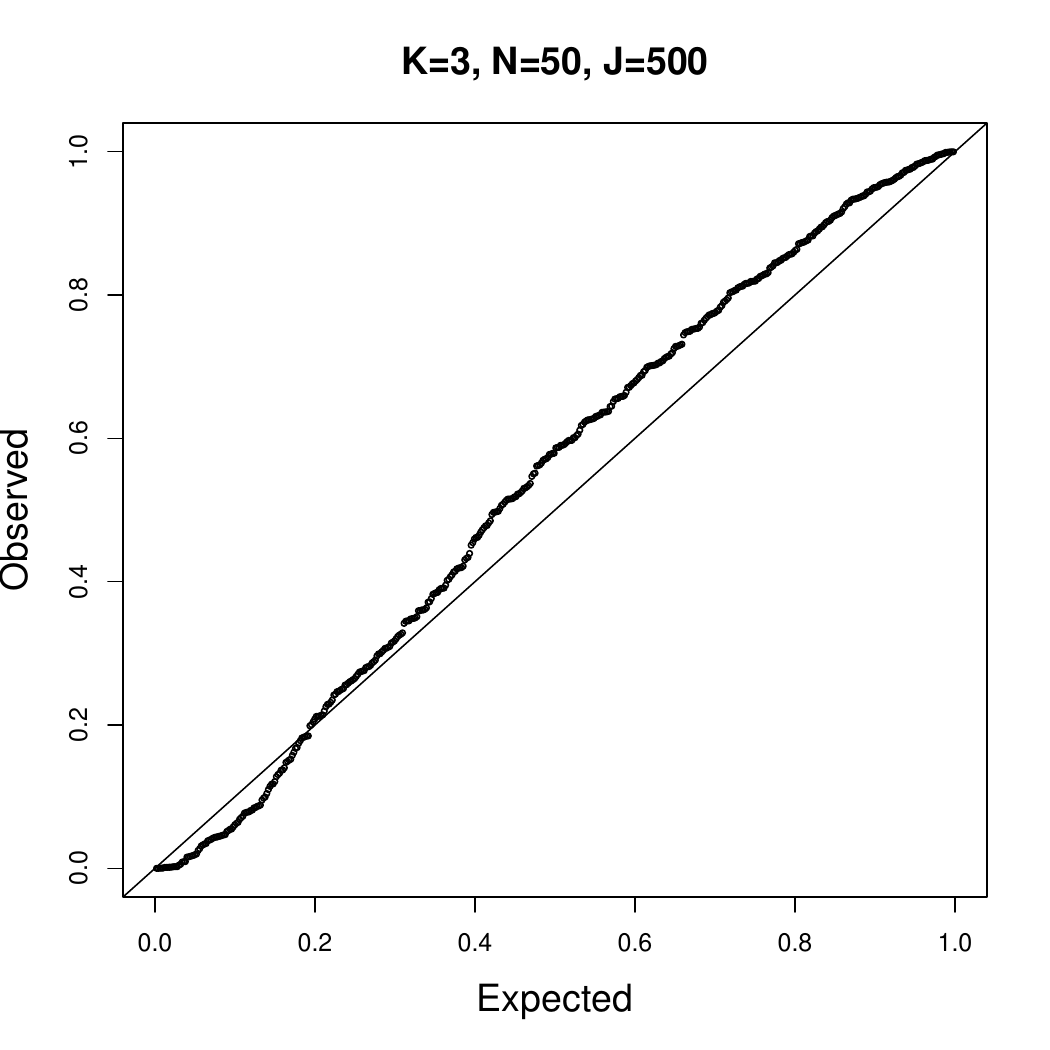}
     \end{subfigure}
    \hfill
     \begin{subfigure}[b]{0.24\textwidth}
         \centering
         \includegraphics[width=0.95\textwidth]{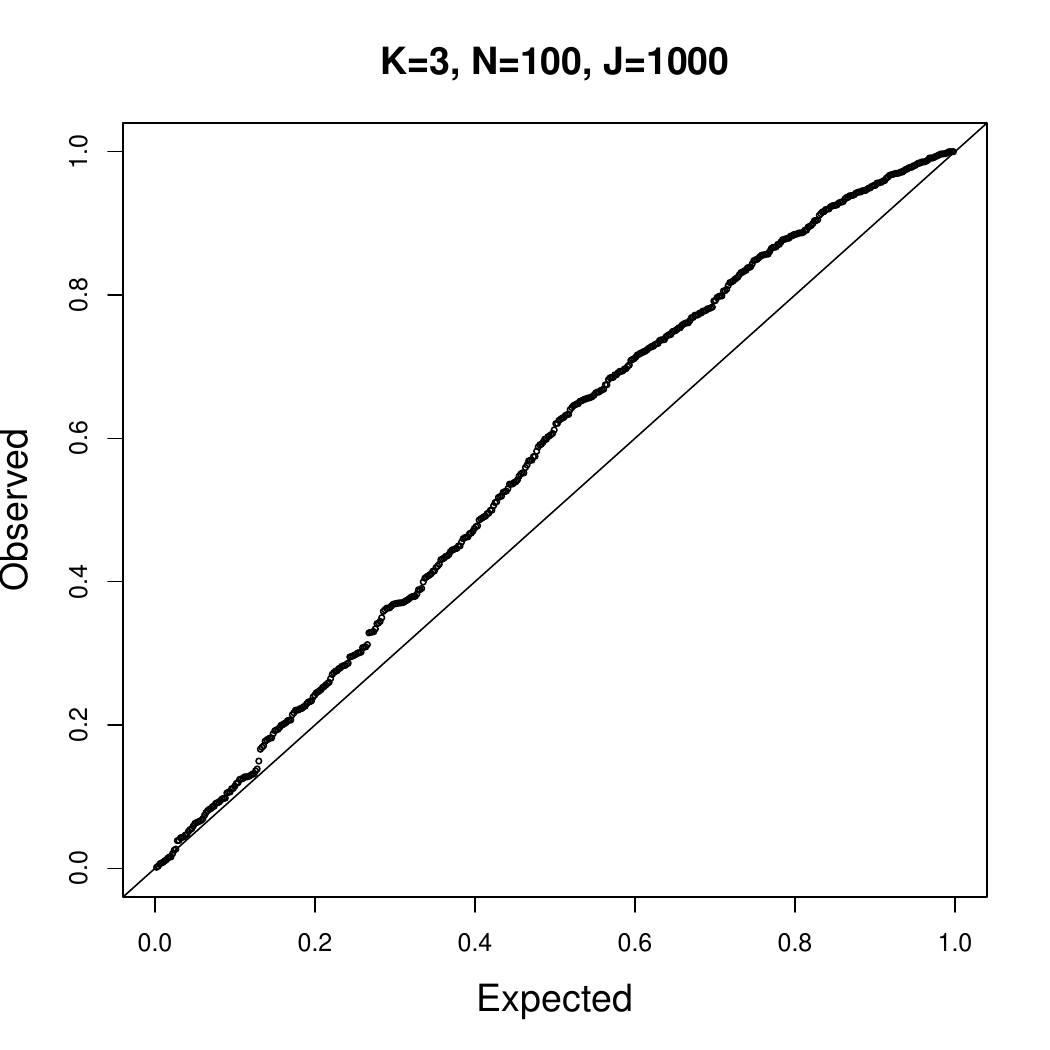}
     \end{subfigure}
     \hfill
     \begin{subfigure}[b]{0.24\textwidth}
         \centering
         \includegraphics[width=0.95\textwidth]{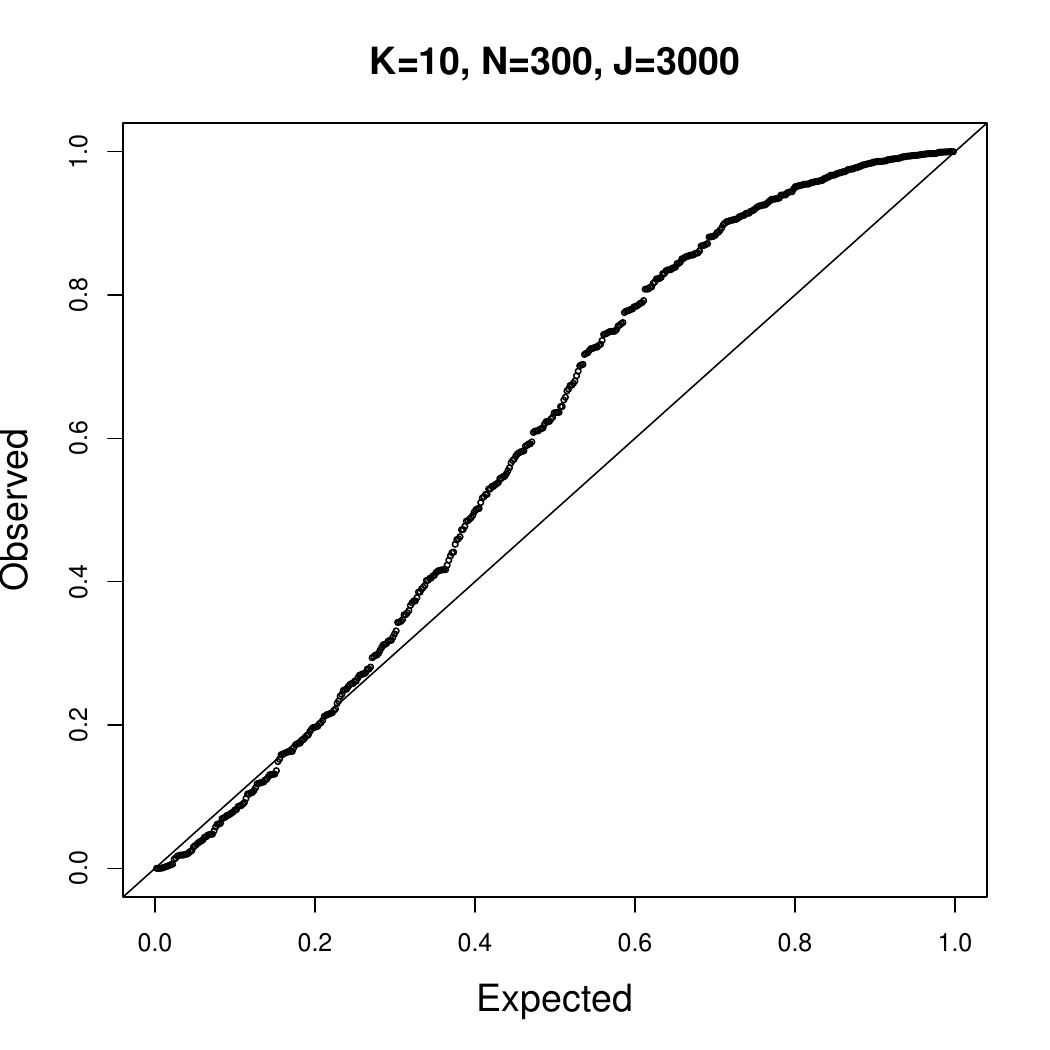}
     \end{subfigure}
     \hfill
     \begin{subfigure}[b]{0.24\textwidth}
         \centering
         \includegraphics[width=0.95\textwidth]{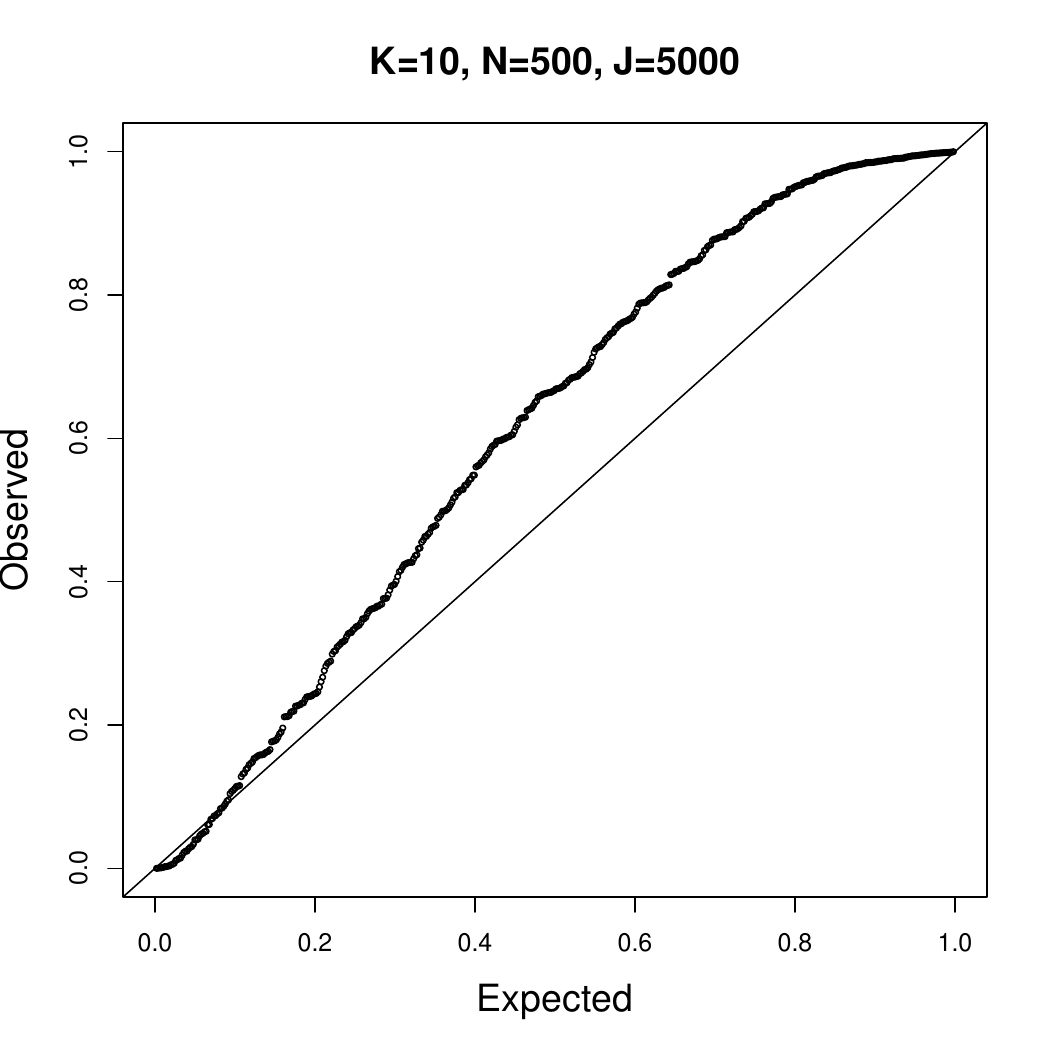}
     \end{subfigure}
     \begin{subfigure}[b]{0.24\textwidth}
         \centering
         \includegraphics[width=0.95\textwidth]{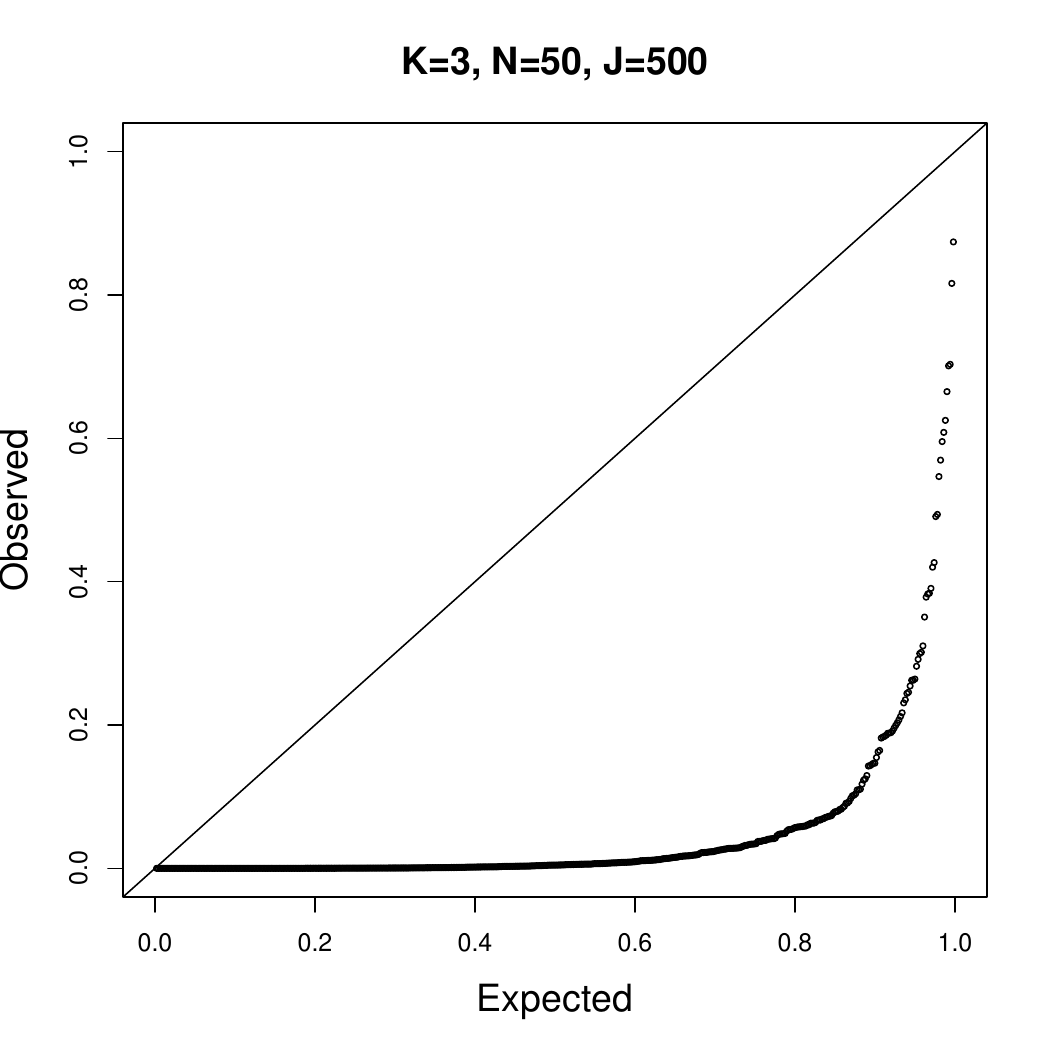}
     \end{subfigure}
    \hfill
     \begin{subfigure}[b]{0.24\textwidth}
         \centering
         \includegraphics[width=0.95\textwidth]{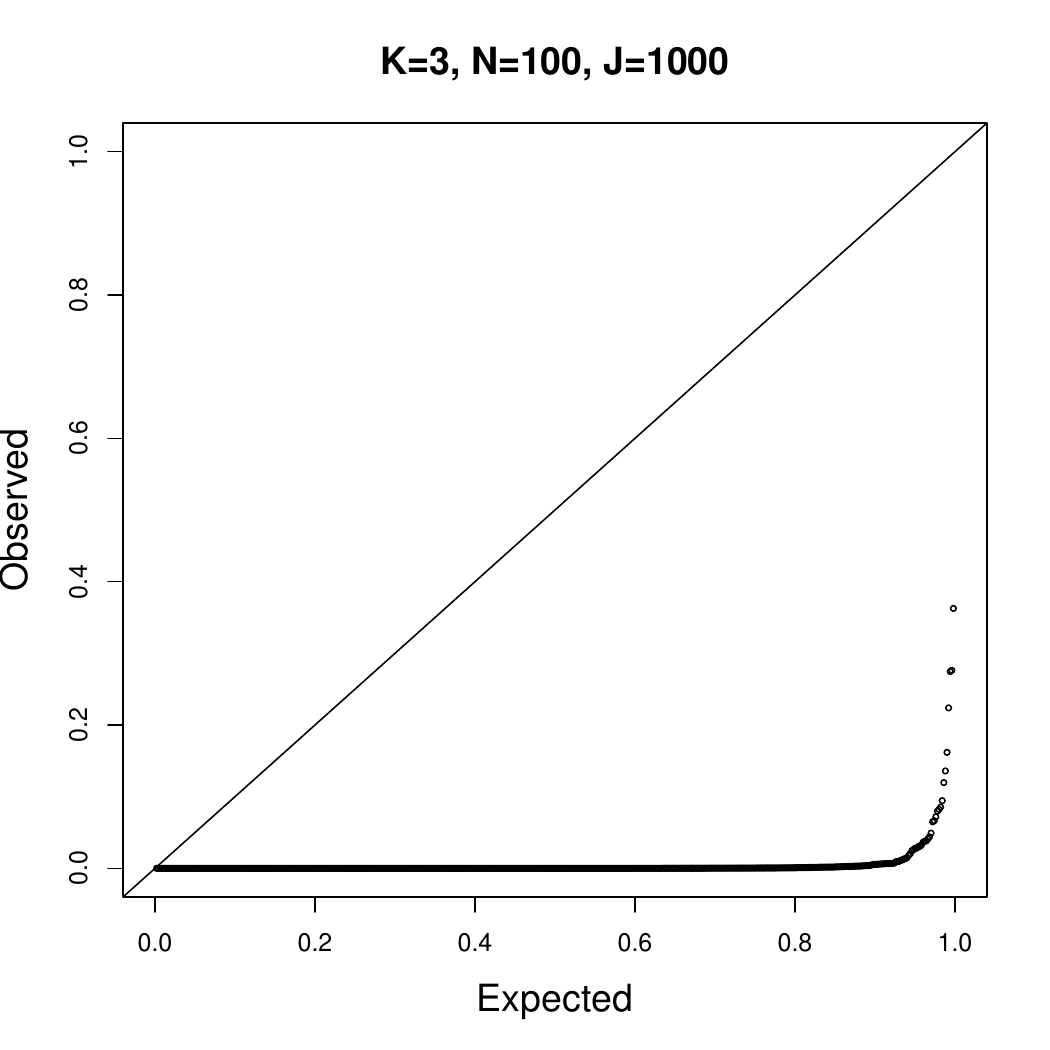}
     \end{subfigure}
     \hfill
     \begin{subfigure}[b]{0.24\textwidth}
         \centering
         \includegraphics[width=0.95\textwidth]{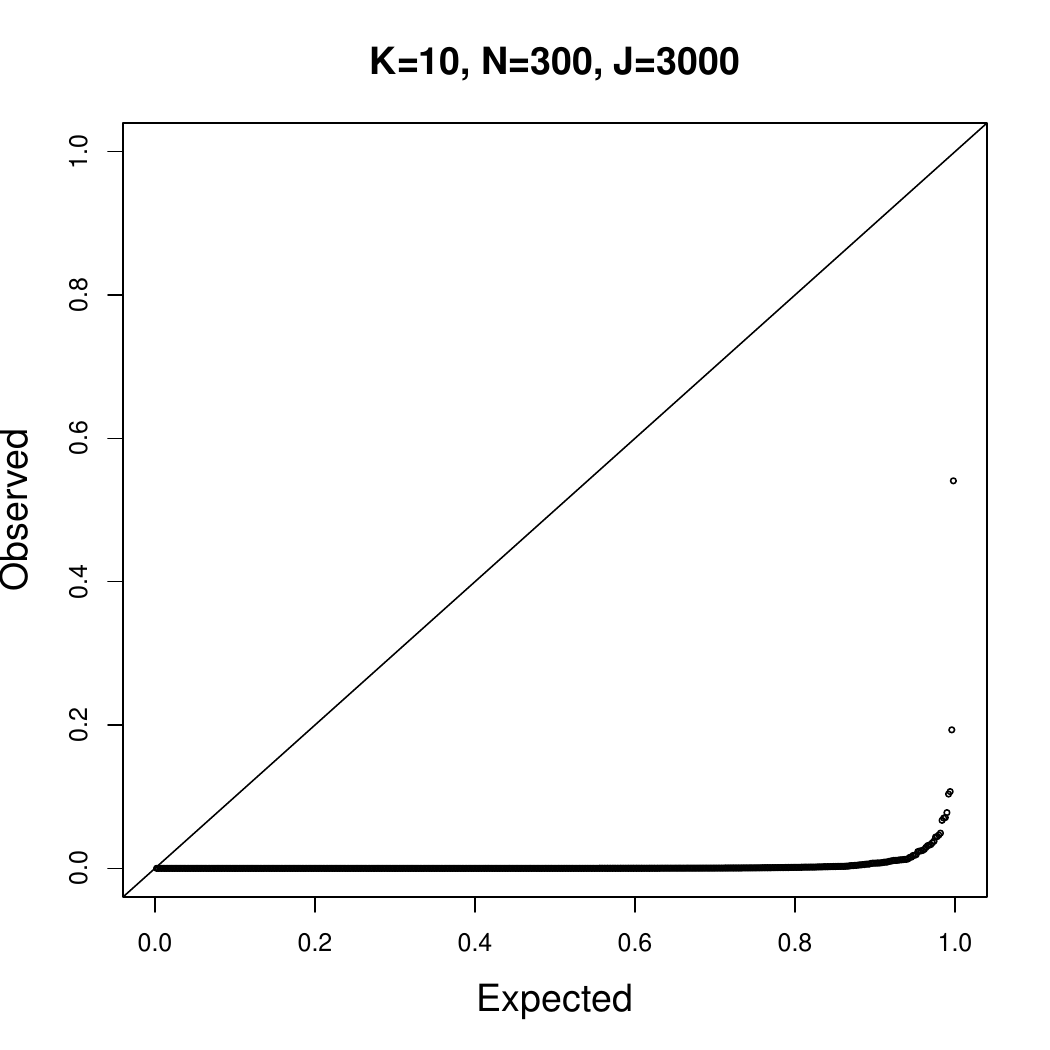}
     \end{subfigure}
     \hfill
     \begin{subfigure}[b]{0.24\textwidth}
         \centering
         \includegraphics[width=0.95\textwidth]{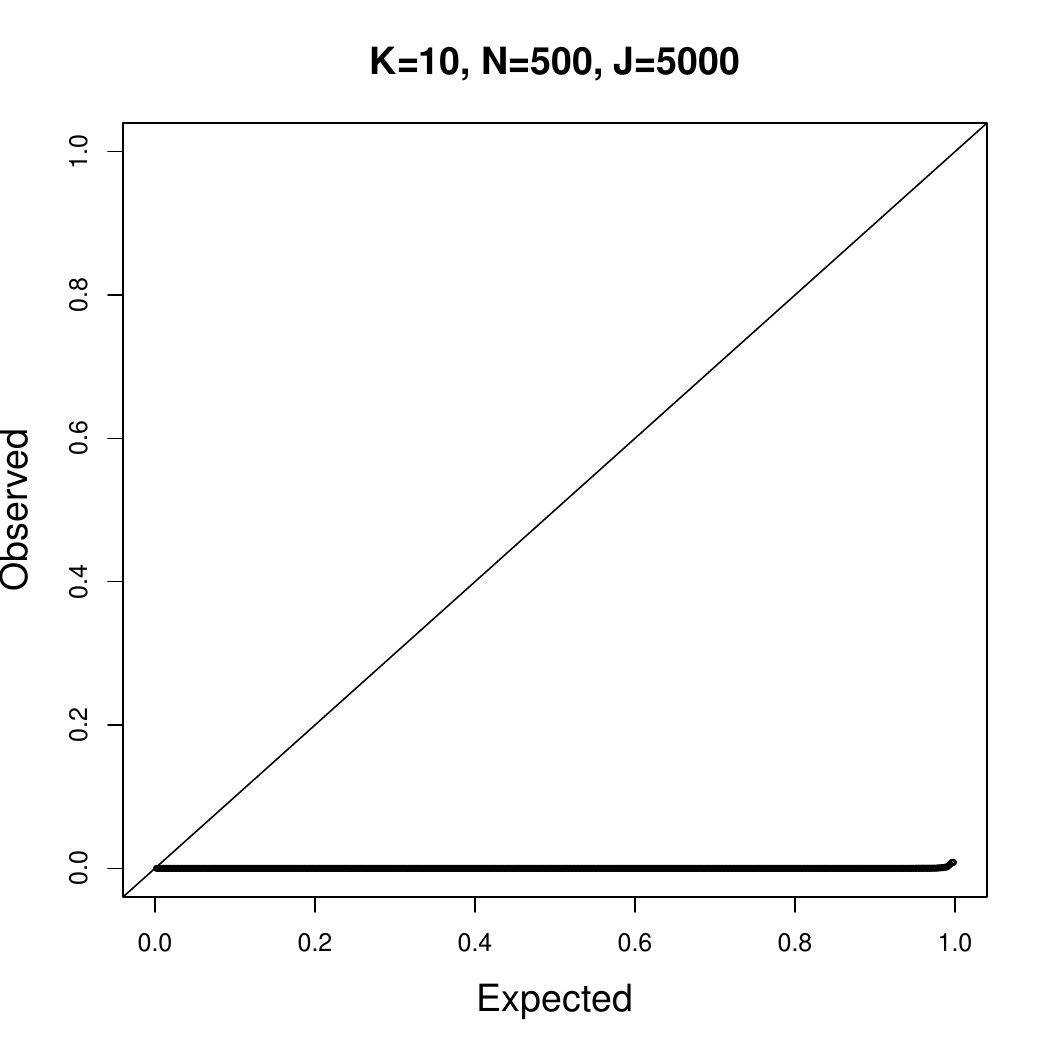}
     \end{subfigure}
\caption{Q-Q plots of $p$-values for testing the null hypothesis $H_0:\theta_{j,1}=\theta_{j,2}=\cdots \theta_{j,K}$ in the Bernoulli model. $H_0$ is true for feature $1$ (upper row) and false for feature $2$ (lower row).}
\label{fig:bernoulli}
\end{figure}

\vspace{-5mm}
\paragraph{Simulation Study 2: Statistical Inference.}
We empirically evaluate the theoretical results concerning the inference on $\bTheta$. Consider $J\in\{500, 1000\}$ when $K=3$ and $J\in\{3000, 5000\}$ when $K=10$. The sample size $N$ is set to $J/10$, $J/5$, and  $J$. 
After generating $\bTheta$ according to the simulation scheme mentioned earlier, we further set $\bTheta_{1,:}=0.5\cdot \mathbf 1^\top_{K}$ so that the entries of the first row of $\bTheta$ have the same value. In addition, $\bTheta_{2,:}=(0.1, 0.3, 0.6)$ for $K=3$ and $\bTheta_{2,:}=0.06\cdot (1,2,\ldots,10)$ for $K=10$, for the entries of the second row to have well-separated values. So, the null hypothesis of $H_0: \theta_{j,1}=\theta_{j,2}=\cdots=\theta_{j,K}$ is  true for feature $j=1$ and false for feature $j=2$.
We reject the null if the $p$-value is smaller than $0.05$.
We calculate the proportion of rejecting the null across the 500 simulation replicates for feature $1$ as Type-I error and the proportion of rejections for feature $2$ as power. 
Table \ref{tab:type_I_power} summarizes the Type-I error and power in various simulation settings. We have the following observations. First, the Type-I error is controlled under $0.05$ except for the most challenging case $N=J/10$. Second,  the power increases as the sample size increases.
Figure \ref{fig:bernoulli} gives the Q-Q plots of the $p$-values. The upper row corresponds to feature $j=1$, for which the null is true, while the lower row corresponds to feature $j=2$, for which the null is false. According to Theorem \ref{thm:test-theta-gen-simple} when $K=3$, we expect the asymptotic distribution of our test statistic to be the maximum of three independent  $\chi_1^2$ variables; when $K=10$, we use the generalized Gumbel distribution as a reference distribution. The Q-Q lines in the upper row of Figure \ref{fig:bernoulli} for testing feature $1$ are close to and above the 45-degree reference line, suggesting the test is slightly conservative but yields a good Type-I error control. For testing feature $2$, the distributions of the $p$-values are severely right-skewed as desired, indicating high test power. 
To summarize, our inference procedure works well in large-scale data scenarios in terms of Type-I error control and power performance. 
Similar results for the Poisson model are provided in Section \ref{sec:tables_figures} in the Supplementary Material.
\begin{table}[h!]
\footnotesize
\centering
\begin{tabular}{ccccccccc}
\toprule
\multirow{2}{*}{Model}  & \multirow{2}{*}{$K$} & \multirow{2}{*}{$J$} & \multicolumn{3}{c}{Type-I error} & \multicolumn{3}{c}{Power} \\
\cmidrule(lr){4-6} \cmidrule(lr){7-9} & & & $N=J/10$ & $N=J/5$ & $N=J$ & $N=J/10$ & $N=J/5$ & $N=J$ \\
\midrule
\multirow{5}{*}{Bernoulli}  & \multirow{2}{*}{3} & 500 & 0.088 & 0.028 & 0.032 & 0.790 & 0.958 & 1 \\
\cmidrule(lr){3-9} & & 1000 & 0.040 & 0.050 & 0.036 & 0.972 & 1 & 1 \\
\cmidrule(lr){2-9}
 & \multirow{2}{*}{10} & 3000 & 0.072 & 0.050 & 0.048 & 0.984 & 1 & 1 \\
\cmidrule(lr){3-9} & & 5000 & 0.060 & 0.050 & 0.042 & 1 & 1 & 1 \\
\midrule 
\multirow{5}{*}{Poisson}  & \multirow{2}{*}{3} & 500 & 0.036 & 0.038 & 0.046 & 0.564 & 0.898 & 1 \\
\cmidrule(lr){3-9} & & 1000 & 0.028 & 0.050 & 0.038 & 0.846 & 0.986 & 1 \\
\cmidrule(lr){2-9}
 & \multirow{2}{*}{10} & 3000 & 0.040 & 0.010 & 0.034 & 0.694 & 1 & 1 \\
\cmidrule(lr){3-9} & & 5000 & 0.040 & 0.032 & 0.048 & 0.996 & 1 & 1 \\
\bottomrule
\end{tabular}
\caption{Type-I error and power for testing $H_0:\theta_{j,1}=\theta_{j,2}=\cdots \theta_{j,K}$ for feature $j=1$ and feature $j=2$, respectively. The null hypothesis is true for feature $1$ and false for feature $2$.}
\label{tab:type_I_power}
\end{table}

We further evaluate our multiple testing procedure in the simulation setting with $K=3, J=1000, N=J/5$. For the first 50 rows of $\bTheta$, the entries of each row are identical and the null hypotheses $H_{0,j}: \theta_{j,1}=\cdots=\theta_{j,K}$ are all true; the values of these rows are independently simulated from the uniform distribution on $[0.2,~ 2/3]$.
The generation scheme of the rest of the parameters is the same as mentioned at the beginning of this section.
Table \ref{table:fdr} gives the mean of the number of discoveries (i.e., the number of rejected null hypotheses among all considered features) and false discovery proportion (FDP) across $500$ replicates. 
The FDP and Type-I errors are well-controlled under all significance levels considered.

\begin{table}[h!]
\centering
\footnotesize
\begin{tabular}{cccccc}
\toprule
Level $\alpha$ & \#False discoveries & \#True discoveries & FDP & Type-I error \\
\midrule
$0.01$ & 0.20 & 27.73 & 0.007 & 0.004 \\
\midrule
$0.05$ & 1.02 & 31.51 & 0.030 & 0.020 \\
\midrule
$0.1$ & 2.03 & 32.79 & 0.057 & 0.041 \\
\midrule
$0.2$ & 4.07 & 34.7 & 0.102 & 0.081 \\
\bottomrule
\end{tabular}
\caption{The numbers of true and false discoveries, FDP, and Type-I error under various levels with simulated data. The values are averaged out of
$500$ independent replications.}
\label{table:fdr}
\end{table}

\vspace{-5mm}
\section{Real Data Applications}\label{sec:real-data}

We apply our methods to three real-world datasets from diverse application fields: political voting records, SNP data in genome-wide association studies (GWAS), and single-cell sequencing data. For all three datasets, we use $100$ different initializations for the $K$-means clustering step in Algorithm \ref{alg:HeteroClustering}. In addition to the clustering error, we also report the rand index \citep{rand1971objective}, which is a commonly used metric to evaluate clustering accuracy.

\subsection{U.S. Senate Roll Call Votes Data}
We consider the United States 112th Senate Roll Call Votes data (publicly available at \href{https://legacy.voteview.com/senate112.htm}{https://legacy.voteview.com/senate112.htm}), containing 102 U.S. senators' voting records for $J=486$ roll calls. The original coding of the votes contains six categories: Yea, Paired Yea, Announced Yea, Announced Nay, Paired Nay, Nay. We convert the responses of Yea, Paired Yea, and Announced Yea to $1$ indicating voting for the roll call, and convert the responses of Nay, Paired Nay, and Announced Nay to $0$ indicating voting against it.
After removing the senators who are neither a Democrat nor a Republican and also senators with over 10\% missing votes, there are $N=94$ senators that we consider. We then randomly assign $0$ or $1$ to the missing entries for each senator with the probability equal to this senator's positive response rate of the non-missing votes. Other approaches to handling missing data are also possible \citep{yan2021inference}.
We again compare six clustering approaches and the results are summarized in Table \ref{tab:real_data} in the ``U.S. Senate data'' columns. Interestingly, with $\ell_2$ normalization, both HeteroPCA and SVD achieve perfect classification.

We also apply our multiple testing method to this dataset. 
Among the $424$ roll calls with positive estimated parameters $(\hat\theta_{j,1},\ldots,\hat\theta_{j,K})$, our BH-based testing procedure rejects the null hypothesis $H_{0,j}: \theta_{j,1}=\cdots=\theta_{j,K}$ for $302$ roll calls under the 0.05 significance level.
The roll call with the smallest adjusted $p$-value of $0$ corresponds to roll 17: ``to exclude employees of the Transportation Security Administration from the collective bargaining rights of Federal employees and provide employment rights and an employee engagement mechanism for passenger and property screeners''. As expected, the Republicans have a much higher probability ($\hat\theta_{17,2}=1$) in voting for this roll as compared to the Democrats ($\hat\theta_{17,1}=0.02$). The second smallest adjusted $p$-value of $9\mathrm{e}{-269}$ corresponds to roll 415: ``to amend the Federal Election Campaign Act of 1971 to provide for additional disclosure requirements for corporations, labor organizations, Super PACs and other entities, and for other purposes''. The Republicans have a much lower probability ($\hat\theta_{415,1}=0.02$) in voting for this roll as compared to the Democrats ($\hat\theta_{415,2}=1$). 
Figure \ref{fig:heatmap} in the Supplementary Material illustrates the heatmap of the $20$ rolls that correspond to the smallest adjusted $p$-values, which serve as markers to best distinguish these two U.S. political parties.

\subsection{SNPs Data of Genetic Variations}
We consider a dataset from the HapMap3 project \citep[publicly available at \href{https://www.broadinstitute.org/medical-and-population-genetics/hapmap-3}{https://www.broadinstitute.org/medical-and-population-genetics/hapmap-3}]{international2010integrating} that comprises high-dimensional SNP data from $K=11$ ethnic sub-populations. This dataset is widely used in GWAS analysis as it encodes the genetic variations of people.
We follow the conventional data preprocessing procedure to remove SNPs with minor allele frequency smaller than 5\% and SNPs with missing data \citep{kranzler2019genome,jiang2024tuning}.
This leads to $N=1115$ individuals and $J=274128$ SNPs. The data are re-coded into $0/1/2$ that represent counts of the minor
alleles at each SNP for each person. The data pre-processing procedures are conducted using the software PLINK \citep{purcell2007plink}. 

\begin{table}[h!]
\footnotesize
\centering
\begin{tabular}{ccccccccccc}
\toprule
\multirow{2}{*}{Normalization}  & \multirow{2}{*}{Spectral method} & \multicolumn{2}{c}{HapMap3 data} & \multicolumn{2}{c}{Single-cell data}  & \multicolumn{2}{c}{U.S. Senate data} \\
\cmidrule(lr){3-4} \cmidrule(lr){5-6} \cmidrule(lr){7-8} & & Error & Rand index & Error & Rand index & Error & Rand index \\
\midrule
\multirow{2}{*}{$\ell_2$} 
  & HeteroPCA & 25.7\% & 0.937 & 11.0\% & 0.917 & 0\% & 1 \\
\cmidrule(lr){2-8} 
    & SVD & 27.3\% & 0.935 & 11.2\% & 0.915 & 0\% & 1 \\
\midrule
\multirow{2}{*}{SCORE} 
  & HeteroPCA & 31.1\% & 0.918 & 13.5\% & 0.896 & 3.19\% & 0.938\\
\cmidrule(lr){2-8} 
    & SVD & 33.5\% & 0.915 & 13.5\% & 0.896 & 3.19\% & 0.938\\
\midrule
\multirow{2}{*}{None} 
  & HeteroPCA & 37.8\% & 0.888 & 16.9\% & 0.865 & 2.13\% & 0.958 \\
\cmidrule(lr){2-8} 
    & SVD & 38.0\% & 0.888 & 21.6\% & 0.830 & 2.13\% & 0.958 \\
\bottomrule
\end{tabular}
\caption{Clustering error and Rand index for HapMap3 data, single-cell data, and U.S. Senate roll call voting data for six clustering methods.
}
\label{tab:real_data}
\end{table}

Figure \ref{fig:hapmap} in the Supplementary Material shows the streak structures in the top singular subspace for the HapMap3 dataset, indicating the existence of degree heterogeneity.
Table \ref{tab:real_data} summarizes the results of the six clustering approaches in the ``HapMap3 data'' columns. It shows normalization in the singular subspace significantly improves the clustering accuracy. Specifically,  $\ell_2$ normalization yields lower clustering error and higher Rand index compared to SCORE normalization. Furthermore, HeteroPCA leads to slightly better results compared to SVD. This result justifies the superior performance of HeteroClustering.

\subsection{Single-cell Sequencing Data}
We consider the single-cell 10x scATAC-seq data \citep[publicly available at \href{https://cellxgene.cziscience.com/collections/d36ca85c-3e8b-444c-ba3e-a645040c6185}{https://cellxgene.cziscience.com/collections/d36ca85c-3e8b-444c-ba3e-a645040c6185}]{lengyel2022molecular}.
There are five annotated cell types in the data: endothelial cell, smooth muscle cell, stromal cell, pericyte, and leukocyte, with the number of cells in each cell type being $156$, $165$, $15607$, $2104$, $283$, respectively. Since the numbers are not balanced across clusters, we randomly select $500$ samples from the stromal cell type and $500$ samples from the pericyte cell type. This leads to $N=1604$ samples and $J=19298$ genes in total. 
This dataset consists of the nonnegative counts of gene expressions in each cell, with the largest count being 212.
Figure \ref{fig:single} in the Supplementary Material demonstrates the streak structures in the the singular subspace, indicating clear degree-heterogeneity within latent classes. 

Table \ref{tab:real_data} summarizes the results of the six spectral clustering approaches in the ``Single-cell data'' columns. Similarly to the HapMap data, we observe that normalizations help reduce the clustering errors, and $\ell_2$ normalization outperforms  SCORE normalization. In addition, HeteroPCA performs better than SVD, especially when no normalization is used. Our HeteroClustering algorithm still achieves the best performance among the six methods.

We apply the multiple testing procedure for the Poisson model; see Section \ref{sec:gen-type} in the Supplementary Material.
Among the $17300$ genes with positive estimated item parameters, our BH-based testing procedure rejects the null $H_{0,j}: \theta_{j,1}=\cdots=\theta_{j,K}$ for $12045$ genes under the 0.05 significance level. This implies these $12045$ genes are found by our method to be useful for differentiating the cell types.
We have also used the popular R package \texttt{Seurat} \citep{seraut} to find the differentially expressed genes. We use the function \texttt{FindAllMarkers} in \texttt{Seurat} with default arguments and found $14277$ genes with BH-adjusted $p$-values below $0.05$. 
Among these markers and those discovered by our procedure, there are $8974$ common ones.
We would like to emphasize that, our preliminary analyses reported here mainly serve as a demonstration of our proposed  method. The substantive interpretation and validity of these discoveries would require further investigation and scrutiny from domain experts.

\section{Discussion}\label{sec-discussion}

In summary, we have proposed {theoretically sound} and {computationally efficient} methods motivated by the following phenomena and need in real-world applications: (a) the wide presence of individual-level quantitative heterogeneity nested within latent classes, and (b) the need for clustering and statistical inference methods with theoretical guarantees for high-dimensional, sparse, heteroskedastic discrete data. 
Our easy-to-implement HeteroClustering algorithm demonstrates broad applicability and superior performance in simulations and real data analyses.
We also offer a suite of estimation and inference results for the high-dimensional continuous parameters in DhLCMs. We borrow the technical tools (Lemma \ref{lem:two-inf-bound}) developed in \cite{yan2021inference} and propose a complete framework for clustering, estimation and inference under DhLCMs, but our theoretical analysis requires careful treatment rather than being an immediate consequence of their results.
Recently, \cite{gao2022selective} and \cite{chen2023selective} considered testing whether there are significant differences in means between clusters. An insight in \cite{gao2022selective} is that naively performing hypothesis testing after clustering can lead to inflated Type-I errors, because the clustering event is not appropriately accounted for. This insight leads the authors in \cite{gao2022selective} and \cite{chen2023selective} to propose novel \emph{selective inference} procedures for clustering. Interestingly, our statistical inference results provide a complementary insight that, in certain high-dimensional mixture models where exact clustering is achievable, we may directly perform valid global testing and multiple testing.

There are several promising directions for future research. \emph{First}, the high power of dependency on $\kappa$ in the general theorems in the Supplementary materials (e.g., Theorem \ref{thm:Thetaerr-gen-inf} and \ref{thm:test-theta-gen}) may be improvable. It would be worthwhile to investigate whether this dependency arises solely from the proof techniques used or if it is inherent in HeteroPCA, necessitating significant algorithmic modifications akin to those proposed in \cite{zhou2023deflated, zhou2023heteroskedastic}.
\emph{Second}, 
Algorithm \ref{alg:HeteroClustering} requires an input of $K$, which can be unknown in real applications. To estimate $K$, a simple yet common strategy is the scree plot method \citep{cattell1966scree}, 
which shall serve as a reliable estimate of $K$ under a suitable signal-to-noise condition. Other approaches include gap statistics \citep{tibshirani2001estimating} and eigen selection \citep{han2023eigen},
among many others. 
Practitioners may first estimate $K$ using those methods, and apply our procedure for clustering and inference. It is intriguing to consider whether such a combined approach yields theoretical guarantees akin to those in \cite{fan2022simplea}. Further investigation along this direction is beyond the scope of this paper and left for future research.
\emph{Third}, it would be interesting to extend the DhLCM to accommodate polytomous responses.
Binary responses are prevalent in numerous applications and pose unique challenges, especially in the sparse regime considered here. Yet polytomous responses with $R_{i,j}\in[d]$ for $d>2$, like those found in Likert-scale questionnaires, are also common \citep{formann1992linear}.
One potential approach is to treat these responses as a $N\times J \times d$ tensor and use tensor-based spectral methods. 
However, the tensor structure's complexity raises nontrivial questions about the optimal guarantees of such approaches, which warrant further investigations. \emph{Third}, in this paper we have considered the case where the $N\times J$ noise matrix $\mathbf E$ has independent entries. In practice, locally dependent errors can occur in educational assessments and survey questionnaires,  due to the design of the items \citep{berzofsky2014local}. 
Our clustering method may still be valid under certain extent of local dependence. Developing clustering error rates and valid inference procedures in such scenarios are intriguing future directions.

\spacingset{1}

\paragraph{Supplementary Material.}
The Supplementary Material contains all proofs of the theoretical results, extensions
to the Binomial and Poisson models, and additional simulation and real data analysis details.

\paragraph{Acknowledgments.}
This research is partially supported by NSF grant DMS-2210796.
The authors thank the Editor Professor Annie Qu, an anonymous Associate Editor, and two anonymous reviewers for constructive comments that helped to greatly improve the quality of this manuscript. The authors thank Chengzhu Huang for helpful discussion.

\bibliographystyle{apalike}
\bibliography{references}

\begin{thebibliography}{}

\bibitem[Balakrishnan et~al., 2017]{balakrishnan2017em}
Balakrishnan, S., Wainwright, M.~J., and Yu, B. (2017).
\newblock Statistical guarantees for the {EM} algorithm: From population to
  sample-based analysis.
\newblock {\em Annals of Statistics}, 45(1):77--120.

\bibitem[Bandeira and Van~Handel, 2016]{bandeira2016sharp}
Bandeira, A.~S. and Van~Handel, R. (2016).
\newblock Sharp nonasymptotic bounds on the norm of random matrices with
  independent entries.

\bibitem[Benjamini and Hochberg, 1995]{benjamini1995controlling}
Benjamini, Y. and Hochberg, Y. (1995).
\newblock Controlling the false discovery rate: a practical and powerful
  approach to multiple testing.
\newblock {\em Journal of the Royal Statistical Society: Series B
  (Methodological)}, 57(1):289--300.

\bibitem[Berzofsky et~al., 2014]{berzofsky2014local}
Berzofsky, M.~E., Biemer, P.~P., and Kalsbeek, W.~D. (2014).
\newblock Local dependence in latent class analysis of rare and sensitive
  events.
\newblock {\em Sociological Methods \& Research}, 43(1):137--170.

\bibitem[Bhattacharya et~al., 2023]{bhattacharya2023inferences}
Bhattacharya, S., Fan, J., and Hou, J. (2023).
\newblock Inferences on mixing probabilities and ranking in mixed-membership
  models.
\newblock {\em arXiv preprint arXiv:2308.14988}.

\bibitem[Bishop, 2006]{bishop2006pattern}
Bishop, C.~M. (2006).
\newblock {\em Pattern recognition and machine learning}, volume~4.
\newblock Springer.

\bibitem[Cai et~al., 2021]{cai2021subspace}
Cai, C., Li, G., Chi, Y., Poor, H.~V., and Chen, Y. (2021).
\newblock Subspace estimation from unbalanced and incomplete data matrices:
  $\ell_{2,\infty}$ statistical guarantees.
\newblock {\em Annals of Statistics}, 49(2):944--967.

\bibitem[Cattell, 1966]{cattell1966scree}
Cattell, R.~B. (1966).
\newblock The scree test for the number of factors.
\newblock {\em Multivariate Behavioral Research}, 1(2):245--276.

\bibitem[Chen et~al., 2022]{chen2022global}
Chen, S., Liu, S., and Ma, Z. (2022).
\newblock Global and individualized community detection in inhomogeneous
  multilayer networks.
\newblock {\em The Annals of Statistics}, 50(5):2664--2693.

\bibitem[Chen and Yang, 2021]{chen2021cutoff}
Chen, X. and Yang, Y. (2021).
\newblock Cutoff for exact recovery of {G}aussian mixture models.
\newblock {\em IEEE Transactions on Information Theory}, 67(6):4223--4238.

\bibitem[Chen et~al., 2021a]{chen2021spectral}
Chen, Y., Chi, Y., Fan, J., Ma, C., et~al. (2021a).
\newblock Spectral methods for data science: A statistical perspective.
\newblock {\em Foundations and Trends{\textregistered} in Machine Learning},
  14(5):566--806.

\bibitem[Chen et~al., 2021b]{chen2021bridging}
Chen, Y., Fan, J., Ma, C., and Yan, Y. (2021b).
\newblock Bridging convex and nonconvex optimization in robust {PCA}: Noise,
  outliers, and missing data.
\newblock {\em Annals of Statistics}, 49(5):2948.

\bibitem[Chen et~al., 2020]{chen2020structured}
Chen, Y., Li, X., and Zhang, S. (2020).
\newblock Structured latent factor analysis for large-scale data:
  Identifiability, estimability, and their implications.
\newblock {\em Journal of the American Statistical Association},
  115(532):1756--1770.

\bibitem[Chen et~al., 2021c]{chen2021unfolding}
Chen, Y., Ying, Z., and Zhang, H. (2021c).
\newblock Unfolding-model-based visualization: theory, method and applications.
\newblock {\em The Journal of Machine Learning Research}, 22(1):548--598.

\bibitem[Chen and Witten, 2023]{chen2023selective}
Chen, Y.~T. and Witten, D.~M. (2023).
\newblock Selective inference for {K}-means clustering.
\newblock {\em Journal of Machine Learning Research}, 24:152.

\bibitem[Consortium et~al., 2010]{international2010integrating}
Consortium, I. H.~. et~al. (2010).
\newblock Integrating common and rare genetic variation in diverse human
  populations.
\newblock {\em Nature}, 467(7311):52.

\bibitem[Deng et~al., 2024]{deng2023strong}
Deng, C., Xu, X.-J., and Ying, S. (2024).
\newblock Strong consistency of spectral clustering for the sparse
  degree-corrected hypergraph stochastic block model.
\newblock {\em IEEE Transactions on Information Theory}, 70(3):1962--1977.

\bibitem[Devroye et~al., 2018]{devroye2018total}
Devroye, L., Mehrabian, A., and Reddad, T. (2018).
\newblock The total variation distance between high-dimensional gaussians with
  the same mean.
\newblock {\em arXiv preprint arXiv:1810.08693}.

\bibitem[Embrechts et~al., 2013]{embrechts2013modelling}
Embrechts, P., Kl{\"u}ppelberg, C., and Mikosch, T. (2013).
\newblock {\em Modelling extremal events: for insurance and finance},
  volume~33.
\newblock Springer Science \& Business Media.

\bibitem[Fan et~al., 2022]{fan2022simplea}
Fan, J., Fan, Y., Han, X., and Lv, J. (2022).
\newblock {SIMPLE}: Statistical inference on membership profiles in large
  networks.
\newblock {\em Journal of the Royal Statistical Society Series B: Statistical
  Methodology}, 84(2):630--653.

\bibitem[Fienberg et~al., 2009]{fienberg2007maximum}
Fienberg, S.~E., Hersh, P., Rinaldo, A., and Zhou, Y. (2009).
\newblock {\em Maximum likelihood estimation in latent class models for
  contingency table data}.
\newblock Cambridge University Press.

\bibitem[Formann, 1992]{formann1992linear}
Formann, A.~K. (1992).
\newblock Linear logistic latent class analysis for polytomous data.
\newblock {\em Journal of the American Statistical Association},
  87(418):476--486.

\bibitem[Fraley and Raftery, 2002]{fraley2002model}
Fraley, C. and Raftery, A.~E. (2002).
\newblock Model-based clustering, discriminant analysis, and density
  estimation.
\newblock {\em Journal of the American Statistical Association},
  97(458):611--631.

\bibitem[Gao et~al., 2018]{gao2018community}
Gao, C., Ma, Z., Zhang, A.~Y., and Zhou, H.~H. (2018).
\newblock Community detection in degree-corrected block models.
\newblock {\em The Annals of Statistics}, 46(5):2153--2185.

\bibitem[Gao et~al., 2022]{gao2022selective}
Gao, L.~L., Bien, J., and Witten, D. (2022).
\newblock Selective inference for hierarchical clustering.
\newblock {\em Journal of the American Statistical Association}, pages 1--11.

\bibitem[Goodman, 1974]{goodman1974exploratory}
Goodman, L.~A. (1974).
\newblock Exploratory latent structure analysis using both identifiable and
  unidentifiable models.
\newblock {\em Biometrika}, 61(2):215--231.

\bibitem[Gu and Xu, 2023]{gu2023joint}
Gu, Y. and Xu, G. (2023).
\newblock A joint {MLE} approach to large-scale structured latent attribute
  analysis.
\newblock {\em Journal of the American Statistical Association},
  118(541):746--760.

\bibitem[Hagenaars and McCutcheon, 2002]{hagenaars2002applied}
Hagenaars, J.~A. and McCutcheon, A.~L. (2002).
\newblock {\em Applied latent class analysis}.
\newblock Cambridge University Press.

\bibitem[Han et~al., 2022]{han2022exact}
Han, R., Luo, Y., Wang, M., and Zhang, A.~R. (2022).
\newblock Exact clustering in tensor block model: Statistical optimality and
  computational limit.
\newblock {\em Journal of the Royal Statistical Society Series B: Statistical
  Methodology}, 84(5):1666--1698.

\bibitem[Han et~al., 2023]{han2023eigen}
Han, X., Tong, X., and Fan, Y. (2023).
\newblock Eigen selection in spectral clustering: a theory-guided practice.
\newblock {\em Journal of the American Statistical Association},
  118(541):109--121.

\bibitem[Hao et~al., 2023]{seraut}
Hao, Y., Stuart, T., Kowalski, M.~H., Choudhary, S., Hoffman, P., Hartman, A.,
  Srivastava, A., Molla, G., Madad, S., Fernandez-Granda, C., and Satija, R.
  (2023).
\newblock Dictionary learning for integrative, multimodal and scalable
  single-cell analysis.
\newblock {\em Nature Biotechnology}.

\bibitem[Hu and Wang, 2022]{hu2022multiway}
Hu, J. and Wang, M. (2022).
\newblock Multiway spherical clustering via degree-corrected tensor block
  models.
\newblock In {\em International Conference on Artificial Intelligence and
  Statistics}, pages 1078--1119. PMLR.

\bibitem[Jiang et~al., 2024]{jiang2024tuning}
Jiang, W., Chen, L., Girgenti, M.~J., and Zhao, H. (2024).
\newblock Tuning parameters for polygenic risk score methods using gwas summary
  statistics from training data.
\newblock {\em Nature Communications}, 15(1):24.

\bibitem[Jin, 2015]{jin2015fast}
Jin, J. (2015).
\newblock Fast community detection by {SCORE}.
\newblock {\em The Annals of Statistics}, 43(1):57--89.

\bibitem[Jin et~al., 2021]{jin2021improvements}
Jin, J., Ke, Z.~T., and Luo, S. (2021).
\newblock Improvements on {SCORE}, especially for weak signals.
\newblock {\em Sankhya A}, pages 1--36.

\bibitem[Jin and Wang, 2016]{jin2016ifpca}
Jin, J. and Wang, W. (2016).
\newblock Influential features {PCA} for high dimensional clustering.
\newblock {\em The Annals of Statistics}, 44(6):2323--2359.

\bibitem[Jing et~al., 2021]{jing2021community}
Jing, B.-Y., Li, T., Lyu, Z., and Xia, D. (2021).
\newblock Community detection on mixture multilayer networks via regularized
  tensor decomposition.
\newblock {\em The Annals of Statistics}, 49(6):3181--3205.

\bibitem[Jing et~al., 2022]{jing2022community}
Jing, B.-Y., Li, T., Ying, N., and Yu, X. (2022).
\newblock Community detection in sparse networks using the symmetrized
  laplacian inverse matrix (slim).
\newblock {\em Statistica Sinica}, 32(1).

\bibitem[Johnson et~al., 2005]{johnson2005univariate}
Johnson, N.~L., Kemp, A.~W., and Kotz, S. (2005).
\newblock {\em Univariate discrete distributions}, volume 444.
\newblock John Wiley \& Sons.

\bibitem[Ke and Jin, 2023]{ke2023special}
Ke, Z.~T. and Jin, J. (2023).
\newblock Special invited paper: The score normalization, especially for
  heterogeneous network and text data.
\newblock {\em Stat}, 12(1):e545.

\bibitem[Ke et~al., 2019]{ke2019community}
Ke, Z.~T., Shi, F., and Xia, D. (2019).
\newblock Community detection for hypergraph networks via regularized tensor
  power iteration.
\newblock {\em arXiv preprint arXiv:1909.06503}.

\bibitem[Kiselev et~al., 2019]{kiselev2019challenges}
Kiselev, V.~Y., Andrews, T.~S., and Hemberg, M. (2019).
\newblock Challenges in unsupervised clustering of single-cell {RNA-seq} data.
\newblock {\em Nature Reviews Genetics}, 20(5):273--282.

\bibitem[Kranzler et~al., 2019]{kranzler2019genome}
Kranzler, H.~R., Zhou, H., Kember, R.~L., Vickers~Smith, R., Justice, A.~C.,
  Damrauer, S., Tsao, P.~S., Klarin, D., Baras, A., Reid, J., et~al. (2019).
\newblock Genome-wide association study of alcohol consumption and use disorder
  in 274,424 individuals from multiple populations.
\newblock {\em Nature Communications}, 10(1):1499.

\bibitem[Lazarsfeld, 1950]{lazarsfeld1950logical}
Lazarsfeld, P.~F. (1950).
\newblock The logical and mathematical foundation of latent structure analysis.
\newblock {\em Studies in social psychology in world war II Vol. IV:
  Measurement and prediction}, pages 362--412.

\bibitem[Lengyel et~al., 2022]{lengyel2022molecular}
Lengyel, E., Li, Y., Weigert, M., Zhu, L., Eckart, H., Javellana, M., Ackroyd,
  S., Xiao, J., Olalekan, S., Glass, D., et~al. (2022).
\newblock A molecular atlas of the human postmenopausal fallopian tube and
  ovary from single-cell rna and atac sequencing.
\newblock {\em Cell Reports}, 41(12).

\bibitem[Linzer and Lewis, 2011]{linzer2011polca}
Linzer, D.~A. and Lewis, J.~B. (2011).
\newblock {poLCA}: An {R} package for polytomous variable latent class
  analysis.
\newblock {\em Journal of Statistical Software}, 42:1--29.

\bibitem[L{\"o}ffler et~al., 2021]{loffler2019optimality}
L{\"o}ffler, M., Zhang, A.~Y., and Zhou, H.~H. (2021).
\newblock Optimality of spectral clustering in the {G}aussian mixture model.
\newblock {\em The Annals of Statistics}, 49(5):2506--2530.

\bibitem[Lu and Zhou, 2016]{lu2016statistical}
Lu, Y. and Zhou, H.~H. (2016).
\newblock Statistical and computational guarantees of lloyd's algorithm and its
  variants.
\newblock {\em arXiv preprint arXiv:1612.02099}.

\bibitem[Lyu et~al., 2023]{lyu2023optimal}
Lyu, Z., Li, T., and Xia, D. (2023).
\newblock Optimal clustering of discrete mixtures: Binomial, poisson, block
  models, and multi-layer networks.
\newblock {\em arXiv preprint arXiv:2311.15598}.

\bibitem[Lyu and Xia, 2022]{lyu2022optimal}
Lyu, Z. and Xia, D. (2022).
\newblock Optimal clustering by lloyd algorithm for low-rank mixture model.
\newblock {\em arXiv preprint arXiv:2207.04600}.

\bibitem[Marchal and Arbel, 2017]{marchal2017sub}
Marchal, O. and Arbel, J. (2017).
\newblock On the sub-gaussianity of the beta and dirichlet distributions.

\bibitem[Ndaoud, 2022]{ndaoud2022sharp}
Ndaoud, M. (2022).
\newblock Sharp optimal recovery in the two component gaussian mixture model.
\newblock {\em The Annals of Statistics}, 50(4):2096--2126.

\bibitem[Nylund-Gibson and Choi, 2018]{nylund2018ten}
Nylund-Gibson, K. and Choi, A.~Y. (2018).
\newblock Ten frequently asked questions about latent class analysis.
\newblock {\em Translational Issues in Psychological Science}, 4(4):440.

\bibitem[Purcell et~al., 2007]{purcell2007plink}
Purcell, S., Neale, B., Todd-Brown, K., Thomas, L., Ferreira, M.~A., Bender,
  D., Maller, J., Sklar, P., De~Bakker, P.~I., Daly, M.~J., et~al. (2007).
\newblock {PLINK}: a tool set for whole-genome association and population-based
  linkage analyses.
\newblock {\em The American Journal of Human Genetics}, 81(3):559--575.

\bibitem[Qin and Rohe, 2013]{qin2013regularized}
Qin, T. and Rohe, K. (2013).
\newblock Regularized spectral clustering under the degree-corrected stochastic
  blockmodel.
\newblock {\em Advances in Neural Information Processing Systems}, 26.

\bibitem[Rai${\check{c}}$, 2019]{raivc2019multivariate}
Rai${\check{c}}$, M. (2019).
\newblock A multivariate berry--esseen theorem with explicit constants.

\bibitem[Rand, 1971]{rand1971objective}
Rand, W.~M. (1971).
\newblock Objective criteria for the evaluation of clustering methods.
\newblock {\em Journal of the American Statistical Association},
  66(336):846--850.

\bibitem[Rohe and Zeng, 2023]{rohe2023varimax}
Rohe, K. and Zeng, M. (2023).
\newblock {Vintage factor analysis with Varimax performs statistical
  inference}.
\newblock {\em Journal of the Royal Statistical Society Series B: Statistical
  Methodology}, 85(4):1037--1060.

\bibitem[Sarkar and Bickel, 2015]{sarkar2015role}
Sarkar, P. and Bickel, P.~J. (2015).
\newblock Role of normalization in spectral clustering for stochastic
  blockmodels.
\newblock {\em The Annals of Statistics}, 43(3):962--990.

\bibitem[Scrucca et~al., 2023]{scrucca2023model}
Scrucca, L., Fraley, C., Murphy, T.~B., and Raftery, A.~E. (2023).
\newblock {\em Model-Based Clustering, Classification, and Density Estimation
  Using mclust in R}.
\newblock Chapman and Hall/CRC.

\bibitem[Skorski, 2023]{skorski2023bernstein}
Skorski, M. (2023).
\newblock Bernstein-type bounds for beta distribution.
\newblock {\em Modern Stochastics: Theory and Applications}, 10(2):211--228.

\bibitem[Squair et~al., 2021]{squair2021confronting}
Squair, J.~W., Gautier, M., Kathe, C., Anderson, M.~A., James, N.~D., Hutson,
  T.~H., Hudelle, R., Qaiser, T., Matson, K.~J., Barraud, Q., et~al. (2021).
\newblock Confronting false discoveries in single-cell differential expression.
\newblock {\em Nature Communications}, 12(1):5692.

\bibitem[Stein-O’Brien et~al., 2018]{stein2018enter}
Stein-O’Brien, G.~L., Arora, R., Culhane, A.~C., Favorov, A.~V., Garmire,
  L.~X., Greene, C.~S., Goff, L.~A., Li, Y., Ngom, A., Ochs, M.~F., et~al.
  (2018).
\newblock Enter the matrix: factorization uncovers knowledge from omics.
\newblock {\em Trends in Genetics}, 34(10):790--805.

\bibitem[Tibshirani et~al., 2001]{tibshirani2001estimating}
Tibshirani, R., Walther, G., and Hastie, T. (2001).
\newblock Estimating the number of clusters in a data set via the gap
  statistic.
\newblock {\em Journal of the Royal Statistical Society: Series B (Statistical
  Methodology)}, 63(2):411--423.

\bibitem[Wainwright, 2019]{wainwright2019high}
Wainwright, M.~J. (2019).
\newblock {\em High-dimensional statistics: A non-asymptotic viewpoint},
  volume~48.
\newblock Cambridge University Press.

\bibitem[Yan et~al., 2024]{yan2021inference}
Yan, Y., Chen, Y., and Fan, J. (2024).
\newblock Inference for heteroskedastic {PCA} with missing data.
\newblock {\em Annals of Statistics}, to appear.

\bibitem[Zeng et~al., 2023]{zeng2023tensor}
Zeng, Z., Gu, Y., and Xu, G. (2023).
\newblock A tensor-{EM} method for large-scale latent class analysis with
  binary responses.
\newblock {\em Psychometrika}, 88(2):580--612.

\bibitem[Zhang et~al., 2022]{zhang2022heteroskedastic}
Zhang, A.~R., Cai, T.~T., and Wu, Y. (2022).
\newblock Heteroskedastic {PCA}: Algorithm, optimality, and applications.
\newblock {\em The Annals of Statistics}, 50(1):53--80.

\bibitem[Zhang and Zhou, 2016]{zhang2016minimax}
Zhang, A.~Y. and Zhou, H.~H. (2016).
\newblock Minimax rates of community detection in stochastic block models.
\newblock {\em The Annals of Statistics}, 44(5):2252--2280.

\bibitem[Zhang and Zhou, 2022]{zhang2022leave}
Zhang, A.~Y. and Zhou, H.~H. (2022).
\newblock Leave-one-out singular subspace perturbation analysis for spectral
  clustering.
\newblock {\em arXiv preprint arXiv:2205.14855}.

\bibitem[Zhang et~al., 2012]{zhang2012latent}
Zhang, B., Chen, Z., and Albert, P.~S. (2012).
\newblock Latent class models for joint analysis of disease prevalence and
  high-dimensional semicontinuous biomarker data.
\newblock {\em Biostatistics}, 13(1):74--88.

\bibitem[Zhang et~al., 2020]{zhang2020detecting}
Zhang, Y., Levina, E., and Zhu, J. (2020).
\newblock Detecting overlapping communities in networks using spectral methods.
\newblock {\em SIAM Journal on Mathematics of Data Science}, 2(2):265--283.

\bibitem[Zhou and Chen, 2023a]{zhou2023deflated}
Zhou, Y. and Chen, Y. (2023a).
\newblock Deflated {HeteroPCA}: Overcoming the curse of ill-conditioning in
  heteroskedastic {PCA}.
\newblock {\em arXiv preprint arXiv:2303.06198}.

\bibitem[Zhou and Chen, 2023b]{zhou2023heteroskedastic}
Zhou, Y. and Chen, Y. (2023b).
\newblock Heteroskedastic tensor clustering.
\newblock {\em arXiv preprint arXiv:2311.02306}.

\end{thebibliography}

\renewcommand{\thesection}{S.\arabic{section}}  
\renewcommand{\thetable}{S.\arabic{table}}  
\renewcommand{\thefigure}{S.\arabic{figure}}
\renewcommand{\theequation}{S.\arabic{equation}}
\renewcommand{\theassumption}{S.\arabic{assumption}}
\renewcommand{\thetheorem}{S.\arabic{theorem}}
\renewcommand{\thelemma}{S.\arabic{lemma}}

\setcounter{section}{0}
\setcounter{equation}{0}
\setcounter{figure}{0}
\setcounter{theorem}{0}
\setcounter{lemma}{0}

\newpage 

{\centering\section*{Supplementary Material}}
\addcontentsline{toc}{section}{Supplementary Material} 


This Supplementary Material is organized as follows.
Section \ref{sec:supp-discussion} provides discussions on some key technical quantities in the theoretical analyses.
Section \ref{sec:gen-type} extends our clustering and inference methods to the Binomial and Poisson models.
Section \ref{sec:tables_figures} presents additional figures for the simulation studies and real-data analyses. Section \ref{sec:notations_assumptions} contains additional notations and general assumptions. Section \ref{sec:lemmas} contains the technical lemmas used in our proofs. Sections \ref{sec:proof_clust}, \ref{sec:proof_est}, \ref{sec:proof_gen} respectively lay out the proofs of our theoretical results for the clustering algorithm in Section \ref{sec:clust}, estimation and inference of $\bTheta$ in Section \ref{sec:est-theta}, and extensions to the Binomial and Poisson models in Section \ref{sec:gen-type}. Section \ref{sec:proof_aux} contains some auxiliary proofs of Propositions and Lemmas.

\section{Discussion on Key Quantities in Section \ref{subsec:assump-key}}\label{sec:supp-discussion}
\subsection{Discussion on $\Delta$, $\sigma_\star$ and $\kappa$}
In clustering analysis, $\Delta$ is a key quantity that characterizes the difficulty of clustering, which is essentially the minimum Euclidean distance between different cluster centers \citep{lu2016statistical,zhang2022leave, loffler2019optimality}.
Another related quantity $\sigma_\star $, defined as the smallest singular value of $\bTheta$, is crucial in spectral methods. See a comprehensive introduction to spectral methods in  \cite{chen2021spectral} and references therein. These two quantities will play important roles in our theoretical analyses. 
The following lemma reveals the properties of these quantities in a DhLCM.

\begin{lemma}\label{lem:incoherence} 
\begin{itemize}
	\item[(a)] The signal strength  and $\sigma_\star$ are of the same order up to $\kappa$, i.e.,
	\begin{align}\label{eq:sig-rel}
		{\Delta}/{\sqrt{2}}\ge  \sigma_\star\ge {\Delta}/({\sqrt{2}\kappa}),
	\end{align}
	\item[(b)] Under Assumptions \ref{cond:balanced}-\ref{cond:const-degee}, the smallest singular value of $\bR^*$, denoted by $\sigma_K\brac{\bR^*}$, satisfies
		$\sigma_K(\bR^*)\gtrsim \sqrt{N/K}\sigma_\star.$
	The condition number of $\bR^*$ is bounded by $\kappa$ up to a constant.
\end{itemize}		
	
\end{lemma}

In light of Lemma \ref{lem:incoherence},  $\Delta$ and $\sigma_\star$, the key quantity in clustering and the signal strength in spectral methods,  differ by a multiplicative factor of $\kappa$. Naturally, this implies that 
the performance of our spectral clustering algorithm depends on $\kappa$. Such dependency has recently been resolved by \cite{zhou2023deflated,zhou2023heteroskedastic}. While it is possible to obtain a sharper bound in terms of $\kappa$ by adopting methods therein, this will complicate our procedure and blur the core principles of our method. We thus do not spare additional effort to that end in this paper. Fortunately, the following proposition indicates that it is indeed reasonable to assume $\kappa=O(1)$ under a common generative model for $\bTheta$ in latent class models.

\begin{proposition}\label{prop:theta-beta}
	Assume $\theta_{j,k}\overset{i.i.d.}{\sim}\rho_J\cdot \textsf{Beta}\brac{a,b}$ for all $j\in[J]$ and $k\in[K]$ with constants $a,b>0$ and sparsity parameter $\rho_J\in(0,1]$,  then we have 
	\begin{align*}
	\kappa\brac{\bTheta}\le C\sqrt\frac{a+b+1}{a+b+\frac{a}{a+b}},
\end{align*}
 with probability at least
\begin{align*}
	1-2K\sqbrac{\exp\brac{-C_1\frac{a^2\brac{a+b+\frac{a}{a+b}}^2}{\brac{a+b}^2\brac{a+b+1}^2}JK}+\exp\brac{-C_2\frac{a\brac{a+b+\frac{a}{a+b}}}{\brac{a+b}\brac{a+b+1}}J}},
\end{align*}
for some absolute constants $C,C_1,C_2>0$.
\end{proposition}

\subsection{Discussion on $\mu_{\bTheta}$}

Essentially, $\mu_{\bTheta}$ quantifies the  magnitude of  row-wise accumulation of ``mass" in $\bTheta$ is uniformly spread out, which can be regarded as a notion of {\it incoherence} of $\bTheta$. Similar quantities arise commonly in the context of low-rank matrix recovery with row-wise error control (i.e., $\ell_{2,\infty}$ error control), particularly relevant in the fields of compressed sensing and matrix completion \citep{chen2021spectral, chen2021bridging}. In the context of clustering in mixture models, it turns out that $\mu_{\bTheta}$ is crucial in deriving sharp exponential error rates for clustering.

\begin{proposition}\label{prop:theta-incoherent}
	Assume $\theta_{j,k}\overset{i.i.d.}{\sim}\rho_J\cdot \textsf{Beta}\brac{a,b}$ for all $j\in[J]$ and $k\in[K]$ with constants $a,b>0$ and sparsity parameter $\rho_J\in(0,1]$, then with probability at least $1-O\brac{J^{-20}}$,
	\begin{align*}
 \mu_{\bTheta}
 \le C\brac{1+\frac{a+b}{b}\min\left\{\rho_J^{-2},\frac{\log J}{K}\right\}},
\end{align*}
for some absolute constant $C>0$.
\end{proposition}
Proposition \ref{prop:theta-incoherent} illustrates the property of $\mu_{\bTheta}$ under the reasonable assumption that entries in $\bTheta$ come from a Beta distribution. In this scenario, with high probability we have $\mu_{\bTheta}=O\brac{1}$ if $\rho_J\asymp 1$, and $\mu_{\bTheta}=O\brac{\log J}$ for sparse data with $\rho_J=o\brac{1}$.

\section{Extensions to Binomial and Poisson models}\label{sec:gen-type}

In addition to common high-dimensional binary data, we also adapt the DhLCM framework for analyzing other high-dimensional discrete data.
Examples motivating such extensions include the SNPs data which take values in 0/1/2 and are usually modeled using the Binomial distribution, and the single-cell gene expression data that are nonnegative counts.
See Section \ref{sec:real-data} for examples of binary, Binomial, and count data.
All our results in the previous sections can be extended to (i) Binomial distribution case where $R_{i,j}\sim \text{Bin}\brac{m,p_{ij}}$ with some positive integer $m\in\NN_+$ for $i\in[N]$ and $j\in[J]$, and (ii) Poisson distribution case where $R_{i,j}\sim \text{Poisson}\brac{\lambda_{ij}}$ for $i\in[N]$ and $j\in[J]$.

\paragraph{Binomial model.}
For Binomial responses, we generalize our results in Section \ref{sec:clust} by incorporating an additional positive integer $m$.
In particular, we consider the model
\begin{align}\label{eq:model-bin}
	R_{i,j}\sim \text{Binomial}\brac{m,~\omega_i\theta_{j,s_i}}, \quad \EE[\bR]=m\bOmega\bZ\bTheta^\top,\quad \bR=\EE[\bR]+\bE,
\end{align}
where $\bE$ consists of mean zero independent Binomial noise. 
The noise random variables $\{E_{i,j}\}$'s are bounded by $m$, so all techniques and results can be extended without essential difficulty.  We state the corresponding result regarding the performance of Algorithm \ref{alg:HeteroClustering}.

\begin{theorem}\label{thm:exp-clustering-error-bin}
Consider the Binomial model defined in \eqref{eq:model-bin} with $ m\lesssim\brac{NJ}^{1/4}\sqrt{{\theta_{\sf max}}/{\log J}}$.
Suppose Assumption \ref{cond:balanced}-\ref{cond:const-degee} hold and $\kappa=O(1)$. In addition, assume  $N\gtrsim \log^2 J$, $J\gtrsim \log^4J$, and there exists some absolute constant $C_{\sf clust}>0$ such that
\begin{align}\label{eq:delta-cond-bin}
	\frac{\Delta^2}{\theta_{\sf max}}\ge \frac{ C_{\sf clust}}{m}K^2\sqrt{\frac{J}{N}}\log J,
\end{align}
then HeteroClustering gives exact clustering $h(\hat s,s)=0$ with probability exceeding $1-\brac{N+J}^{-20}$.
\end{theorem}

\paragraph{Poisson model.}
Poisson model is defined the same way as in \eqref{eq:model-original} and \eqref{eq:model-decomp}, except that 
\begin{align}\label{eq:model-poisson}
R_{i,j}\sim \text{Poisson}\brac{\omega_i\theta_{j,s_i}},
 \quad \EE[\bR]=\bOmega\bZ\bTheta^\top,
 \quad 
 \bR=\EE[\bR] + \bE,
\end{align}
where $\bE$ consists of independent mean zero Poisson noise.
Some extra efforts on dealing with a truncated version of the unbounded noise matrix are needed to establish clustering guarantees for the Poisson case. We have the following theoretical  result for HeteroClustering.

\begin{theorem}\label{thm:exp-clustering-error-poisson}
Consider the Poisson model defined in \eqref{eq:model-poisson} with $\min_{j,k}\theta_{j,k}\gtrsim\brac{N+J}^{-C_{\theta}}$ for some absolute constant $C_{\theta}>0$. Suppose Assumption \ref{cond:balanced}-\ref{cond:const-degee} hold and $\kappa=O(1)$. In addition, assume $N\gtrsim \log^2 J$, $J\gtrsim \log^4 J$, and there exists some absolute constant $C_{\sf clust}>0$ such that
\begin{align}\label{eq:delta-cond-poisson}
	\frac{\Delta^2}{\theta_{\sf max}}\ge  C_{\sf clust}K^2\sqrt{\frac{J}{N}}\log^3 J,
\end{align}
then HeteroClustering gives exact clustering $h(\hat s,s)=0$ with probability exceeding $1-\brac{N+J}^{-20}$.
\end{theorem}

\paragraph{Statistical inference on $\bTheta$ under the Binomial or Poisson model.}
After using the HeteroClustering algorithm to recover $\hat s$ under either the Binomial model \eqref{eq:model-bin} or the Poisson model \eqref{eq:model-poisson}, we can obtain estimation and inference results similar to those in Section \ref{sec:est-theta}. Due to the space constraint, we next only briefly present how to adapt the global testing procedure for \eqref{eq:null} and multiple testing procedure for \eqref{eq:multi-null} to Binomial and Poisson models. 

We define the estimator of $\bTheta$ as
\begin{align}\label{eq:thetahat-def-gen}
    \hat\bTheta^\top =\begin{cases}
        \dfrac{1}{m}\brac{\hat\bZ^\top \hat\bZ}^{-1} \hat\bZ^\top \hat\bOmega^{-1}\bR, &\text{for~Binomial~model~}\eqref{eq:model-bin}, \\
        \brac{\hat\bZ^\top \hat\bZ}^{-1} \hat\bZ^\top \hat\bOmega^{-1}\bR, &\text{for~Poisson~model~}\eqref{eq:model-poisson}. 
    \end{cases}
\end{align}
where $\hat\bOmega$ is defined the same way as in Section \ref{subsec:theta}. Note that $\hat\bTheta^\top$ for the Poisson model takes the same form as in the previous Bernoulli case in \eqref{eq:theta-est}. Meanwhile, we also need to modify the asymptotic variance expression in \eqref{eq:theta-variance} to the following,
\begin{align}\label{eq:sigmahat-def-gen}
\hat \sigma^2_{j,k}=\begin{cases}
   \frac{\hat \theta_{j,k}}{m\ab{\hat \calC_k}^2}\sum_{i\in\hat\calC_k}\frac{1-\hat\omega_i\hat \theta_{j,k}}{\hat\omega_i},& \text{for~Binomial~model~}\eqref{eq:model-bin},\\
     \frac{\hat \theta_{j,k}}{\ab{\hat \calC_k}^2}\sum_{i\in\hat\calC_k}\frac{1}{\hat\omega_i},& \text{for~Poisson~model~}\eqref{eq:model-poisson}.
\end{cases}
\end{align}
To perform the global hypothesis testing \eqref{eq:null} and the multiple hypothesis testing \eqref{eq:multi-null},  we use the same test statistic $T$ defined in \eqref{eq:test-statistic} and $T_j$ defined in \eqref{eq:multi-test-statistic} respectively,  but replace the definition of $\hat\bTheta$ and $\hat\sigma_{j,k}^2$ by \eqref{eq:thetahat-def-gen} and \eqref{eq:sigmahat-def-gen}, respectively. 

\begin{theorem}\label{thm:test-theta-bin-poisson}
Suppose the conditions of Theorem \ref{thm:exp-clustering-error-simple} and Assumption \ref{cond:iden-theta} hold, and there exists some absolute constant $C_{\sf inf}>0$ such that 
\begin{align*}
	\frac{\Delta^2}{\theta_{\sf max}}\ge C_{\sf inf}\mu^2_{\bTheta}K^3\brac{\frac{\theta_{\sf max}}{\theta^*_{\sf min}}}\frac{J}{N}\log^3(N+J).
\end{align*}
The conclusions of Theorem \ref{thm:test-theta-gen-simple} and Theorem  \ref{thm:fdr-control}   hold under either of the following scenarios:
\begin{enumerate}
    \item[(i)] \sloppy For the  Binomial model defined in \eqref{eq:model-bin} with $m\lesssim \sqrt{\frac{N\theta_{\sf max}}{\log J}}$, assuming that  $J\gtrsim m\mu^3_{\bTheta}\brac{\frac{\theta_{\sf max}}{\theta^*_{\sf min}}}\log^2J$.
    \item[(ii)] \sloppy For the Poisson model defined in \eqref{eq:model-poisson} with $\min_{j,k}\theta_{j,k}\gtrsim\brac{N+J}^{-C_{\theta}}$, assuming that  $J\gtrsim \mu^3_{\bTheta}\brac{\frac{\theta_{\sf max}}{\theta^*_{\sf min}}}\log^3J$.
\end{enumerate}
\end{theorem}

Theorem \ref{thm:test-theta-bin-poisson} implies that both the global testing and the multiple testing procedures in Section \ref{subsec:test} are still valid for Binomial and Poisson models.

\section{Lower Bound for Exponential Error Rate}
\begin{theorem}\label{thm:lower-bound}
	Assume $\theta_{\sf max}\lesssim \min_{j,k}\theta_{j,k}$. If ${\Delta^2}/({\theta_{\sf max}}\log K)\rightarrow \infty $, then
	\begin{align*}
\inf_{\hat s} \sup_{\calP(s,\bOmega,\bTheta)}\EE h(\hat s,s)&\ge \frac{1}{N}\sum_{i=1}^N\exp\brac{-c_0\omega_i\cdot \frac{\Delta^2}{\theta_{\sf max}}},
\end{align*}
for some absolute constant $c_0>0$. 
\end{theorem}
\begin{remark}
{\color{black}When $\mu_{\bTheta},K=O(1)$, Theorem \ref{thm:lower-bound} complements Theorem \ref{thm:exp-clustering-error} by matching the upper bound  in \eqref{eq:exp-error-rate}  when $\frac{\Delta^2}{\theta_{\sf max}}\gtrsim\frac{J}{N}\log J$ holds, in which case the \textsf{SNR}$^2$ defined in Theorem \ref{thm:exp-clustering-error} has the same order as $\frac{\Delta^2}{\theta_{\sf max}}$.
This result indicates that Algorithm \ref{alg:HeteroClustering} delivers an  optimal error rate in this scenario. Notably,  $\frac{\Delta^2}{\theta_{\sf max}}\gtrsim\frac{J}{N}\log J$ required in Theorem \ref{thm:lower-bound} aligns with the minimal condition for valid statistical inference on $\bTheta$ as required in Section \ref{sec:est-theta}, so it is  not a stringent condition if one focuses on the inferenc of $\bTheta$. 
When $\frac{\Delta^2}{\theta_{\sf max}}\ll \frac{J}{N}\log J$, an additional multiplicative factor $\brac{1+\frac{KJ\theta_{\sf max}\log J}{N\Delta^2}}^{-1}$ arises in \textsf{SNR}$^2$ in Theorem \ref{thm:exp-clustering-error} and it is unclear if one can remove it. }
\end{remark}
\paragraph{Proof of Theorem \ref{thm:lower-bound}.}
Denote $\calC_k(\tilde s):=\{i:\tilde s_i=k\}$ for any $\tilde s\in[K]^N$. Without loss of generality, assume that there exists a $s^*\in[K]^N$ such that $|\calC_1(s^*)|\le |\calS_2(s^*)|\le \cdots \le |\calC_k(s^*)|$ and $\frac{c N}{K}\le |\calC_k(s)|\le \frac{N}{c K}$ for all $k\in[K]$ for some small constant $c\in(0,1)$. Now given $s^*$, for any $k\in[K]$ let $\calI_{k}\subset \{i:s^*_i=k\}$ with cardinality $|\calI_k|=|\calC_k(s^*)|-\frac{\delta N}{K^2}$ collecting indices of the largest $\omega_i$'s in $\{\omega_i:s^*_i=k\}$, for some  $\delta>0$ sufficiently small. Let $\calI=\cup_{k\in[K]}\calI_k$ and define
\begin{align*}
	\calV^*=\left \{s\in[K]^N:s_i=s_i^*,\forall i\in \calI, \frac{c N}{K}\le |\calC_k(s)|\le \frac{N}{c K},\forall k\in[K]\right \}
\end{align*}
By definition, $s^*\in\calV^*$. We can define a parameter space for $ \bR^*$ as
\begin{align*}
	\calP_0\brac{s,\bOmega,\bTheta}=\left\{\wt \bR:s\in\calV^*,  \wt R_{i,j}=\omega_i\theta_{j,s_i} \right \}
\end{align*}
Restricting our attention to this parameter space gives that 
\begin{align*}
	\inf_{\hat s} \sup_{\calP_0(s,\bOmega,\bTheta)}\EE Nh(\hat s,s)=\inf_{\hat s} \sup_{\calP_0(s,\bOmega,\bTheta)}\EE H(\hat s,s)
\end{align*}
where $H(\cdot,\cdot)$ is the hamming distance without permutation, owing to the fact that any two distinct lable vectors in $\calV^*$ will have hamming distance at most $\frac{\delta N}{K}$ and hence we can avoid the permutation in the definition of $h(\cdot,\cdot)$ if $\delta$ is small. Observe that
\begin{align*}
	\inf_{\hat s}\sup_{\calP_0}\EE H(\hat s,s)\ge \inf_{\hat s}\frac{1}{|\calV^*|}\sum_{s\in \calV^*}\EE H(\hat s,s)\ge \sum_{i\in \calI^c}\inf_{\hat s_i}\frac{1}{|\calV^*|}\sum_{s\in \calV^*}\PP(\hat s_i\ne s_i)
\end{align*}
Here the first inequality holds since minimax risk is lower bounded by Bayes risk by assuming a uniform prior on $\calV^*$, where with slight abuse of notation we drop the dependence on $(\bOmega,\bTheta)$. The second inequality holds as all $s\in \cal V^*$ satisfying $s_i=s_i^*$ for $i\in \cal I$, hence it suffices for us to consider those $\hat s$ satisfying $\hat s_i=s_i$ for $i\in \cal I$. Now it suffices to consider a lower bound for $\inf_{\hat s_i}\frac{1}{|\calV^*|}\sum_{s\in \calV^*}\PP(\hat s_i\ne s_i)$ for $i\in \calI^c$. Without loss of generality, assume $1\in \calI^c$, we can further partition $\calV^*=\cup_{k\in[K]}\calV^*_k$ with
\begin{align*}
	\calV_k^*=\{s\in \calV^*:s_1=k\}
\end{align*}
For any $k_1\ne k_2$, $|\calV_{k_1}^*|=|\calV_{k_2}^*|$ by symmetry and we can index all $\{\calV_k^*\}_{k\in[K]}$ by the second to the last coordinates of $s$ contained in them, denoted by $(k,s_{-1})$. We collect all $s_{-1}$'s into a set $\calV_{-1}$, it is readily seen from definition that $K|\calV_{-1}|=|\calV^*|$. Then 
\begin{align*}
	\inf_{\hat s_1}\frac{1}{|\calV^*|}\sum_{s\in \calV^*}\PP(\hat s_1\ne s_1)&=\inf_{\hat s_1}\frac{1}{|\calV^*|}\sum_{k\in[K]}\sum_{s\in \calV_{k}^*}\PP(\hat s_1\ne k)\\
	&\ge \inf_{\hat s_1}\frac{1}{|\calV^*|}\frac{1}{K}\sum_{k_1< k_2}\brac{\sum_{s\in \calV_{k_1}^*}\PP(\hat s_1\ne k_1)+\sum_{s\in \calV_{k_2}^*}\PP(\hat s_1\ne k_2)}\\
	&\ge \inf_{\hat s_1}\frac{1}{|\calV^*|}\frac{1}{K}\brac{\sum_{s\in \calV_{1}^*}\PP(\hat s_1\ne 1)+\sum_{s\in \calV_{2}^*}\PP(\hat s_1\ne 2)}\\
	&\ge\frac{1}{K}\frac{1}{|\calV^*|}\sum_{s_{-1}\in \calV_{-1}} \inf_{\hat s_1}\brac{\PP_{s=(1,s_{-1})}(\hat s_1\ne 1)+\PP_{s=(2,s_{-1})}(\hat s_1\ne 2)}\\
	&\ge\frac{1}{K^2}\inf_{\hat s_1}\brac{\PP_{H_0}(\hat s_1=2 )+\PP_{H_1}(\hat s_1=1)}
\end{align*}
where we define 
\begin{align*}
	H_0:\bR_{1,:}\sim \bigotimes_{j=1}^J\text{Bern}(\omega_1\theta_{j,1}) \quad \text{vs.} \quad H_1:\bR_{1,:}\sim \bigotimes_{j=1}^J\text{Bern}(\omega_1\theta_{j,2})
\end{align*}
By Neyman-Pearson Lemma (c.f., Lemma A.2 in \cite{chen2022global}), the optimal test that minimizes the Type-I plus Type-II error of the above simple versus simple hypothesis test is given by the likelihood ratio test. In particular, LRT reject $H_0$ if 
\begin{align*}
	\prod_{j=1}^J(\omega_1\theta_{j,2})^{R_{1,j}}(1-\omega_1\theta_{j,2})^{1-R_{1,j}}> \prod_{j=1}^J(\omega_1\theta_{j,1})^{R_{1,j}}(1-\omega_1\theta_{j,1})^{1-R_{1,j}}
\end{align*}
By simple algebra we obtain that 
\begin{align*}
	\sum_{j=1}^JR_{1,j}\log \frac{\omega_1\theta_{j,2}(1-\omega_1\theta_{j,1})}{\omega_1\theta_{j,1}(1-\omega_1\theta_{j,2})}> \sum_{j=1}^J\log\frac{1-\omega_1\theta_{j,1}}{1-\omega_1\theta_{j,2}}
\end{align*}
We need the following lemma to establish the minimax lower bound.
\begin{lemma}\label{lem:lower-bound}
 Suppose $X_1,\cdots,X_J$ are independent Bernoulli random variables with $X_j\sim \text{Bern}(p_{j,1})$, where $p_{j,1}\asymp p_{j,2}=o(1)$ for all $j\in[J]$, then we have
	\begin{align*}
		\PP\brac{\sum_{j=1}^J X_j \log \frac{p_{j,2}(1-p_{j,1})}{p_{j,1}(1-p_{j,2})}> \sum_{j=1}^J\log\frac{1-p_{j,1}}{1-p_{j,2}}}\ge \exp\brac{-(1+o(1))\sum_{j=1}^J I_j}
	\end{align*}
	as $\sum_{j=1}^JI_j\rightarrow \infty$, where $I_j:=-2\log\brac{\sqrt{p_{j,1}p_{j,2}}+\sqrt{(1-p_{j,1})(1-p_{j,2})}}$ for all $j\in[J]$.
\end{lemma}
By Lemma \ref{lem:lower-bound}, we obtain that 
\begin{align*}
	\PP\brac{\sum_{j=1}^JR_{1,j}\log \frac{\omega_1\theta_{j,2}(1-\omega_1\theta_{j,1})}{\omega_1\theta_{j,1}(1-\omega_1\theta_{j,2})}> \sum_{j=1}^J\log\frac{1-\omega_1\theta_{j,1}}{1-\omega_1\theta_{j,2}}}\ge  \exp\brac{-c_0\omega_1\cdot \frac{\sum_{j=1}^J(\theta_{j,1}-\theta_{j,2})^2}{\theta_{\sf max}}}
\end{align*}
where we have used Lemma B.1 in \cite{zhang2016minimax}. Therefore, we arrive at
\begin{align*}
\inf_{\hat s} \sup_{\calP_0(s,\bOmega,\bTheta)}\EE h(\hat s,s)&\ge \frac{ 1}{NK^2}\sum_{i\in \calI^c}\exp\brac{-c_0\omega_i\cdot \frac{\Delta^2}{\theta_{\sf max}}}\ge  \frac{ \delta}{K^3}\frac{1}{|\calI^c|}\sum_{i\in \calI^c}\exp\brac{-c_0\omega_i\cdot \frac{\Delta^2}{\theta_{\sf max}}}\\
&\ge \frac{ \delta}{K^3}\frac{1}{N}\sum_{i=1}^N\exp\brac{-c_0\omega_i\cdot \frac{\Delta^2}{\theta_{\sf max}}}=\frac{1}{N}\sum_{i=1}^N\exp\brac{-c_0\omega_i\cdot \frac{\Delta^2}{\theta_{\sf max}}(1+o(1))}
\end{align*}
where the last inequality holds provided that $\omega_{\sf min}{\Delta^2}/{\theta_{\sf max}}\gg \log K$.\hfill$\square$

\section{Additional Figures for the Numerical Results}\label{sec:tables_figures}

\subsection{Comparisons with JML and MML}
\begin{figure}[h!]
    \centering
    \includegraphics[width=0.7\textwidth]{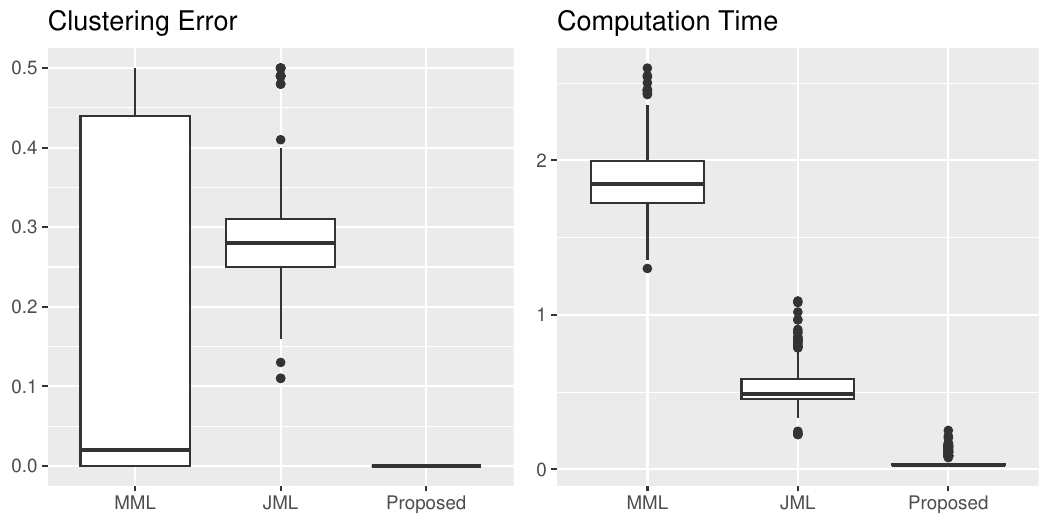}
    \caption{Clustering error and computation time in seconds comparing JML, MML, and our proposed method, when the ground truth model does not have degree heterogeneity.}
    \label{fig:mle}
\end{figure}

\begin{figure}[h!]
    \centering
    \includegraphics[width=0.8\textwidth]{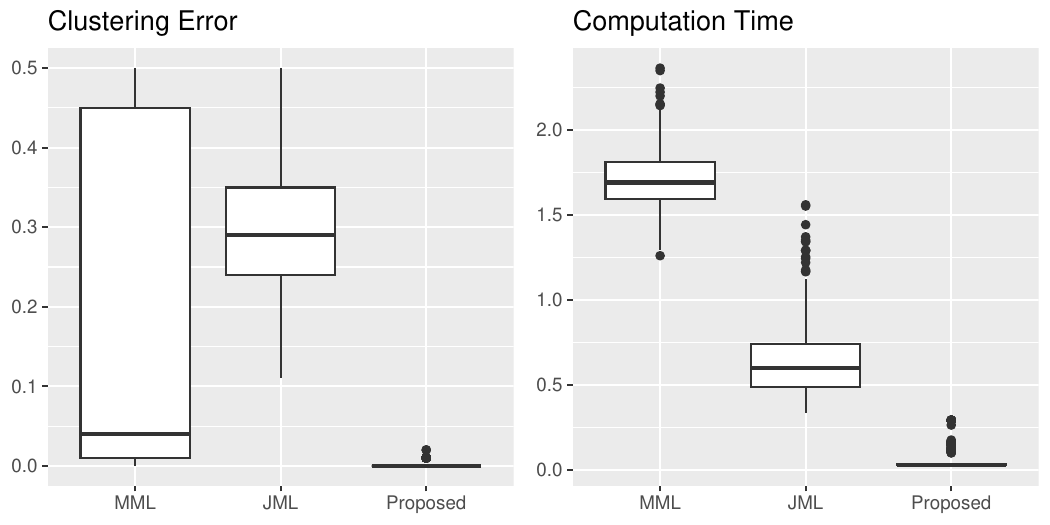}
    \caption{Clustering error and computation time in seconds comparing JML, MML, and our proposed method, when the ground truth model has degree heterogeneity.}
    \label{fig:mle_w_degree}
\end{figure}

We extend our comparison to include two popular likelihood-based clustering methods for LCA, namely, joint maximum likelihood (JML) and marginal maximum likelihood (MML). Consider the simulation setting with $K=2$, $J=500$, and $N=100$.
For JML, we use an iterative algorithm that monotonically increases the joint likelihood function;
see additional details for this algorithm  in Section \ref{sec:JMLE}.
For MML, we use the EM algorithm implemented in the function \texttt{poLCA} in the popular R package \texttt{poLCA} \citep{linzer2011polca},
where each data point is assigned to the latent class with the largest posterior probability after convergence of the EM algorithm. Random initialization is used for both JML and MML.
Figure \ref{fig:mle} presents the clustering error and computation time of our HeteroClustering algorithm and the above two likelihood methods under the classical LCM without degree heterogeneity. Note that both JML and MML are only developed for this degree-free model setting. 
We can see that JML gives the largest clustering error on average, while MML clustering results are not robust. Notably, our spectral method not only excels in clustering accuracy but also is the most computationally efficient compared to the iterative MML and JML algorithms.
We also conduct simulations in the challenging setting with $K=10$. In this setting, our proposed method still works well, whereas JML usually gets stuck at local minimums and MML takes a long time to converge. 
Figure \ref{fig:mle_w_degree} demonstrates the clustering error and computation time comparisons for DhLCM with degree heterogeneity, in contrast to the setting in Figure \ref{fig:mle}.


\subsection{Statistical Inference Simulation under the Poisson Model}
\begin{figure}[h!]
\centering
     \begin{subfigure}[b]{0.24\textwidth}
         \centering
         \includegraphics[width=0.9\textwidth]{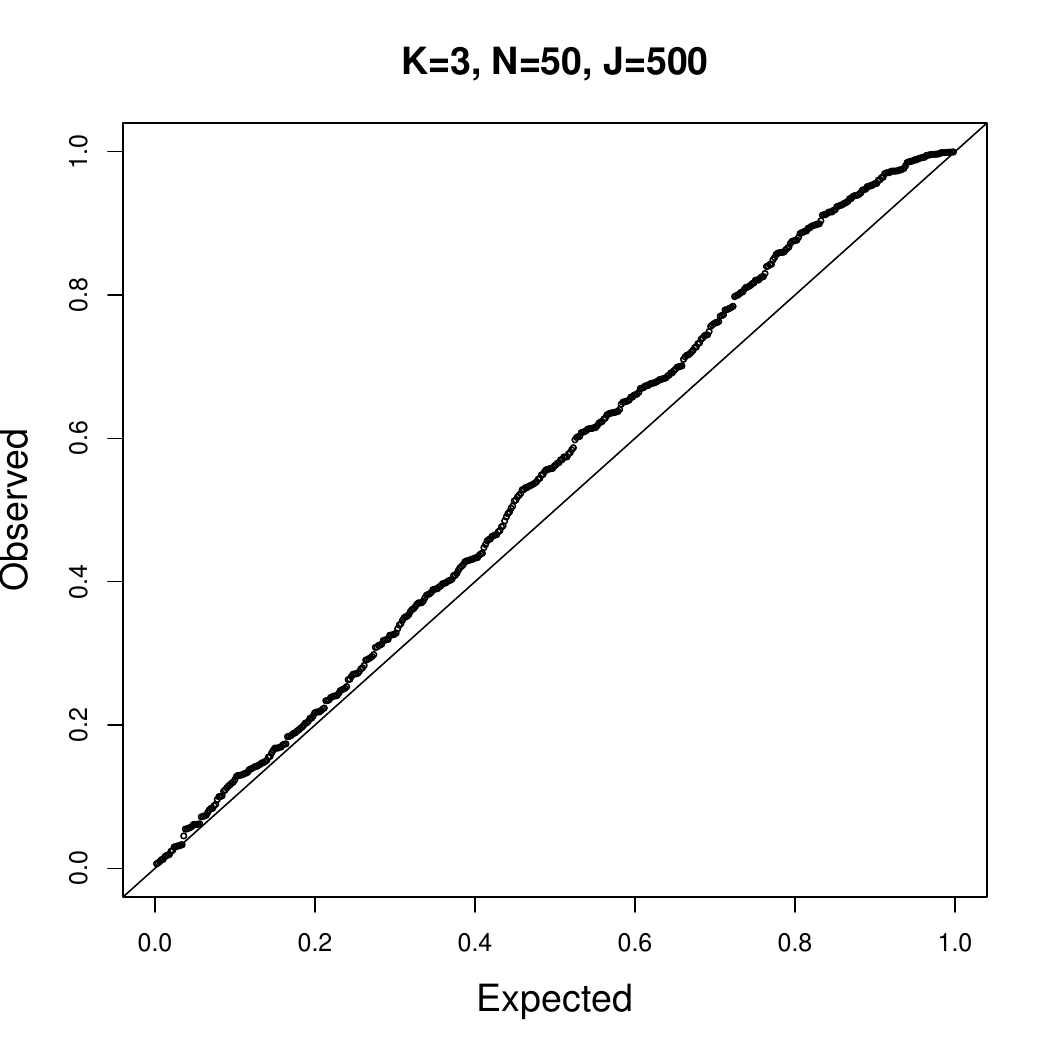}
     \end{subfigure}
    \hfill
     \begin{subfigure}[b]{0.24\textwidth}
         \centering
         \includegraphics[width=0.9\textwidth]{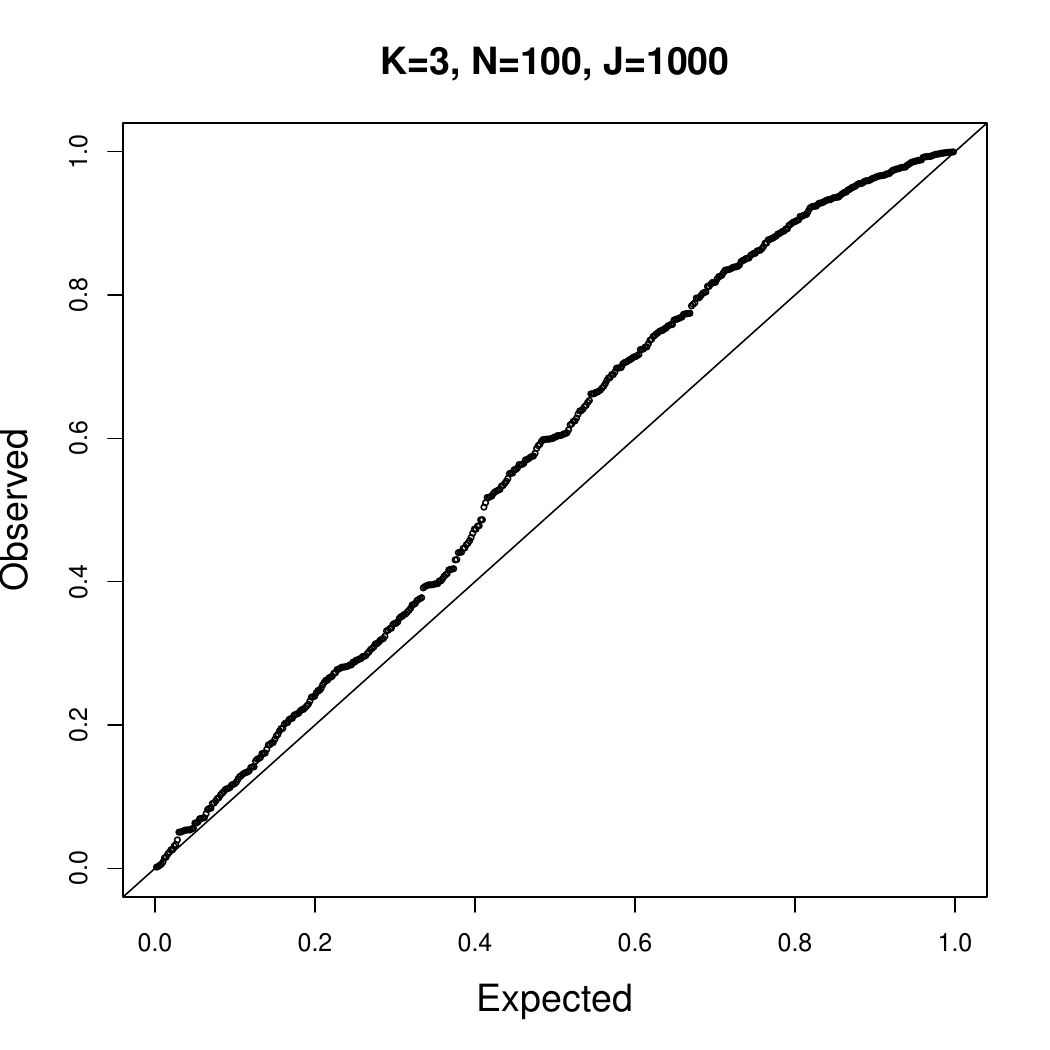}
     \end{subfigure}
     \hfill
     \begin{subfigure}[b]{0.24\textwidth}
         \centering
         \includegraphics[width=0.9\textwidth]{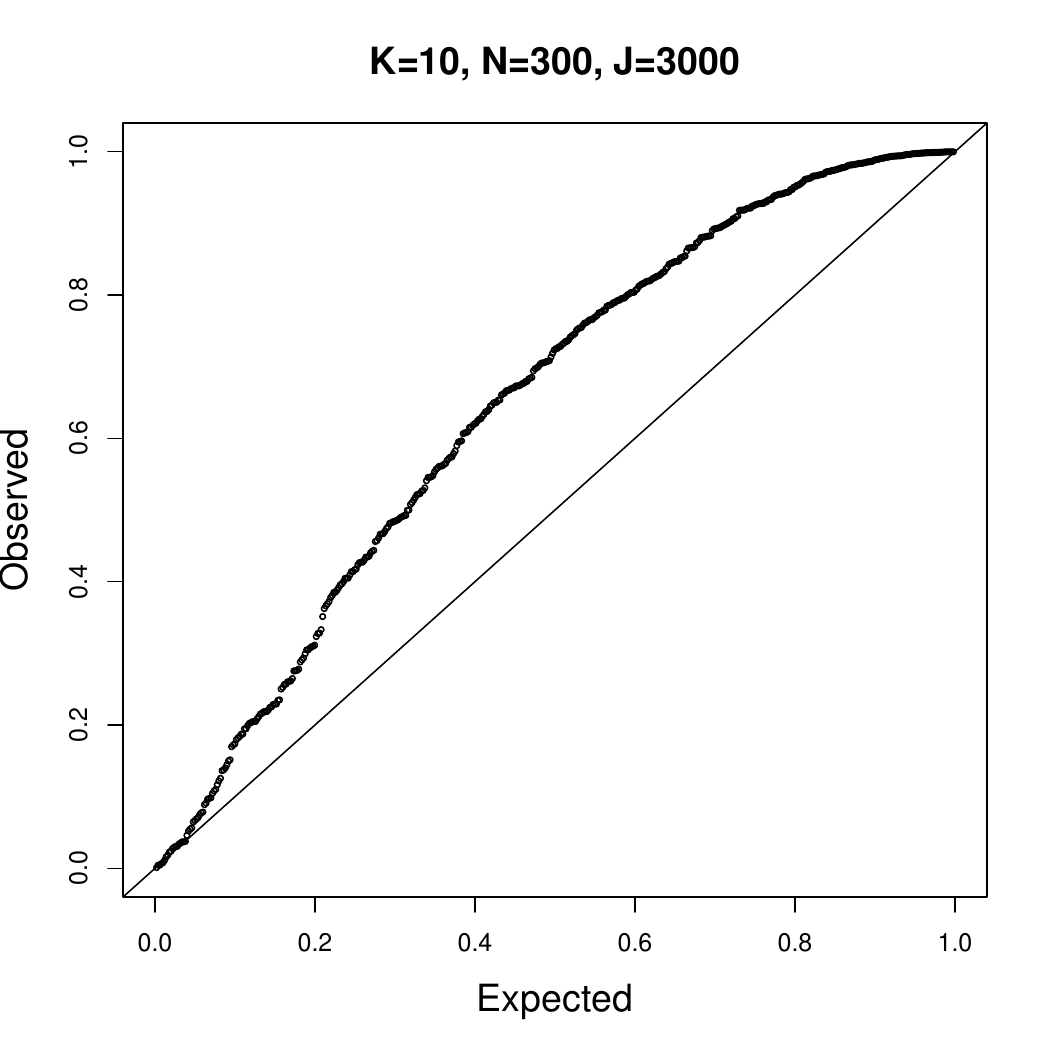}
     \end{subfigure}
     \hfill
     \begin{subfigure}[b]{0.24\textwidth}
         \centering
         \includegraphics[width=0.9\textwidth]{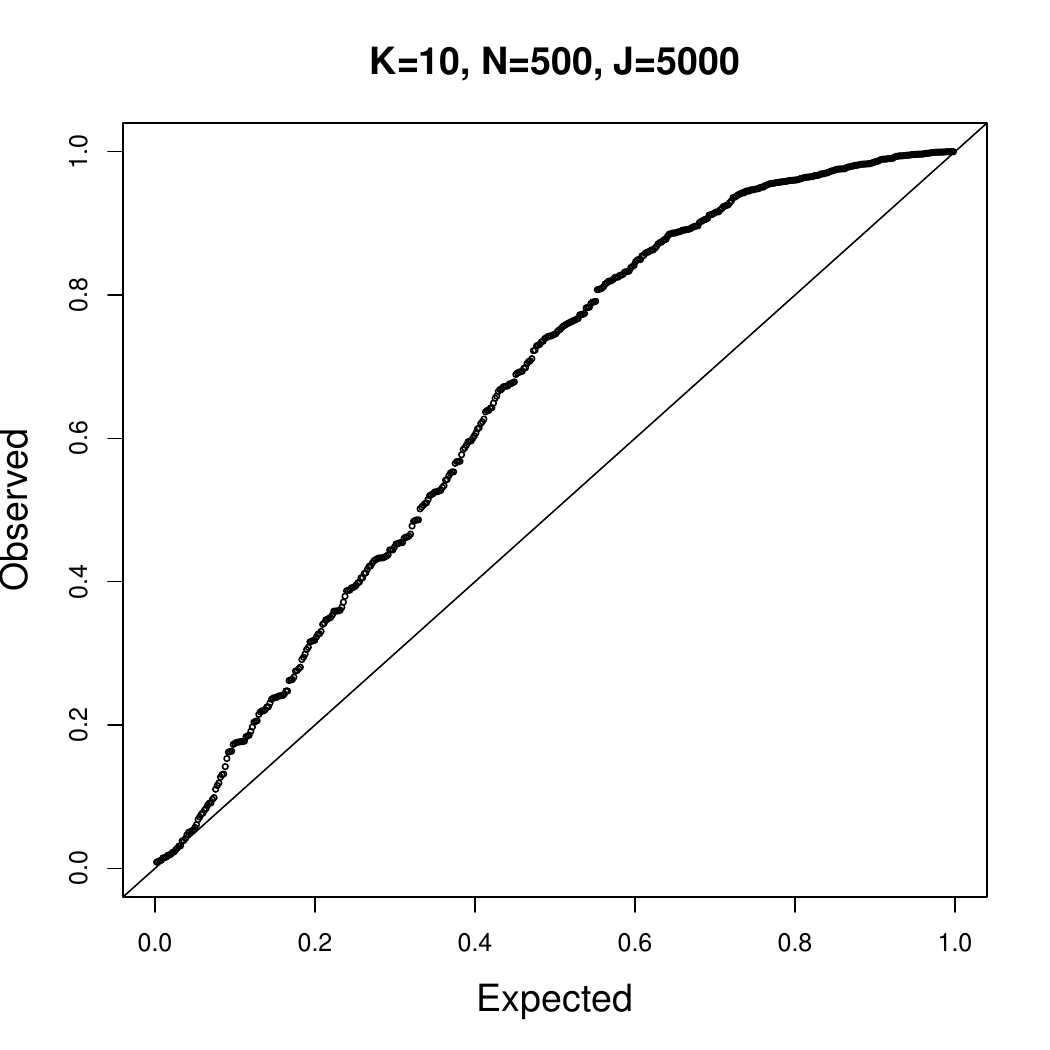}
     \end{subfigure}
     \begin{subfigure}[b]{0.24\textwidth}
         \centering
         \includegraphics[width=0.9\textwidth]{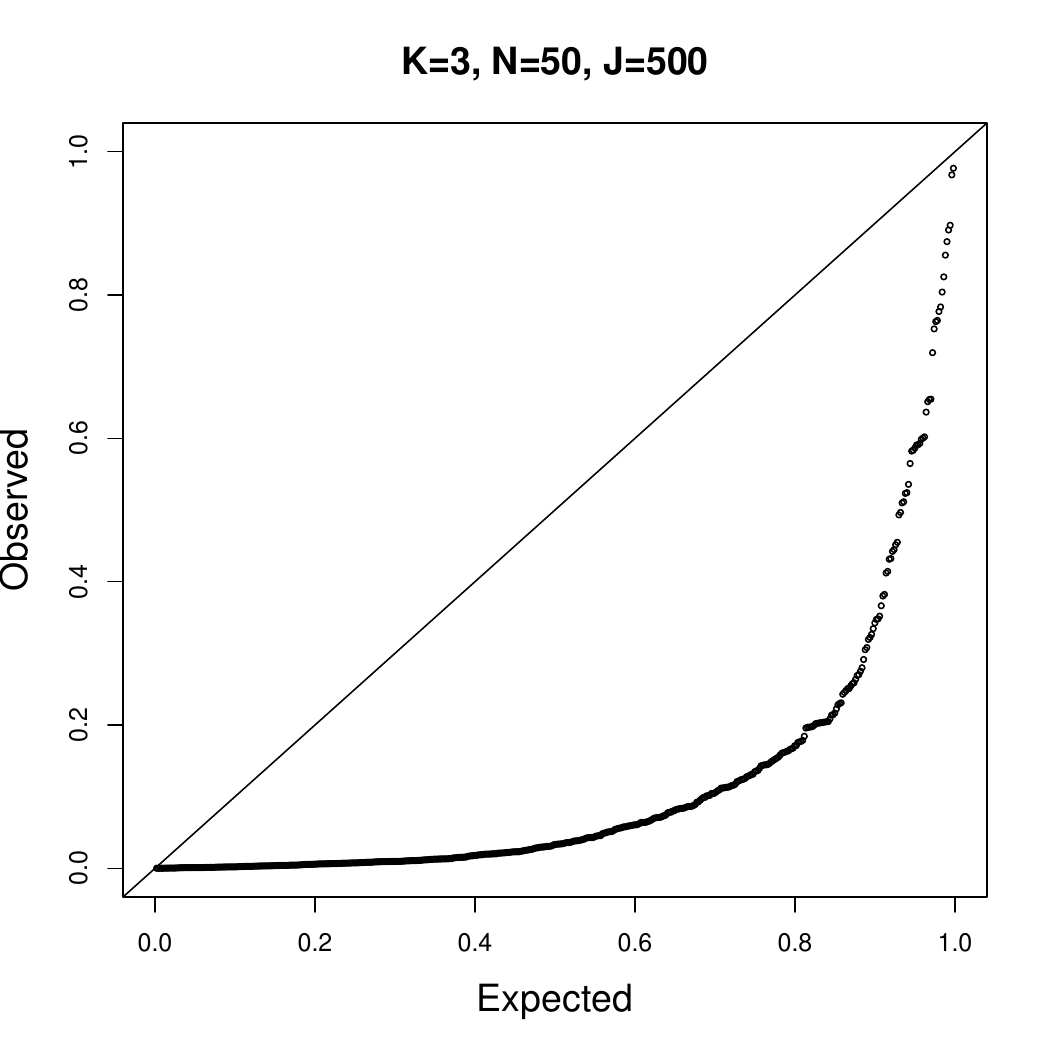}
     \end{subfigure}
    \hfill
     \begin{subfigure}[b]{0.24\textwidth}
         \centering
         \includegraphics[width=0.9\textwidth]{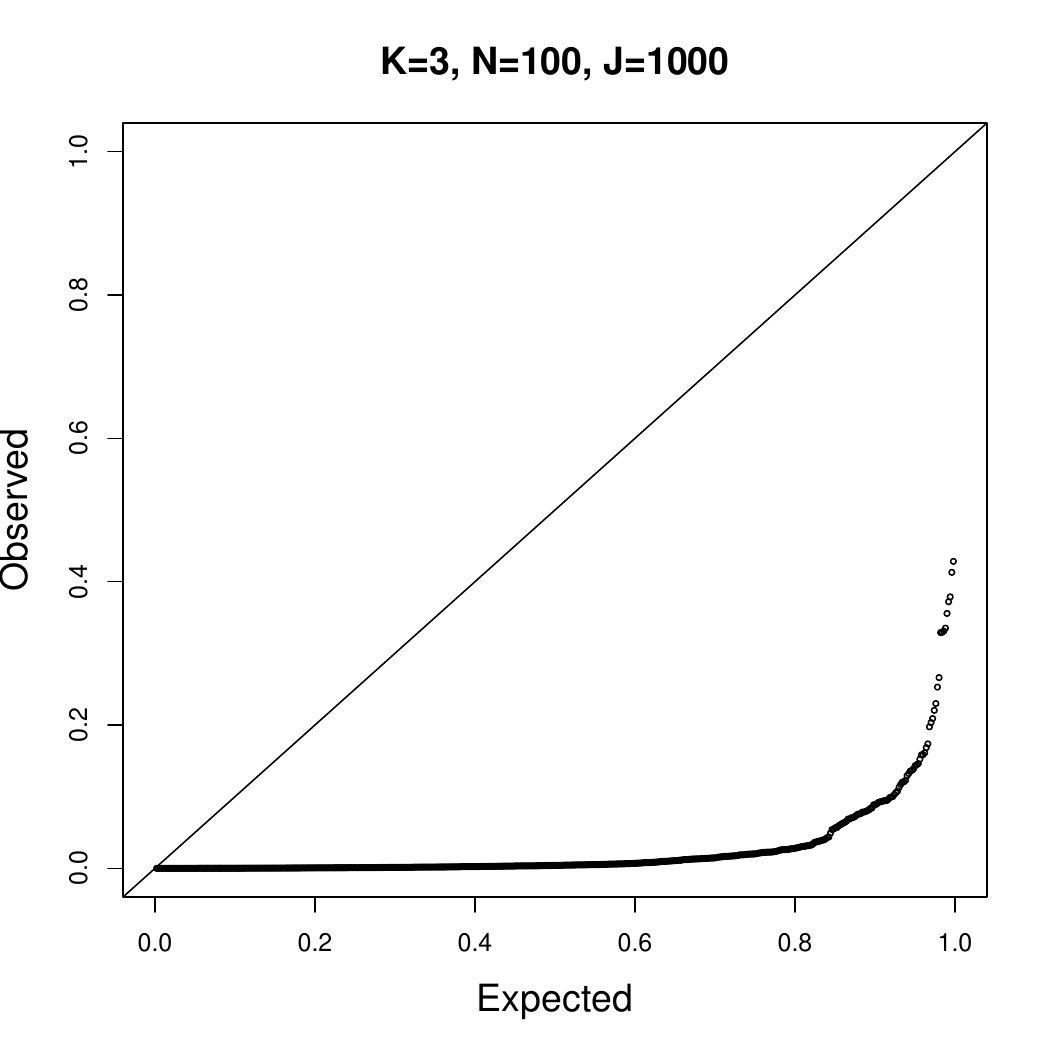}
     \end{subfigure}
     \hfill
     \begin{subfigure}[b]{0.24\textwidth}
         \centering
         \includegraphics[width=0.9\textwidth]{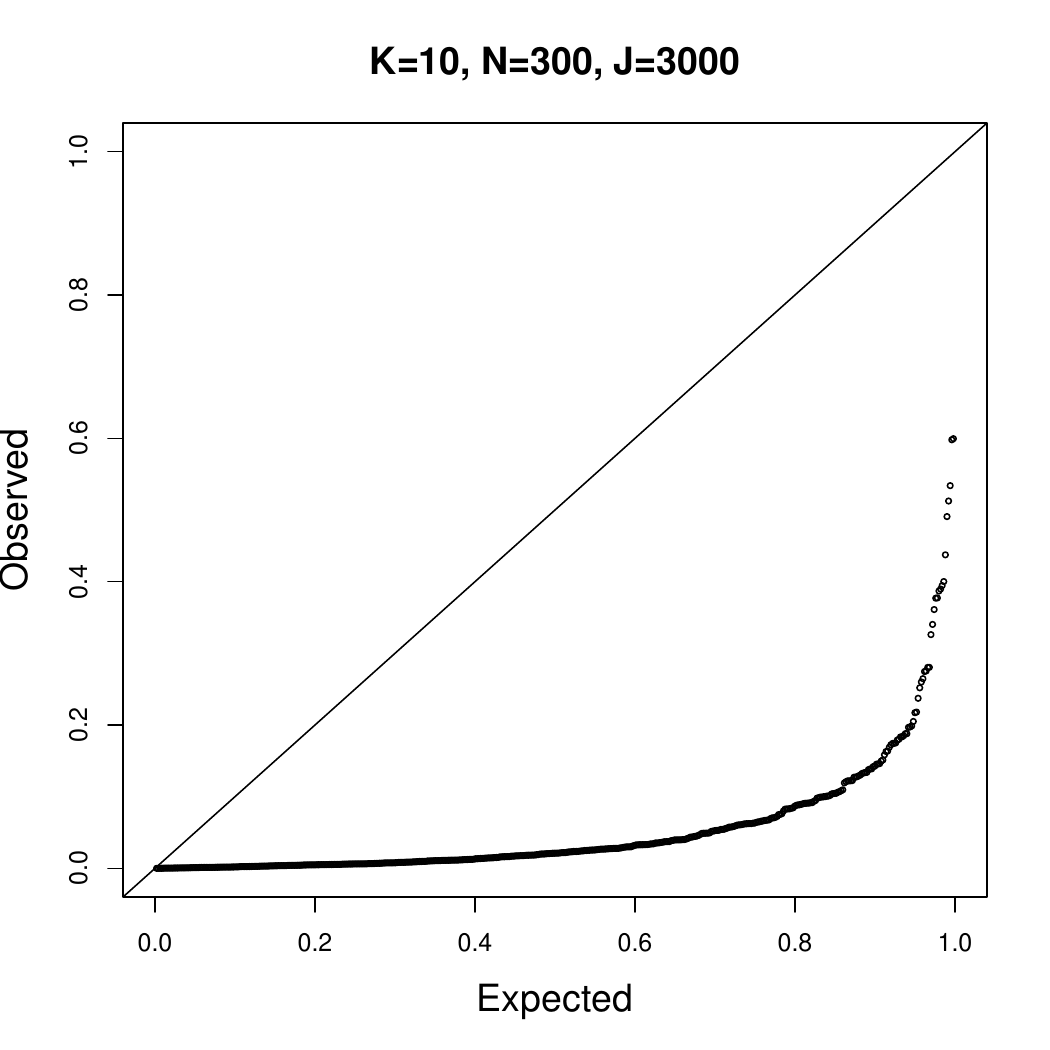}
     \end{subfigure}
     \hfill
     \begin{subfigure}[b]{0.24\textwidth}
         \centering
         \includegraphics[width=0.9\textwidth]{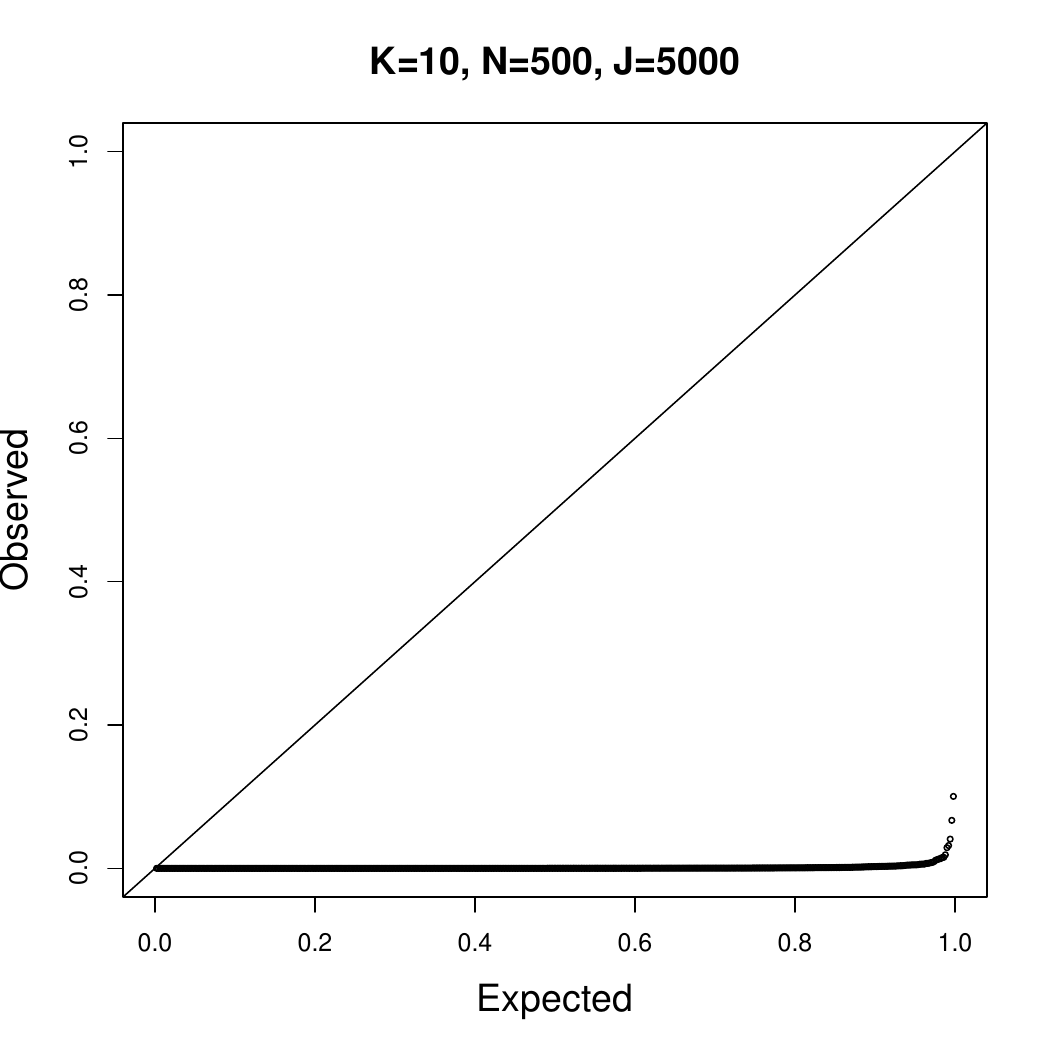}
     \end{subfigure}
\caption{Q-Q plots of the $p$-values for testing$H_0:\theta_{j,1}=\theta_{j,2}=\cdots \theta_{j,K}$ under the Poisson model. The null hypothesis is true for feature $1$ (upper row) and false for feature $2$ (lower row).}
\label{fig:poisson}
\end{figure}

Figure \ref{fig:poisson} demonstrates the Q-Q pots of the $p$-values for testing the null hypothesis $H_0:\theta_{j,1}=\theta_{j,2}=\cdots \theta_{j,K}$ under the Poisson model. The upper row corresponds to feature $1$, for which the null hypothesis is correct; the lower row corresponds to feature $2$, for which the null hypothesis is incorrect. We use the same simulation settings stated in Section \ref{sec:sim}, except for that the Poisson model is used now. This figure parallels with Figure \ref{fig:bernoulli} that is under the Bernoulli model.

{\color{black}
\subsection{Comparisons of SVD and HeteroPCA}

We have conducted additional simulation studies to compare HeteroPCA and SVD 
in terms of (a) \emph{recovering the top singular subspace}, (b) \emph{clustering}, and (c) \emph{estmating the continuous item parameters} in our proposed DhLCMs.
We set $J=3000, N=500, K=2, \bTheta_{:, 1}=(0.3\cdot \mathbf{1}_{J/2}^\top, 0.5\cdot 
\mathbf{1}_{J/2}^\top)^\top, \bTheta_{:, 2}=(0.1\cdot \mathbf{1}_{J/2}^\top, 0.06\cdot 
\mathbf{1}_{J/2}^\top)^\top$ for both the Bernoulli and the Poisson models, and generate $100$ independent replications. Figure \ref{fig:svd_hetero_compare_supp} (i.e. Figure \ref{fig:svd_hetero_compare} in the main paper) displays the boxplots of the Frobenius norm error of $\hat\bU$ (defined as $\|\bU\bU^\top - \hat\bU \hat\bU^\top\|_F$) on the left, clustering error in the middle, and the maximum absolute error of $\hat\bTheta$ on the right. The upper and lower rows correspond to the Bernoulli model and the Poisson model, respectively. Results show that HeteroPCA, on average, yields higher accuracy in terms of singular space estimation, clustering, as well as item parameter estimation compared to regular SVD for both models. More specifically, the out-performance is the most obvious for the Poisson model due to its severe heteroskedasticity, which aligns with the theoretical insight.

\begin{figure}
    \centering
    \includegraphics[width=\linewidth]{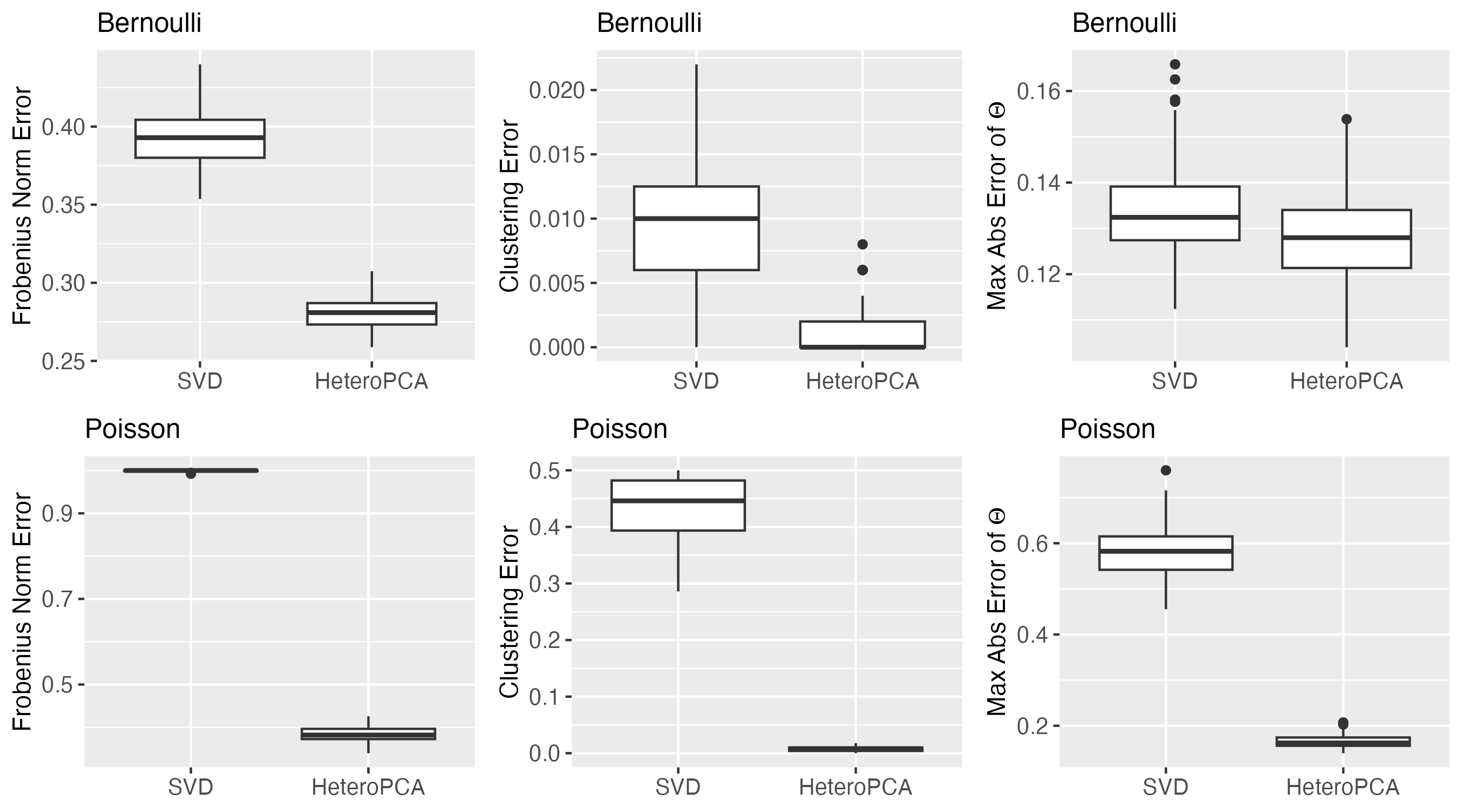}
    \caption{Boxplots comparing SVD and HeteroPCA in terms of the Frobenius norm error of $\hat \bU$ (left), clustering error (middle), and the maximum absolute error of $\hat\bTheta$ (right) for the Bernoulli model (upper row) and the Poisson model (lower row).}
    \label{fig:svd_hetero_compare_supp}
\end{figure}

Figure \ref{fig:inf_compare_supp} demonstrates the inference results of the two methods. The null hypothesis is true in the upper row and false in the lower row. The distribution of $p$-values yielded by HeteroPCA are much closer to the 45-degree reference line compared to SVD, when the null hypothesis is true. This suggests that the test with HeteroPCA yields better Type-I error control. In the second row, the distributions of the p-values from HeteroPCA indicate higher test power compared to SVD for the Poisson model.
More specifically, our test fails for the Poisson model with SVD. This is because the heteroskedasticity is quite severe in that case and SVD fails to accurately estimates the singular space, sequentially leading to poor clustering and inference results.

\begin{figure}[h!]
    \centering
         \begin{subfigure}[b]{0.24\textwidth}
             \centering
             \includegraphics[width=\textwidth]{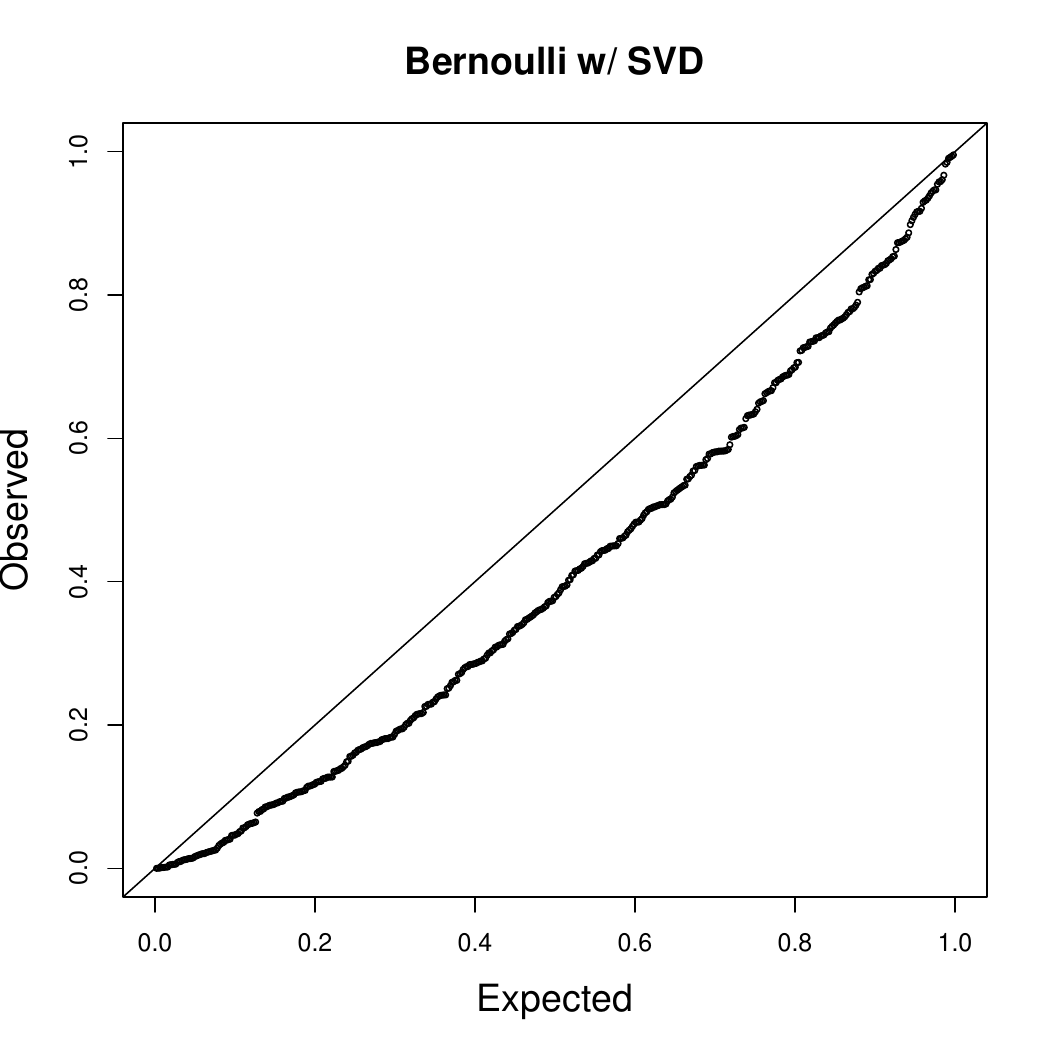}
         \end{subfigure}
        \hfill
         \begin{subfigure}[b]{0.24\textwidth}
             \centering
             \includegraphics[width=\textwidth]{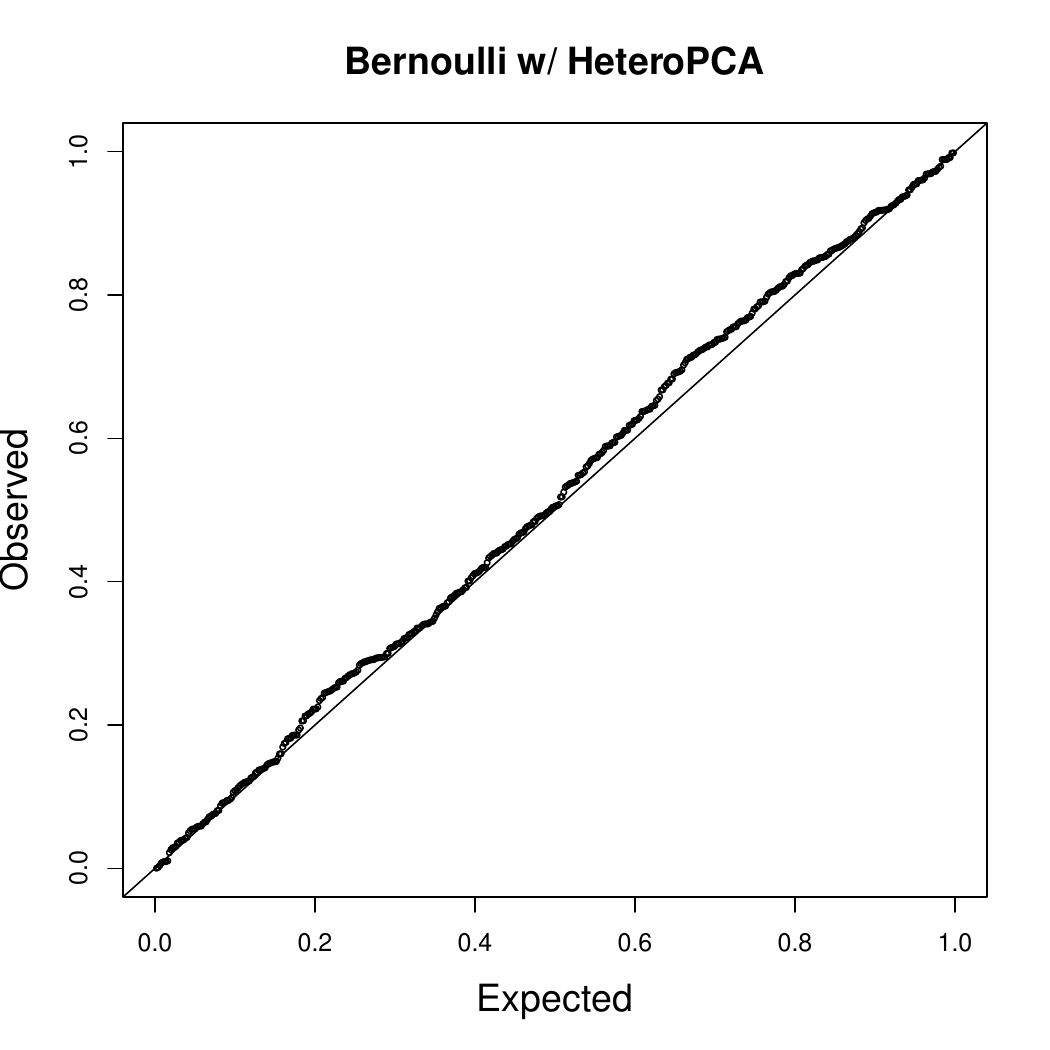}
         \end{subfigure}
         \hfill
         \begin{subfigure}[b]{0.24\textwidth}
             \centering
             \includegraphics[width=\textwidth]{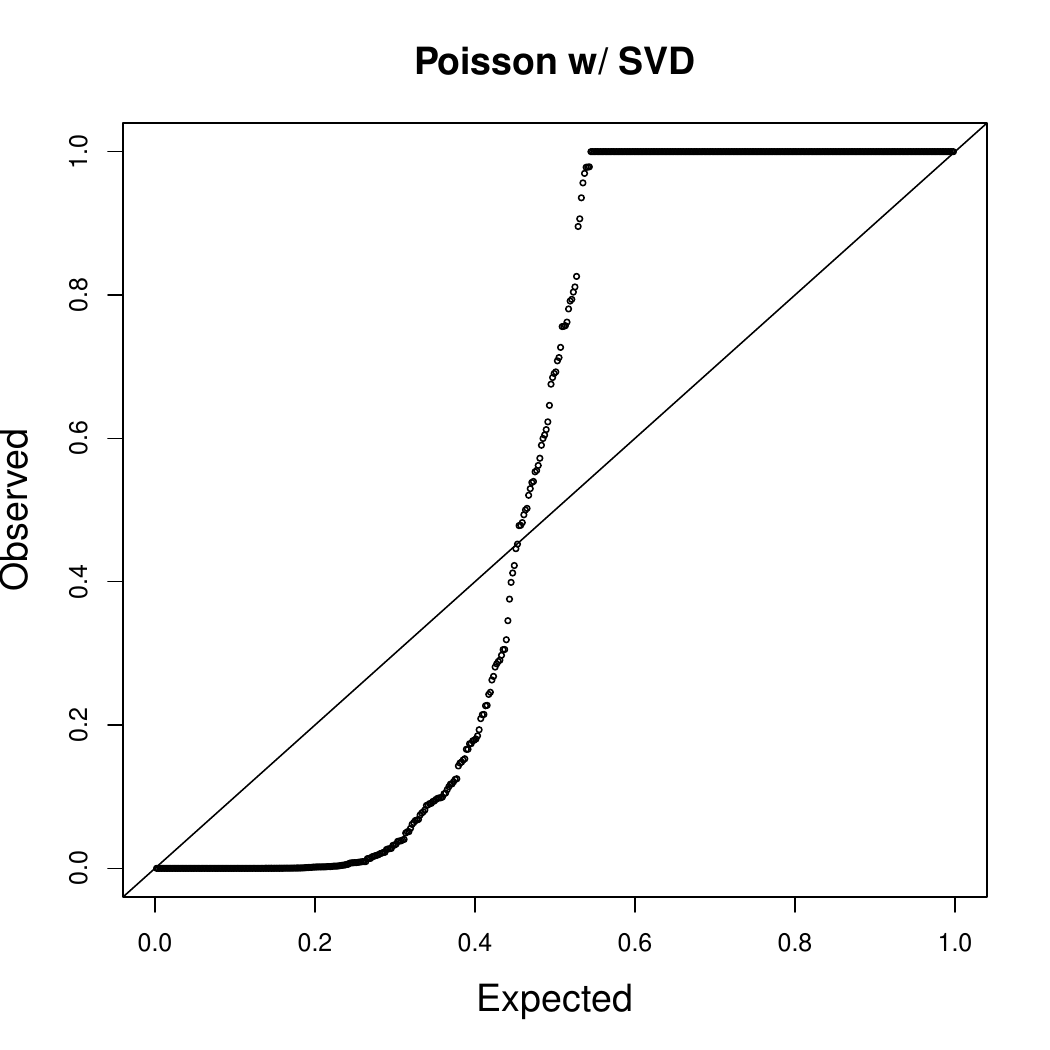}
         \end{subfigure}
         \begin{subfigure}[b]{0.24\textwidth}
             \centering
             \includegraphics[width=\textwidth]{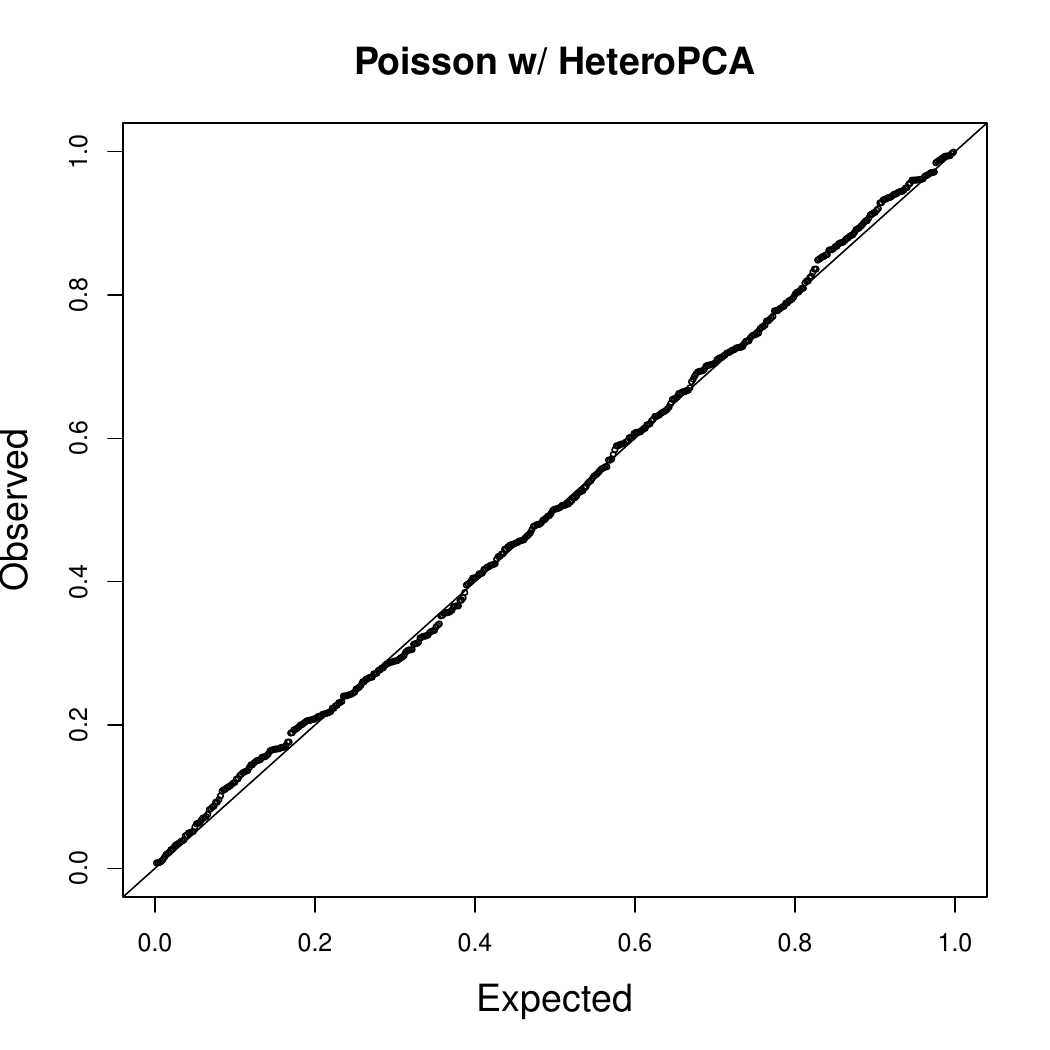}
         \end{subfigure}
        \hfill
         \begin{subfigure}[b]{0.24\textwidth}
             \centering
             \includegraphics[width=\textwidth]{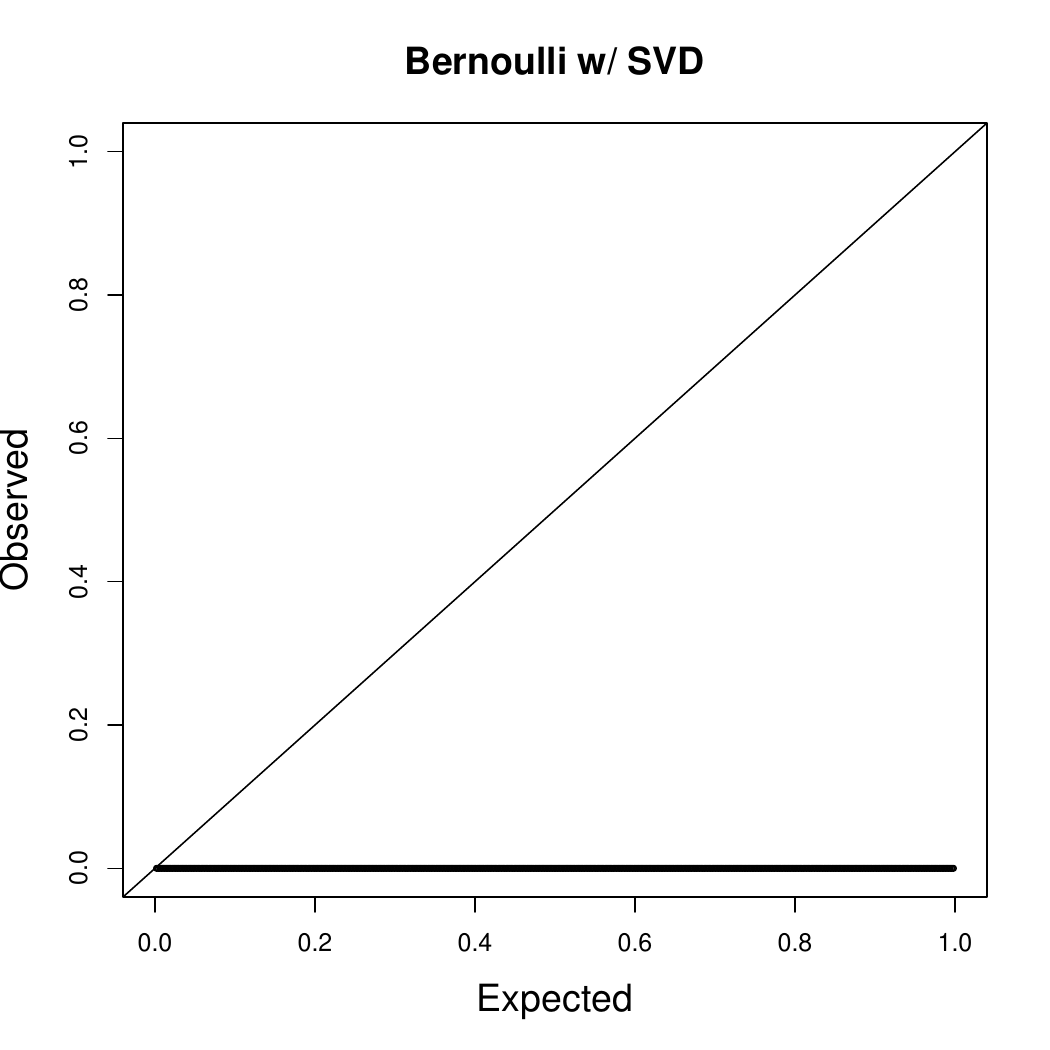}
         \end{subfigure}
         \hfill
         \begin{subfigure}[b]{0.24\textwidth}
             \centering
             \includegraphics[width=\textwidth]{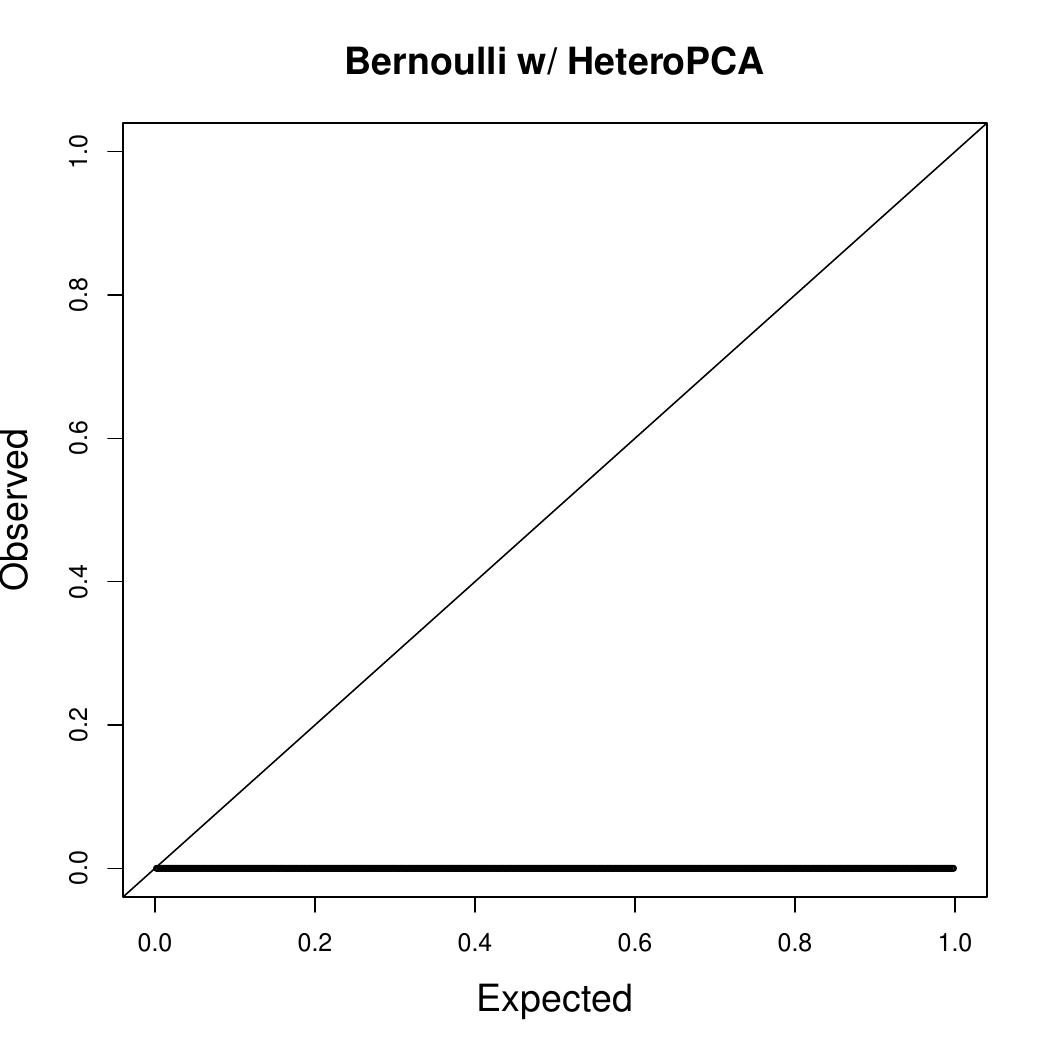}
         \end{subfigure}
         \hfill
         \begin{subfigure}[b]{0.24\textwidth}
             \centering
             \includegraphics[width=\textwidth]{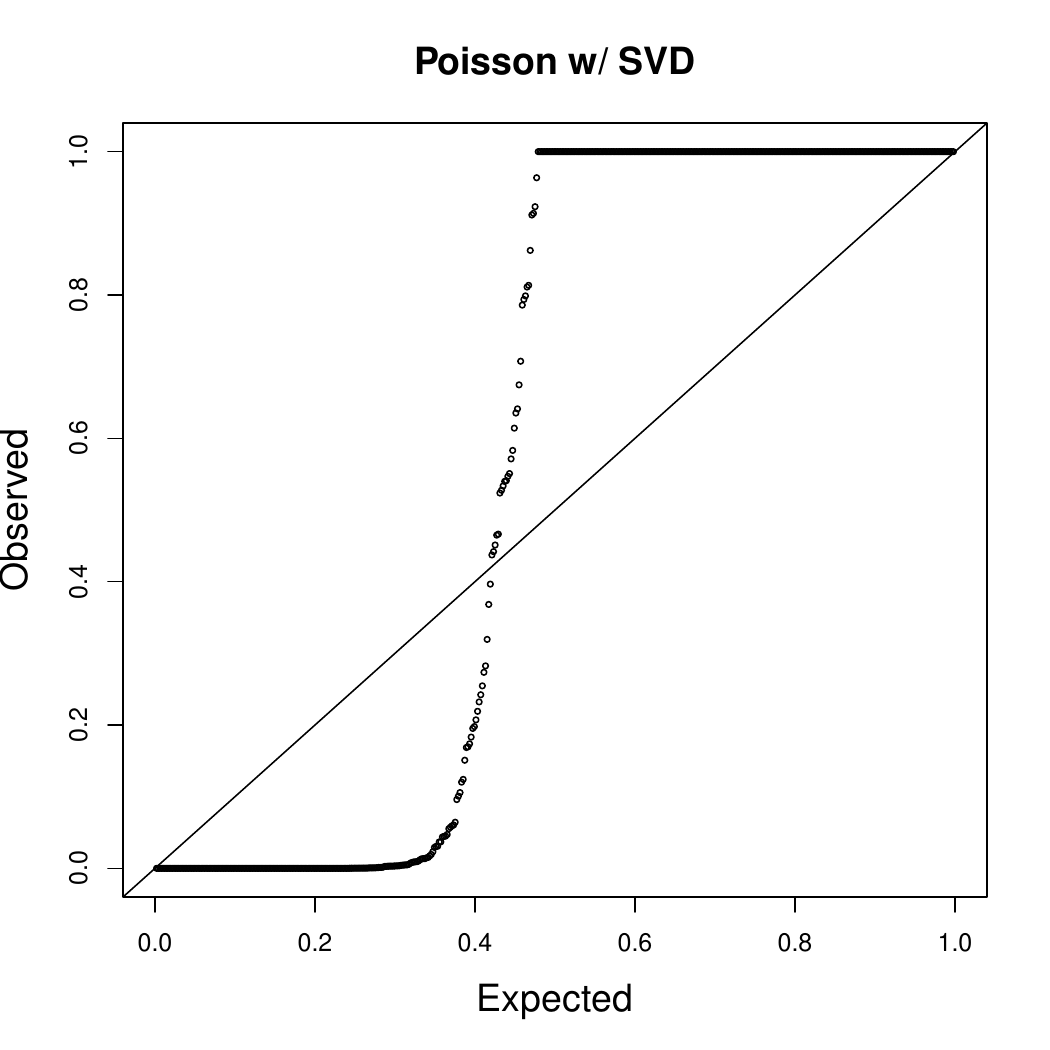}
         \end{subfigure}
         \hfill
         \begin{subfigure}[b]{0.24\textwidth}
             \centering
             \includegraphics[width=\textwidth]{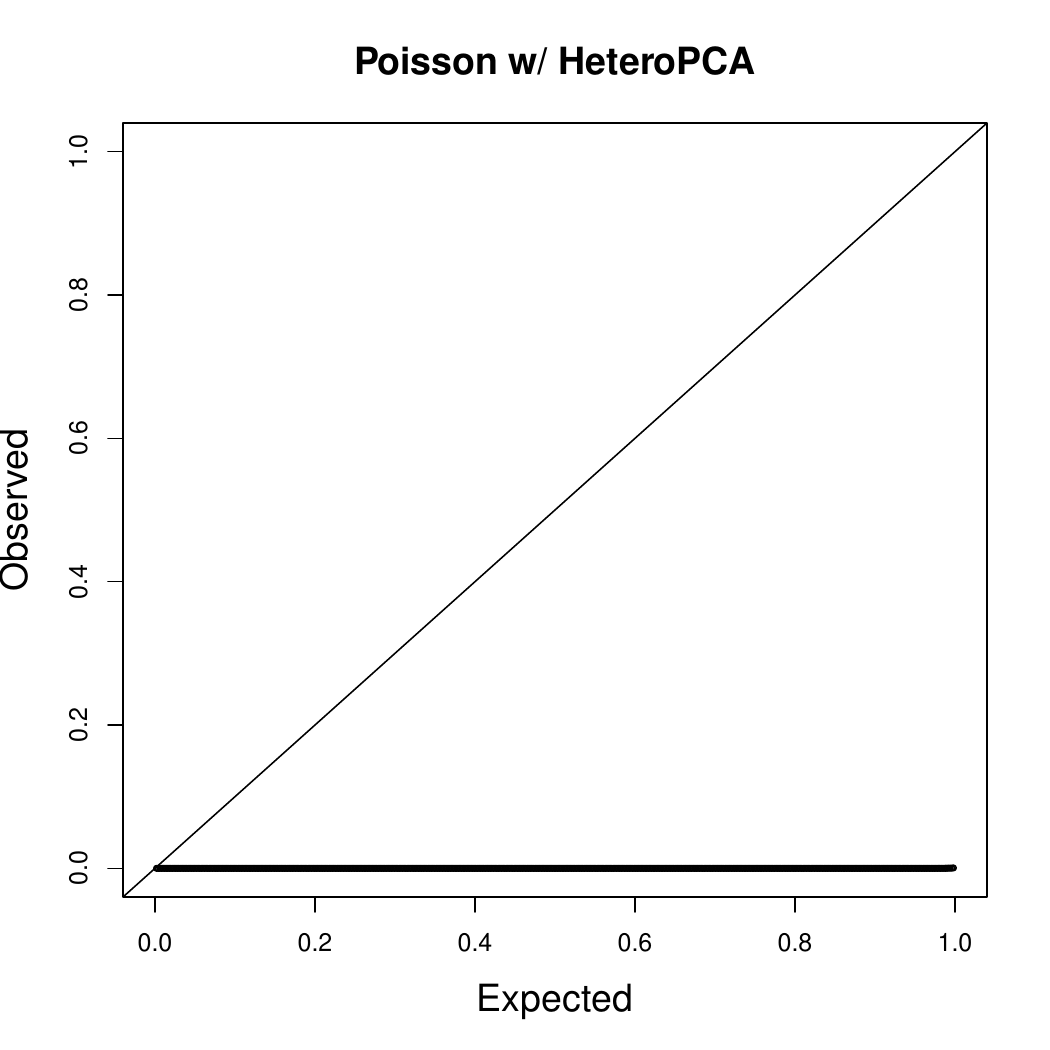}
         \end{subfigure}
    \caption{Q-Q plots of $p$-values for testing the null hypothesis $H_0:\theta_{j,1}=\theta_{j,2}=\cdots \theta_{j,K}$ in the Bernoulli and the Poisson model. Regular SVD or HeteroPCA is used to estimate the singular subspace $\bU$. $H_0$ is true for feature $1$ (upper row) and false for feature $2$ (lower row).}
    \label{fig:inf_compare_supp}
\end{figure}

\subsection{Relaxation on Assumption 3}
One way to relax Assumption 3 is to adjust the constant in front of $\ab{\calC_k}$. Specifically, we could assume  $\sum_{i\in\calC_k}\omega^2_i=\alpha_k\ab{\calC_k}$ for $k\in[K]$, where $\ebrac{\alpha_k}_{k\in[K]}$ are some pre-specified known constants. Under this modified assumption, our theoretical results still hold by incorporating the appropriate normalization constant when constructing $\hat\bTheta$.  In particular, we  adjust our estimator for $\bOmega$ as $\hat\bOmega=\textsf{diag}\brac{\ab{\alpha_{\hat s_1}\calC_{\hat s_1}}^{1/2}
\op{\hat\bU_{1,:}},\cdots,\ab{\alpha_{\hat s_N}\calC_{\hat s_N}}^{1/2}\op{\hat\bU_{N,:}}}$. In practice, $\ebrac{\alpha_k}_{k\in[K]}$ should be determined by some prior information on the degree assumptions of clusters based on different application contexts. 

If $\ebrac{\alpha_k}_{k\in[K]}$ are unknown, it is generally difficult to derive some meaningful results, except for estimation of $\bTheta$ (Theorem 4). In this case,  we can instead assume $\ab{\sum_{i\in\calC_k}\omega^2_i-\ab{\calC_k}}/\ab{\calC_k}\le \gamma$ for $k\in[K]$ with $$\gamma=o\brac{\mu_{\bTheta}^{1/2}K^{1/2}\brac{\frac{K\log\brac{N+J}}{\Delta^2/\theta_{\sf max}}\sqrt{\frac{J}{N}}+\sqrt{\frac{K\log\brac{N+J}}{\Delta^2/\theta_{\sf max}}}}}.$$
    To illustrate this point, we have added some additional numerical experiments as follows. We set $K=3, N=J/5$ with $J$ varying in $\{200, 500, 1000, 2000\}$, and fix $\alpha_1=1, \alpha_2=0.95, \alpha_3=1.05$. Other simulation settings follow those in Section 5 in the manuscript and $100$ independent replications are generated. We compare the item parameter estimation results to those of the case without perturbation ($\alpha_k=1$), using the same estimation procedure. Figure \ref{fig:assumption3_supp} demonstrates the comparisons of the two cases with respect to the mean absolute error of $\hat\bTheta$ for the Bernoulli model. We can see that the item parameter estimation with some violation of Assumption 3 is comparable to when the assumption holds. Similar results are also observed for the Poisson model.
    
    \begin{figure}[h!]
        \centering
        \includegraphics[width=0.65\linewidth]{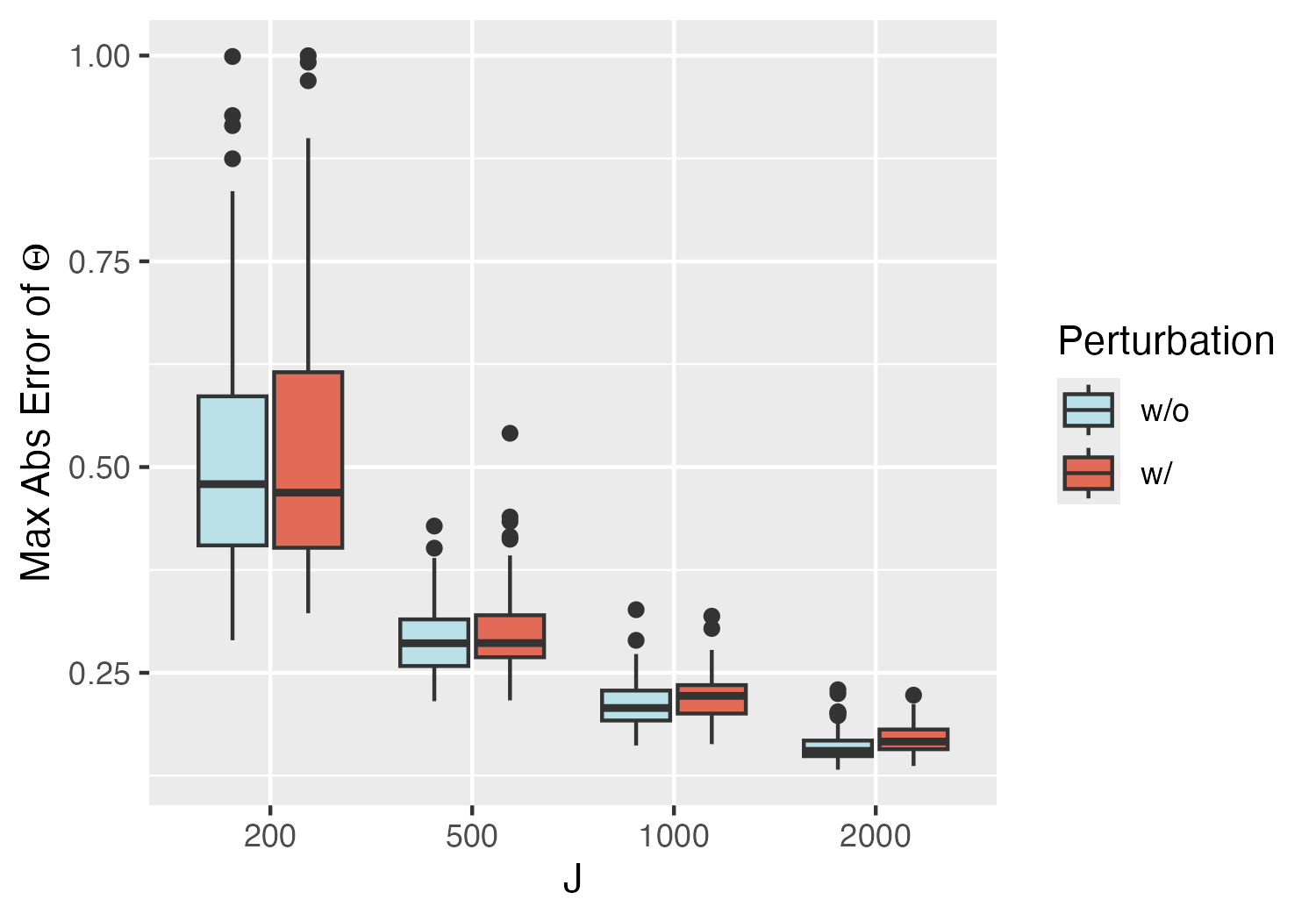}
        \caption{Maximum absolute error of $\hat\bTheta$ comparing the case when $\alpha_1=1, \alpha_2=0.95, \alpha_3=1.05$ (w/ perturbation) and the case when $\alpha_k\equiv 1$ (w/o perturbation) for the Bernoulli model. We set $K=3, N=J/5$ with $100$ independent replications.}
        \label{fig:assumption3_supp}
    \end{figure}

\subsection{Method Robustness to $\kappa$}
Empirically, we have illustrated the robustness of our method to variations in $\kappa$ through additional numerical experiments. These experiments evaluate the performance of our method under $J=1000, N=200, K=3$. We right multiply $\bTheta$ by $\diag(1, 1, c)$ and tune the constant $c\ (c<1)$ so that the $\kappa$ is set as desired. The type-I error and power for $H_0 : \theta_{j,1}=\theta_{j,2}=\dots \theta_{j,K}$ are calculated with $500$ replications. Table \ref{tab:cn_supp} summarizes the results for the two models and various values of $\kappa$. As $\kappa$ increases up to $4$, the Type-I error is well-controlled and the power is close to $1$. This empirically demonstrate the robustness of our method to $\kappa$.

\begin{table}[h!]
\footnotesize
\centering
\begin{tabular}{ccccccccc}
\toprule
\multirow{2}{*}{Model} & \multirow{2}{*}{Metric} & \multicolumn{7}{c}{$\kappa$} \\
\cmidrule(lr){3-9} & & 1.5 & 2 & 2.5 & 3 & 3.5 & 4 \\
\midrule
\multirow{2}{*}{Bernoulli} & Type-I error & 0.05 & 0.056 & 0.034 & 0.056 & 0.024 & 0.026 & \\
\cmidrule(lr){2-9}
 & Power & 1 & 1 & 1 & 1 & 1 & 1 & \\
\midrule 
\multirow{2}{*}{Poisson} & Type-I error & 0.04 & 0.028 & 0.04 & 0.054 & 0.036 & 0.048 \\
\cmidrule(lr){2-9}
 & Power & 0.99 & 0.994 & 0.996 & 0.994 & 0.992 & 0.994 \\
\bottomrule
\end{tabular}
\caption{Type-I error and power for testing $H_0 : \theta_{j,1}=\theta_{j,2}=\dots \theta_{j,K}$ with the Bernoulli and Poisson models. We consider various values of $\kappa$, and set $J=1000, N=200, K=3$. Type-I error and power are calculated with $500$ replications.}
\label{tab:cn_supp}
\end{table}
}
\subsection{Real Data Figures}
\begin{figure}[h!]
    \centering
    \includegraphics[width=0.9\textwidth]{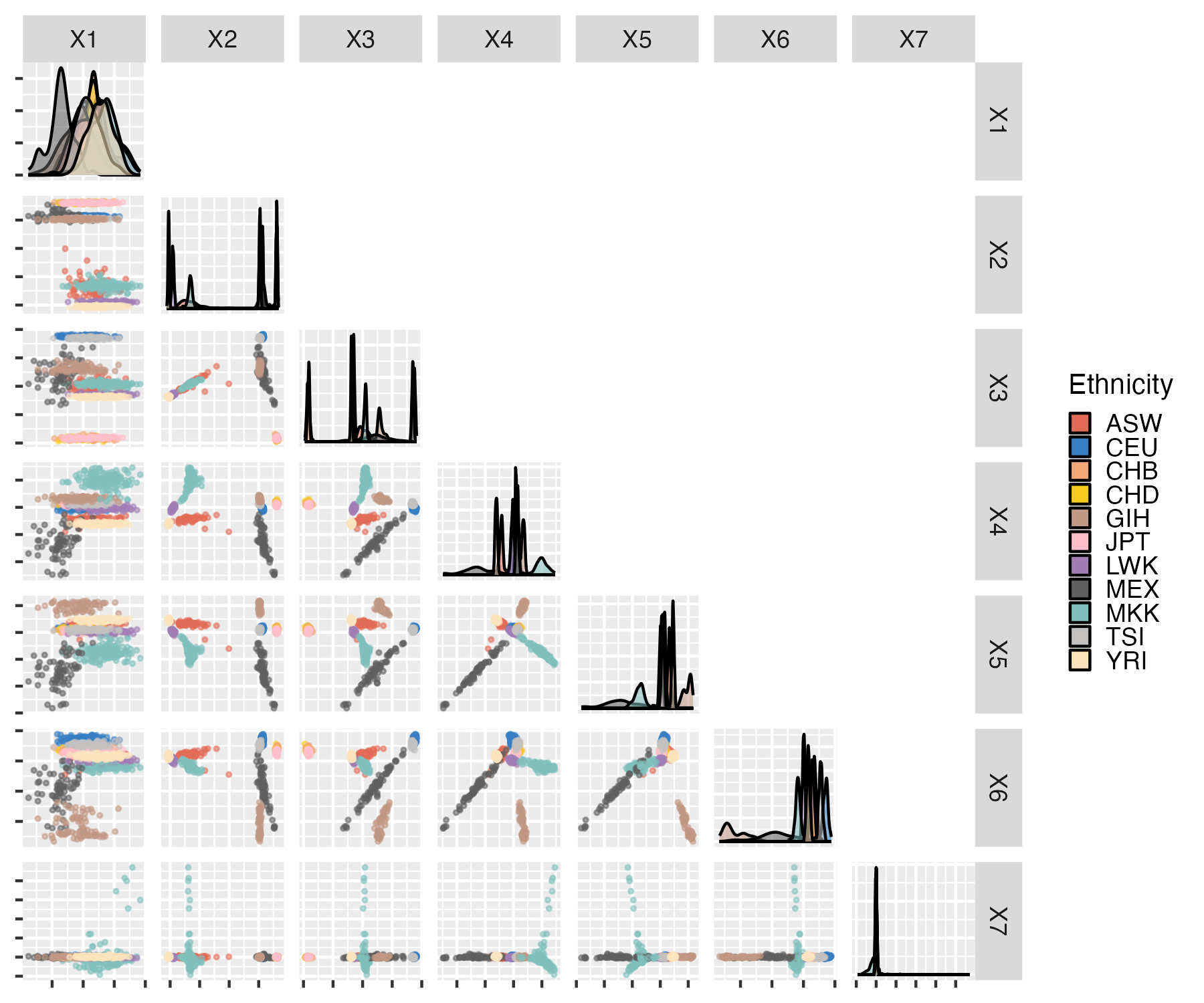}
    \caption{Pairplot of the first $7$ singular vectors given by HeteroPCA for the HapMap3 data. We observe streak structures in the singular embeddings. For the label information, refer to \url{https://www.broadinstitute.org/medical-and-population-genetics/hapmap-3}.}
    \label{fig:hapmap}
\end{figure}

\begin{figure}[h!]
    \centering
    \includegraphics[width=0.9\textwidth]{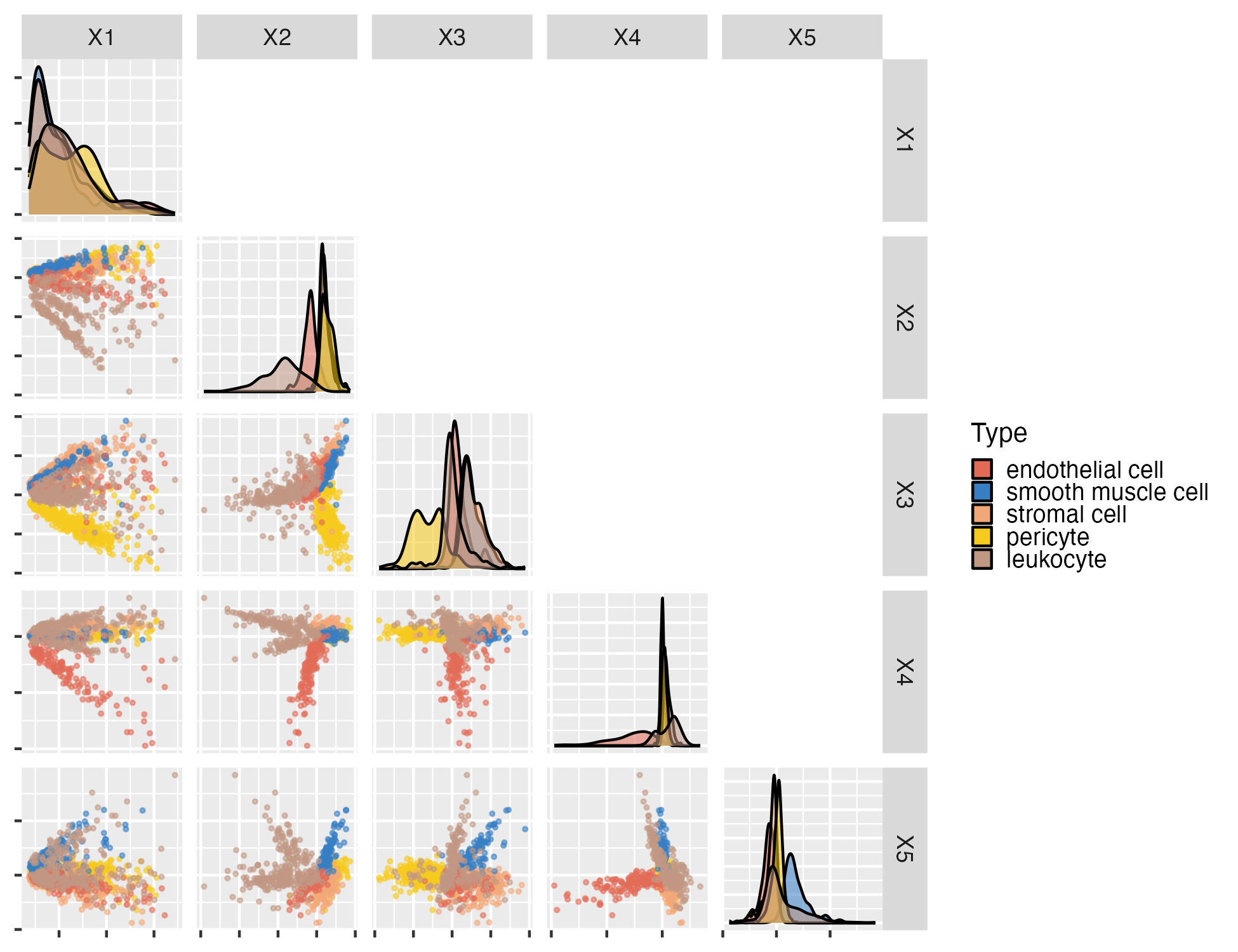}
    \caption{Pairplot of the first $5$ singular vectors given by HeteroPCA for the single cell data. We observe streak structures in the singular embeddings.}
    \label{fig:single}
\end{figure}

Figure \ref{fig:hapmap} demonstrates the pairplot of the first five singular vectors given by HeteroPCA for the HapMap3 data. We observe radial streak structures in the singular subspace embeddings. Figure \ref{fig:single} demonstrates the pairplot of the first $5$ singular vectors given by HeteroPCA for the single cell data \citep{lengyel2022molecular}. We again observe radial streak structures in the singular subspace embeddings.

\vspace*{\fill}

\begin{figure}[h!]
    \centering
    \includegraphics[width=\textwidth]{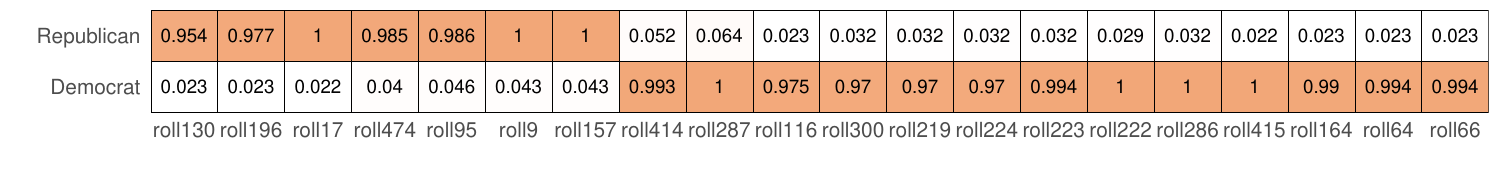}
    \caption{Heatmap of $\hat{\bTheta}$ of a subset of 20 rolls with the smallest adjusted $p$-values for testing $H_0:\theta_{j,1}=\theta_{j,2}=\cdots=\theta_{j,K}$. }
    \label{fig:heatmap}
\end{figure}

Figure \ref{fig:heatmap} illustrates the heatmap of the $20$ rolls that correspond to the smallest BH-adjusted $p$-values, which are all extremely small and serve as markers to best distinguish the two U.S. political parties. 
As we can see in the figure, the two parties can be easily distinguished based on their respective response probabilities towards these $20$ roll calls.

\section{Additional Notations and General Assumptions}\label{sec:notations_assumptions}
For completeness, all results in our proofs will be stated without the constant degree assumption  and  the high-dimensional regime assumption (i.e., $J\gtrsim N$). Instead, we replace Assumption \ref{cond:const-degee} in the main paper with the following condition, which can be viewed as a more general version of Assumption \ref{cond:const-degee}.

\begin{assumption}[Balanced degrees across clusters]\label{cond:general-degree}
\begin{align*}
\max_{k\in[K]}\sqrt{\sum_{i\in \calC_k}\omega_i^2}\lesssim \min_{k\in[K]}\sqrt{\sum_{i\in \calC_k}\omega_i^2}.
\end{align*}
\end{assumption}
We start with defining some additional incoherence parameters in terms of $\bR^* := \mathbb E[\bR] = \bU \bSigma \bV^\top$. Define $\mu_0,\mu_1,\mu_2>0$ to be
\begin{align*}
	\mu_1:=\frac{N\op{\bU}_{2,\infty}^2}{K}, \quad \mu_2:=\frac{J\op{\bV}_{2,\infty}^2}{K},\quad \text{and~}\mu_0:=\frac{NJ\op{\bR^*}_{\max}^2}{\fro{\bR^*}^2}
\end{align*}
\sloppy For convenience, we denote  $\omega_{\sf max}=\max_{i\in[N]}\omega_i$, $\omega_{\sf min}=\max_{i\in[N]}\omega_i$, $\omega_*=\sqrt{N^{-1}\sum_{i=1}^N\omega_i^2}$, $\omega_{**}:=\max_{k\in [K]}\brac{\frac{1}{|\calC_k|}\sum_{i\in\calC_k}\omega_i^{-1}}^{-1}$ and $\mu_{\omega}:=\omega_{\sf max}/\omega_*$. Moreover, we define the following quantity:
\begin{align*}
	&\xi_{\sf err}:=\omega_{\sf max}\theta_{\sf max}\sqrt{NJ}\log(N+J)+\sigma_1(\bR^*)\sqrt{\omega_{\sf max}\theta_{\sf max}N\log(N+J)}.
\end{align*}
Lemma \ref{lem:incoherence-par} entails that under Assumption \ref{cond:general-degree},
\begin{align*}
	\sigma_K(\bR^*)\gtrsim \sqrt{\frac{N}{K}}\omega_*\sigma_\star,\quad \mu_1=O\brac{\mu_\omega^2},\quad \mu_2=O\brac{\mu_{\bTheta}\kappa^2},\quad \mu_0=O\brac{\mu_\omega^2\mu_{\bTheta}\kappa^4K},
\end{align*}
which will be used throughout our proofs. 

\section{Technical Lemmas}\label{sec:lemmas}
\begin{lemma}\label{lem:incoherence-par}
	Under Assumption \ref{cond:general-degree}, the incoherence parameters satisfy $\mu_1=O\left(1\right ),\mu_2=O(\mu_{\bTheta}\kappa^2), \mu_0=O\left(\kappa^4K\right )$.
\end{lemma}
\begin{lemma}\label{lem:two-inf-bound}
	Suppose Assumptions \ref{cond:balanced} and \ref{cond:general-degree} hold. Assume $N\gtrsim \mu_\omega^4\mu_{\bTheta}\kappa^8K^3\log^2(N+J)$, $J\gtrsim K\log^4(N+J)$, 
	\begin{align}\label{eq:signal-cond}
		\sigma^2_K(\bR^*)&\ge  C_{\sf gap}\kappa^2\xi_{\sf err},
	\end{align}
and that Algorithm \ref{alg:HeteroPCA} is run for $T_0\ge \log\left(\frac{\sigma^2_1\brac{\bR^*}}{\xi_{\sf err}}\right)$ iterations. Then with probability at least $1-(N+J)^{-20}$, we have that 
\begin{align}\label{eq:Uhat-decomp}
	\hat \bU\bO^\top   -\bU=\bE\bV\bSigma^{-1}+\calH(\bE\bE^\top)\bU\bSigma^{-2}+\mPsi,
\end{align}
where
\begin{align*}
	&\op{\mPsi}_{2,\infty}\lesssim \kappa^2\frac{\xi_{\sf err}}{\sigma_K^2(\bR^*)}\frac{\mu K}{N}+\kappa^2\frac{\xi^2_{\sf err}}{\sigma_K^4(\bR^*)}\sqrt{\frac{\mu K}{N}}\\
	&\op{\hat\bU-\bU\bO}\lesssim \frac{\xi_{\sf err}}{\sigma^2_K(\bR^*)},\quad \op{\hat\bU-\bU\bO}_{2,\infty}\lesssim \kappa^2\frac{\xi_{\sf err}}{\sigma^2_K(\bR^*)}\sqrt{\frac{\mu K}{N}},
\end{align*} 
where $\bO:=\arg\min_{\wt\bO\in\OO_K}\fro{\hat\bU-\bU\wt\bO}^2$.  Furthermore, if 
$\sigma^2_K(\bR^*)\ge  C_{\sf gap}\brac{\mu_{\bTheta}K}^{1/2}\kappa^4\xi_{\sf err}$, then $$\infn{\hat\bU-\bU\bO}\lesssim \infn{\bU}.$$
\end{lemma}

\begin{lemma}\label{lem:trunc}
	There exists some absolute constants $C_1,C_2$ such that for any centered Poisson random variable $X$ with parameter $\lambda>0$, consider any $\delta\in\brac{0,C_1\sqrt{\lambda}}$ if $\lambda\ge 1$ and $\delta\in\brac{0,C_1\lambda^2}$ if $0<\lambda<1$, we can construct a random variable $\wt X$ satisfying the following properties:
\begin{enumerate}
	\item $\Prob\brac{\wt X=X}\ge 1-\delta$.
	\item $\EE \wt X=0$.
	\item $\wt X$ is a bounded random variable such that 
$$\ab{\wt X}\le C_2\brac{\sqrt{\lambda\brac{\log \brac{\delta^{-1}}+\log \brac{\lambda^{-1}}}}+\log\brac{\delta^{-1}} +\log \brac{\lambda^{-1}}}.$$
	\item The variance of $\wt X$ satisfies $\textsf{Var}\brac{\wt X}=\brac{1+O\brac{\sqrt{\delta}}}\lambda$.
\end{enumerate}
\end{lemma}
The following lemma serves as the building block for the proof of Theorem \ref{thm:Thetaerr-gen-inf}.
 \begin{lemma}\label{lem:Thetaerr-inf}
	Suppose the conditions of Theorem \ref{thm:exp-clustering-error} for exact recovery  and Assumption \ref{cond:iden-theta} hold.  In addition, assume that $J\gtrsim \mu_\omega^4\mu^3_{\bTheta}\kappa^{18}K^4\brac{\frac{\omega_{\sf max}}{\omega_{\sf min}}}^2\brac{\frac{\theta_{\sf max}}{\theta^*_{\sf min}}}\log^2(N+J)$ and there exist some absolute constant $C_{\sf inf}>0$ such that 
\begin{align}\label{eq:theta-inf-cond-lem}
	\frac{\Delta^2}{\theta_{\sf max}}\ge \frac{C_{\sf inf}\mu_\omega^4\mu^2_{\bTheta}\kappa^{18}K^3}{\omega_{\sf max}} \brac{\frac{\omega_{\sf max}}{\omega_{\sf min}}}^6\brac{\frac{\theta_{\sf max}}{\theta^*_{\sf min}}}\brac{\frac{J}{N}+\frac{N}{J}}\log^3(N+J),
\end{align}
then we have
\begin{align*}
	\sup_{t\in \RR}\ab{\PP\brac{\sigma_{j,k}^{-1}\brac{\hat \theta_{j,k}-\theta_{j,k}}\le t}-\Phi(t)}\lesssim \frac{1}{\sqrt{\log(N+J)}},
\end{align*}
	where $\sigma_{j,k}^2:=\theta_{j,k}\frac{1}{\ab{\calC_k}^2}\sum_{i\in\calC_k}\frac{1-\omega_i\theta_{j,k}}{\omega_i}$.
\end{lemma}
\begin{remark}
	In its full generality, there is an additional term $\frac{N}{J}$ showing up in Lemma \ref{lem:Thetaerr-inf} in comparison to the condition for estimation of $\bTheta$ in Theorem \ref{thm:Thetaerr-ave}. Indeed, Theorem \ref{thm:Thetaerr-ave} conveys that the estimation error of $\theta_{j,k}$ is of order $\sqrt{\frac{\theta_{\sf max}\log(N+J)}{N\wedge J}}$ and the standard deviation $\sigma_{j,k}$ is of order $\sqrt{\frac{\theta_{\sf max}}{N}}$. Therefore, the inference result would not be valid in general  if $J$ is much smaller than $N$, due to the inflation of the error term in $\sigma_{j,k}^{-1}\brac{\hat\theta_{j,k}-\theta_{j,k}}$. 
\end{remark}
The following proposition indicates the consistency of $\hat\sigma_{j,k}$. 
\begin{lemma}\label{lem:consistent-sigma}
Suppose the conditions of Theorem  \ref{thm:exp-clustering-error} for exact clustering and  Assumption \ref{cond:iden-theta} hold. Assume that there exists some absolute constant $C_{\sf sigma}>0$ such that
\begin{align*}
	\frac{\Delta^2}{\theta_{\sf max}}\ge  C_{\sf sigma} \frac{\mu_\omega^4\mu_{\bTheta}\kappa^{14}K^3}{\omega_{\sf max}} \brac{\frac{\omega_{\sf max}}{\omega_{\sf min}}}^6\brac{\frac{\theta_{\sf max}}{\theta^*_{\sf min}}}\brac{\frac{J}{N}+1}\log^2(N+J),
\end{align*}
	then we have for any $j\in[J]$ and $k\in[K]$,  
	\begin{align*}
	&\Prob\brac{\ab{\frac{\hat \sigma^2_{j,k}}{\sigma_{j,k}^2}-1}\ge \frac{C_1}{\sqrt{\log(N+J)}}}\le C_2\brac{N+J}^{-20},
\end{align*}
for some absolute constant $C_1,C_2>0$.
\end{lemma}

\section{Proofs of Results in Section \ref{sec:clust}}\label{sec:proof_clust}
\subsection{General Versions of Results in Section \ref{sec:clust}}
\begin{theorem}\label{thm:exp-clustering-error}
Suppose Assumption \ref{cond:balanced} and \ref{cond:general-degree} hold. Assume $N\gtrsim \mu_\omega^4\mu_{\bTheta}\kappa^8K\brac{K^2\log^2(N+J)+\left({\omega_*}/{\omega_{\sf min}}\right)^2}$, $J\gtrsim K\log^4(N+J)$, $T_0\ge \log\left(\frac{\sigma^2_1\brac{\bR^*}}{\xi_{\sf err}}\right)$, and there exist some absolute constant $C_{\sf clust}>0$ such that
\begin{align*}
\frac{\Delta^2}{\theta_{\sf max}}&\ge  C_{\sf clust}\sqbrac{\frac{\mu_\omega\mu_{\bTheta}^{1/4}\kappa^4 K^{3/2}}{\omega_*}\left(\frac{\omega_{\sf max}}{\omega_{\sf min}}\right)^{2}\sqrt{\frac{J}{N}}\log \brac{N+J}+\frac{\mu_\omega\mu_{\bTheta}^{1/2}\kappa^8 K^2}{\omega_{\sf min}}\left(\frac{\omega_{\sf max}}{\omega_{\sf min}}\right)\log \brac{N+J}}.
\end{align*}
Then there exists some absolute constant $c_0>0$ such that  $\hat s$, the output of Algorithm \ref{alg:HeteroClustering}, satisfies 
\begin{align}\label{eq:exp-error-rate}
		\E h(\hat s,s)\le \frac{2K}{N}\sum_{i=1}^N&\exp\left(-c_0\omega_i\cdot \textsf{SNR}^2\right)+O\brac{\brac{N+J}^{-20}},
\end{align}
where 
\begin{align*}
	\textsf{SNR}^2:=\frac{\Delta^2}{\theta_{\sf max}}\cdot \frac{ 1}{\kappa^2 K\brac{\mu_{\omega}+\kappa \mu^{1/2}_{\bTheta} }\left(1+\frac{\kappa KJ\theta_{\sf max}\log \brac{N+J}}{N\omega_*\Delta^2}\right)}.
\end{align*}  
Moreover, we have with probability exceeding $1-\brac{N+J}^{-20}$ that $h(\hat s,s)=0$ provided that 
	\begin{align}\label{eq:delta-cond-exact-gen}
		\frac{\Delta^2}{\theta_{\sf max}}&\ge C_{\sf exact}\sqbrac{\frac{\mu_{\omega}\mu^{1/2}_{\bTheta}\kappa^6K^{{3/2}}}{\omega_*}\brac{\frac{\omega_{\sf max}}{\omega_{\sf min}}}\sqrt{\frac{J}{N}}\log(N+J)+\frac{\mu_\omega\mu_{\bTheta}\kappa^{12}K^2}{\omega_*}\brac{\frac{\omega_{\sf max}}{\omega_{\sf min}}}^2\log\brac{N+J}}.
	\end{align}
for some absolute constant $C_{\sf exact}>0$.
\end{theorem}
\begin{remark}
{\color{black}In contrast to Theorem \ref{thm:exp-clustering-error-simple}, in Theorem \ref{thm:exp-clustering-error} we established an  exponential error rate in addition to exact clustering, which is achievable under the general degree assumption \ref{cond:general-degree}. In particular, if a constant fraction $\delta\in(0,1)$ of subjects in each cluster has a degree  of $\omega_{\sf max}\asymp 1$ and the remaining has a  degree of $\omega_{\sf min}=o(1/\log\brac{N+J})$,  then Assumption \ref{cond:general-degree} holds with $\omega_{\sf max},\omega_*,\mu_{\omega}\asymp 1 $. In this scenario, the clustering proportion given by Theorem \ref{thm:exp-clustering-error} would be dominated by $\frac{2K}{N}\delta \exp\brac{-c_0\omega_{\sf min}\cdot \textsf{SNR}^2}$ even if $\textsf{SNR}^2\ge C \log\brac{N+J}$ for some large constant $C>0$. We have chosen to retain the exponential error rate in the general theorem  to distinguish it clearly from a polynomial one. 
}
\end{remark}
\subsection{Preliminary Results for Section \ref{sec:clust}}
\paragraph{Event $\calB_{\sf good}$.}

By Lemma \ref{lem:two-inf-bound}, there exists an event $\calB_{\sf good}$ with $\Prob\brac{\calB_{\sf good}}\ge 1-(N+J)^{-20}$ such that on $\calB_{\sf good}$, we have
\begin{align*}
	\hat \bU\bO^\top   -\bU=\bE\bV\bSigma^{-1}+\calH(\bE\bE^\top)\bU\bSigma^{-2}+\mPsi,
\end{align*}
where
\begin{align*}
	&\op{\mPsi}_{2,\infty}\lesssim \kappa^2\frac{\xi_{\sf err}}{\sigma_K^2(\bR^*)}\frac{\mu K}{N}+\kappa^2\frac{\xi^2_{\sf err}}{\sigma_K^4(\bR^*)}\sqrt{\frac{\mu K}{N}}\\
	&\fro{\hat\bU-\bU\bO}\lesssim \frac{\xi_{\sf err}K^{1/2}}{\sigma^2_K(\bR^*)},\quad \op{\hat\bU-\bU\bO}_{2,\infty}\lesssim \kappa^2\frac{\xi_{\sf err}}{\sigma^2_K(\bR^*)}\sqrt{\frac{\mu K}{N}},
\end{align*} 
provided that $\sigma^2_K(\bR^*)\gtrsim \kappa^2\xi_{\sf err}$.

\subsection{Proof of Theorem \ref{thm:exp-clustering-error}}
\paragraph{Exponential Rate in Expectation.}
Our analysis is conducted on the event $\calB_{\sf good}$. For any $k\in[K]$ and $i\in\calC_k$, we have
\begin{align*}
	{\wt\bU_{i,:}-\bar\bU_{i,:}\bO}=\frac{\hat \bU_{i,:}}{\op{\hat \bU_{i,:}}}-\frac{\bU_{i,:}\bO}{\op{\bU_{i,:}}}=\frac{1}{\op{\bU_{i,:}}}(\hat \bU_{i,:}- \bU_{i,:}\bO)+\left(\frac{1}{\op{\hat \bU_{i,:}}}-\frac{1}{\op{ \bU_{i,:}}}\right)\hat \bU_{i,:}.
\end{align*}
Therefore, we have
\begin{align*}
	\op{\wt\bU_{i,:}-\bar\bU_{i,:}\bO}&\le \frac{1}{\op{\bU_{i,:}}}\left(\op{\hat \bU_{i,:}- \bU_{i,:}\bO}+\left|\op{\bU_{i,:}\bO}-\op{\hat \bU_{i,:}}\right|\right)\\
	&\le \frac{2}{\op{\bU_{i,:}}}\op{\hat \bU_{i,:}- \bU_{i,:}\bO}.
\end{align*}
On the other hand,
\begin{equation*}
	\op{\bU_{i,:}}=\frac{\omega_i}{\sqrt{\sum_{j\in S_k}\omega_j^2}}\gtrsim \sqrt{\frac{K}{N}}\frac{\omega_{\sf min}}{\omega_*}.
\end{equation*}
We arrive at
\begin{align*}
	\fro{\wt\bU-\bar\bU\bO}\lesssim \sqrt{\frac{N}{K}}\frac{\omega_*}{\omega_{\sf min}}\fro{\hat \bU- \bU\bO},\quad \op{\wt\bU-\bar\bU\bO}_{2,\infty}\lesssim \sqrt{\frac{N}{K}}\frac{\omega_*}{\omega_{\sf min}}\op{\hat \bU- \bU\bO}_{2,\infty}.
\end{align*}
On event $\calB_{\sf good}$, we obtain that,
\begin{align*}
	\fro{\wt\bU-\bar\bU\bO}\lesssim  \brac{\frac{\omega_*}{\omega_{\sf min}}}\frac{\sqrt{N}\xi_{\sf err}}{\sigma^2_K(\bR^*)}.
\end{align*}
Notice that $\sum_{i=1}^N\fro{\wt\bU_{i,:}-\bar\bU_{i,:}\bO}^2=\fro{\wt\bU-\bar\bU\bO}^2$ and $\op{\bar\bU_{i,:}-\bar\bU_{i,:}}=\sqrt{2}$ for any $i,j$ such that $s_i\ne s_j$, then by  the proof of Theorem 5.2 in \cite{jing2021community}, we can obtain $h(\hat s,s)\le C_0 \zeta$ with
\begin{align*}
	\zeta:= \brac{\frac{\omega_*}{\omega_{\sf min}}}^2\frac{\xi^2_{\sf err}}{\sigma^4_K(\bR^*)},
\end{align*}
and $C_0>0$ is some absolute constant. As a result, we can conclude that $\zeta\le c_{0}/K$ for some sufficiently small constant $c_0>0$, if the following condition holds:
\begin{align}\label{eq:exp-err-init-cond}
	\frac{\Delta^2}{\theta_{\sf max}}&\gtrsim\frac{\kappa^2K^{3/2}}{\omega_*} \brac{\frac{\omega_*}{\omega_{\sf min}}}^2\sqrt{\frac{J}{N}}\log(N+J)+\frac{\kappa^4K^2}{\omega_{\sf min}} \brac{\frac{\omega_{\sf max}}{\omega_{\sf min}}}\log \brac{N+J}.
\end{align}
where we have used Lemma \ref{lem:incoherence}. Let $\hat\calC_k$ be the k-means cluster associated with $\calC_k$ for all $k\in[K]$. Under Assumption \ref{cond:balanced-cluster}, for each $k\in[K]$ we have 
\begin{align*}
	|\hat\calC_k \backslash \calC_k|\vee |\calC_k\backslash\hat\calC_k|\le C_0N\zeta.
\end{align*}
In addition by k-means, the cluster center of $\hat\calC_k$, denoted by $\hat\bU_{k,:}^\dag$, satisfies $	\hat\bU^\dag_{k,:}=\frac{1}{|\hat\calC_k|}\sum_{i\in \hat\calC_k}\wt\bU_{i,:}$.
Hence we can arrive at
\begin{align*}
	\op{\hat\bU^\dag_{k,:}-\bU^\dag_{k,:}\bO}
	&\le \frac{1}{|\hat\calC_k|}\left(\op{\sum_{i\in\hat\calC_k}(\wt\bU_{i,:}-\bar\bU_{i,:}\bO)}+\op{\sum_{i\in\hat\calC_k\backslash\calC_k}(\bar\bU_{i,:}-\bU^\dag_{k,:}\bO)}\right)\\
	&\le \frac{1}{|\hat\calC_k|}\sqrt{|\hat\calC_k|\sum_{i\in\hat\calC_k}\op{\wt\bU_{i,:}-\bar\bU_{i,:}\bO}^2}+CK\zeta\\
	&\le \sqrt{\frac{C^\prime K}{N}\fro{\wt\bU-\bar\bU\bO}^2}+CK\zeta\\
    &=O(\sqrt{K\zeta}).
\end{align*}
Combined with the assumption that $\zeta\le c_0/K$ with $c_0$ sufficiently small, we have 
\begin{align}\label{eq:cluster-close}
	\op{\hat\bU^\dag_{k,:}-\bU^\dag_{k,:}\bO}\le \frac{\sqrt{2}}{8},\quad \forall k\in[K].
\end{align}
Notice that
\begin{align}\label{eq:h-exp-bound}
	\E h(\hat s,s)&=\frac{1}{N}\sum_{i=1}^N\Prob\brac{\hat s_i\ne s_i}\le \frac{1}{N}\sum_{i=1}^N\Prob\left(\left\{\hat s_i\ne s_i\right\}\cap\calB_{\sf good}\right )+\brac{N+J}^{-20}.
\end{align}
For any $i\in\calC_k$, if $\op{\wt\bU_{i,:}-\bar\bU_{i,:}\bO}\le \sqrt{2}/4$, then $\op{\wt\bU_{i,:}-\hat\bU_{k,:}^\dag}\le \op{\wt\bU_{i,:}-\bar\bU_{i,:}\bO}+\op{\hat\bU^\dag_{k,:}-\bar\bU_{i,:}\bO}\le \frac{3}{8}\sqrt{2}$ due to \eqref{eq:cluster-close}. For any $l\ne k$, $\op{\wt\bU_{i,:}-\hat\bU_{l,:}^\dag}\ge \op{\bU^\dag_{k,:}-\bU^\dag_{l,:}}-\op{\wt\bU_{i,:}-\bU_{k,:}^\dag\bO}-\op{\hat \bU^\dag_{l,:}-\bU_{l,:}^\dag\bO}\ge \frac{5}{8}\sqrt{2}$. This implies we must have $\hat s_i=s_i$. In other words, we have 
\begin{align*}
	\Prob\left(\left\{\hat s_i\ne s_i\right\}\cap\calB_{\sf good}\right )\le \Prob\left(\op{\wt\bU_{i,:}-\bar\bU_{i,:}\bO}>\frac{\sqrt{2}}{4}\right )\le \Prob\left(\op{\hat \bU_{i,:}- \bU_{i,:}\bO}>\frac{\sqrt{2}}{8}\op{\bU_{i,:}}\right ),
\end{align*}
where the last inequality we have used the relation between $\op{\wt\bU_{i,:}-\bar\bU_{i,:}\bO}$ and  $\op{\hat \bU_{i,:}- \bU_{i,:}\bO}$. By Theorem 10 in \cite{yan2021inference}, we obtain that with probability at least $1-(N+J)^{-20}$,
\begin{align*}
	\hat \bU\bO^\top  -\bU=\bE\bV\bSigma^{-1}+\calH(\bE\bE^\top)\bU\bSigma^{-2}+\mPsi,
\end{align*}
where
\begin{align*}
	\op{\mPsi}_{2,\infty}\lesssim \kappa^2\frac{\xi_{\sf err}}{\sigma_K^2(\bR^*)}\frac{\mu K}{N}+\kappa^2\frac{\xi^2_{\sf err}}{\sigma_K^4(\bR^*)}\sqrt{\frac{\mu K}{N}}.
\end{align*}
Notice that
\begin{align*}
	\op{e_i^\top \left(\bE\bV\bSigma^{-1}+\calH(\bE\bE^\top)\bU\bSigma^{-2}+\mPsi\right)}\le \op{e_i^\top \left(\bE\bV\bSigma^{-1}+\calH(\bE\bE^\top)\bU\bSigma^{-2}\right)}+\op{\mPsi}_{2,\infty}.
\end{align*}
We need  $\op{\mPsi}_{2,\infty}\le c\min_i\op{\bU_{i,:}}$ for some sufficiently small constant $c>0$, which is fulfilled by 
\begin{align*}
	N\gtrsim \mu_\omega^4\mu_{\bTheta}\kappa^8K^3\left(\frac{\omega_*}{\omega_{\sf min}}\right)^2,
\end{align*}
and 
\begin{align*}
\frac{\Delta^2}{\theta_{\sf max}}&\gtrsim \frac{\mu_\omega\mu_{\bTheta}^{1/2}\kappa^8 K^{3/2}}{\omega_*}\left(\frac{\omega_{\sf max}}{\omega_{\sf min}}\right)\log \brac{N+J}+\frac{\mu_\omega\mu_{\bTheta}^{1/4}\kappa^4 K^{5/4}}{\omega_*}\left(\frac{\omega_{\sf max}}{\omega_{\sf min}}\right)^{1/2}\sqrt{\frac{J}{N}}\log \brac{N+J}.
\end{align*}
Then we can have
\begin{align*}
	&~\Prob\left(\op{\hat \bU_{i,:}- \bU_{i,:}\bO}>\frac{\sqrt{2}}{8}\op{\bU_{i,:}}\right )\\
    \le &~ \Prob\left( \op{e_i^\top \left(\bE\bV\bSigma^{-1}+\calH(\bE\bE^\top)\bU\bSigma^{-2}\right)}>c_0\op{\bU_{i,:}}\right )+O\brac{\brac{N+J}^{-20}},
\end{align*}
for some small absolute constant $c_0>0$. Let $\bE^{-i}$ be the matrix by zeroing out entries in $i$-th row of $\bE$, then 
\begin{align*}
	 \op{e_i^\top \left(\bE\bV\bSigma^{-1}+\calH(\bE\bE^\top)\bU\bSigma^{-2}\right)}&=\op{\sum_{j=1}^JE_{i,j}\left (\bV_{j,:}\bSigma^{-1}+\brac{\bE_{:,j}^{-i}}^\top \bU\bSigma^{-2}\right)}\\
	 &\le \max_{k\in[K]}\sqrt{K}\ab{\sum_{j=1}^JE_{i,j}\left (\bV_{j,:}\bSigma^{-1}+\brac{\bE_{:,j}^{-i}}^\top \bU\bSigma^{-2}\right)e_k}.
\end{align*}
Conditional on $\bE^{-i}$, this is a sum of independent zero-mean random variables. Now in order to apply Bernstein's inequality for  fixed $k\in[K]$, it suffices to calculate
\begin{align*}
	L=\max_{j\in[J]}\left|E_{i,j}\left (\bV_{j,:}\bSigma^{-1}+\brac{\bE_{:,j}^{-i}}^\top \bU\bSigma^{-2}\right)e_k\right|\le \op{\bV}_{2,\infty}\sigma^{-1}_{K}(\bR^*)+\op{(\bE^{-i})^{\top} \bU\bSigma^{-2}}_{2,\infty},
\end{align*}
and 
\begin{align*}
	V=\sum_{j=1}^J\E\brac{E_{i,j}^2}\ab{\bV_{j,:}\bSigma^{-1}e_k+\brac{\bE_{:,j}^{-i}}^\top \bU\bSigma^{-2}e_k}^2\le  \omega_i\theta_{\sf max}\left(\sigma^{-2}_{K}(\bR^*)+\op{(\bE^{-i})^\top \bU_{:,k}}^2\sigma^{-4}_{K}(\bR^*)\right).
\end{align*}
Following the proofs of Lemma 12 and Lemma 14 in \cite{cai2021subspace}, we can obtain with probability at least $1-(N+J)^{-20}$,
\begin{align}\label{eq:L-V-bound}
	&\op{(\bE^{-i})^{\top} \bU\bSigma^{-2}}_{2,\infty}\lesssim \frac{\op{\bU}_{2,\infty}\left(\log(N+J)+\sqrt{K^{-1/2}N\omega_*\infn{\bTheta}\log(N+J)}\right)}{\sigma^2_K(\bR^*)},\notag\\
	&\op{(\bE^{-i})^\top \bU_{:,k}}^2
	\lesssim \omega_{\sf max}\max_k\op{\bTheta_{:,k}}_1+\frac{\mu_1 K}{N}\log^3(N+J)\lesssim J\omega_{\sf max}\theta_{\sf max},
\end{align}
where we have used Cauchy-Schwarz inequality on $\max_{j\in[J]}\sum_{i\in[N]}\omega_i\theta_{j,s_i}$, and that $N\gtrsim \mu K\log^2(N+J)$. Denote $\calB_0:=\{\eqref{eq:L-V-bound} \text{~holds}\}$, our analysis shall proceed on $\calB_0\cap\calB_{\sf good}$ with $\Prob(\calB_0\cap\calB_{\sf good})\ge 1-2(N+J)^{-20}$. Hence using $\sigma_{K}(\bR^*)\gtrsim (N/K)^{1/2}\omega_*\sigma_\star$ and \eqref{eq:L-V-bound}, we can further arrive at
\begin{align*}
	&L
	\lesssim \frac{K}{\sqrt{N}\omega_*\sigma_\star}\left(\sqrt{\frac{\mu_2}{J}}+\sqrt{\frac{\mu_1K}{N}}\frac{\log(N+J)+\sqrt{K^{-1/2}N\omega_*\infn{\bTheta}\log(N+J)}}{\sqrt{N}\omega_*\sigma_\star}\right)\\
	&V
	\lesssim \frac{\omega_i \theta_{\sf max}K}{N\omega_*^2\sigma^2_\star}\left(1+\frac{KJ\omega_{\sf max}\theta_{\sf max}}{N\omega_*^2\sigma^2_\star}\right).
\end{align*}

Due to the independence between $\{E_{i,j}\}_{j=1,\cdots,J}$ and $\bE^{-i}$, we can then apply Bernstein's inequality with $t=c_0K^{-1/2}\op{\bU_{i,:}}=\tilde c_0\omega_i/(N^{1/2}\omega_*)$ to obtain that
\begin{align*}
	\Prob &\left(\sqrt{K}\ab{\sum_{j=1}^JE_{i,j}\left (\bV_{j,:}\bSigma^{-1}+\brac{\bE_{:,j}^{-i}}^\top \bU\bSigma^{-2}\right)e_k}\ge c_0\op{\bU_{i,:}}\right)\le 2\exp\left(-\frac{t^2/2}{V+tL/3}\right)\\
	&\overset{(a)}{\lesssim} \exp\left[-\omega_i\cdot\min\left\{\frac{ \sigma_\star^2}{\theta_{\sf max}K\left(1+\frac{K(\omega_{\sf max}/\omega_*)J\theta_{\sf max}}{N\omega_*\sigma^2_\star}\right)},\frac{\sigma_\star^2}{\sqrt{\mu_1\vee \mu_2}K\left(\theta_{\sf max}+\frac{\sqrt{K}\log(N+J)+\sqrt{KN\omega_*\theta_{\sf max}\log(N+J)}}{N\omega_*}\right)}\right\}\right]\\
	&\overset{(b)}{\lesssim} \exp\left(-\frac{ \omega_i\cdot \sigma_\star^2}{\theta_{\sf max}\brac{\mu_{\omega}+\kappa \mu^{1/2}_{\bTheta} }K\left(1+\frac{KJ\theta_{\sf max}\log(N+J)}{N\omega_*\sigma^2_\star}+\sqrt{\frac{K\log(N+J)}{N\theta_{\sf max}\omega_*}}\right)}\right)\\
	&\overset{(c)}{\lesssim} \exp\left(-\frac{ \omega_i\cdot \sigma_\star^2}{\theta_{\sf max}\brac{\mu_{\omega}+\kappa \mu^{1/2}_{\bTheta} }K\left(1+\frac{KJ\theta_{\sf max}\log(N+J)}{N\omega_*\sigma^2_\star}\right)}\right),
\end{align*}
where in (a) we have used that $\sigma_\star\le  \frac{\fro{\bTheta}}{\sqrt{K}}\le \theta_{\sf max} \sqrt{J}$ and $\infn{\bTheta}\le \theta_{\sf max}\sqrt{K}$, (b) holds due to $\mu_1\vee \mu_2\lesssim \mu^2_{\omega}+\kappa^2$ and $\sigma_\star^2\le J\theta^2_{\sf max}$, (c) holds since if $\sqrt{\frac{K\log(N+J)}{N\theta_{\sf max}\omega_*}}>1$, then we must have $\sqrt{\frac{K\log(N+J)}{N\theta_{\sf max}\omega_*}}<\frac{K\log(N+J)}{N\theta_{\sf max}\omega_*}<\frac{KJ\theta_{\sf max}\log(N+J)}{N\omega_*\sigma_\star^2}$, implying that $\sqrt{\frac{K\log(N+J)}{N\theta_{\sf max}\omega_*}}\le 1\vee \frac{KJ\theta_{\sf max}\log(N+J)}{N\omega_*\sigma_\star^2}$. Finally, combining a union bound over $[K]$ and \eqref{eq:h-exp-bound} gives the following bound in expectation:
\begin{align}\label{eq:expec-bound}
	\E h(\hat s,s)\lesssim \frac{1}{N}\sum_{i=1}^N\exp\left(-\omega_i\cdot \textsf{SNR}^2\right)+\brac{N+J}^{-20},
\end{align}
where 
\begin{align*}
	\textsf{SNR}^2=\frac{\sigma_\star^2}{\theta_{\sf max}\brac{\mu_{\omega}+\kappa \mu^{1/2}_{\bTheta} }K\left(1+\frac{KJ\theta_{\sf max}\log(N+J)}{N\omega_*\sigma^2_\star}\right)}.
\end{align*}
provided that $\min_{i\in[n]}\omega_i\cdot \textsf{SNR}^2>C\log K$ for some sufficiently large $C>0$. The proof of this part is completed by noticing the relation between $\sigma_\star$ and $\Delta$ by Lemma \ref{lem:incoherence}.
\paragraph{Exact Recovery with High Probability.}
On event $\calB_{\sf good}$, we know that
\begin{align*}
	  \op{\wt\bU-\bar\bU\bO}_{2,\infty}\le \frac{2}{\min_{i\in[N]}\op{\bU_{i,:}}}\op{\hat\bU-\bU\bO}_{2,\infty}\lesssim \frac{\mu_\omega\mu^{1/2}_{\bTheta}\kappa^4K^{1/2}}{\omega_{\sf min}}\cdot \frac{\xi_{\sf err}}{\sigma^2_K(\bR^*)}\le \frac{\sqrt{2}}{8}
\end{align*}
provided that $\sigma^2_K(\bR^*)\ge C\kappa^4\brac{\mu_{\bTheta}K}^{1/2}\brac{\omega_{\sf max}/\omega_{\sf min}}^{-1}\xi_{\sf err}$ for some large constant $C>0$. On the other hand, we have
\begin{align*}
	&\Prob\brac{\bigcup_{i\in[N]}\left\{\hat s_i\ne s_i\right\}}\le \Prob\left(\bigcup_{i\in[N]}\left\{\hat s_i\ne s_i\right\}\cap\calB_{\sf good}\right )+\Prob\left(\calB_{\sf good}^c\right )\\
	&\le  \Prob\left(\bigcup_{i\in[N]}\left\{\op{\wt\bU_{i,:}-\bar\bU_{i,:}\bO}>\frac{\sqrt{2}}{4}\right\}\right )+\brac{N+J}^{-20}= \brac{N+J}^{-20}
\end{align*}
Hence we conclude that with probability exceed $1-{\brac{N+J}^{-20}}$ we can have exact recovery if the following holds
		\begin{align*}
		\frac{\Delta^2}{\theta_{\sf max}}&\gtrsim\frac{\mu_{\omega}\mu^{1/2}_{\bTheta}\kappa^6K^{{3/2}}}{\omega_*}\brac{\frac{\omega_{\sf max}}{\omega_{\sf min}}}\sqrt{\frac{J}{N}}\log(N+J)+\frac{\mu_\omega\mu_{\bTheta}\kappa^{12}K^2}{\omega_*}\brac{\frac{\omega_{\sf max}}{\omega_{\sf min}}}^2\log\brac{N+J}
	\end{align*}
\hfill$\square$
\subsection{Proof of Theorem \ref{thm:exact-lower-bound}}
Denote $\theta_k=\bTheta_{:,k}$ for $k\in\ebrac{1,2}$. Let $\pi=\pi_{s}\times \pi_{\bTheta}$ be a prior on $\brac{s,\bTheta}$ and following the proof of Theorem 1  in \cite{ndaoud2022sharp} line by line, we obtain  there exists an absolute constant  $c>0$ that
\begin{align}\label{eq:lb-general}
	\inf_{\hat s}\sup_{\calP\brac{s,\bTheta}}\EE h\brac{\hat s,s}\ge c\brac{\frac{1}{n/2}\sum_{i=1}^{n/2}\inf_{\hat T_i\in[1,2]}\EE_\pi\EE_{\bR}\ab{\hat T_i-s_i}-\pi_\bTheta\brac{\op{\theta_1-\theta_2}<\Delta}}.
\end{align} 
where $\inf_{\hat T_i\in[1,2]}$ is the infimum over all estimators $\hat T_i$ with values in $[1,2]$, $\EE_\pi$ and $\EE_{\bR}$ is taking expectation with respect to $\brac{s,\bTheta}$ with prior $\pi$ and data $\bR$ conditional on $\brac{s,\bTheta}$ respectively.

We start with bounding the first term in \eqref{eq:lb-general}. To this end, our goal is to derive a result similar to Proposition 2 in \cite{ndaoud2022sharp}. In particular, let	 $\pi_{\bTheta}=\pi^{\alpha_1}_{\theta}\times \pi^{\alpha_2}_{\theta}$ be a product prior on $\brac{\theta_1,\theta_2}\in \brac{0,1}^{J}\times \brac{0,1}^{J}$, where $\pi_\theta^{\alpha}$ is the distribution of the Beta$\brac{\alpha,N-\alpha}$ random vector with i.i.d.  entries, let $\pi_s$ be the distribution of the vector with i.i.d. uniform distribution in $\ebrac{1,2}$. Let $\pi^{i}_{s}$ and   $\pi^{-i}_{s}$  denote the  prior on $s_i$ and $s_{-i}:=\brac{s_1,\cdots,s_{i-1},s_{i+1},\cdots,s_N}$ respectively. Notice that for any $i\in[N]$ any $\hat T_i\in[1,2]$, we have
\begin{align*}
	\EE_\pi\EE_{\bR}\ab{\hat T_i-s_i}=\EE_{\pi_s^{-i}}\EE_{\pi_s^{i}}\EE_{\pi_\bTheta}\EE_{\bR}\ab{\hat T_i-s_i}=\EE_{\pi_s^{-i}}\EE \brac{\ab{\hat T_i-s_i}\mid s_{-i}}.
\end{align*}
Denote $\beta_k:=N-\alpha_k$, $\calC_k^{-i}:=\ebrac{l\in[N]\backslash\ebrac{i}:s_l=k}$ and $n_k^{-i}:=\ab{\calC_k^{-i}}$. We define, for $k\in\ebrac{1,2}$, $\wt f^i_{k}$ the density of the observation $\bR$ given $\ebrac{s_l,l\ne i}$ and $s_i=k$, then by Neyman–Pearson lemma, we get that for each $i\in[N ]$,
\begin{align*}
	s^{**}_i=\begin{cases}
		1,&\text{if~}f^i_{1}\brac{\bR}>f^i_{2}\brac{\bR},\\
		2,& \text{o.w.}
	\end{cases}
\end{align*}
minimizes $\EE \brac{\ab{\hat T_i-s_i}\mid s_{-i}}$ over all functions of $s_{-i}$  and $\bR$, hence we arrive at 
\begin{align}\label{eq:lb-reduce-np}
	\inf_{\hat T_i\in[1,2]}\EE_\pi\EE_{\bR}\ab{\hat T_i-s_i}=\inf_{\hat T_i\in[1,2]}\EE_{\pi_s^{-i}}\EE \brac{\ab{\hat T_i-s_i}\mid s_{-i}}\ge \EE_\pi\EE_{\bR}\brac{\ab{s_i^{**}-s_i}}=\PP\brac{s_i^{**}\ne s_i}.
\end{align}
where the last $\PP$ is taken over $\bR$ and $\brac{s,\bTheta}$. Notice that 
\begin{align*}
	\wt f^i_{1}\brac{\bR}&=\prod_{j=1}^J\int\theta_{j1}^{R_{ij}}\brac{1-\theta_{j1}}^{1-R_{ij}}\prod_{k\in\ebrac{1,2}}\frac{\theta_{jk}^{\sum_{l\in \calC_k^{-i}}R_{lj}}\brac{1-\theta_{jk}}^{\sum_{l\in \calC_k^{-i}}\brac{1-R_{lj}}}\theta_{jk}^{\alpha_k-1}\brac{1-\theta_{jk}}^{\beta_k-1}}{\textsf{B}\brac{\alpha_k,\beta_k}}d\bTheta\\
	&=\prod_{j=1}^J\int\frac{\theta_{j1}^{R_{ij}+\sum_{l\in \calC_1^{-i}}R_{lj}+\alpha_1-1}\brac{1-\theta_{j1}}^{1-R_{ij}+\sum_{l\in \calC_1^{-i}}\brac{1-R_{lj}}+\beta_1-1}}{\textsf{B}\brac{\alpha_1,\beta_1}}d\theta_{j1}\\
	&\times \int\frac{\theta_{j2}^{\sum_{l\in \calC_2^{-i}}R_{lj}+\alpha_2-1}\brac{1-\theta_{j2}}^{\sum_{l\in \calC_2^{-i}}\brac{1-R_{lj}}+\beta_2-1}}{\textsf{B}\brac{\alpha_2,\beta_2}}d\theta_{j2}\\
	&=\prod_{j=1}^J\frac{\textsf{B}\brac{\alpha_1+R_{ij}+\sum_{l\in \calC_1^{-i}}R_{lj},\beta_1+1-R_{ij}+\sum_{l\in \calC_1^{-i}}\brac{1-R_{lj}}}}{\textsf{B}\brac{\alpha_1,\beta_1}}\\
	&\times \frac{\textsf{B}\brac{\alpha_2+\sum_{l\in \calC_2^{-i}}R_{lj},\beta_2+\sum_{l\in \calC_2^{-i}}\brac{1-R_{lj}}}}{\textsf{B}\brac{\alpha_2,\beta_2}},
\end{align*}
where
\begin{align*}
	&\textsf{B}\brac{\alpha_1+R_{ij}+\sum_{l\in \calC_1^{-i}}R_{lj},\beta_1+1-R_{ij}+\sum_{l\in \calC_1^{-i}}\brac{1-R_{lj}}}\\
	&=\begin{cases}
		\frac{\alpha_1+\sum_{l\in \calC_1^{-i}}R_{lj}}{\alpha_1+\beta_1+n_1^{-i}}\textsf{B}\brac{\alpha_1+\sum_{l\in \calC_1^{-i}}R_{lj},\beta_1+\sum_{l\in \calC_1^{-i}}\brac{1-R_{lj}}}, & \text{if~} R_{ij}=1\\
		\frac{\beta_1+\sum_{l\in \calC_1^{-i}}\brac{1-R_{lj}}}{\alpha_1+\beta_1+n_1^{-i}}\textsf{B}\brac{\alpha_1+\sum_{l\in \calC_1^{-i}}R_{lj},\beta_1+\sum_{l\in \calC_1^{-i}}\brac{1-R_{lj}}}, & \text{if~} R_{ij}=0
	\end{cases}.
\end{align*}
We thereby have
\begin{align*}
	\wt f^i_{1}\brac{\bR}&=\frac{1}{\sqbrac{\textsf{B}\brac{\alpha_1,\beta_1}\textsf{B}\brac{\alpha_2,\beta_2}}^J}\prod_{j=1}^J\brac{\frac{\alpha_1+\sum_{l\in \calC_1^{-i}}R_{lj}}{\alpha_1+\beta_1+n_1^{-i}}}^{R_{ij}}\brac{\frac{\beta_1+\sum_{l\in \calC_1^{-i}}\brac{1-R_{lj}}}{\alpha_1+\beta_1+n_1^{-i}}}^{1-R_{ij}}\\
	&\times \textsf{B}\brac{\alpha_1+\sum_{l\in \calC_1^{-i}}R_{lj},\beta_1+\sum_{l\in \calC_1^{-i}}\brac{1-R_{lj}}}\textsf{B}\brac{\alpha_2+\sum_{l\in \calC_2^{-i}}R_{lj},\beta_2+\sum_{l\in \calC_2^{-i}}\brac{1-R_{lj}}}.
\end{align*}
A similar result holds for $\wt f^i_{2}\brac{\bR}$. Denote $p_{kj}:=\frac{\alpha_k+\sum_{l\in \calC_k^{-i}}R_{lj}}{\alpha_k+\beta_k+n_k^{-i}}$ for $k\in\ebrac{1,2}$, we then have
\begin{align*}
	\wt f^i_{k}\brac{\bR}&=\frac{1}{\sqbrac{\textsf{B}\brac{\alpha_1,\beta_1}\textsf{B}\brac{\alpha_2,\beta_2}}^J}\prod_{j=1}^Jp_{kj}^{R_{ij}}\brac{1-p_{kj}}^{1-R_{ij}}\\
	&\times \textsf{B}\brac{\alpha_1+\sum_{l\in \calC_1^{-i}}R_{lj},\beta_1+\sum_{l\in \calC_1^{-i}}\brac{1-R_{lj}}}\textsf{B}\brac{\alpha_2+\sum_{l\in \calC_2^{-i}}R_{lj},\beta_2+\sum_{l\in \calC_2^{-i}}\brac{1-R_{lj}}}.
\end{align*}
Hence we have
\begin{align*}
	\frac{\wt f^i_{1}\brac{\bR}}{\wt f^i_{2}\brac{\bR}}=\prod_{j=1}^J\frac{p_{1j}^{R_{ij}}\brac{1-p_{1j}}^{1-R_{ij}}}{p_{2j}^{R_{ij}}\brac{1-p_{2j}}^{1-R_{ij}}}.
\end{align*}
As a result, we have
\begin{align*}
	s^{**}_i=\begin{cases}
		1,&\text{if~}\sum_{j=1}^JR_{ij}\log \frac{p_{1j}\brac{1-p_{2j}}}{p_{2j}\brac{1-p_{1j}}}\ge \sum_{j=1}^J\log\frac{1-p_{2j}}{1-p_{1j}},\\
		2,& \text{o.w.}
	\end{cases}
\end{align*}
Now we temporarily assume that $p_{1j}\asymp p_{2j}=o(1)$, which will be verified in the end, then by independence structure between $R_{ij}$ and $\ebrac{p_{1j},p_{2j}}$ and Lemma \ref{lem:lower-bound}  we have 
\begin{align*}
	\PP\brac{s_i^{**}\ne s_i}=\PP\brac{\sum_{j=1}^JR_{ij}\log \frac{p_{1j}\brac{1-p_{2j}}}{p_{2j}\brac{1-p_{1j}}}\ge \sum_{j=1}^J\log\frac{1-p_{2j}}{1-p_{1j}}}\gtrsim\EE\sqbrac{\exp\brac{-c\frac{\sum_{j=1}^J\brac{p_{1j}-p_{2j}}^2}{\max_{j,k}p_{kj}}}}.
\end{align*}
It suffices to investigate $p_{1j}$ and $p_{2j}$. Without loss of generality, we can assume $n_1^{-i}=n_2^{-i}=\brac{N-1}/2$ and set $\alpha_k+\beta_k=\brac{N+1}/2$ and $\alpha_k=o\brac{N}$. First of all, by Bernstein-type inequality for Beta distribution in \cite{skorski2023bernstein} we get that
\begin{align*}
	\PP\brac{\ab{\theta_{jk}-\frac{\alpha_k}{\alpha_k+\beta_k}}\ge t}\le \exp\brac{-\frac{t^2}{2\brac{v_k+c_kt/3}}},
\end{align*}
where $v_k\asymp\frac{\alpha_k\beta_k}{N^3},c_k\asymp\frac{\beta_k-\alpha_k}{N^2}$. It suffices to take $t=\sqrt{v_k\log J}+c_k\log J$, we obtain that
\begin{align*}
\ab{\theta_{jk}-\frac{\alpha_k}{\alpha_k+\beta_k}}\lesssim\frac{\sqrt{\alpha_k\log J}+\log J}{N}
\end{align*}
Hence we can take $\alpha_k\asymp N^{\epsilon}$ for $\epsilon\in(1/2,1)$ and conclude that there exists an event $\calA_1$ with $\PP\brac{\calA_1}\ge 1-\exp\brac{-c\log J}$,
\begin{align*}
	\theta_{jk}=\frac{\alpha_k}{\alpha_k+\beta_k}\brac{1+O\brac{\sqrt{\frac{\log J}{N^{\epsilon}}}}},\quad \forall j\in[J],k\in\ebrac{1,2}.
\end{align*}
This implies that $\theta_{jk}\asymp 1/N^{1-\epsilon}$ and $\max_{j,k}\theta_{j,k}=\frac{2\max_{k\in\ebrac{1,2}}\alpha_k}{N+1}\brac{1+O\brac{\sqrt{\frac{\log J}{N^{\epsilon}}}}}$ on $\calA_1$. Denote $\theta_{\sf max}:=3\alpha_1/N$. Moreover, we have
\begin{align*}
	\EE\sum_{j=1}^J\brac{\theta_{j1}-\theta_{j2}}^2&=\sum_{j=1}^J\brac{\EE\theta_{j1}^2+\EE\theta_{j2}^2-2\EE\theta_{j1}\EE\theta_{j2}}\\
	&=\sum_{j=1}^J\brac{\frac{\alpha_1\beta_1+\alpha_2\beta_2}{\brac{\alpha_1+\beta_1}^2\brac{\alpha_1+\beta_1+1}}+\frac{\alpha_1^2+\alpha_2^2}{\brac{\alpha_1+\beta_1}^2}-\frac{2\alpha_1\alpha_2}{\brac{\alpha_1+\beta_1}^2}}\\
	&=\sum_{j=1}^J\frac{\brac{\alpha_1-\alpha_2}^2+\alpha_1\frac{\beta_1}{\alpha_1+\beta_1+1}+\alpha_2\frac{\beta_2}{\alpha_1+\beta_1+1}}{\brac{\alpha_1+\beta_1}^2}.
\end{align*}
Hence we have $\EE\sum_{j=1}^J\brac{\theta_{j1}-\theta_{j2}}^2\asymp  J\brac{\frac{\brac{\alpha_1-\alpha_2}^2}{N^2}+\frac{1}{N^{2-\epsilon}}}$. 
By \cite{marchal2017sub}, $\op{\theta_{jk}}_{\psi_2}\lesssim 1/\sqrt{N}$ for $k\in\ebrac{1,2}$. This implies that $\op{\theta_{j1}-\theta_{j2}}_{\psi_2}\lesssim 1/\sqrt{N}$ and hence
\begin{align*}
	\sum_{j=1}^J\op{\brac{\theta_{j1}-\theta_{j2}}^2}_{\psi_1}^2\lesssim \frac{J}{N^2}, \quad \max_{j\in[J]}\op{\brac{\theta_{j1}-\theta_{j2}}^2}_{\psi_1}\lesssim \frac{1}{N}.
\end{align*}
By Bernstein's inequality, we obtain that
\begin{align*}
	&\PP\brac{\ab{\sum_{j=1}^J\brac{\theta_{j1}-\theta_{j2}}^2-\EE\sum_{j=1}^J\brac{\theta_{j1}-\theta_{j2}}^2}\ge t,~\calA_2\text{~occurs}}\\
	&\le 2\exp\brac{-c\min\ebrac{\frac{t^2}{\sum_{j=1}^J\op{\brac{\theta_{j1}-\theta_{j2}}^2}_{\psi_1}^2},\frac{t}{\max_{j\in[J]}\op{\brac{\theta_{j1}-\theta_{j2}}^2}_{\psi_1}}}}.
\end{align*}
Hence we can take $\alpha_1-\alpha_2= C_0\Delta\cdot N/\sqrt{J\brac{1+\gamma}}$ with $\gamma:=J\theta_{\sf max}\log J/\brac{N\Delta^2}$ and $t=c_0\Delta^2$ for some constants $C_0,c_0>0$ such that
\begin{align}\label{eq:diffleDelta-small-prob}
	\PP\brac{\sum_{j=1}^J\brac{\theta_{j1}-\theta_{j2}}^2\ge  \Delta^2}\ge 1-2\exp\brac{-c\min\ebrac{\Delta^2N,\frac{\Delta^4N^2}{J}}}-\exp\brac{-c\log J}.
\end{align}
Denote the above event as $\calA_2$.

On the other hand, by Bernstein's inequality we have
\begin{align*}
	\PP\brac{\ab{\sum_{l\in\calC_k^{-i}}R_{lj}-n_k^{-i}\theta_{jk}}\le C\brac{\sqrt{n_k^{-i}\theta_{jk}\log J}+\log J}\mid \bTheta}\ge 1-\exp\brac{-c\log J}.
\end{align*}
Thus we can conclude that
\begin{align*}
	\PP\brac{\ab{\sum_{l\in\calC_k^{-i}}R_{lj}-n_k^{-i}\frac{\alpha_k}{\alpha_k+\beta_k}}\le C^\prime \sqrt{N^{\epsilon}\log J}}\ge 1-2\exp\brac{-c\log J}.
\end{align*}
We can further take a union bound over $j\in[J]$ to obtain that 
\begin{align*}
	\PP\brac{\bigcap_{j\in[J]}\ebrac{\ab{\sum_{l\in\calC_k^{-i}}R_{lj}-n_k^{-i}\frac{\alpha_k}{\alpha_k+\beta_k}}\le C^\prime \sqrt{N^{\epsilon}\log J}}}\ge 1-2\exp\brac{-c^\prime\log J}.
\end{align*}
Denote the above event as $\calA_3$,  on which we have  that 
\begin{align*}
	p_{1j}-p_{2j}=\frac{1}{N}\sqbrac{\alpha_1-\alpha_2+n_1^{-i}\frac{\alpha_1-\alpha_2}{\brac{N+1}/2}\brac{1+O\brac{\sqrt{\frac{\log J}{N^{\epsilon}}}}}}=\frac{2\brac{\alpha_1-\alpha_2}}{N+1}\brac{1+O\brac{\sqrt{\frac{\log J}{N^{\epsilon}}}}}
\end{align*}
Moreover, on $\calA_3$ we have
\begin{align*}
	\max_{j,k}p_{kj}=\max_{j,k}\frac{1}{N}\sqbrac{\alpha_k+n_k^{-i}\frac{\alpha_k}{\brac{N+1}/2}\brac{1+O\brac{\sqrt{\frac{\log J}{N^{\epsilon}}}}}}=\frac{2\max_{k\in\ebrac{1,2}}\alpha_k}{N+1}\brac{1+O\brac{\sqrt{\frac{\log J}{N^{\epsilon}}}}}.
\end{align*}
Hence we have
\begin{align*}
	\PP\brac{s_i^{**}\ne s_i}\gtrsim\EE\sqbrac{\exp\brac{-c\frac{\sum_{j=1}^J\brac{p_{1j}-p_{2j}}^2}{\max_{j,k}p_{kj}}}\II\brac{\bigcap_{l=1}^3\calA_l}}\gtrsim \exp\brac{-c_1\frac{J\brac{\alpha_1-\alpha_2}^2}{N\max_{k}\alpha_k}}.
\end{align*}
for some absolute constant $c_1>0$. We take
\begin{align*}
	\alpha_1=\frac{\brac{N-1}\theta_{\sf max}}{6}+\frac{C_0\Delta N}{\sqrt{J\brac{1+\frac{J\log J}{N\Delta^2/\theta_{\sf max}}}}},\quad \alpha_2=\frac{\brac{N-1}\theta_{\sf max}}{6},
\end{align*}
and can obtain that
\begin{align}\label{eq:lb-exp}
	\PP\brac{s_i^{**}\ne s_i}\gtrsim \exp\brac{-c_1\frac{\Delta^2}{\theta_{\sf max}\brac{1+\frac{J\log J}{N\Delta^2/\theta_{\sf max}}}}}.
\end{align}
Notice that from ${\Delta^2}/{\theta_{\sf max}}\le J\brac{\log J}/{N}$ we get that
\begin{align*}
	\frac{\Delta N}{\sqrt{J\brac{1+\frac{J\log J}{N\Delta^2/\theta_{\sf max}}}}}=o\brac{N\theta_{\sf max}}, \quad \frac{J\log J}{N\Delta^2/\theta_{\sf max}}\ge 1,
\end{align*}
which leads to  $p_{1j}\asymp p_{2j}=o(1)$ for $j\in[J]$.

To finish the proof, it suffices to bound the second term in \eqref{eq:lb-general}.   Notice that \eqref{eq:diffleDelta-small-prob} implies that 
\begin{align*}
	\pi_\bTheta\brac{\op{\theta_1-\theta_2}<\Delta}\le 2\exp\brac{-c\min\ebrac{\Delta^2N,\frac{\Delta^4N^2}{J}}}+\exp\brac{-c\log J}.
\end{align*}
Notice that
\begin{align*}
	\min\ebrac{\Delta^2N,\frac{\Delta^4N^2}{J},\log J}\ge C\frac{\Delta^2}{\theta_{\sf max}\brac{1+\frac{J\log J}{N\Delta^2/\theta_{\sf max}}}}, 
\end{align*}
for some sufficiently large constant $C>0$, which is implied by $\theta_{\sf max}\asymp N^{1-\epsilon}$, $J=\omega\brac{N}$ and
\begin{align*}
	\frac{\Delta^2}{\theta_{\sf max}}\lesssim \sqrt{\frac{J}{N}}\log J,\quad \theta^2_{\sf max}\gtrsim \frac{1}{N\log J}.
\end{align*}
Hence we can obtain that if 
\begin{align*}
	\frac{\Delta^2}{\theta_{\sf max}}\le  c_2\sqrt{\frac{J}{N}\log J},
\end{align*}
for some sufficiently small  absolute constant $c_2>0$, by \eqref{eq:lb-reduce-np} and \eqref{eq:lb-exp} we have
\begin{align*}
	\inf_{\hat T_i\in[1,2]}\EE_\pi\EE_{\bR}\ab{\hat T_i-s_i}\ge \PP\brac{s_i^{**}\ne s_i}\gtrsim \exp\brac{-c_1\frac{\Delta^2}{\theta_{\sf max}\brac{1+\frac{J\log J}{N\Delta^2/\theta_{\sf max}}}}}\ge c_3.
\end{align*}
for some absolute constant $c_3\in(0,1)$.

\section{Proofs in Section \ref{sec:est-theta}}\label{sec:proof_est}
\subsection{General Versions of Results in Section \ref{sec:est-theta}}
 \begin{theorem}\label{thm:Thetaerr-ave}
	Suppose the conditions of Theorem \ref{thm:exp-clustering-error} for exact recovery and Assumption \ref{cond:iden-theta} hold,  
then we have with probability exceeding $1-O\brac{\brac{N+J}^{-20}}$,  
	\begin{align}\label{eq:thetaer-event-gen}
		\min_{\bPi\in \SS_{K}}\op{\hat\bTheta-\bTheta\bPi}_{\sf max}&\le C\mu_\omega\mu^{1/2}_{\bTheta}\kappa^6K^{3/2}\brac{\frac{\omega_*}{\omega_{\sf min}}}^2\sqrt{\frac{\omega_{\sf max}\theta_{\sf max}\log(N+J)}{N\wedge J}}+\frac{K\log(N+J)}{N\omega_{\sf min}},
	\end{align}
	for some large constant $C>0$, where $\SS_K$ stands for the set of $K\times K$ permutation matrices.
\end{theorem}

\begin{theorem}\label{thm:Thetaerr-gen-inf}
Suppose the conditions of Theorem \ref{thm:exp-clustering-error} for exact recovery  and Assumption \ref{cond:iden-theta} hold.  In addition, assume that $MK=o\brac{\brac{\log\brac{N+J}}^{2/5}}$,  $J\gtrsim \mu_\omega^4\mu^3_{\bTheta}\kappa^{18}K^4\brac{\frac{\omega_{\sf max}}{\omega_{\sf min}}}^2\brac{\frac{\theta_{\sf max}}{\theta^*_{\sf min}}}\log^2(N+J)$ and there exist some absolute constant $C_{\sf inf}>0$ such that 
\begin{align*}
	\frac{\Delta^2}{\theta_{\sf max}}\ge \frac{C_{\sf inf}\mu_\omega^4\mu^2_{\bTheta}\kappa^{18}K^3}{\omega_{\sf max}} \brac{\frac{\omega_{\sf max}}{\omega_{\sf min}}}^6\brac{\frac{\theta_{\sf max}}{\theta^*_{\sf min}}}\brac{\frac{J}{N}+\frac{N}{J}}\log^3(N+J),
\end{align*}
then we have
\begin{align*}
    \sup_{\calC\in\scrC^{MK}}\ab{\PP\brac{\bSigma^{-1/2}_{\bTheta,\calJ_0}\textsf{vec}\brac{\hat \bTheta^\top_{\calJ_0,:}-\bTheta^\top _{\calJ_0,:}}\in \calC}-   \PP\brac{\calN\brac{0,\bI_{MK}}\in \calC}}\lesssim \frac{\brac{MK}^{5/4}}{\sqrt{\log\brac{N+J}}},
\end{align*}
where $\scrC^{MK}$ is the set of all convex sets in $\RR^{MK}$ and  $\bSigma_{\bTheta,\calJ_0}$ is a diagonal matrix defined as
\begin{align*}
	\bSigma_{\bTheta,\calJ_0}:=\begin{bmatrix}
		\bSigma_{\bTheta,j_1}&&\\
		&\ddots&\\
		&&\bSigma_{\bTheta,j_{M}}
	\end{bmatrix}\in \RR^{MK\times MK}
\end{align*}
and $\bSigma_{\bTheta,j}:=\textsf{diag}\brac{\left\{\sigma^2_{j,k}\right\}_{k=1,\cdots,K}}$ for $j\in\calJ_+$.  Moreover, the same conclusion continues to hold if we replace $\bSigma_{\bTheta,\calJ_0}$ by its plug-in estimator $\hat \bSigma_{\bTheta,\calJ_0}:=\textsf{diag}\brac{\left\{\hat \sigma^2_{j_m,k}\right\}_{m\in[M],k\in[K]}}$.
\end{theorem}

\begin{theorem}\label{thm:test-theta-gen}
Suppose the conditions of Theorem \ref{thm:exp-clustering-error} for exact recovery  and Assumption \ref{cond:iden-theta} hold.  In addition, assume that $J\gtrsim \mu_\omega^4\mu^3_{\bTheta}\kappa^{18}K^4\brac{\frac{\omega_{\sf max}}{\omega_{\sf min}}}^2\brac{\frac{\theta_{\sf max}}{\theta^*_{\sf min}}}\log^2(N+J)$ and there exist some absolute constant $C_{\sf inf}>0$ such that 
\begin{align}\label{eq:theta-test-cond-gen}
	\frac{\Delta^2}{\theta_{\sf max}}\ge \frac{C_{\sf inf}\mu_\omega^4\mu^2_{\bTheta}\kappa^{18}K^3}{\omega_{\sf max}} \brac{\frac{\omega_{\sf max}}{\omega_{\sf min}}}^6\brac{\frac{\theta_{\sf max}}{\theta^*_{\sf min}}}\brac{\frac{J}{N}+\frac{N}{J}}\log^3(N+J),
\end{align}
then 
\begin{enumerate}
	\item[(i)] under the null hypothesis $H_0$, 
 \begin{itemize}
     \item if $MK^2=O(1)$, we have 
\begin{align*}
	&\sup_{t\in \RR}\ab{\Prob\brac{T\le t}-\sqbrac{\Prob\brac{\chi_1^2\le t}}^{M{K\choose 2}}}=o(1).
\end{align*}
\item  if $MK^2\rightarrow \infty$ and $MK^2=o\brac{\log^{1/2}\brac{N+J}}$, we have  
	\begin{align*}
	\sup_{t\in\RR}\ab{\Prob\brac{\frac{T-c_{M,K}}{2}\le t}-\calG(t)}=o(1),
\end{align*}
where $\calG(x):=\exp(-e^{-x})$ is the Gumbel distribution and 
\begin{align*}
	c_{M,K}:=2\brac{\log M+\log {K\choose 2}}-\log\brac{\log M+\log {K\choose 2}}-\log \pi.
\end{align*}
 \end{itemize}
	\item[(ii)] under the alternative hypothesis $H_a$ such that 
	\begin{align}\label{eq:cn-cond-gen}
		d_N\gg \sqrt{\frac{\theta_{\sf max}\log \log(N+J)}{N}},
	\end{align}
	then we have for any constant $C>0$,
	\begin{align*}
		\Prob\brac{T>C}=1-o(1).
	\end{align*}
\end{enumerate}
\end{theorem}
{\color{black}{
\begin{remark}
    The high power dependency in Theorems \ref{thm:Thetaerr-gen-inf} and \ref{thm:test-theta-gen} on $\kappa$ arises from two main factors. First, the singular space perturbation developed in \cite{yan2021inference} contributes significantly to the power dependency on $\kappa$. Second, repeated applications of Lemma \ref{lem:incoherence} to translate conditions on $\sigma_\star$ to those on $\Delta$, and Lemma \ref{lem:incoherence-par} to translate bounds in terms of  $\mu_1,\mu_2,\mu_3$ to those of $\mu_{\bTheta},\kappa,K$ contribute to the high power dependency on $\kappa$. While it is possible to present our theoretical results in a more sophisticated form with lower power dependency on $\kappa$ by using $\sigma_\star$ and $\mu_1,\mu_2,\mu_3$, it might blur the interpretability of our result.
\end{remark}
}}
\subsection{Preliminary Results for  Section \ref{sec:est-theta}}

\paragraph{Event $\calB_{\sf exact}$.} By Theorem \ref{thm:exp-clustering-error}, there exists an event $\calB_{\sf exact}$ such that $h(\hat s,s)=0$ with $\PP\brac{\calB_{\sf exact}}\ge 1-\brac{N+J}^{-20}$.

\paragraph{Event $\wt \calB_{\sf good}$.}
For convenience, we first state a sufficient condition that guarantee all SNR requirements in the following discussion:
\begin{align}\label{eq:signal-cond-strong}
	N\theta_{\sf max}\omega_{\sf max}\gtrsim \brac{\frac{\omega_{\sf max}}{\omega_{\sf min}}}\brac{\frac{\theta_{\sf max}}{\theta^*_{\sf min}}}K\log(N+J)
\end{align}
\begin{align*}
	\frac{\Delta^2}{\theta_{\sf max}}\gtrsim  \frac{\mu_{\omega}\mu^2_{\bTheta}\kappa^6K}{\omega_*\log\brac{N+J}}\brac{\frac{J}{N}+\frac{N}{J}}+\frac{\mu^3_{\omega}\mu_{\bTheta}\kappa^{12}K^2}{\omega_*}\log\brac{N+J}.
\end{align*}
\begin{itemize}
	\item By Lemma \ref{lem:two-inf-bound}, we have  with probability at least $1-(N+J)^{-20}$ that
\begin{align}\label{eq:event-wt-good-decomp}
	\hat \bU\bO^\top   -\bU=\bE\bV\bSigma^{-1}+\calH(\bE\bE^\top)\bU\bSigma^{-2}+\mPsi,
\end{align}
where
\begin{align*}
	&\op{\mPsi}_{2,\infty}\lesssim \kappa^2\frac{\xi_{\sf err}}{\sigma_K^2(\bR^*)}\frac{\mu K}{N}+\kappa^2\frac{\xi^2_{\sf err}}{\sigma_K^4(\bR^*)}\sqrt{\frac{\mu K}{N}},\\
	&\op{\hat\bU-\bU\bO}_{2,\infty}\lesssim \kappa^2\frac{\xi_{\sf err}}{\sigma^2_K(\bR^*)}\sqrt{\frac{\mu K}{N}}.
\end{align*} 
Note that  
\begin{align}\label{eq:xi-err-bound}
	\xi_{\sf err}\lesssim \sigma_1(\bR^*)\sqrt{\omega_{\sf max}\theta_{\sf max}N\log(N+J)}
\end{align}
under the condition  
\begin{align*}
	\frac{\Delta^2}{\theta_{\sf max}}\gtrsim \frac{\mu_{\sf \omega}K}{\omega_*}\frac{J}{N}\log(N+J).
\end{align*}
This leads to 
\begin{align*}
	\op{\hat\bU-\bU\bO}_{2,\infty}\lesssim \mu_\omega\mu^{1/2}_{\bTheta}\kappa^5K\frac{\sqrt{\omega_{\sf max}\theta_{\sf max}\log(N+J)}}{\sigma_K(\bR^*)}
\end{align*}
\begin{align*}
	\op{\mPsi}_{2,\infty}\lesssim \kappa^3\frac{\sqrt{N\omega_{\sf max}\theta_{\sf max}\log(N+J)}}{\sigma_K(\bR^*)}\frac{\mu K}{N}+\kappa^4\frac{N\omega_{\sf max}\theta_{\sf max}\log(N+J)}{\sigma_K^2(\bR^*)}\sqrt{\frac{\mu K}{N}}
\end{align*}
Furthermore, we can see from the proof of Theorem \ref{thm:exp-clustering-error} that with probability at least $1-(N+J)^{-20}$,
\begin{align*}
	\infn{\calH(\bE\bE^\top)\bU\bSigma^{-2}}&\lesssim \infn{\bU}\frac{\sqrt{N\omega_{\sf max}\theta_{\sf max}}}{\sigma^2_K(\bR^*)}\log^{{3/2}}(N+J)+\frac{K\sqrt{J}\omega_{\sf max}\theta_{\sf max}}{\sigma^2_K(\bR^*)}\log^{1/2}(N+J)
\end{align*}

Thereby we can redefine the residual matrix as  $\wt\mPsi:=\calH(\bE\bE^\top)\bU\bSigma^{-2}+\mPsi$ which satisfies
\begin{align}\label{eq:sec-order-bound}
	\infn{\wt\mPsi}&\le\kappa^3\mu^{1/2}K^{1/2}\frac{\sqrt{\omega_{\sf max}\theta_{\sf max}\log(N+J)}}{\sigma_K(\bR^*)}\brac{\sqrt{\frac{\mu K}{N}}+\frac{\log \brac{N+J}}{\sigma_K(\bR^*)}}\notag\\
	&+\kappa^4{\mu}^{1/2} K\frac{\sqrt{N+J}\omega_{\sf max}\theta_{\sf max}\log(N+J)}{\sigma_K^2(\bR^*)}\notag\\
	&\lesssim\frac{\kappa^3 \mu K}{\sigma_K\brac{\bR^*}}\sqrt{\frac{\omega_{\sf max}\theta_{\sf max}\log(N+J)}{N}}+\kappa^4{\mu}^{1/2} K\frac{\sqrt{N+J}\omega_{\sf max}\theta_{\sf max}\log(N+J)}{\sigma_K^2(\bR^*)}
\end{align}
where the second inequality holds due to 
\begin{align*}
    \sqrt{\brac{N+J}\omega_{\sf max}\theta_{\sf max}}\gtrsim \sqrt{\log\brac{N+J}}
\end{align*}
under the SNR condition \eqref{eq:signal-cond-strong}.
Thus we have the following  decomposition
\begin{align}\label{eq:explicit-sim-decomp}
	\hat \bU\bO^\top   -\bU=\bE\bV\bSigma^{-1}+\wt\mPsi
\end{align}
	\item Note that for any $i\in[N]$,
\begin{align}\label{eq:omega-crude-bound}
	\ab{\frac{1}{\hat \omega_i}-\frac{1}{\omega_i}}&=\frac{1}{\ab{ \calC_k}^{1/2}}\frac{1}{\op{\hat \bU_{i,:}}\op{\bU_{i,:}}}\ab{\op{\hat\bU_{i,:}}-\op{\bU_{i,:}}}\lesssim \brac{\frac{\omega_*}{\omega_{\sf min}}}^2\frac{\mu_\omega \mu^{1/2}_{\bTheta}\kappa^4K^{1/2}\xi_{\sf err}}{\sigma^2_K(\bR^*)}
\end{align}
	\item By random matrix theory, we have that the noise matrix $\bE$ satisfies  
\begin{align}\label{eq:prop-Eop}
	\op{\bE}\lesssim \sqrt{(J+N)\omega_{\sf max}\theta_{\sf max}}
\end{align}
with probability at least $1-(N+J)^{-20}$.
By Wedin's sin theorem and the proof of Theorem 9 in  \citep{yan2021inference}, we can directly obtain the following facts:
\begin{align}\label{eq:prop-Vop}
	\op{\hat\bV\hat\bV^\top-\bV\bV^\top }\lesssim\xi^\bV_{\sf op}:=\frac{ \sqrt{(J+N)\omega_{\sf max}\theta_{\sf max}}}{\sigma_K(\bR^*)}
\end{align}
and
\begin{align}\label{eq:prop-V2}
	&\max\left\{\infn{\hat\bV\bH_{\bV}-\bV},\infn{\hat\bV-\bV\bO_{\bV}}\right\}\notag\\
	&\lesssim\frac{\sqrt{\omega_{\sf max}\theta_{\sf max}K\log (N+J)}}{\sigma_K(\bR^*)}+\frac{\mu_{\bTheta}\kappa^2 (J+N)\omega_{\sf max}\theta_{\sf max}}{\sigma^2_K(\bR^*)}\sqrt{\frac{K}{J}}\notag\\
	&\lesssim \xi^{\bV}_{2,\infty }:=\frac{\sqrt{\omega_{\sf max}\theta_{\sf max}K\log (N+J)}}{\sigma_K(\bR^*)}
\end{align}
where the last inequality holds provided that 
\begin{align*}
	\frac{\Delta^2}{\theta_{\sf max}}\gtrsim \frac{\mu_{\omega}\mu^2_{\bTheta}\kappa^6K}{\omega_*\log\brac{N+J}}\brac{\frac{J}{N}+\frac{N}{J}}.
\end{align*}
For notational consistency, we denote 
\begin{align*}
	\xi^\bU_{\sf op}:=\frac{\xi_{\sf err}}{\sigma^2_K(\bR^*)},\quad \xi^{\bU}_{2,\infty }:=\kappa^2\frac{\xi_{\sf err}}{\sigma^2_K(\bR^*)}\sqrt{\frac{\mu K}{N}}
\end{align*}
It is readily seen that by definition of $\xi^{\bU}_{\sf op}$  and  $\xi^{\bV}_{\sf op}$ that 
\begin{align}\label{eq:prop-UVop}
	\max\left\{\xi^{\bU}_{\sf op},\xi^{\bV}_{\sf op}\right \}\lesssim \xi_{\sf op}:=\frac{ \sqrt{\omega_{\sf max}\theta_{\sf max}(J+\kappa^2N\log(N+J))}}{\sigma_{K}(\bR^*)}
\end{align}
	\item In addition, we have
\begin{align*}
	\infn{\hat \bU}\le \infn{\hat \bU-\bU\bO_{\bU}}+\infn{\bU}\lesssim \infn{\bU}\le \sqrt{\frac{\mu_1 K}{N}}
\end{align*}
provided that 
\begin{align*}
	\frac{\Delta^2}{\theta_{\sf max}}\gtrsim \frac{\mu^3_{\omega}\mu_{\bTheta}\kappa^{12}K^2}{\omega_*}\log\brac{N+J},
\end{align*}
and 
\begin{align*}
	\infn{\hat \bV}\le \infn{\hat \bV-\bV\bO_\bV}+\infn{\bV}\lesssim  \sqrt{\frac{\mu_2 K}{J}}
\end{align*}
provided that 
\begin{align*}
	\frac{\Delta^2}{\theta_{\sf max}}\gtrsim \frac{\mu_{\omega}\kappa^2K}{\omega_*}\frac{J}{N}\log\brac{N+J}.
\end{align*} 
\end{itemize}

The event $\wt \calB_{\sf good}$ is  defined as $\wt \calB_{\sf good}:=\left\{\eqref{eq:sec-order-bound}, \eqref{eq:explicit-sim-decomp},\eqref{eq:prop-Eop}\text{~hold}\right \}$ with $\PP\brac{\wt \calB_{\sf good}}\ge 1-3\brac{N+J}^{-20}$.
\subsection{Proof of Theorem \ref{thm:Thetaerr-ave}}
Our analysis is conducted on the event $\calB_{\sf exact}\cap \wt \calB_{\sf good}$. 
 Without loss of generality, we can assume the permutation that achieved minimum in $h(\hat s,s)$ is the identity map.  By \eqref{eq:omega-crude-bound}, we have for any $i\in[N]$,
\begin{align*}
	\ab{\frac{1}{\hat \omega_i}-\frac{1}{\omega_i}}\lesssim \brac{\frac{\omega_*}{\omega_{\sf min}}}^2\frac{\mu_\omega \mu^{1/2}_{\bTheta}\kappa^4K^{1/2}\xi_{\sf err}}{\sigma^2_K(\bR^*)}
\end{align*}
where the last inequality holds due to  $\wt \calB_{\sf good}$. Then for any $j\in[J]$ and $k\in[K]$, we have 
\begin{align}\notag
	&\quad \ab{e_k^\top \brac{\hat \bTheta-\bTheta}^\top e_j}\\
 &=\ab{e_k^\top\brac{ \bZ^\top  \bZ}^{-1} \bZ^\top \brac{{\hat \bOmega}^{-1}\bR-\bOmega^{-1}\bR^*} e_j}=\frac{1}{\ab{\calC_k}}\ab{\sum_{i\in\calC_k}\brac{\frac{R_{i,j}}{\hat\omega_i}-\frac{R^*_{i,j}}{\omega_i}}}\notag\\
	&=\frac{1}{\ab{\calC_k}}\ab{\sum_{i\in\calC_k}\left[\brac{\frac{1}{\hat\omega_i}-\frac{1}{\omega_i}}R_{i,j}+\frac{R_{i,j}-R^*_{i,j}}{\omega_i}\right]}\notag\\
 \label{eq:theta-decomp}
	&\le \max_{l\in[N]}\ab{\frac{1}{\hat\omega_l}-\frac{1}{\omega_l}}\ab{\frac{1}{\ab{\calC_k}}\sum_{i\in\calC_k}R^*_{i,j}}+\max_{l\in[N]}\ab{\frac{1}{\hat\omega_l}-\frac{1}{\omega_l}}\ab{\frac{1}{\ab{\calC_k}}\sum_{i\in\calC_k}E_{i,j}}+\ab{\frac{1}{\ab{\calC_k}}\sum_{i\in\calC_k}\frac{E_{i,j}}{\omega_i}}
\end{align}
We will bound each term in \eqref{eq:theta-decomp}. The first term can be bounded as 
\begin{align*}
	 \max_{l\in[N]}&\ab{\frac{1}{\hat\omega_l}-\frac{1}{\omega_l}}\ab{\frac{1}{\ab{\calC_k}}\sum_{i\in\calC_k}R^*_{i,j}}\overset{\eqref{eq:omega-crude-bound}}{\lesssim} \brac{\frac{\omega_*}{\omega_{\sf min}}}^2\frac{\mu_\omega\mu^{1/2}_{\bTheta}\kappa^4 K^{1/2}\xi_{\sf err}}{\sigma^2_K(\bR^*)}\infn{\bU}\infn{\bV}\sigma_1(\bR^*)\\
	 &\overset{\eqref{eq:xi-err-bound}}{\lesssim} \brac{\frac{\omega_*}{\omega_{\sf min}}}^2\frac{\mu_\omega^2\mu^{1/2}_{\bTheta}\kappa^5K^{3/2}}{\sigma_K(\bR^*)}\sqrt{\frac{1}{NJ}}\sigma_1(\bR^*)\sqrt{\omega_{\sf max}\theta_{\sf max}N\log(N+J)}\\
	 &\lesssim \brac{\frac{\omega_*}{\omega_{\sf min}}}^2\mu_\omega^2\mu^{1/2}_{\bTheta}\kappa^6K^{3/2}\sqrt{\frac{\omega_{\sf max}\theta_{\sf max}\log(N+J)}{J}}
\end{align*}
The second term in \eqref{eq:theta-decomp} can be bounded as 
\begin{align*}
	\max_{l\in[N]}\ab{\frac{1}{\hat\omega_l}-\frac{1}{\omega_l}}\ab{\frac{1}{\ab{\calC_k}}\sum_{i\in\calC_k}E_{i,j}}&\lesssim\brac{\frac{\omega_*}{\omega_{\sf min}}}^2\frac{\mu_\omega\mu^{1/2}_{\bTheta}\kappa^4K^{1/2}\xi_{\sf err}}{\sigma^2_K(\bR^*)}\brac{\sqrt{\frac{\omega_{\sf max}\theta_{\sf max}\log(N+J)}{N/K}}+\frac{K\log(N+J)}{N}}.
\end{align*}
where the first inequality holds with probability at least $1-\brac{N+J}^{-20}$ by Bernstein's inequality and \eqref{eq:omega-crude-bound}. 
It remains to bound the third term in \eqref{eq:theta-decomp} by Bernstein's inequality such that with probability at least $1-(N+J)^{-20}$, 
\begin{align*}
	\ab{\frac{1}{\ab{\calC_k}}\sum_{i\in\calC_k}\frac{E_{i,j}}{\omega_i}}&\lesssim  \sqrt{\frac{\theta_{\sf max}\log(N+J)}{\omega_{**}N/K}}+\frac{\log(N+J)}{\omega_{\sf min}N/K}\\
 &\lesssim K^{1/2}\brac{\frac{\omega_{*}}{\omega_{**}}}\sqrt{\frac{\omega_{\sf max}\theta_{\sf max}\log(N+J)}{N}}+\frac{K\log(N+J)}{N\omega_{\sf min}}
\end{align*}
 Here, $\omega_{**}:=\max_{k\in [K]}\brac{\frac{1}{|\calC_k|}\sum_{i\in\calC_k}\omega_i^{-1}}^{-1}$, i.e., the maximum harmonic mean of $\omega_i$ across all clusters.
 Collecting all pieces we obtain the desired bound:
 \begin{align*}
 	\ab{e_k^\top \brac{\hat \bTheta-\bTheta}^\top e_j}\lesssim  \brac{\frac{\omega_*}{\omega_{\sf min}}}^2\mu_\omega^2\mu^{1/2}_{\bTheta}\kappa^6K^{3/2}\sqrt{\frac{\omega_{\sf max}\theta_{\sf max}\log(N+J)}{N\wedge J}}+\frac{K\log(N+J)}{N\omega_{\sf min}}.
 \end{align*}
The proof is completed by a standard union bound argument on $\calB_{\sf exact}\cap \wt \calB_{\sf good}$. \hfill$\square$

\subsection{Proof of Theorem \ref{thm:est-theta-lb}}
It suffices to consider $\calP_1\brac{s,\bI_N,\bTheta}$ (Bernoulli location model) where $\bTheta\in[0,1]^{J\times 1}$, as the rate for $\calP_1\brac{s,\bI_N,\bTheta}$  constitutes a lower bound for $\calP_2\brac{s,\bOmega,\bTheta}$ (DhLCM). Let $\delta\in(0,1)$ and define
\begin{align*}
	\bTheta^{(1)}=\theta_*\boldsymbol{1}+\frac{1}{2}\sqrt{\frac{\theta_*}{N}\log\brac{\frac{1}{4\delta\brac{1-\delta}}}}\be_1,\quad 	\bTheta^{(2)}=\theta_*\boldsymbol{1}.
\end{align*}
where $\theta_*=o(1)$. We thus have $\op{\bTheta^{(1)}-\bTheta^{(2)}}_{\sf max}\ge \sqrt{\frac{\theta_*}{N}\log\brac{\frac{1}{4\delta\brac{1-\delta}}}}$ and 
\begin{align*}
	\textsf{D}_{\sf KL}\brac{\bP_{\bTheta^{(1)}}\|\bP_{\bTheta^{(2)}}}&=\sum_{l\in[N]}\sum_{j\in[J]}\textsf{D}_{\sf KL}\brac{\textsf{Bern}\brac{\theta^{(1)}_{j}}\|\textsf{Bern}\brac{\theta^{(2)}_{j}}}\\
 &\le\frac{N\fro{\bTheta^{(1)}-\bTheta^{(2)}}^2}{\theta_*\brac{1-\theta_*}}\le  \log\brac{\frac{1}{4\delta\brac{1-\delta}}}.
\end{align*}
Applying Bretagnolle–Huber inequality leads to 
\begin{align}\label{eq:TV-KL-upper}
	\textsf{D}_{\sf TV}\brac{\bP_{\bTheta^{(1)}}\|\bP_{\bTheta^{(2)}}}\le \sqrt{1-\exp\brac{-\textsf{D}_{\sf KL}\brac{\bP_{\bTheta^{(1)}}\|\bP_{\bTheta^{(2)}}}}}\le 1-2\delta.
\end{align}
On the other hand, define
\begin{align*}
	\wt\bTheta^{(1)}=\wt \theta_*\boldsymbol{1}+\frac{1-2\delta}{2N}\be_1,\quad 	\wt\bTheta^{(2)}=\wt \theta_*\boldsymbol{1}.
\end{align*}
where $\wt \theta_*=(0,1/2)$, we have $\op{\bTheta^{(1)}-\bTheta^{(2)}}_{\sf max}\ge \frac{1-2\delta}{2N}$ and
\begin{align}\label{eq:TV-upper}
	\textsf{D}_{\sf TV}\brac{\bP_{\wt\bTheta^{(1)}}\|\bP_{\wt\bTheta^{(2)}}}=\sum_{l\in[N]}\sum_{j\in[J]}\textsf{D}_{\sf TV}\brac{\textsf{Bern}\brac{\wt\theta^{(1)}_{j}}\|\textsf{Bern}\brac{\wt\theta^{(2)}_{j}}}\le 1-2\delta.
\end{align}
Collecting the bounds for \eqref{eq:TV-KL-upper} and \eqref{eq:TV-upper} with $\delta=1/4$ and applying two-point Le Cam bound (e.g., eq. (15.14) in \cite{wainwright2019high}) we can complete the proof. \hfill$\square$

\subsection{Proof of Theorem \ref{thm:Thetaerr-gen-inf}}
For any $j\in \calJ_0$, denote $\calI_{j}:=\brac{\hat \bTheta-\bTheta}^\top e_j$. We then  have 
\begin{align*}
	\calI_{j}\II_{\calB_{\sf exact}\cap \wt\calB_{\sf good}}&=\underbrace{\brac{\frac{1}{\ab{\calC_1}}\sum_{i\in\calC_1}\frac{E_{i,j}}{\omega_i},\cdots,\frac{1}{\ab{\calC_K}}\sum_{i\in\calC_K}\frac{E_{i,j}}{\omega_i}}^\top}_{=:\calI_{j,0}}\\
	&+\underbrace{\brac{\frac{1}{\ab{\calC_1}}\sum_{i\in\calC_1}\brac{\frac{1}{\hat\omega_i}-\frac{1}{\omega_i}}\brac{R^*_{i,j}+E_{i,j}},\cdots,\frac{1}{\ab{\calC_K}}\sum_{i\in\calC_K}\brac{\frac{1}{\hat\omega_i}-\frac{1}{\omega_i}}\brac{R^*_{i,j}+E_{i,j}}}^\top}_{=:\calI_{j,1}}.
\end{align*}
Define
\begin{align*}
	\bar\bE_{i,j}:=\brac{\boldsymbol{0}^\top_{s_i-1},\frac{1}{\ab{\calC_{s_i}}}\frac{E_{i,j}}{\omega_i},\boldsymbol{0}^\top_{K-s_i}}^\top\in\RR^{K},\quad i\in[N],j\in\calJ_0.
\end{align*} 
It is readily seen that  $\calI_{j,0}=\sum_{i\in[N]}\bar\bE_{i,j}$ for $j\in\calJ_0$. Moreover, we have the following decomposition:
\begin{align*}
	\textsf{vec}\brac{\hat \bTheta^\top_{\calJ_0,:}-\bTheta^\top _{\calJ_0,:}}&=\textsf{vec}\brac{\hat \bTheta^\top_{\calJ_0,:}-\bTheta^\top _{\calJ_0,:}}\II_{\calB_{\sf exact}\cap \wt\calB_{\sf good}}+\textsf{vec}\brac{\hat \bTheta^\top_{\calJ_0,:}-\bTheta^\top _{\calJ_0,:}}\II_{\calB^c_{\sf exact}\cup \wt\calB^c_{\sf good}}\\
	&=\bcalI_0+\bcalI_1+\textsf{vec}\brac{\hat \bTheta^\top_{\calJ_0,:}-\bTheta^\top _{\calJ_0,:}}\II_{\calB^c_{\sf exact}\cup \wt\calB^c_{\sf good}},
\end{align*}
where we define $\bcalI_l:=\brac{\calI^\top _{j_1,l},\calI^\top _{j_2,l},\cdots,\calI^\top _{j_{M,l}}}^\top\in \RR^{MK}$ for $l=0,1$. We thereby can write $\bcalI_0=\sum_{i\in[N]}\bB_i$ where $\bB_i:=\brac{\bar\bE^\top_{i,j_1},\cdots,\bar\bE^\top_{i,j_M}}^\top\in\RR^{MK}$. By construction, $\bB_i$'s are independent, $\EE\bB_i=0$ and $\textsf{Var}\brac{\bSigma_{\bTheta,\calJ_0}^{-1/2}\bB_i}=\bI_{MK}$. Moreover, we have
\begin{align*}
	\sum_{i\in[N]}\EE{\op{\bSigma_{\bTheta,\calJ_0}^{-1/2}\bB_i}^3}=\sum_{m\in[M]}\sum_{i
\in[N]}\EE\ab{\frac{E_{i,j_m}}{\omega_i\ab{\calC_{s_i}}\sigma_{j_m,s_i}}}^3=\sum_{m\in[M]}\sum_{k\in[K]}\sum_{i
\in\calC_k}\EE\ab{\frac{E_{i,j_m}}{\omega_i\ab{\calC_k}\sigma_{j_m,k}}}^3\lesssim\frac{MK}{\sqrt{\log\brac{N+J}}}.
\end{align*}
where the last inequality holds due to \eqref{eq:theta-sup-error}  in the proof of Lemma \ref{lem:Thetaerr-inf}. By Theorem 1 in \cite{raivc2019multivariate}, we can conclude that 
\begin{align}\label{eq:I0-sup-bound}
	\sup_{\calC\in\scrC^{MK}}\ab{\PP\brac{\bSigma_{\bTheta,\calJ_0}^{-1/2}\bcalI_0\in\calC}-\PP\brac{\calN\brac{0,\bI_{MK}}\in \calC}}\le \frac{C_0\brac{MK}^{5/4}}{\sqrt{\log\brac{N+J}}}.
\end{align}
for some absolute constant $C_0>0$.

Observe that 
\begin{align*}
	\op{\bSigma_{\bTheta,\calJ_0}^{-1/2}\bcalI_1}&\le \sum_{m=1}^{M}\op{\bSigma^{-1/2}_{\bTheta,j_m}\calI^\top _{j_m,1}}=\sum_{m=1}^{M}\sqrt{\sum_{k=1}^K\sigma^{-2}_{j_{m},k}\brac{\frac{1}{\ab{\calC_k}}\sum_{i\in\calC_k}\brac{\frac{1}{\hat\omega_i}-\frac{1}{\omega_i}}\brac{R^*_{i,j}+E_{i,j}}}^2}.
\end{align*}
By \eqref{eq:theta-res-bound}, we obtain that
\begin{align*}
	\PP\brac{\op{\bSigma_{\bTheta,\calJ_0}^{-1/2}\bcalI_1}\ge \frac{C_1MK^{1/2}}{\sqrt{\log \brac{N+J}}}}=O\brac{\brac{N+J}^{-20}}.
\end{align*}
for some absolute constant $C_1>0$. On the other hand, let $\bcalI_2:=\bSigma_{\bTheta,\calJ_0}^{-1/2}\textsf{vec}\brac{\hat \bTheta^\top_{\calJ_0,:}-\bTheta^\top _{\calJ_0,:}}\II_{\calB^c_{\sf exact}\cup \wt\calB^c_{\sf good}}$, by properties of $\calB_{\sf exact}$ and $\calB_{\sf good}$ we have
\begin{align*}
\PP\brac{\op{\bSigma_{\bTheta,\calJ_0}^{-1/2}\bcalI_2}=0}=O\brac{\brac{N+J}^{-20}}.
\end{align*} 
In the following, we introduce some additional notation that used in \cite{raivc2019multivariate}.
For any point $x\in\RR^{MK}$ and any non-empty convex set $\calC\in\scrC^{MK}$ with $\calC\ne \RR^{MK}$, define
\begin{align*}
	\delta_{\calC}\brac{x}:=\begin{cases}
		-\textsf{dist}\brac{x,\RR^{MK}\backslash\calC},& \text{if~}x\in\calC;\\
		\textsf{dist}\brac{x,\calC},& \text{if~}x\notin\calC.
	\end{cases}
\end{align*}
where $\textsf{dist}\brac{x,\calC}$ is the Euclidean distance between $x$ and $\calC$. In addition, for any non-empty convex set $\calC\in\scrC^{MK}$ with $\calC\ne \RR^{MK}$ and $\varepsilon$, let $\calC^{\varepsilon}:=\ebrac{x\in\RR^{MK}:\delta_{\calC}\brac{x}\le \varepsilon}$ and define $\emptyset^{\varepsilon}:=\emptyset$ and $\brac{\RR^{MK}}^{\varepsilon}=\RR^{MK}$.

Now define $\varepsilon_0:=\frac{C_1MK^{1/2}}{\sqrt{\log \brac{N+J}}}$. For any convex set $\calC\in\RR^{MK}$, we have
\begin{align}\label{eq:C-eps-bound}
	\PP\brac{\bSigma_{\bTheta,\calJ_0}^{-1/2}\bcalI_0\in\calC^{-\varepsilon_0}}&=\PP\brac{\bSigma_{\bTheta,\calJ_0}^{-1/2}\bcalI_0\in\calC^{-\varepsilon_0}, \op{\bSigma_{\bTheta,\calJ_0}^{-1/2}\brac{\bcalI_1+\bcalI_2}}\le \varepsilon_0}\notag\\
	&+\PP\brac{\bSigma_{\bTheta,\calJ_0}^{-1/2}\bcalI_0\in\calC^{-\varepsilon_0},\op{\bSigma_{\bTheta,\calJ_0}^{-1/2}\brac{\bcalI_1+\bcalI_2}}> \varepsilon_0}\notag\\
	&\le \PP\brac{\bSigma_{\bTheta,\calJ_0}^{-1/2}\textsf{vec}\brac{\hat \bTheta^\top_{\calJ_0,:}-\bTheta^\top _{\calJ_0,:}}\in \calC}+O\brac{\brac{N+J}^{-20}}.
\end{align}
On the other hand, we have
\begin{align}\label{eq:C-bound}
		\PP\brac{\bSigma_{\bTheta,\calJ_0}^{-1/2}\textsf{vec}\brac{\hat \bTheta^\top_{\calJ_0,:}-\bTheta^\top _{\calJ_0,:}}\in \calC}&=\PP\brac{\bSigma_{\bTheta,\calJ_0}^{-1/2}\textsf{vec}\brac{\hat \bTheta^\top_{\calJ_0,:}-\bTheta^\top _{\calJ_0,:}}\in \calC,\op{\bSigma_{\bTheta,\calJ_0}^{-1/2}\brac{\bcalI_1+\bcalI_2}}\le \varepsilon_0 }\notag\\
	&+\PP\brac{\bSigma_{\bTheta,\calJ_0}^{-1/2}\textsf{vec}\brac{\hat \bTheta^\top_{\calJ_0,:}-\bTheta^\top _{\calJ_0,:}}\in \calC,	\op{\bSigma_{\bTheta,\calJ_0}^{-1/2}\brac{\bcalI_1+\bcalI_2}}> \varepsilon_0}\notag\\
	&\le \PP\brac{\bSigma_{\bTheta,\calJ_0}^{-1/2}\bcalI_0\in \calC^{\varepsilon_0}}+O\brac{\brac{N+J}^{-20}}.
\end{align}
In addition, we have
\begin{align}\label{eq:I0-upper-bound}
	 \PP\brac{\bSigma_{\bTheta,\calJ_0}^{-1/2}\bcalI_0\in \calC^{\varepsilon_0}}&\le \PP\brac{\calN\brac{0,\bI_{MK}}\in \calC^\varepsilon_0}+\frac{C_0\brac{MK}^{5/4}}{\sqrt{\log\brac{N+J}}}\notag\\
	 &\le \PP\brac{\calN\brac{0,\bI_{MK}}\in \calC}+\varepsilon_0\brac{0.59\brac{MK}^{1/4}+0.21}+\frac{C_0\brac{MK}^{5/4}}{\sqrt{\log\brac{N+J}}}\notag\\
	 &\le \PP\brac{\calN\brac{0,\bI_{MK}}\in \calC}+\frac{C_2\brac{MK}^{5/4}}{\sqrt{\log\brac{N+J}}}.
\end{align}
where the first inequality holds due to \eqref{eq:I0-sup-bound} and the second inequality holds due to Theorem 2 in \cite{raivc2019multivariate}. Similarly, we could obtain that 
\begin{align}\label{eq:I0-lower-bound}
	 \PP\brac{\bSigma_{\bTheta,\calJ_0}^{-1/2}\bcalI_0\in \calC^{-\varepsilon_0}}&\ge  \PP\brac{\calN\brac{0,\bI_{MK}}\in \calC}-\frac{C_2\brac{MK}^{5/4}}{\sqrt{\log\brac{N+J}}}.
\end{align}
Combining \eqref{eq:C-eps-bound}, \eqref{eq:C-bound}, \eqref{eq:I0-upper-bound} and \eqref{eq:I0-lower-bound} and taking supremum over $\calC\in\scrC^{MK}$, we can conclude that 
\begin{align}\label{eq:first-claim}
	\sup_{\calC\in\scrC^{MK}}\ab{\PP\brac{\bSigma_{\bTheta,\calJ_0}^{-1/2}\textsf{vec}\brac{\hat \bTheta^\top_{\calJ_0,:}-\bTheta^\top _{\calJ_0,:}}\in \calC}-  \PP\brac{\calN\brac{0,\bI_{MK}}\in \calC}}\le  \frac{C_3\brac{MK}^{5/4}}{\sqrt{\log\brac{N+J}}}.
\end{align}
It remains to show that the above inequality holds up to constants by substituting $\bSigma_{\bTheta,\calJ_0}$ with $\hat \bSigma_{\bTheta,\calJ_0}$. By construction and Lemma \ref{lem:consistent-sigma}, we have that 
\begin{align}\label{eq:bSigmahat-error}
	\PP\brac{\fro{ \bSigma_{\bTheta,\calJ_0}^{-1} \hat\bSigma_{\bTheta,\calJ_0}-\bI_{MK}}\ge  \frac{C_4\brac{MK}^{1/2}}{\sqrt{\log\brac{N+J}}}}\le O\brac{\brac{N+J}^{-19}}.
\end{align}
Let $\calE_\bTheta:=\ebrac{\fro{ \bSigma_{\bTheta,\calJ_0}^{-1} \hat\bSigma_{\bTheta,\calJ_0}-\bI_{MK}}\ge  \frac{C_4\brac{MK}^{1/2}}{\sqrt{\log\brac{N+J}}}}$. By $MK=o\brac{\log\brac{N+J}}$, we can conclude that $\hat \bSigma_{\bTheta,\calJ_0}\succ 0$ on $\calE_{\bTheta}$. By Theorem 21 in \cite{yan2021inference} (Theorem 1.1 in \cite{devroye2018total}), we obtain that 
\begin{align*}
	\sup_{\calC\in \scrC^{MK}}&\ab{\PP\brac{\calN\brac{0,\hat \bSigma_{\bTheta,\calJ_0}}\in \calC\big |\hat \bSigma_{\bTheta,\calJ_0}}-\PP\brac{\calN\brac{0,\bSigma_{\bTheta,\calJ_0}}\in \calC\big |\hat \bSigma_{\bTheta,\calJ_0}}}\\
	&\lesssim \fro{ \bSigma_{\bTheta,\calJ_0}^{-1} \hat\bSigma_{\bTheta,\calJ_0}-\bI_{MK}}\wedge 1+2\II_{\calE^c_\bTheta}.
\end{align*}
Using the above inequality, we could derive that 
\begin{align*}
	\PP\brac{\calN\brac{0,\bSigma_{\bTheta,\calJ_0}}\in \calC}
	&= \EE\sqbrac{\PP\brac{\calN\brac{0,\bSigma_{\bTheta,\calJ_0}}\in \calC\Big|\hat \bSigma_{\bTheta,\calJ_0}}}\\
	&\le \EE\sqbrac{\PP\brac{\calN\brac{0,\hat \bSigma_{\bTheta,\calJ_0}}\in \calC\Big|\hat \bSigma_{\bTheta,\calJ_0}}+\fro{ \bSigma_{\bTheta,\calJ_0}^{-1} \hat\bSigma_{\bTheta,\calJ_0}-\bI_{MK}}\wedge 1+2\II_{\calE^c_\bTheta}}\\
	&\le \PP\brac{\calN\brac{0,\hat \bSigma_{\bTheta,\calJ_0}}\in \calC}+\frac{C_5\brac{MK}^{1/2}}{\sqrt{\log\brac{N+J}}}.
\end{align*}
and similarly, that 
\begin{align*}
	\PP\brac{\calN\brac{0,\hat \bSigma_{\bTheta,\calJ_0}}\in \calC}
	&\ge  \PP\brac{\calN\brac{0, \bSigma_{\bTheta,\calJ_0}}\in \calC}-\frac{C_5\brac{MK}^{1/2}}{\sqrt{\log\brac{N+J}}}.
\end{align*}
We thus obtain that 
\begin{align}\label{eq:normal-bSigma-bound}
	\sup_{\calC\in\scrC^{MK}}\ab{\PP\brac{\calN\brac{0,\bSigma_{\bTheta,\calJ_0}}\in \calC}-	\PP\brac{\calN\brac{0,\hat \bSigma_{\bTheta,\calJ_0}}\in \calC}}\le \frac{C_5\brac{MK}^{1/2}}{\sqrt{\log\brac{N+J}}}.
\end{align}
In addition, by $MK=o\brac{\log\brac{N+J}}$ and \eqref{eq:bSigmahat-error}, we can conclude that 
\begin{align}\label{eq:rank-def-bound}
	\PP\brac{\textsf{rank}\brac{\hat \bSigma_{\bTheta,\calJ_0}}<MK}\le O\brac{\brac{N+J}^{-19}}.
\end{align}
Now for any $\calC\in\scrC^{MK}$, it follows that 
\begin{align*}
	&\ab{\PP\brac{\hat \bSigma^{-1/2}_{\bTheta,\calJ_0}\textsf{vec}\brac{\hat \bTheta^\top_{\calJ_0,:}-\bTheta^\top _{\calJ_0,:}}\in \calC}-   \PP\brac{\calN\brac{0,\bI_{MK}}\in \calC}}\\
	&\overset{\eqref{eq:bSigmahat-error}}{\le}\left|\PP\brac{\textsf{vec}\brac{\hat \bTheta^\top_{\calJ_0,:}-\bTheta^\top _{\calJ_0,:}}\in \hat \bSigma^{1/2}_{\bTheta,\calJ_0}\calC,\calE_\bTheta\text{~holds}}-  \PP\brac{\calN\brac{0,\hat \bSigma_{\bTheta,\calJ_0}}\in\hat \bSigma^{1/2}_{\bTheta,\calJ_0}\calC,\calE_\bTheta\text{~holds}}\right|\\
 &+O\brac{\brac{N+J}^{-19}}\\
	&\le\sup_{\calC\in\scrC^{MK}}\left|\PP\brac{\textsf{vec}\brac{\hat \bTheta^\top_{\calJ_0,:}-\bTheta^\top _{\calJ_0,:}}\in \calC}-  \PP\brac{\calN\brac{0, \hat \bSigma_{\bTheta,\calJ_0}}\in\calC}\right|+O\brac{\brac{N+J}^{-19}}\\
	&\overset{\eqref{eq:normal-bSigma-bound}}{\le} \sup_{\calC\in\scrC^{MK}}\left|\PP\brac{\textsf{vec}\brac{\hat \bTheta^\top_{\calJ_0,:}-\bTheta^\top _{\calJ_0,:}}\in \calC}-  \PP\brac{\calN\brac{0,  \bSigma_{\bTheta,\calJ_0}}\in\calC}\right|+\frac{C_5\brac{MK}^{1/2}}{\sqrt{\log\brac{N+J}}}+O\brac{\brac{N+J}^{-19}}\\
	&\le \sup_{\calC\in\scrC^{MK}}\ab{\PP\brac{\bSigma_{\bTheta,\calJ_0}^{-1/2}\textsf{vec}\brac{\hat \bTheta^\top_{\calJ_0,:}-\bTheta^\top _{\calJ_0,:}}\in \calC}-  \PP\brac{\calN\brac{0,\bI_{MK}}\in \calC}}+\frac{C_6\brac{MK}^{1/2}}{\sqrt{\log\brac{N+J}}}\\
	&\overset{\eqref{eq:first-claim}}{\le}  \frac{C_7\brac{MK}^{5/4}}{\sqrt{\log\brac{N+J}}}.
\end{align*}
The proof is completed by taking supremum over $\calC\in\scrC^{MK}$.
\hfill$\square$

\subsection{Proof of Theorem \ref{thm:test-theta-gen}}
We first show part (i) with $MK^2\rightarrow \infty$ and $MK^2=o\brac{\log^{1/2}\brac{N+J}}$. Note that for any $j\in\calJ_0$ and $k_1,k_2\in[K]$, we have
\begin{align}\label{eq:k1k2-decomp}    &\brac{\sigma_{j,k_1}^{2}+\sigma_{j,k_2}^{2}}^{-1/2}\brac{\hat\theta_{j,k_1}-\hat\theta_{j,k_2}}\II_{\calB_{\sf exact}\cap \wt\calB_{\sf good}}\notag\\
&=\frac{\brac{\theta_{j,k_1}-\theta_{j,k_2}}\II_{\calB_{\sf exact}\cap \wt\calB_{\sf good}}}{\sqrt{\sigma_{j,k_1}^{2}+\sigma_{j,k_2}^{2}}}+\frac{1}{\sqrt{\sigma_{j,k_1}^{2}+\sigma_{j,k_2}^{2}}}\sum_{l=1,2}\frac{(-1)^{l-1}}{\ab{\calC_{k_l}}}\sum_{i\in\calC_{k_l}}\left[\brac{\frac{1}{\hat\omega_i}-\frac{1}{\omega_i}}\brac{R^*_{i,j}+E_{i,j}}+\frac{E_{i,j}}{\omega_i}\right]\notag\\
	&=\frac{\brac{\theta_{j,k_1}-\theta_{j,k_2}}\II_{\calB_{\sf exact}\cap \wt\calB_{\sf good}}}{\sqrt{\sigma_{j,k_1}^{2}+\sigma_{j,k_2}^{2}}}\notag\\
 &+\frac{1}{\sqrt{\sigma_{j,k_1}^{2}+\sigma_{j,k_2}^{2}}}\sum_{l=1,2}(-1)^{l-1}\sqbrac{\frac{1}{\ab{\calC_{k_l}}}\sum_{i\in\calC_{k_l}}\frac{E_{i,j}}{\omega_i}+\frac{1}{\ab{\calC_{k_l}}}\sum_{i\in\calC_{k_l}}\brac{\frac{1}{\hat\omega_i}-\frac{1}{\omega_i}}\brac{R^*_{i,j}+E_{i,j}}}
\end{align}

Under $H_0$, \eqref{eq:k1k2-decomp} can be simplified as 
\begin{align}\label{eq:theta-exact-decomp-gen}
&\brac{\sigma_{j,k_1}^{2}+\sigma_{j,k_2}^{2}}^{-1/2}\brac{\hat\theta_{j,k_1}-\hat\theta_{j,k_2}}\II_{\calB_{\sf exact}\cap \wt\calB_{\sf good}}\notag\\
&=\frac{1}{\sqrt{\sigma_{j,k_1}^{2}+\sigma_{j,k_2}^{2}}}\sum_{l=1,2}(-1)^{l-1}\sqbrac{\frac{1}{\ab{\calC_{k_l}}}\sum_{i\in\calC_{k_l}}\frac{E_{i,j}}{\omega_i}+\frac{1}{\ab{\calC_{k_l}}}\sum_{i\in\calC_{k_l}}\brac{\frac{1}{\hat\omega_i}-\frac{1}{\omega_i}}\brac{R^*_{i,j}+E_{i,j}}}
\end{align}
Moreover, by definition of $\sigma_{j,k}^2$ and Assumption \ref{cond:balanced}
we have
\begin{align}\label{eq:sigma2-lb-gen}
    \sigma_{j,k_1}^2+\sigma_{j,k_2}^2\gtrsim\frac{1}{\omega_{**}}\brac{\frac{\theta_{j,k_1}}{\ab{\calC_{k_1}}}+\frac{\theta_{j,k_2}}{\ab{\calC_{k_2}}}}\gtrsim\frac{\theta^*_{\sf min }K}{N\omega_{**}}
\end{align}
Following the same arguments in the proof of Lemma \ref{lem:Thetaerr-inf} and utilizing \eqref{eq:sigma2-lb-gen}, we arrive at
\begin{align*}
    \frac{1}{\sqrt{\sigma_{j,k_1}^{2}+\sigma_{j,k_2}^{2}}}\ab{\sum_{l=1,2}\frac{(-1)^{l-1}}{\ab{\calC_{k_l}}}\sum_{i\in\calC_{k_l}}\brac{\frac{1}{\hat\omega_i}-\frac{1}{\omega_i}}\brac{R^*_{i,j}+E_{i,j}}}\lesssim\frac{1}{\sqrt{\log \brac{N+J}}}
\end{align*}
with probability at least $1-O\brac{\brac{N+J}^{-20}}$ provided that \eqref{eq:theta-test-cond-gen} holds. We thus obtain that 
\begin{align*}
&\brac{\sigma_{j,k_1}^{2}+\sigma_{j,k_2}^{2}}^{-1/2}\brac{\hat\theta_{j,k_1}-\hat\theta_{j,k_2}}\II_{\calB_{\sf exact}\cap \wt\calB_{\sf good}}\notag\\
&=\frac{1}{\sqrt{\sigma_{j,k_1}^{2}+\sigma_{j,k_2}^{2}}}\sum_{l=1,2}\frac{(-1)^{l-1}}{\ab{\calC_{k_l}}}\sum_{i\in\calC_{k_l}}\frac{E_{i,j}}{\omega_i}+O\brac{\frac{1}{\sqrt{\log\brac{N+J}}}}
\end{align*}
with probability at least $1-O\brac{\brac{N+J}^{-20}}$. This gives that 
\begin{align}\label{eq:Tj-decomp-gen}
T\II_{\calB_{\sf exact}\cap \wt\calB_{\sf good}}=\max_{j\in\calJ_0}\max_{k_1<k_2\in[K]}\sqbrac{\frac{1}{\sqrt{\sigma_{j,k_1}^{2}+\sigma_{j,k_2}^{2}}}\sum_{l=1,2}\frac{(-1)^{l-1}}{\ab{\calC_{k_l}}}\sum_{i\in\calC_{k_l}}\frac{E_{i,j}}{\omega_i}}^2+O\brac{\frac{1}{\sqrt{\log\brac{N+J}}}}
\end{align}
with probability at least $1-O\brac{\brac{N+J}^{-20}}$. Moreover, by Berry-Esseen Theorem we have  
\begin{align*}
    \sup_{t\in\RR}&\ab{\PP\brac{\sqbrac{\frac{1}{\sqrt{\sigma_{j,k_1}^{2}+\sigma_{j,k_2}^{2}}}\sum_{l=1,2}\frac{(-1)^{l-1}}{\ab{\calC_{k_l}}}\sum_{i\in\calC_{k_l}}\frac{E_{i,j}}{\omega_i}}^2\le t}-\PP\brac{\chi_1^2\le  t}}\\
&\lesssim\sum_{l=1,2}\sum_{i\in\calC_{k_l}}\EE\ab{\frac{E_{i,j}}{\omega_i\ab{\calC_{k_l}}\brac{\sigma_{j,k_1}^{2}+\sigma_{j,k_2}^{2}}^{1/2}}}^3\\
&\lesssim\frac{1}{\omega_{\sf min}\brac{\sigma_{j,k_1}^{2}+\sigma_{j,k_2}^{2}}^{1/2}\min_{l=1,2}\ab{\calC_{k_l}}}\\
&\overset{\eqref{eq:sigma2-lb-gen}}{\lesssim}\sqrt{\frac{\omega_{**}}{\omega_{\sf min}}}\sqrt{\frac{K}{N\theta_{\sf min}\omega_{\sf min}}}\lesssim\frac{1}{\sqrt{\log\brac{N+J}}}
\end{align*}

This implies the random variables $\frac{1}{\sqrt{\sigma_{j,k_1}^{2}+\sigma_{j,k_2}^{2}}}\sum_{l=1,2}\frac{(-1)^{l-1}}{\ab{\calC_{k_l}}}\sum_{i\in\calC_{k_l}}\frac{E_{i,j}}{\omega_i}$ in \eqref{eq:Tj-decomp-gen} with $j\in \calJ_0$ and $k_1\ne k_2\in[K]$ are independent and asymptotically distributed as $\chi_1^2$. We thereby have the following for any $t\in\RR$,
\begin{align}\label{eq:gumbel-approx}
	&\Prob\brac{\max_{j\in\calJ_0}\max_{k_1< k_2\in[K]}\sqbrac{\frac{1}{\sqrt{\sigma_{j,k_1}^{2}+\sigma_{j,k_2}^{2}}}\sum_{l=1,2}\frac{(-1)^{l-1}}{\ab{\calC_{k_l}}}\sum_{i\in\calC_{k_l}}\frac{E_{i,j}}{\omega_i}}^2\le 2t+c_{M,K}}\notag\\
	&=\sqbrac{\Prob\brac{\chi_1^2\le 2t+c_{M,K}}+O\brac{\frac{1}{\sqrt{\log\brac{N+J}}}}}^{M{K\choose 2}}\notag\\
 &=\sqbrac{\Prob\brac{\chi_1^2\le 2t+c_{M,K}}}^{M{K\choose 2}}+O\brac{\frac{MK^2}{\sqrt{\log\brac{N+J}}}}\notag\\
 &\rightarrow \exp\brac{-e^{-t}},
\end{align}
where the last step is due to the fact that the maximum of i.i.d. $\chi^2_1$ random variables converges weakly to the Gumbel distribution $\calG$ (c.f. Table 3.4.4 of \cite{embrechts2013modelling}), provided that $MK^2\rightarrow \infty$. On the other hand, we have
\begin{align}\label{eq:T-exact-event-small}
	\ab{\PP\brac{T\le  2t+c_{M,K}}-\PP\brac{T\II_{\calB_{\sf exact}\cap \wt\calB_{\sf good}}\le  2t+c_{M,K}}}=O\brac{\brac{N+J}^{-20}}
\end{align}
Combining \eqref{eq:Tj-decomp-gen},  \eqref{eq:gumbel-approx} and \eqref{eq:T-exact-event-small}, we obtain that the desired result under $H_0$ when $MK^2\rightarrow\infty$ and $MK^2=o\brac{\log^{1/2}\brac{N+J}}$. When $MK=O(1)$, the proof is identical except that we only need the following intermediate result instead of \eqref{eq:gumbel-approx}:
\begin{align}\label{eq:finite-approx}
	&\Prob\brac{\max_{j\in\calJ_0}\max_{k_1< k_2\in[K]}\sqbrac{\frac{1}{\sqrt{\sigma_{j,k_1}^{2}+\sigma_{j,k_2}^{2}}}\sum_{l=1,2}\frac{(-1)^{l-1}}{\ab{\calC_{k_l}}}\sum_{i\in\calC_{k_l}}\frac{E_{i,j}}{\omega_i}}^2\le t}\notag\\
 &=\sqbrac{\Prob\brac{\chi_1^2\le t}}^{M{K\choose 2}}+O\brac{\frac{1}{\sqrt{\log\brac{N+J}}}}
\end{align}
Combining \eqref{eq:Tj-decomp-gen}, \eqref{eq:T-exact-event-small} and \eqref{eq:finite-approx}, we finish the proof for part (i).

Under the alternative $H_a$, \eqref{eq:k1k2-decomp} still holds  for any $j\in\calJ_+$ and $k_1,k_2\in[K]$. Now fix $j\in\calJ_0$ and $k_1\ne k_2\in[K]$ such that $\ab{\theta_{j,k_1}-\theta_{j,k_2}}\ge d_N$, then we have
\begin{align}\label{eq:theta-power-sec-gen}
	&\frac{1}{\sqrt{\hat\sigma^2_{j,k_1}+\hat\sigma^2_{j,k_2}}}\ab{\frac{1}{\ab{\calC_{k_1}}}\sum_{i\in\calC_{k_1}}\frac{E_{i,j}}{\omega_i}+\frac{1}{\ab{\calC_{k_1}}}\sum_{i\in\calC_{k_1}}\brac{\frac{1}{\hat\omega_i}-\frac{1}{\omega_i}}\brac{R^*_{i,j}+E_{i,j}}}\notag\\
	&=\frac{\sqrt{\sigma^2_{j,k_1}+\sigma^2_{j,k_2}}}{\sqrt{\hat\sigma^2_{j,k_1}+\hat\sigma^2_{j,k_2}}}\frac{\sigma_{j,k_1}}{\sqrt{\sigma^2_{j,k_1}+\sigma^2_{j,k_2}}}\frac{1}{\sigma_{j,k_1}}\ab{\frac{1}{\ab{\calC_{k_1}}}\sum_{i\in\calC_{k_1}}\frac{E_{i,j}}{\omega_i}+\frac{1}{\ab{\calC_{k_1}}}\sum_{i\in\calC_{k_1}}\brac{\frac{1}{\hat\omega_i}-\frac{1}{\omega_i}}\brac{R^*_{i,j}+E_{i,j}}}
\end{align}
Combining Lemma \ref{lem:consistent-sigma} and the proof of Lemma \ref{lem:Thetaerr-inf}, in particular \eqref{eq:theta-main-dist} and \eqref{eq:theta-res-bound}, we can deduce that under SNR condition \eqref{eq:theta-test-cond-gen}, the term of \eqref{eq:theta-power-sec-gen} is upper bounded by $C\sqrt{\log \log(N+J)}$ with probability at least $1-O\brac{1/\sqrt{\log \brac{N+J}}}$ for some absolute constant $C>0$. It suffices to show the first term in  \eqref{eq:k1k2-decomp} diverge faster than $\sqrt{\log\log\brac{N+J}}$. Observe that 
\begin{align}\label{eq:theta-power-first-gen}
	\frac{1}{\sqrt{\hat\sigma^2_{j,k_1}+\hat\sigma^2_{j,k_2}}}\brac{\theta_{j,k_1}-\theta_{j,k_2}}=\frac{\sqrt{\sigma^2_{j,k_1}+\sigma^2_{j,k_2}}}{\sqrt{\hat\sigma^2_{j,k_1}+\hat\sigma^2_{j,k_2}}}\frac{1}{\sqrt{\sigma^2_{j,k_1}+\sigma^2_{j,k_2}}}\brac{\theta_{j,k_1}-\theta_{j,k_2}}
\end{align}
By \eqref{eq:theta-power-first-gen} and Lemma  \ref{lem:consistent-sigma}, we obtain that 
\begin{align*}
\frac{1}{\sqrt{\hat\sigma^2_{j,k_1}+\hat\sigma^2_{j,k_2}}}\ab{\theta_{j,k_1}-\theta_{j,k_2}}\ge 3C \sqrt{\log \log(N+J)}
\end{align*}
with probability exceeding $1-O\brac{\brac{N+J}^{-20}}$, provided that 
\begin{align*}
	\frac{N\omega_{\sf max}\brac{\theta_{j,k_1}-\theta_{j,k_2}}^2}{\theta_{j,k_1}\vee \theta_{j,k_2}}\ge \wt C \brac{\frac{\omega_{\sf max}}{\sf \omega_{**}}}K\log \log(N+J)
\end{align*}
for some sufficiently large constant $\wt C$ depending only on $C$, which is further guaranteed by \eqref{eq:cn-cond-gen}. In conclusion, we obtain that 
\begin{align*}
		&\Prob\brac{T^{1/2}\ge \sqrt{\log\log\brac{N+J}}}\ge \Prob\brac{T^{1/2}_j(k_1,k_2)\ge \sqrt{\log\log\brac{N+J}}}\\
  &\ge \Prob\brac{T^{1/2}_j(k_1,k_2)\II_{\calB_{\sf exact}\cap \wt\calB_{\sf good}}\ge \sqrt{\log\log\brac{N+J}}}=1-O\brac{\frac{1}{\sqrt{\log (N+J)}}}.
\end{align*}
\hfill$\square$

\subsection{Proof of Theorem \ref{thm:fdr-control}}
To show the first part of the lemma, it suffices to consider $\ab{\calN_{\calJ_0}}\ge 1$. Let $r$ be  the number of rejections, then we have
\begin{align*}
\textsf{FDP}&=\sum_{j\in\calN_{\calJ_0}}\frac{\II\brac{H_{0,j}\text{~is~rejected}}}{1\vee r}=\sum_{j\in\calN_{\calJ_0}}\sum_{q=1}^{M}\frac{\II\brac{T_j>\chi^2_{1,\beta_1\brac{\alpha_0\cdot q/M}}}}{q}\II\brac{r=q}
\end{align*}
Notice that from the proof of Theorem \ref{thm:test-theta-gen}, we have with probability exceeding $1-O\brac{\brac{N+J}^{-20}}$ that 
\begin{align}\label{eq:Tj-decomp}
    T_j=\max_{k_1<k_2\in[K]}\sqbrac{\frac{1}{\sqrt{\sigma_{j,k_1}^{2}+\sigma_{j,k_2}^{2}}}\sum_{l=1,2}\frac{(-1)^{l-1}}{\ab{\calC_{k_l}}}\sum_{i\in\calC_{k_l}}\frac{E_{i,j}}{\omega_i}}^2+O\brac{\frac{1}{\sqrt{\log\brac{N+J}}}}
\end{align}
To keep notation's simplicity,  we denote the first maximum term in \eqref{eq:Tj-decomp} as $T^*_j$. Using \eqref{eq:Tj-decomp}, we can continue as with probability exceeding $1-O\brac{\brac{N+J}^{-19}}$,
\begin{align}\label{eq:FDP-hp-decomp}
\textsf{FDP}&=\sum_{j\in\calN_{\calJ_0}}\sum_{q=1}^{M}\frac{\II\brac{T^*_j>\chi^2_{1,\beta_1\brac{\alpha_0\cdot q/M}} \brac{1-O\brac{\frac{1}{\chi^2_{1,\beta_1\brac{\alpha_0\cdot q/M}}\sqrt{\log\brac{N+J}}}}}}}{ q}\II\brac{r=q}\notag\\
&=\sum_{j\in\calN_{\calJ_0}}\sum_{q=1}^{M}\frac{\II\brac{T^*_j>\chi^2_{1,\beta_1\brac{\alpha_0\cdot q/M}} \brac{1-O\brac{\frac{1}{\chi^2_{1,\beta_1\brac{\alpha_0\cdot q/M}}\sqrt{\log\brac{N+J}}}}}}}{q}\II\brac{r\brac{T^*_j}=q}
\end{align}
\sloppy where $r\brac{T^*_j}$ is the number of rejections we get if we set $T^*_j=\infty$ and the rest of $T^*_{j^\prime}$'s unchanged.  The last equality holds since for $r=q$, if  $T^*_j>\chi^2_{1,\beta_1\brac{\alpha_0\cdot q/M}} \brac{1-O\brac{\frac{1}{\chi^2_{1,\beta_1\brac{\alpha_0\cdot q/M}}\sqrt{\log\brac{N+J}}}}}$, then  $r\brac{T_j^*}=r=q$ and hence
\begin{align*}
    \II\brac{r=q}=\II\brac{r\brac{T_j^*}=q}.
\end{align*}
On the other hand, if  $T^*_j\le \chi^2_{1,\beta_1\brac{\alpha_0\cdot q/M}} \brac{1-O\brac{\frac{1}{\chi^2_{1,\beta_1\brac{\alpha_0\cdot q/M}}\sqrt{\log\brac{N+J}}}}}$, then 
\begin{align*}
    \II\brac{T^*_j> \chi^2_{1,\beta_1\brac{\alpha_0\cdot q/M}} \brac{1-O\brac{\frac{1}{\chi^2_{1,\beta_1\brac{\alpha_0\cdot q/M}}\sqrt{\log\brac{N+J}}}}}}=0.
\end{align*}
Therefore, in both cases we have \eqref{eq:FDP-hp-decomp} holds. It suffices to calculate the  expectation on the term in \eqref{eq:FDP-hp-decomp}. For any $j\in\calN_{\calJ_0}$, we have
\begin{align}\label{eq:each-j-expect}
&\EE\sum_{q=1}^{M}\frac{\II\brac{T^*_j>\chi^2_{1,\beta_1\brac{\alpha_0\cdot q/M}} \brac{1-O\brac{\frac{1}{\chi^2_{1,\beta_1\brac{\alpha_0\cdot q/M}}\sqrt{\log\brac{N+J}}}}}}}{ q}\II\brac{r\brac{T^*_j}=q}\notag\\
&=\EE\sum_{q=1}^{M}\frac{\PP\brac{T^*_j>\chi^2_{1,\beta_1\brac{\alpha_0\cdot q/M}} \brac{1-O\brac{\frac{1}{\chi^2_{1,\beta_1\brac{\alpha_0\cdot q/M}}\sqrt{\log\brac{N+J}}}}}\Bigg| T^*_{j^\prime},j^\prime\ne j}}{q}\II\brac{r\brac{T^*_j}=q}\notag\\
&=\EE\sum_{q=1}^{M}\frac{\PP\brac{T^*_j>\chi^2_{1,\beta_1\brac{\alpha_0\cdot q/M}} \brac{1-O\brac{\frac{1}{\chi^2_{1,\beta_1\brac{\alpha_0\cdot q/M}}\sqrt{\log\brac{N+J}}}}}}}{q}\II\brac{r\brac{T^*_j}=q}
\end{align}
Let $\ebrac{Z_{k_1,k_2}, k_1<k_2\in[K]}$ be ${K\choose 2}$ independent standard normal random variables,  we then have
\begin{align*}
\PP\brac{\max_{k_1<k_2\in[K]}Z^2_{k_1,k_2}>\chi^2_{1,\beta_{1}\brac{\alpha_0\cdot q/M}}}= \alpha_0\cdot\frac{q}{M}
\end{align*}
Note that the c.d.f. of $\max_{k_1<k_2\in[K]}Z^2_{k_1,k_2}$ is continuous, we thereby conclude that
\begin{align}\label{eq:max-chi-control}
\PP\brac{\max_{k_1<k_2\in[K]}Z^2_{k_1,k_2}>\chi^2_{1,\beta_{1}\brac{\alpha_0\cdot q/M}}\brac{1-O\brac{\frac{1}{\chi^2_{1,\beta_1\brac{\alpha_0\cdot q/M}}\sqrt{\log\brac{N+J}}}}}}= \alpha_0\cdot\frac{q}{M}\brac{1+o\brac{1}}
\end{align}
On the other hand, 
\begin{align}\label{eq:max-chi-Tj-close}
\sup_{t\in\RR}\ab{\PP\brac{\max_{k_1<k_2\in[K]}Z^2_{k_1,k_2}>t}-\PP\brac{T^*_j>t}}= O\brac{\frac{K^2}{\sqrt{\log\brac{N+J}}}}
\end{align}
Using \eqref{eq:max-chi-control} and \eqref{eq:max-chi-Tj-close}, we can proceed from \eqref{eq:each-j-expect} as 
\begin{align*}
&\EE\sum_{q=1}^{M}\frac{\II\brac{T^*_j>\chi^2_{1,\beta_1\brac{\alpha_0\cdot q/M}} \brac{1-O\brac{\frac{1}{\chi^2_{1,\beta_1\brac{\alpha_0\cdot q/M}}\sqrt{\log\brac{N+J}}}}}}}{ q}\II\brac{r\brac{T^*_j}=q}\notag\\
&=\EE\frac{\alpha_0}{M}\brac{1+o\brac{1}}\sum_{q=1}^{M}\II\brac{r\brac{T^*_j}=q}=\frac{\alpha_0}{M}\brac{1+o\brac{1}}
\end{align*}
Hence we obtain that
\begin{align*}
&\textsf{FDR}=\EE\brac{\textsf{FDR}}\\
&=\EE\sum_{j\in\calN_{\calJ_0}}\sum_{q=1}^{M}\frac{\II\brac{T^*_j>\chi^2_{1,\beta_1\brac{\alpha_0\cdot q/M}} \brac{1-O\brac{\frac{1}{\chi^2_{1,\beta_1\brac{\alpha_0\cdot q/M}}\sqrt{\log\brac{N+J}}}}}}}{q}\II\brac{r\brac{T^*_j}=q}\\
&+O\brac{\brac{N+J}^{-18}}\\
&=\frac{\ab{\calN_{\calJ_0}}}{M}\cdot \alpha_0\brac{1+o\brac{1}}.
\end{align*}
\hfill$\square$

\section{Proofs in Section \ref{sec:gen-type}}\label{sec:proof_gen}
\subsection{Proof of Theorem \ref{thm:exp-clustering-error-bin}}
The proof is essentially the same as the second part in proof of Theorem \ref{thm:exp-clustering-error}. We only outline the necessary modifications here. First notice that Lemma \ref{lem:two-inf-bound} still holds under the Binomial model. Hence a sufficient condition for the SNR condition in Lemma \ref{lem:two-inf-bound} to hold shall be cast as
\begin{align*}
m^2N\sigma_\star^2\gtrsim m\theta_{\sf max}\sqrt{NJ}\log\brac{N+J}+m\sqrt{N}\sigma_\star\sqrt{mN\theta_{\sf max}\log\brac{N+J}}
\end{align*}
which is equivalent to \eqref{eq:delta-cond-bin} under Assumption \ref{cond:balanced}-\ref{cond:const-degee} and $J\gtrsim N$.
Following the arguments line by line in the second part of the proof for Theorem \ref{thm:exp-clustering-error}, we can obtain the desired result.\hfill$\square$

\subsection{Proof of Theorem \ref{thm:exp-clustering-error-poisson}}

First of all, we can apply Lemma \ref{lem:trunc} to each entry of $\bE$ with $\delta=C\brac{N+J}^{-100}$ for some sufficiently large absolute constant $C>0$, we can produce an auxiliary noise matrix $\wt \bE$ such that 
\begin{itemize}
	\item $\wt E_{i,j}$’s are independent, obey  that $\EE \wt E_{i,j}=0$ and 
	\begin{align*}
		\textsf{Var}\brac{\wt E^2_{i,j}}&=\brac{1+O\brac{{\brac{N+J}^{-50}}}}\omega_i\theta_{j,s_i}, \\
       \ab{\wt E_{i,j}}\lesssim B:&=\brac{\sqrt{\omega_i\theta_{j,s_i}\log \brac{N+J}}+\log\brac{N+J}}
	\end{align*}
 where in the bound on $\ab{E_{i,j}}$ we have used  $\theta_{j,k}\gtrsim \brac{N+J}^{-C_\theta}$ and Assumption \ref{cond:const-degee}.
	\item $\wt \bE$  is identical to  $\bE$ with high probability  in the sense that
	\begin{align*}
		\PP\brac{E_{i,j}=\wt E_{i,j}\text{~for~all~}i\in[N]\text{~and~}j\in[J]}\ge 1-O\brac{\brac{N+J}^{-98}}.
	\end{align*} 
\end{itemize}
Then notice that we have the following condition holds 
\begin{align*}
	\sqrt{\omega_i\theta_{j,s_i}\log \brac{N+J}}+\log\brac{N+J}\lesssim\frac{\sqrt{\omega_{\sf max}\theta_{\sf max}}\min\left\{J^{1/2},\brac{NJ}^{1/4}\right\}}{\sqrt{\log\brac{N+J}}},
\end{align*}
provided that
\begin{align*}
	\frac{\Delta^2}{\theta_{\sf max}}\gtrsim \brac{\sqrt{\frac{J}{N}}+1}\log^{3}\brac{N+J}.
\end{align*}
This guarantees that $\calB_{\sf good}$  holds with high probability by Theorem 10 in \cite{yan2021inference}, and the desired result follows by the same the arguments in the second part of the proof for Theorem \ref{thm:exp-clustering-error}.\hfill$\square$

\subsection{Proof of Theorem \ref{thm:test-theta-bin-poisson}}
The proof of Theorem \ref{thm:test-theta-bin-poisson} is almost identical to that of Theorem \ref{thm:test-theta-gen-simple} and Theorem \ref{thm:fdr-control}, and we only sketch the necessary modifications here. 

First, our general condition for Binomial model is
\begin{align}\label{eq:m-bound-inf}
    m\lesssim \min\ebrac{N^{1/2}, J^{1/2}}\sqrt{\frac{\omega_{\sf max}\theta_{\sf max}}{\log \brac{N+J}}}
\end{align}
For any $j\in\calJ_0$ and $k_1,k_2\in[K]$, similar to Theorem \ref{thm:test-theta-gen} we have under $H_0$,
\begin{align}\label{eq:theta-exact-decomp-gen-ext}
&\brac{\sigma_{j,k_1}^{2}+\sigma_{j,k_2}^{2}}^{-1/2}\brac{\hat\theta_{j,k_1}-\hat\theta_{j,k_2}}\II_{\calB_{\sf exact}\cap \wt\calB_{\sf good}}\notag\\
&=\frac{1}{\sqrt{\sigma_{j,k_1}^{2}+\sigma_{j,k_2}^{2}}}\sum_{l=1,2}(-1)^{l-1}\sqbrac{\frac{1}{\ab{\calC_{k_l}}}\sum_{i\in\calC_{k_l}}\frac{E_{i,j}}{\omega_i}+\frac{1}{\ab{\calC_{k_l}}}\sum_{i\in\calC_{k_l}}\brac{\frac{1}{\hat\omega_i}-\frac{1}{\omega_i}}\brac{R^*_{i,j}+E_{i,j}}}
\end{align}
It remains to modify the  arguments in the proof of Lemma \ref{lem:Thetaerr-inf} to arrive at
\begin{align}\label{eq:high-order-bound-ext}
    \frac{1}{\sqrt{\sigma_{j,k_1}^{2}+\sigma_{j,k_2}^{2}}}\ab{\sum_{l=1,2}\frac{(-1)^{l-1}}{\ab{\calC_{k_l}}}\sum_{i\in\calC_{k_l}}\brac{\frac{1}{\hat\omega_i}-\frac{1}{\omega_i}}\brac{R^*_{i,j}+E_{i,j}}}\lesssim\frac{1}{\sqrt{\log \brac{N+J}}}
\end{align}
with probability at least $1-O\brac{\brac{N+J}^{-20}}$. To this end, we split our discussion into two cases as follows.
\begin{enumerate}
    \item [(a)] Under the Binomial model \eqref{eq:model-bin}, by standard matrix tail bound (c.f. Remark 3.13 in \cite{bandeira2016sharp} together with a standard device called the ``symmetric dilation trick", Theorem 3.1.4 in \cite{chen2021spectral}) we obtain that with probability at least $1-O\brac{\brac{N+J}^{-20}}$,
    \begin{align*}
        \op{\bE}\lesssim \sqrt{m\brac{N+J}\omega_{\sf max}\theta_{\sf max}}+ m\sqrt{\log\brac{N+J}}\lesssim \sqrt{m\brac{N+J}\omega_{\sf max}\theta_{\sf max}}
    \end{align*}
    provided that \eqref{eq:m-bound-inf} holds. 
    Similar to the proof of Lemma \ref{lem:Thetaerr-inf} and notice that the bound on $\sigma_{j,k}^{-1}$ becomes
    $\sigma_{j,k}^{-1}\le \sqrt{\frac{mN\omega_{**}}{\theta^*_{\sf min }K}}$, the condition for \eqref{eq:high-order-bound-ext} to be hold becomes
    \begin{align*}
        \frac{\Delta^2}{\theta_{\sf max}}\ge \frac{C_{\sf inf}\mu_\omega^4\mu^2_{\bTheta}\kappa^{18}K^3}{\omega_{\sf max}} \brac{\frac{\omega_{\sf max}}{\omega_{\sf min}}}^6\brac{\frac{\theta_{\sf max}}{\theta^*_{\sf min}}}\brac{\frac{J}{N}+\frac{N}{J}}\log^3(N+J),
    \end{align*}
    and
    \begin{align*}
        J\gtrsim \mu_\omega^4\mu^3_{\bTheta}\kappa^{18}m K^4\brac{\frac{\omega_{\sf max}}{\omega_{\sf min}}}^2\brac{\frac{\theta_{\sf max}}{\theta^*_{\sf min}}}\log^2(N+J).
    \end{align*}
    \item[(b)] Under the Poisson model \eqref{eq:model-poisson}, we first apply  Lemma \ref{lem:trunc}, similar to the proof of Theorem \ref{thm:exp-clustering-error-poisson}, to each entry of $\bE$ with $\delta=C\brac{N+J}^{-100}$ for some sufficiently large absolute constant $C>0$, we can produce an auxiliary noise matrix $\wt \bE$ we shall focus on. Again, standard matrix tail bound gives that with probability exceeding  $1-O\brac{(N+J)^{-20}}$,
    \begin{align*}
        \op{\wt\bE}\lesssim \sqrt{\brac{N+J}\omega_{\sf max}\theta_{\sf max}}+ \log^{3/2}\brac{N+J}\lesssim \sqrt{\brac{N+J}\omega_{\sf max}\theta_{\sf max}}
    \end{align*}
    provided that $\brac{N+J}\omega_{\sf max}\theta_{\sf max}\gtrsim \log^3\brac{N+J}$.
    Moreover, we can obtain the following result, as an analogue to the bound regarding $\infn{\calH(\bE\bE^\top)\bU\bSigma^{-2}}$ in Preliminary results in Section \ref{sec:est-theta}:
    \begin{align*}
	&\infn{\calH(\wt \bE\wt \bE^\top)\bU\bSigma^{-2}}\\
 &\lesssim \infn{\bU}\frac{\sqrt{N\omega_{\sf max}\theta_{\sf max}}\brac{1+\sqrt{\omega_{\sf max}\theta_{\sf max}}}}{\sigma^2_K(\bR^*)}\log^{2}(N+J)+\frac{K\sqrt{J}\omega_{\sf max}\theta_{\sf max}}{\sigma^2_K(\bR^*)}\log^{1/2}(N+J)
    \end{align*}
    This leads to the same bound for $\infn{\wt\mPsi}$ in \eqref{eq:sec-order-bound}   if 
    \begin{align*}
       \brac{N+J}\omega_{\sf max}\theta_{\sf max}\gtrsim \log^2\brac{N+J}.
    \end{align*}
It remains to follow the proof of Lemma \ref{lem:Thetaerr-inf} to reach \eqref{eq:high-order-bound-ext}, where the condition for \eqref{eq:high-order-bound-ext} to be hold becomes
    \begin{align*}
        \frac{\Delta^2}{\theta_{\sf max}}\ge \frac{C_{\sf inf}\mu_\omega^4\mu^2_{\bTheta}\kappa^{18}K^3}{\omega_{\sf max}} \brac{\frac{\omega_{\sf max}}{\omega_{\sf min}}}^6\brac{\frac{\theta_{\sf max}}{\theta^*_{\sf min}}}\brac{\frac{J}{N}+\frac{N}{J}}\log^3(N+J),
    \end{align*}
    and
    \begin{align*}
        J\gtrsim \mu_\omega^4\mu^3_{\bTheta}\kappa^{18} K^4\brac{\frac{\omega_{\sf max}}{\omega_{\sf min}}}^2\brac{\frac{\theta_{\sf max}}{\theta^*_{\sf min}}}\log^3(N+J).
    \end{align*}
\end{enumerate}
The remaining proof is almost identical to that in the proof of Theorem \ref{thm:test-theta-gen-simple} and Theorem \ref{thm:fdr-control} and hence omitted.\hfill$\square$

\section{Details of the Iterative Algorithm for Joint MLE}\label{sec:JMLE}
Under a traditional LCM without degree heterogeneity, the joint likelihood function of latent class memberships $\mathbf{Z}$ and item parameters $\mathbf\Theta$ given binary response matrix $\mathbf R$ can be written as follows \citep{zeng2023tensor}:
\begin{align*}
    L(\mathbf Z,\mathbf \Theta\mid \mathbf R) = \sum_{i=1}^N \sum_{j=1}^J R_{i,j} \log\left(\sum_{k=1}^K Z_{i,k}\theta_{j,k}\right) + 
    \sum_{i=1}^N \sum_{j=1}^J (1-R_{i,j}) \log\left(1-\sum_{k=1}^K Z_{i,k}\theta_{j,k}\right).
\end{align*}
We next describe an iterative algorithm to maximize the above $L(\mathbf Z,\mathbf \Theta\mid \mathbf R)$.
Given $(\mathbf Z^{(t)}, \mathbf\Theta^{(t)})$ at iteration $t$, we update $\bTheta^{(t+1)} = (\theta_{j,k}^{(t+1)})$ to be: 
$$
\theta_{j,k}^{(t+1)} = \frac{\sum_{i=1}^N R_{i,j} Z_{i,k}^{(t)}}{\sum_{i=1}^N Z_{i,k}^{(t)}},\quad j\in[J],~ k\in[K],
$$
which is the maximizer of $L(\mathbf Z^{(t)},\mathbf \Theta\mid \mathbf R)$.
Then, given $\bTheta^{(t+1)}$, we further update $\mb Z^{(t)}$ to $\mb Z^{(t+1)}$ with $Z_{i,k_i}^{(t+1)} = 1$ where $k_i$ is obtained from
$$
k_i = \argmax_{k\in[K]}
\sum_{j=1}^J \Big[R_{i,j} \log\Big(\theta_{j,k}^{(t+1)}\Big) + 
    (1-R_{i,j}) \log\Big(1-\theta_{j,k}^{(t+1)}\Big)\Big],\quad \forall i\in[N].
$$
It is not hard to see that iterating the above two updates monotonically increases the function value of the joint likelihood $L(\mathbf Z,\mathbf \Theta\mid \mathbf R)$.

\section{Auxiliary Proofs of Propositions and Lemmas}\label{sec:proof_aux}
\subsection{Proof of Proposition \ref{prop:theta-beta}}
By definition, we have 
\begin{align}\label{eq:thetaTtheta-explicit}
	\EE\bTheta^\top _{j,:}\bTheta _{j,:}=\begin{bmatrix}
		\rho_J^2\brac{\frac{a}{a+b}}\brac{\frac{a+1}{a+b+1}}&\cdots& 	\rho_J^2\brac{\frac{a}{a+b}}\brac{\frac{a}{a+b}}\\
			\vdots&\ddots&\vdots\\
				\rho_J^2\brac{\frac{a}{a+b}}\brac{\frac{a}{a+b}}&\cdots &	\rho_J^2\brac{\frac{a}{a+b}}\brac{\frac{a+1}{a+b+1}}
	\end{bmatrix}=\rho_J^2\frac{a}{a+b}\brac{\frac{}{}\mathbf{1}_K\mathbf{1}_K^\top  + \frac{b}{\brac{a+b}\brac{a+b+1}}\bI_K}
\end{align}
Denote $\bD_j:=\bTheta_{j,:}^\top \bTheta_{j,:}-\EE \bTheta^\top _{j,:}\bTheta _{j,:}$ for $j\in[J]$, then $\bD_j$'s are independent mean-zero symmetric $K\times K$ matrices. In particular, 
\begin{align*}
	\op{\bD_j}\le \op{\bTheta_{j,:}^\top \bTheta_{j,:}}+\op{\EE \bTheta^\top _{j,:}\bTheta _{j,:}}&=\op{\bTheta_{j,:}}^2+\frac{a}{a+b}\cdot \brac{K+\frac{b}{\brac{a+b}\brac{a+b+1}}}\\
	&= \rho_J^2K\sqbrac{1+\frac{a}{a+b}\brac{1+\frac{b}{K\brac{a+b}\brac{a+b+1}}}}\\
	&\le 2\rho_J^2K 
\end{align*}
where the last inequality holds as 
\begin{align*}
	1+\frac{b}{K\brac{a+b}\brac{a+b+1}}<\frac{a+b}{a}
\end{align*}
Moreover, we have
\begin{align*}
	\op{\EE \bD_j^2}\le\EE  \op{\bD_j^2}\le \EE  \op{\bD_j}^2\le 4\rho_J^4 K^2
\end{align*}
Standard matrix Bernstein's inequality gives that for any $t>0$,
\begin{align*}
	\PP\brac{\op{\bTheta^\top \bTheta-\EE\bTheta^\top \bTheta}\ge t}\le K\exp\brac{-\frac{t^2}{8\rho_J^4 JK^2+4\rho_J^2Kt/3}}=:\delta_t
\end{align*}
On the other hand, \eqref{eq:thetaTtheta-explicit} implies that
\begin{align*}
	\lambda_1\brac{\EE\bTheta^\top \bTheta}\le \frac{2a}{a+b}\cdot \rho_J^2JK,\quad \lambda_K\brac{\EE\bTheta^\top \bTheta}\ge  \frac{a}{a+b}\cdot \frac{a+b+\frac{a}{a+b}}{a+b+1}\cdot \rho_J^2JK
\end{align*}
For any $t_1,t_2>0$, Weyl’s inequality gives that with probability at least $1-\delta_{t_1}-\delta_{t_2}$,  
\begin{align*}
	\ab{\lambda_1\brac{\bTheta^\top \bTheta}-\lambda_1\brac{\EE\bTheta^\top \bTheta}}\le t_1, \quad 	\ab{\lambda_K\brac{\bTheta^\top \bTheta}-\lambda_K\brac{\EE\bTheta^\top \bTheta}}\le t_2
\end{align*}
It suffices to take $t_1= \frac{a}{a+b}\cdot \rho_J^2JK$ and $t_2=  \frac{a}{2\brac{a+b}}\cdot \frac{a+b+\frac{a}{a+b}}{a+b+1}\cdot \rho_J^2JK$, which leads to
\begin{align*}
	\kappa\brac{\bTheta^\top \bTheta}=\frac{\lambda_1\brac{\bTheta^\top \bTheta}}{\lambda_K\brac{\bTheta^\top \bTheta}}\le \frac{6\brac{a+b+1}}{a+b+\frac{a}{a+b}}
\end{align*}
with probability at least
\begin{align*}
	1-2K\sqbrac{\exp\brac{-\frac{a^2\brac{a+b+\frac{a}{a+b}}^2JK}{64\brac{a+b}^2\brac{a+b+1}^2}}+\exp\brac{-\frac{3a\brac{a+b+\frac{a}{a+b}}J}{16\brac{a+b}\brac{a+b+1}}}}
\end{align*}\hfill$\square$

\subsection{Proof of Proposition \ref{prop:theta-incoherent}}
Let $B$ be some r.v. such that $B\sim \textsf{Beta}\brac{a,b}$. Notice that for any $j\in[J]$,
\begin{align*}
	\EE\op{\bTheta_{j,:}}^2=\EE\sum_{k=1}^K\theta^2_{j,k}=\rho_J^2K\EE B^2=\rho_J^2K\frac{a\brac{a+1}}{\brac{a+b}\brac{a+b+1}}
\end{align*}
and
\begin{align*}
	\textsf{Var}\brac{\sum_{k=1}^K\theta^2_{j,k}}&=\rho_J^4K\textsf{Var}\brac{B^2}=\rho_J^4K\sqbrac{\EE B^4-\brac{\EE B^2}^2}\\
	&=\rho_J^4K\sqbrac{\frac{a\brac{a+1}}{\brac{a+b}\brac{a+b+1}}\brac{\frac{\brac{a+2}\brac{a+3}}{\brac{a+b+2}\brac{a+b+3}}-\frac{a\brac{a+1}}{\brac{a+b}\brac{a+b+1}}}}\\
	&\le \rho_J^4K \frac{a}{a+b}
\end{align*}
Hence we obtain that
\begin{align*}
	\Prob\brac{\ab{\op{\bTheta_{j,:}}^2-\EE\op{\bTheta_{j,:}}^2}\ge t}\le \exp\brac{-\frac{t^2}{\rho_J^4K \frac{a}{a+b}+\rho_J^2t}}
\end{align*}
Taking $t=C\brac{\rho_J^2\sqrt{\frac{a}{a+b}K\log J}+\rho_J^2\log J}$ for some large constant $C>0$ and applying a union bound over $j\in[J]$, we obtain that with probability at least $1-O\brac{J^{-20}}$,
\begin{align*}
	\infn{\bTheta}^2\lesssim  \rho_J^2\brac{\frac{a\brac{a+1}}{\brac{a+b}\brac{a+b+1}}K+\sqrt{\frac{a}{a+b}K\log J}+\log J}\lesssim \rho_J^2\brac{\frac{a}{a+b}K+\log J}
\end{align*}
On the other hand, we have $\infn{\bTheta}^2\le K$. We thereby get
\begin{align*}
	\infn{\bTheta}^2\lesssim  \rho_J^2\brac{\frac{a}{a+b}K+\frac{K}{\rho_J^2}\wedge \log J}
\end{align*}
We can similarly obtain that 
\begin{align*}
	\EE\fro{\bTheta}^2=\EE\sum_{j=1}^J\sum_{k=1}^K\theta^2_{j,k}=\rho_J^2JK\EE B^2=\rho_J^2JK\frac{a\brac{a+1}}{\brac{a+b}\brac{a+b+1}}
\end{align*}
and
\begin{align*}
	\textsf{Var}\brac{\sum_{j=1}^J\sum_{k=1}^K\theta^2_{j,k}}&=\rho_J^4JK\textsf{Var}\brac{B^2}\le \rho_J^4JK \frac{a}{a+b}
\end{align*}
Hence we reach with probability at least $1-O\brac{J^{-20}}$,
\begin{align*}
	\fro{\bTheta}^2\gtrsim  \rho_J^2\brac{\frac{a\brac{a+1}}{\brac{a+b}\brac{a+b+1}}JK-\sqrt{\frac{a}{a+b}JK\log J}-\log J}\gtrsim \rho_J^2JK\frac{a}{a+b}
\end{align*}
Therefore, we conclude that
\begin{align*}
	\frac{J\infn{\bTheta}^2}{\fro{\bTheta}^2}\lesssim \frac{\frac{a}{a+b}JK+J\min\left\{K\rho_J^{-2},\log J\right\}}{\frac{a}{a+b}JK}=1+\frac{a+b}{b}\min\left\{\rho_J^{-2},\frac{\log J}{K}\right\}
\end{align*}
with probability at least $1-O\brac{J^{-20}}$. \hfill$\square$

\subsection{Proof of Proposition \ref{prop:identifiability}}
Suppose there is another parameter set $\brac{\wt\bOmega,\wt\bTheta}$ with corresponding $\wt \bR^*$ such that  $$\wt \omega_i\wt \theta_{j, s_i}=\omega_i\theta_{j,s_i},\quad \forall i\in[N], j\in[J]$$
Then we must have $\wt \bR^*$ has exactly the same SVD as $\bR^*$ since $\wt \bR^*= \bR^*$. By Lemma \ref{lem:U-prop} and Assumption \ref{cond:iden-theta}, we obtain
\begin{align*}
	\frac{\omega_i}{\sqrt{\ab{\calC_{s_i}}}}=\frac{\wt \omega_i}{\sqrt{\ab{\calC_{ s_i}}}},\quad \forall i\in[N]
\end{align*}
and hence $\omega_i=\wt \omega_i$ for all $i\in[N]$. We thereby have $\theta_{j,s_i}=\wt \theta_{j,s_i}$ for all $i\in[N]$ and $j\in[J]$, which collectively leads to $\brac{\wt \bOmega,\wt \bTheta}=\brac{\bOmega,\bTheta}$.\hfill$\square$

\subsection{Proof of Lemma \ref{lem:U-prop}}
By definition, Eq.~\eqref{eq:model-original} can be rewritten as 
\begin{align*}
	\bR^*=\bOmega\bZ\bW^{-1}\bW\bTheta^\top =\bOmega\bZ\bW^{-1}\bU^\dag\bSigma^\dag\bV^{\dag\top}.
\end{align*}
As a consequence, we have that 
\begin{align*}
	(\bOmega\bZ\bW^{-1}\bU^\dag)^\top \bOmega\bZ\bW^{-1}\bU^\dag=\bU^{\dag\top}\bW^{-1}\bZ^\top \bOmega^2\bZ \bW^{-1}\bU^\dag=\bI_K.
\end{align*}
This implies that $\bU=\bOmega\bZ\bW^{-1}\bU^\dag$, $\bSigma=\bSigma^\dag, \bV=\bV^\dag$. Then for any $i\in\calC_k$  we can write
\begin{align*}
	\bU_{i,:}=\frac{\omega_i}{\sqrt{\sum_{j\in\calC_k}\omega_j^2}}\bU^\dag_{k,:}.
\end{align*}
Since $\bU^\dag\in\OO_{K,K}$, we have  $\bU^\dag\bU^{\dag\top }=\bU^{\dag\top}\bU^\dag=\bI_K$. Hence we can conclude that $\bar\bU$  has $K$ distinct rows and in particular,
\begin{align*}
	\bar\bU_{i,:}=\bU^\dag_{k,:},\quad\quad  \forall k\in[K],\forall i\in\calC_k.
\end{align*}
For any $k\ne l\in[K]$, $\bar\bU_{i,:}=\bar\bU_{j,:}$ if $i,j\in\calC_k$ and $\bar\bU_{i,:}\ne \bar\bU_{j,:}$ if $i\in\calC_k$ and $j\in\calC_l$; moreover, we have
\begin{align*}
	\op{\bar\bU_{i,:}-\bar\bU_{j,:}}=\op{(e_k-e_l)^\top \bU^\dag}=\op{e_k-e_l}=\sqrt{2}.
\end{align*}
\hfill$\square$

\subsection{Proof of Lemma \ref{lem:incoherence}}
\begin{itemize}
	\item The first claim directly follows by noting that 
\begin{align*}
	\frac{\Delta}{\sqrt{2}}\ge  \sigma_\star=\kappa^{-1}\op{\bTheta}\ge \frac{\Delta}{\sqrt{2}\kappa}
\end{align*}
	\item For the second claim, by definition  and Assumption \ref{cond:balanced}-\ref{cond:const-degee}, we can immediately obtain that
\begin{align*}
	\op{\bR^*}=\op{\bW\bTheta^\top }\le  \op{\bW}\op{\bTheta}\lesssim\sqrt{\frac{N}{K}}\op{\bTheta}
\end{align*}
and
\begin{align*}
	\sigma_K(\bR^*)=\sigma_K(\bW\bTheta^\top )\ge \sigma_K(\bW)\sigma_K(\bTheta)\gtrsim \sqrt{\frac{N}{K}}\sigma_\star
\end{align*}
\sloppy On the other hand, we directly deduce  that the condition number of $\bR^*$ is also bounded by $\kappa$ up to constant. 
\hfill$\square$
\end{itemize}

\subsection{Proof of Lemma \ref{lem:incoherence-par}}
	First, eq. \eqref{eq:Ui-Udagi} and Assumption \ref{cond:const-degee} imply that $\mu_1=O\left(\mu_\omega^2\right)$.  Next, we have
\begin{align*}
	\op{\bV}_{2,\infty}&=\op{\bV^\dag}_{2,\infty}=\op{\bTheta\bW\bU^\dag(\bSigma^{\dag})^{-1}}_{2,\infty}\lesssim\op{\bTheta}_{2,\infty}\op{\bW}\sigma^{-1}_K(\bR^*)\\
	&\lesssim\frac{\op{\bTheta}_{2,\infty}}{\sigma_\star}\le\frac{\op{\bTheta}_{2,\infty}}{\kappa^{-1}K^{-1/2}\fro{\bTheta}}\lesssim\kappa\sqrt{\frac{\mu_{\bTheta}K}{J}}
\end{align*}
implying that $\mu_2=O(\mu_{\bTheta}\kappa^2)$. Finally, observe that
\begin{align*}
	\mu_0\leq \frac{NJ\op{\bR^*}^2\op{\bU}_{2,\infty}^2\op{\bV}_{2,\infty}^2}{K\sigma^2_{K}(\bR^*)}\lesssim\frac{NJ\op{\bTheta}^2\frac{\mu_1\mu_2K^2}{NJ}}{K\sigma_\star^2} =\mu_1\mu_2\kappa^2K
\end{align*}
Hence $\mu_0=O\left(\mu_\omega^2\mu_{\bTheta}\kappa^4K\right )$. \hfill$\square$

\subsection{Proof of Lemma \ref{lem:two-inf-bound}}
The decomposition \eqref{eq:Uhat-decomp} and the residual bound on $\mPsi$ follows directly from Theorem 10 in \cite{yan2021inference}. The operator/two-to-infinity bound on $\hat\bU-\bU\bO$ follows by applying Lemma 24 and Lemma 28 in \cite{yan2021inference}. Notice that the eigen-gap condition in \cite{yan2021inference} is stated as 
	\begin{align}
		\frac{\sigma^2_K(\bR^*)}{\kappa^2}\gg \xi_{\sf err}.
	\end{align}
By speculating the proofs therein, we can substitute ``$\gg$'' in the above condition with ``$\ge C_{\sf gap}$'', where $C_{\sf gap}>0$ is a sufficiently large absolute constant.

If the SNR condition is stronger by a factor of $\kappa^2\brac{\mu_{\bTheta}K}^{1/2}$, we can guarantee that 
\begin{align*}
	\op{\hat\bU-\bU\bO}_{2,\infty}\lesssim \kappa^2\frac{\xi_{\sf err}}{\sigma^2_K(\bR^*)}\sqrt{\frac{\mu K}{N}}\lesssim \kappa^{-2}\brac{\mu_{\bTheta}K}^{-1/2}\sqrt{\frac{\mu K}{N}}\lesssim \sqrt{\frac{\mu_1 K}{N}}=\infn{\bU },
\end{align*}
where Lemma \ref{lem:incoherence} is used.\hfill$\square$

\subsection{Proof of Lemma \ref{lem:trunc}}
The rationale of the proof  follows Lemma 51 in \cite{yan2021inference}, while there are some subtleties regarding the Poisson distribution which we treat differently.   
\paragraph{Step 1: lower bounding $\EE \ab{X}$.} 
By equation (4.19) in \cite{johnson2005univariate}, we have that
\begin{align*}
	\EE \ab{X}=\frac{2\lambda^{\lfloor \lambda\rfloor+1}e^{-\lambda}}{\lfloor\lambda\rfloor!}.
\end{align*}
If $\lambda\ge 1$, by Stirling's approximation we have that 
\begin{align*}
	\EE \ab{X}=\frac{2\lambda^{\lfloor \lambda\rfloor+1}e^{-\lambda}}{\lfloor\lambda\rfloor!}\ge \frac{2{\lfloor\lambda\rfloor}^{\lfloor \lambda\rfloor+1}e^{-\lambda}}{\sqrt{2\pi \lfloor\lambda\rfloor}\brac{\frac{\lfloor\lambda\rfloor}{e}}^{\lfloor\lambda\rfloor}e^{\frac{1}{12\lfloor\lambda\rfloor}}}=\sqrt{\lfloor\lambda\rfloor}\cdot \sqrt{\frac{2}{\pi}}e^{-\brac{\lambda-\lfloor\lambda\rfloor}}e^{-\frac{1}{12\lfloor\lambda\rfloor}}\ge C_{\sf lb,1}\sqrt{\lambda}
\end{align*}
If $0<\lambda< 1$, by Stirling's approximation we have that 
\begin{align*}
	\EE \ab{X}=2\lambda e^{-\lambda}\ge C_{\sf lb,2}\lambda
\end{align*}
Thus we can conclude that there exists some absolute constant $C_{\sf lb}>0$ such that
\begin{align*}
	\EE \ab{X}\ge \frac{C_{\sf lb}\lambda}{\sqrt{\lambda}\vee 1}
\end{align*} 
\paragraph{Step 2: constructing $\wt X$ by truncating $X$ randomly.} Define $X^+=X\vee 0$ and $X^-=\brac{-X}\vee 0$. Since $\EE X=0$, we get that
\begin{align}\label{eq:step2-EX-lower}
	\EE X^+=\EE X^-=\frac{1}{2}\EE \ab{X}\ge \frac{C_{\sf lb}\lambda}{\sqrt{\lambda}\vee 1}.
\end{align}
Define the function $f:\RR^+\mapsto\RR^+$ as
\begin{align*}
	f\brac{x}:=\EE\brac{X\II_{X\ge x}}.
\end{align*}
Notice that $f(x)$ is non-increasing in $[0,\infty)$ and left-continuous. Moreover, we can conclude that $\lim_{x\searrow 0}f(x)=\EE X^+$ by monotone convergence theorem via $X\II_{X\ge 1/n}\overset{a.s.}{\rightarrow} X^+$, and that $\lim_{x\rightarrow +\infty}f(x)=0$ by dominated convergence theorem via $X\II_{X\ge n}\overset{a.s.}{\rightarrow} 0$ and $\ab{X\II_{X\ge n}}\le \ab{X}$.   For any $x\in \RR^+$, we can get
\begin{align}\label{eq:step2-upper}
	f(x)=\EE\brac{X\II_{\ab{X}>x}}&\le \brac{\EE X^2}^{1/2}\sqbrac{\PP\brac{\ab{X}>x}}^{1/2}\le 2\sqrt{\lambda}\exp\brac{-\frac{x^2}{C\brac{\lambda+x}}},
\end{align}
for some absolute constant $C>0$, where the first inequality comes from Cauchy-Schwarz inequality and the second follows from standard tail bound for Poisson distribution.

For any given $\epsilon\in \brac{0, \frac{C_{\sf lb}\lambda}{2\brac{\sqrt{\lambda}\vee 1}}}$, we define
\begin{align*}
	x_\epsilon:=\sup\left\{x\in \RR^+:f\brac{x}\ge \epsilon\right \}.
\end{align*}
Since  $f$ is left-continuous, we obtain that 
\begin{align}\label{eq:step2-left-cont}
	\lim_{x\rightarrow x^-_\epsilon}f(x)=f\brac{x_\epsilon}\ge \epsilon\ge \lim_{x\rightarrow x_\epsilon^+}f(x).
\end{align}
By \eqref{eq:step2-upper}, we further have
\begin{align*}
	\epsilon\le f\brac{x_\epsilon}\le 2\sqrt{\lambda}\exp\brac{-\frac{x_\epsilon ^2}{C\brac{\lambda+x_\epsilon}}},
\end{align*}
which implies  there exists  some sufficiently large constant $C_{\sf ub}>0$ such that 
\begin{align*}
	x_\epsilon\le \sqrt{2C\lambda\log \brac{\frac{2\sqrt{\lambda}}{\epsilon}}}\vee \brac{2C\log \brac{\frac{2\sqrt{\lambda}}{\epsilon}}}\le  C_{\sf ub}\sqbrac{\sqrt{\lambda\log \brac{\frac{\sqrt{\lambda}}{\epsilon}}}+ \log \brac{\frac{\sqrt{\lambda}}{\epsilon}}}.
\end{align*}
On the other hand, a lower bound on $x_\epsilon$ can be obtained by observing that
\begin{align*}
	\EE X^+&=\EE\brac{X^+\II_{X\le x_\epsilon	}}+\EE\brac{X\II_{X>x_\epsilon}}=\EE\brac{X^+\II_{X\le x_\epsilon}}+\lim_{x\rightarrow x_\epsilon^+}f\brac{x}\\
	&\le x_\epsilon+\epsilon\le x_\epsilon+\frac{C_{\sf lb}\lambda}{2\brac{\sqrt{\lambda}\vee 1}}.
\end{align*}
Combined with \eqref{eq:step2-EX-lower}, we arrive at
\begin{align}\label{eq:step2-xe-lower}
	x_\epsilon\ge \frac{C_{\sf lb}\lambda}{2\brac{\sqrt{\lambda}\vee 1}}.
\end{align}
Then we can construct $\wt X$ as follows:
\begin{itemize}
	\item If $\lim_{x\rightarrow x_\epsilon^+}f(x)=f\brac{x_\epsilon}$, then $ f\brac{x_\epsilon}=\epsilon$ by \eqref{eq:step2-left-cont}. We then set
	\begin{align*}
		\wt X^+:=X^+\II_{X^+<x_\epsilon}.
	\end{align*}
	This gives us $\wt X^+< x_\epsilon\le C_{\sf ub}\sqbrac{\sqrt{\lambda\log \brac{\frac{\sqrt{\lambda}}{\epsilon}}}+ \log \brac{\frac{\sqrt{\lambda}}{\epsilon}}}$ and
	\begin{align*}
		\EE \wt X^+=\EE X^+-f\brac{x_\epsilon}=\EE X^+-\epsilon
	\end{align*}
	We can deduce from \eqref{eq:step2-xe-lower} and Markov's inequality that 
	\begin{align*}
		\PP\brac{\wt X^+\ne X^+}=\PP\brac{X^+\ge x_\epsilon}\le \frac{f\brac{x_\epsilon}}{x_\epsilon}=\frac{\epsilon}{x_\epsilon}\le \frac{2C_{\sf lb}\epsilon\brac{\sqrt{\lambda}\vee 1}}{\lambda}
	\end{align*}
	\item If $\lim_{x\rightarrow x_\epsilon^+}f(x)<f\brac{x_\epsilon}$, we get that
	\begin{align*}
		\EE \brac{X^+\II_{X^+=x_\epsilon}}=\EE \brac{X^+\II_{X^+\ge x_\epsilon}}-\EE \brac{X^+\II_{X^+>x_\epsilon}}=f\brac{x_\epsilon}-\lim_{x\rightarrow x_\epsilon^+}f(x)>0
	\end{align*}  
	We then set
	\begin{align*}
		\wt X^+:=X^+\II_{X^+<x_\epsilon}+X^+\II_{X^+=x_\epsilon}Q
	\end{align*}
	where  $Q\sim \text{Ber}(q)$ (independent of $X$) with 
	\begin{align*}
		q=\frac{f\brac{x_\epsilon}-\epsilon}{f\brac{x_\epsilon}-\lim_{x\rightarrow x_\epsilon^+}f\brac{x}}
	\end{align*}
	This construction still gives us $\wt X^+< x_\epsilon\le C_{\sf ub}\sqbrac{\sqrt{\lambda\log \brac{\frac{\lambda}{\epsilon}}}+ \log \brac{\frac{\lambda}{\epsilon}}}$, and
	\begin{align*}
		\EE \wt X^+&=\EE\brac{X^+\II_{X^+<x_\epsilon}}+q\EE\brac{X^+\II_{X^+=x_\epsilon}}\\
		&=\EE X^+-f\brac{x_\epsilon}+q\brac{f\brac{x_\epsilon}-\lim_{x\rightarrow x_\epsilon^+}f(x)}\\
		&=\EE X^+-\epsilon
	\end{align*}
	Moreover, we have that
	\begin{align*}
		\PP\brac{\wt X^+\ne X^+}&=\PP\brac{X^+> x_\epsilon}+\PP\brac{X^+= x_\epsilon,Q=0}\\
		&\le  \frac{\EE\brac{X^+\II_{X^+>x_\epsilon}}}{x_\epsilon}+\brac{1-q}\frac{\EE\brac{X^+\II_{X^+=x_\epsilon}}}{x_\epsilon}\\
		&=\frac{\EE X^+-\EE\brac{X^+\II_{X^+<x_\epsilon}}-q\EE\brac{X^+\II_{X^+=x_\epsilon}}}{x_\epsilon}\\
		&=\frac{\EE X^+-\EE \wt X^+}{x_\epsilon}\le \frac{\epsilon}{x_\epsilon}\le \frac{2C_{\sf lb}\epsilon\brac{\sqrt{\lambda}\vee 1}}{\lambda}.
	\end{align*}
\end{itemize}
So far, we have constructed $\wt X^+$ such that: (i) $\wt X^+$ equals either $X^+$ or 0; (ii) $\PP\brac{\wt X^+\ne X^+}\le \frac{2C_{\sf lb}\epsilon\brac{\sqrt{\lambda}\vee 1}}{\lambda}$; (iii) $\EE \wt X^+=\EE X^+-\epsilon$; and (iv) $0\le \wt X^+\le C_{\sf ub}\sqbrac{\sqrt{\lambda\log \brac{\frac{\sqrt{\lambda}}{\epsilon}}}+ \log \brac{\frac{\sqrt{\lambda}}{\epsilon}}}$. Similarly, we can also construct another random variable $X^-$ satisfying the same properties. Then we can construct
\begin{align*}
	\wt X:=\wt X^+-\wt X^-.
\end{align*}
\paragraph{Step 3: verifying the advertised properties of $\wt X$.} It suffices to check the following properties are satisfied for $\wt X$:
\begin{itemize}
	\item $\wt X$ has mean zero, i.e.,
	\begin{align*}
		\EE \wt X=\EE \wt X^+-\EE \wt X^-=\EE X^+-\epsilon -\EE X^-+\epsilon=\EE X=0.
	\end{align*}
	\item $\wt X$ is identical to $X$ with high probability, i.e.,
	\begin{align*}
		\PP\brac{X\ne \wt X}\le \PP\brac{X^+\ne \wt X^+}+\PP\brac{X^-\ne \wt X^-}\le \frac{4C_{\sf lb}\epsilon\brac{\sqrt{\lambda}\vee 1}}{\lambda}
	\end{align*}
	\item $\wt X$ is a bounded random variable such that $\ab{\wt X}\le C_{\sf ub}\sqbrac{\sqrt{\lambda\log \brac{\frac{\sqrt{\lambda}}{\epsilon}}}+ \log \brac{\frac{\sqrt{\lambda}}{\epsilon}}}$.
	\item The variance of $\wt X$ is close to $\lambda$ in the sense that
	\begin{align*}
		\textsf{Var}\brac{\wt X}=\EE \wt X^2= \EE X^2-\EE \brac{X^2\II_{X\ne \wt X}}=\lambda\brac{1-O\brac{\sqrt{1\vee \frac{1}{\lambda}}\brac{\frac{\epsilon\brac{\sqrt{\lambda}\vee 1}}{\lambda}}^{1/2}}}
	\end{align*}
	where we have used the fact that 
	\begin{align*}
		\EE \brac{X^2\II_{X\ne \wt X}}\le \brac{\EE X^4}^{1/2}\brac{\PP\brac{X\ne \wt X}}^{1/2}\lesssim\sqrt{\lambda\brac{1+\lambda}} \brac{\frac{\epsilon\brac{\sqrt{\lambda}\vee 1}}{\lambda}}^{1/2}.
	\end{align*}
\end{itemize}
For $\lambda>1$, we can take $\epsilon=\delta\sqrt{\lambda}/\brac{4C_{\sf lb}}$ for any $\delta\in(0,1)$. For $0<\lambda<1$, we can take $\epsilon={\delta\lambda^2}/\brac{4C_{\sf lb}}$ for any $\delta\in(0,1)$. Hence  we establish the desired result.\hfill$\square$

\subsection{Proof of Lemma \ref{lem:Thetaerr-inf}}
The SNR condition implies that
\begin{align}\label{eq:snr-thetaerr-inf}
	N\omega_{\sf max}\theta_{\sf max}\ge \brac{\frac{\omega_{\sf max}}{\omega_{\sf min}}}^{{3/2}}\brac{\frac{\theta_{\sf max}}{\theta^*_{\sf min}}}K\log(N+J)
\end{align}
By scrutinizing the proof of Theorem \ref{thm:Thetaerr-ave}, we can obtain that the variance of signal part in $\hat\theta_{j,k}-\theta_{j,k}$ admits 
\begin{align}\label{eq:sigmajk-formula}
	\sigma_{j,k}^2:=\textsf{Var}\brac{\frac{1}{\ab{\calC_k}}\sum_{i\in\calC_k}\frac{E_{i,j}}{\omega_i}}=\theta_{j,k}\frac{1}{\ab{\calC_k}^2}\sum_{i\in\calC_k}\frac{1-\omega_i\theta_{j,k}}{\omega_i}
\end{align}
Note that for any $j\in[J]$, $k\in[K]$ and any $t\in\RR$,
\begin{align*}
	&\PP\brac{\sigma_{j,k}^{-1}\brac{\hat \theta_{j,k}-\theta_{j,k}}\le  t}\\
	&=\PP\brac{\left\{\sigma_{j,k}^{-1}\brac{\hat \theta_{j,k}-\theta_{j,k}}\le  t\right\}\bigcap \calB_{\sf exact}\bigcap \wt\calB_{\sf good}}+\PP\brac{\left\{\sigma_{j,k}^{-1}\brac{\hat \theta_{j,k}-\theta_{j,k}}\ge t\right\}\bigcap\left\{ \calB^c_{\sf exact} \bigcup \wt\calB^c_{\sf good}\right \}}\\
	&\le \PP\brac{\sigma_{j,k}^{-1}\brac{\hat \theta_{j,k}-\theta_{j,k}}\II_{\calB_{\sf exact}\cap \wt\calB_{\sf good}}\le  t}+O\brac{\brac{N+J}^{-20}}
\end{align*}
implying that
\begin{align}\label{eq:theta-event-small}
	\ab{\PP\brac{\sigma_{j,k}^{-1}\brac{\hat \theta_{j,k}-\theta_{j,k}}\le  t}-\PP\brac{\sigma_{j,k}^{-1}\brac{\hat \theta_{j,k}-\theta_{j,k}}\II_{\calB_{\sf exact}\cap \wt\calB_{\sf good}}\le  t}}=O\brac{\brac{N+J}^{-20}}
\end{align}
Then it suffices for us to consider the following term 
\begin{align}\label{eq:theta-exact-decomp}
	\sigma_{j,k}^{-1}\brac{\hat \theta_{j,k}-\theta_{j,k}}\II_{\calB_{\sf exact}\cap \wt\calB_{\sf good}}&=\sigma_{j,k}^{-1}\frac{1}{\ab{\calC_k}}\sum_{i\in\calC_k}\left[\brac{\frac{1}{\hat\omega_i}-\frac{1}{\omega_i}}\brac{R^*_{i,j}+E_{i,j}}+\frac{E_{i,j}}{\omega_i}\right]\notag\\
	&=\sigma_{j,k}^{-1}\frac{1}{\ab{\calC_k}}\sum_{i\in\calC_k}\frac{E_{i,j}}{\omega_i}+\sigma_{j,k}^{-1}\frac{1}{\ab{\calC_k}}\sum_{i\in\calC_k}\brac{\frac{1}{\hat\omega_i}-\frac{1}{\omega_i}}\brac{R^*_{i,j}+E_{i,j}}
\end{align}
which consists of two parts. 
\paragraph{Main term in \eqref{eq:theta-exact-decomp}}
We can apply Berry-Esseen Theorem to the first part in \eqref{eq:theta-exact-decomp} to have 
\begin{align}\label{eq:theta-main-dist}
	\sup_{t\in \RR}\ab{\PP\brac{\sigma_{j,k}^{-1}\frac{1}{\ab{\calC_k}}\sum_{i\in\calC_k}\frac{E_{i,j}}{\omega_i}\le t}-\Phi(t)}\lesssim \sum_{i\in\calC_k}\EE \ab{\frac{E_{i,j}}{\omega_i\ab{\calC_k}\sigma_{j,k}}}^3
\end{align}
It thus boils down to bound $\sum_{i\in\calC_k}\EE \ab{\frac{E_{i,j}}{\omega_i\ab{\calC_k}\sigma_{j,k}}}^3$. Observe that
\begin{align}\label{eq:theta-sup-error}
	\sum_{i\in\calC_k}\EE \ab{\frac{E_{i,j}}{\omega_i\ab{\calC_k}\sigma_{j,k}}}^3&\le \frac{1}{\sigma_{j,k}}\max_{i\in\calC_k}\frac{1}{\omega_i\ab{\calC_k}}\lesssim \frac{1}{\omega_{\sf min}\ab{\calC_k}\sqrt{\theta_{j,k}\frac{1}{\ab{\calC_k}^2}\sum_{i\in\calC_k}\frac{1}{\omega_i}}}\notag\\
	&\lesssim \sqrt{\frac{\omega_{**}}{\omega_{\sf min}}}\sqrt{\frac{K}{N\theta_{\sf min }\omega_{\sf min}}}\lesssim \frac{1}{\sqrt{\log (N+J)}},
\end{align}
where the last inequality holds due to \eqref{eq:snr-thetaerr-inf}.
\paragraph{High order term in \eqref{eq:theta-exact-decomp}}
Recall that on $\wt\calB_{\sf good}$ that $\hat \bU\bO^\top   -\bU=\bE\bV\bSigma^{-1}+\wt\mPsi$. Denote $\bDelta=\bE\bV\bSigma^{-1}+\wt\mPsi$ with $\bDelta_i$ being the $i$-th row of $\bDelta$. Observe that 
\begin{align*}
	\hat\omega_i&=\ab{\calC_k}^{1/2}\op{\hat\bU_{i,:}\bO^\top }=\ab{\calC_k}^{1/2}\op{\bU_{i,:}+\bDelta_i}=\ab{\calC_k}^{1/2}\op{\bU_{i,:}}\sqrt{1+\underbrace{\frac{\op{\bDelta_i}^2}{\op{\bU_{i,:}}^2}+\frac{2\inp{\bU_{i,:} }{\bDelta_i}}{\op{\bU_{i,:}}^2}}_{=:2\epsilon_i}}
\end{align*}
We claim that $\epsilon_i=o(1)$ for all $i\in[N]$ which will be verified later. Then Taylor expansion of $1/\sqrt{1+x}$ gives that 
\begin{align*}
	\frac{1}{\hat\omega_i}-\frac{1}{\omega_i}=\frac{1-\brac{\epsilon_i-\delta_i}}{\ab{\calC_k}^{1/2}\op{\bU_{i,:}}}-\frac{1}{\ab{\calC_k}^{1/2}\op{\bU_{i,:}}}=-\frac{\tilde\epsilon_i}{\ab{\calC_k}^{1/2}\op{\bU_{i,:}}}
\end{align*}
for some $\delta_i=O\brac{\epsilon_i^2}=o(\epsilon_i)$ and $\tilde\epsilon_i:=\epsilon_i-\delta_i$. Therefore, we have that
\begin{align}\label{eq:hatw-w-R}
	&\ab{\frac{1}{\ab{\calC_k}}\sum_{i\in\calC_k}\brac{\frac{1}{\hat\omega_i}-\frac{1}{\omega_i}}R^*_{i,j}}=\ab{\frac{1}{\ab{\calC_k}}\sum_{i\in\calC_k}\frac{\epsilon_i}{\ab{\calC_k}^{1/2}\op{\bU_{i,:}}}R^*_{i,j}}+\ab{\frac{1}{\ab{\calC_k}}\sum_{i\in\calC_k}\frac{\delta_i}{\ab{\calC_k}^{1/2}\op{\bU_{i,:}}}R^*_{i,j}}\notag\\
	&\lesssim  \ab{\frac{1}{\ab{\calC_k}}\sum_{i\in\calC_k}\frac{\inp{\bU_{i,:}}{\bDelta_i}}{\ab{\calC_k}^{1/2}\op{\bU_{i,:}}^3}R^*_{i,j}}+\ab{\frac{1}{\ab{\calC_k}}\sum_{i\in\calC_k}\frac{\op{\bDelta_i}^2}{\ab{\calC_k}^{1/2}\op{\bU_{i,:}}^3}R^*_{i,j}}+\ab{\frac{1}{\ab{\calC_k}}\sum_{i\in\calC_k}\frac{\delta_i}{\ab{\calC_k}^{1/2}\op{\bU_{i,:}}}R^*_{i,j}}
\end{align}
First by the the definition of $\bDelta_i$,  the leading term in the first term of \eqref{eq:hatw-w-R} can be written as
\begin{align*}
	&\frac{1}{\ab{\calC_k}}\sum_{i\in\calC_k}\frac{\inp{\bU_{i,:}}{\bE_{i,:}\bV\bSigma^{-1}}}{\ab{\calC_k}^{1/2}\op{\bU_{i,:}}^3}R^*_{i,j}=\frac{1}{\ab{\calC_k}}\sum_{l=1}^J\sum_{i\in\calC_k}\frac{R^*_{i,j}}{\ab{\calC_k}^{1/2}\op{\bU_{i,:}}^3}E_{i,l}\brac{\bV_{l,:}\bSigma^{-1}\bU^\top e_i}
\end{align*}
To apply Bernstein's inequality, observe that
\begin{align*}
	L=\max_{l\in[J],i\in\calC_k}\frac{R^*_{i,j}}{\ab{\calC_k}^{1/2}\op{\bU_{i,:}}^3}\ab{\bV_{l,:}\bSigma^{-1}\bU^\top e_i}\lesssim\frac{\infn{\bV}^2\sigma_1(\bR^*)}{\ab{\calC_k}^{1/2}\op{\bU_{i,:}}}\sigma^{-1}_K(\bR^*)\lesssim \kappa\frac{\infn{\bV}^2}{\omega_{\sf min}}
\end{align*}
and
\begin{align*}
	V&\le \omega_{\sf max}\theta_{\sf max}\sum_{l=1}^J\sum_{i\in\calC_k}\frac{\op{\bV_{j,:}}^2\sigma^2_1(\bR^*)}{\ab{\calC_k}\op{\bU_{i,:}}^4}\ab{\bV_{l,:}\bSigma^{-1}\bU^\top e_i}^2\\
	&\le \omega_{\sf max}\theta_{\sf max}\sum_{l=1}^J\sum_{i\in\calC_k}\frac{\op{\bV_{j,:}}^2\sigma^2_1(\bR^*)}{\ab{\calC_k}\op{\bU_{i,:}}^2}\op{\bV_{l,:}}^2\sigma^{-2}_K(\bR^*)\le \kappa^2K\omega_{\sf max}\theta_{\sf max}\ab{\calC_k}\frac{\infn{\bV}^2}{\omega^2_{\sf min}}
\end{align*}
Hence we obtain that with probability at least $1-(N+J)^{-20}$,
\begin{align*}
	\ab{\frac{1}{\ab{\calC_k}}\sum_{i\in\calC_k}\frac{\inp{\bU_{i,:}}{\bE_{i,:}\bV\bSigma^{-1}}}{\ab{\calC_k}^{1/2}\op{\bU_{i,:}}^3}R^*_{i,j}}&\le \kappa\frac{\infn{\bV}^2}{\omega_{\sf min}\ab{\calC_k}}\log(N+J)+\kappa K^{1/2}\frac{\infn{\bV}\sqrt{\omega_{\sf max}\theta_{\sf max}}}{\omega_{\sf min}\ab{\calC_k}^{1/2}}\log^{1/2}(N+J)
\end{align*}
On the other hand, by \eqref{eq:sigmajk-formula} we have $\sigma^{-1}_{j,k}\le \ab{\calC_k}\sqrt{K\omega_{**}/N\theta^*_{\sf min}}$, which yields that 
\begin{align*}
	&\ab{\sigma^{-1}_{j,k}\frac{1}{\ab{\calC_k}}\sum_{i\in\calC_k}\frac{\inp{\bU_{i,:}}{\bE_{i,:}\bV\bSigma^{-1}}}{\ab{\calC_k}^{1/2}\op{\bU_{i,:}}^3}R^*_{i,j}}\\
	&\lesssim  \kappa^2 K^{{3/2}}\frac{\sqrt{\omega_{**}/\omega_{\sf min }}}{J\sqrt{N\theta^*_{\sf min}\omega_{\sf min}}}\log(N+J)+\kappa^2 K\sqrt{\frac{\theta_{\sf max}}{\theta^*_{\sf min}}}\frac{\sqrt{\brac{\omega_{**}/\omega_{\sf min}}\brac{\omega_{\sf max}/\omega_{\sf min}}}}{J}\log^{1/2}(N+J)\\
	&\lesssim  \kappa^2 K^{{3/2}}\brac{\frac{\omega_{\sf max}}{\omega_{\sf min}}}\brac{\frac{\theta_{\sf max}}{\theta^*_{\sf min}}}^{1/2}\frac{\log(N+J)}{J\sqrt{N\omega_{\sf max}\theta_{\sf max}}}+\kappa^2 K\brac{\frac{\omega_{\sf max}}{\omega_{\sf min}}}\brac{\frac{\theta_{\sf max}}{\theta^*_{\sf min}}}^{1/2}\frac{\log^{1/2}(N+J)}{J}
\end{align*}
with probability exceeding $1-(N+J)^{-20}$. It remains to treat the residual term in the first term of \eqref{eq:hatw-w-R} as 
\begin{align*}
	&\ab{\sigma_{j,k}^{-1}\frac{1}{\ab{\calC_k}}\sum_{i\in\calC_k}\frac{\inp{\bU_{i,:}}{\wt \mPsi_i}}{\ab{\calC_k}^{1/2}\op{\bU_{i,:}}^3}R^*_{i,j}}\le \sqrt{\frac{N}{K}}\sqrt{\frac{\omega_{**}}{\theta^*_{\sf min}}}\frac{\infn{\bV}\sigma_1(\bR^*)}{\omega_{\sf min}} \infn{\wt\mPsi}\\
	&\overset{\eqref{eq:sec-order-bound}}{\lesssim}\kappa^4\mu_2^{1/2}{\mu K}\brac{\frac{\omega_{\sf max}}{\omega_{\sf min}}}\brac{\frac{\theta_{\sf max}}{\theta^*_{\sf min}}}^{1/2}\sqrt{\frac{\log(N+J)}{J}}\\
	&+\mu^2_\omega\kappa^5\mu_2^{1/2}\mu^{1/2}K^{3/2}\brac{\frac{\omega_{\sf max}}{\omega_{\sf min}}}\brac{\frac{\theta_{\sf max}}{\theta^*_{\sf min}}}^{1/2}\frac{\brac{1+\sqrt{\frac{N}{J}}}\sqrt{\theta_{\sf max}}\log(N+J)}{\sqrt{\omega_{\sf max}}\sigma_\star}\\
	&\lesssim \kappa^9\mu^{3/2}_{\bTheta}{K^2}\brac{\frac{\omega_{\sf max}}{\omega_{\sf min}}}\brac{\frac{\theta_{\sf max}}{\theta^*_{\sf min}}}^{1/2}\sqrt{\frac{\log(N+J)}{J}}\\
	&+\mu^2_\omega\kappa^8\mu_{\bTheta}K^{2}\brac{\frac{\omega_{\sf max}}{\omega_{\sf min}}}\brac{\frac{\theta_{\sf max}}{\theta^*_{\sf min}}}^{1/2}\frac{\brac{1+\sqrt{\frac{N}{J}}}\sqrt{\theta_{\sf max}}\log(N+J)}{\sqrt{\omega_{\sf max}}\sigma_\star}\\
\end{align*}
Collecting the above three bounds we can conclude  that the first term in \eqref{eq:hatw-w-R} is bounded by 
\begin{align*}
	\ab{\frac{1}{\ab{\calC_k}}\sum_{i\in\calC_k}\frac{\inp{\bU_{i,:}}{\bDelta_i}}{\ab{\calC_k}^{1/2}\op{\bU_{i,:}}^3}R^*_{i,j}}\lesssim\frac{1}{\sqrt{\log(N+J)}}
\end{align*}
provided that
\begin{align*}
	\frac{\Delta^2}{\theta_{\sf max}}\gtrsim\frac{\mu_\omega^4\kappa^{18}\mu^2_{\bTheta}K^4}{\omega_{\sf max}} \brac{\frac{\omega_{\sf max}}{\omega_{\sf min}}}^2\brac{\frac{\theta_{\sf max}}{\theta^*_{\sf min}}}\brac{1+\frac{N}{J}}\log^3(N+J)
\end{align*}
and $J\gtrsim \mu_\omega^4\mu^3_{\bTheta}\kappa^{18}K^4\brac{\frac{\omega_{\sf max}}{\omega_{\sf min}}}^2\brac{\frac{\theta_{\sf max}}{\theta^*_{\sf min}}}\log^2(N+J)$.
\\Next, we consider the following components in the second term in \eqref{eq:hatw-w-R}, which can be seperately bounded as
 \begin{align*}
	&\ab{\sigma_{j,k}^{-1}\frac{1}{\ab{\calC_k}}\sum_{i\in\calC_k}\frac{\op{\bE_{i,:}\bV\bSigma^{-1}}^2}{\ab{\calC_k}^{1/2}\op{\bU_{i,:}}^3}R^*_{i,j}}\le \sqrt{\frac{N\omega_{**}}{K\theta^*_{\sf min}}}\frac{\omega_{*}K}{\ab{\calC_k}^{1/2}}\frac{\op{\bE\bV\bSigma^{-1}}^2}{\omega_{\sf min}^2}\infn{\bV}\sigma_1(\bR^*)\\
	&\lesssim\mu^{1/2}_{\bTheta}\kappa^2 K^{2}\sqrt{\frac{\omega_{**}}{\theta^*_{\sf min}}}\frac{\op{\bE}^2}{\omega_{\sf min}^2\sqrt{NJ}\sigma_\star}\lesssim\mu^{1/2}_{\bTheta}\kappa^2 K^{2}\brac{\frac{\theta_{\sf max}}{\theta^*_{\sf min}}}^{1/2}\brac{\frac{\omega_{\sf max}}{\omega_{\sf min}}}^2\frac{\brac{\sqrt{\frac{J}{N}}+\sqrt{\frac{N}{J}}}\sqrt{\theta_{\sf max}}}{\sqrt{\omega_{\sf max}}\sigma_\star}
\end{align*}
and 
 \begin{align*}
	&\ab{\sigma_{j,k}^{-1}\frac{1}{\ab{\calC_k}}\sum_{i\in\calC_k}\frac{\inp{\bE_{i,:}\bV\bSigma^{-1}}{\wt\mPsi_i}}{\ab{\calC_k}^{1/2}\op{\bU_{i,:}}^3}R^*_{i,j}}\le \sigma_{j,k}^{-1}\frac{1}{\ab{\calC_k}}\sum_{i\in\calC_k}\frac{\infn{\bE\bV\bSigma^{-1}}\infn{\wt\mPsi}}{\omega_{\sf min}\op{\bU_{i,:}}}\infn{\bV}\sigma_1(\bR^*)\\
	&\lesssim \kappa \sqrt{\frac{\omega_{**}}{\theta^*_{\sf min}}}\frac{\omega_*}{\omega^2_{\sf min}}{\frac{N}{K}}\infn{\bE\bV}\infn{\wt\mPsi}\infn{\bV}\lesssim \mu_{\bTheta}\kappa^3 \sqrt{\frac{\omega_{**}}{\theta^*_{\sf min}}}\frac{\omega_*}{\omega^2_{\sf min}}N\sqrt{\frac{\omega_{\sf max}\theta_{\sf max}\log(N+J)}{J}}\infn{\wt\mPsi}\\
	&\lesssim \mu^3_\omega\mu^2_{\bTheta}\kappa^{10}K^{5/2}\brac{\frac{\theta_{\sf max}}{\theta^*_{\sf min}}}^{1/2}\brac{\frac{\omega_{\sf max}}{\omega_{\sf min}}}^2 \sqrt{\frac{\log(N+J)}{J}}\frac{\sqrt{\theta_{\sf max}\log(N+J)}}{\sqrt{\omega_{\sf max}}\sigma_\star}\\
	&+\mu_\omega^3\mu^{3/2}_{\bTheta}\kappa^9K^2 \brac{\frac{\theta_{\sf max}}{\theta^*_{\sf min}}}^{1/2}\brac{\frac{\omega_{\sf max}}{\omega_{\sf min}}}^2\frac{\brac{1+\sqrt{\frac{N}{J}}}\theta_{\sf max}\log^{{3/2}}(N+J)}{\omega_{\sf max}\sigma_\star^2}
\end{align*}
and
 \begin{align*}
	&\ab{\sigma_{j,k}^{-1}\frac{1}{\ab{\calC_k}}\sum_{i\in\calC_k}\frac{\op{\wt\mPsi_i}^2}{\ab{\calC_k}^{1/2}\op{\bU_{i,:}}^3}R^*_{i,j}}\lesssim  \sqrt{\frac{\omega_{**}}{\theta^*_{\sf min}}}\frac{\omega_*}{\omega^2_{\sf min}}{\frac{N}{K}}\infn{\wt\mPsi}^2\infn{\bV}\sigma_1(\bR^*)\\
	&\lesssim \brac{\frac{\theta_{\sf max}}{\theta^*_{\sf min}}}^{1/2}\brac{\frac{\omega_{\sf max}}{\omega_{\sf min}}}^2  \mu^4_\omega\mu^{5/2}_{\bTheta}\kappa^{16}K^{{7/2}}\frac{\sqrt{\theta_{\sf max}}\log(N+J)}{\sqrt{NJ}\sqrt{\omega_{\sf max}}\sigma_\star}\\
	&+\brac{\frac{\theta_{\sf max}}{\theta^*_{\sf min}}}^{1/2}\brac{\frac{\omega_{\sf max}}{\omega_{\sf min}}}^2\mu_\omega^{{11/2}}\mu^{3/2}_{\bTheta}\kappa^{14}K^3\frac{\brac{\sqrt{\frac{J}{N}}+\sqrt{\frac{N}{J}}}\theta^{{3/2}}_{\sf max}\log^2(N+J)}{\omega^{{3/2}}_{\sf max}\sigma_\star^3}
\end{align*}
Collecting the above three bounds we can conclude  that the second term in \eqref{eq:hatw-w-R} is bounded by 
\begin{align*}
	\ab{\sigma_{j,k}^{-1}\frac{1}{\ab{\calC_k}}\sum_{i\in\calC_k}\frac{\op{\bDelta_i}^2}{\ab{\calC_k}^{1/2}\op{\bU_{i,:}}^3}R^*_{i,j}}\lesssim\frac{1}{\sqrt{\log(N+J)}}
\end{align*}
provided that
\begin{align*}
	\frac{\Delta^2}{\theta_{\sf max}}\gtrsim\frac{\mu_\omega^4\mu^{3/2}_{\bTheta}\kappa^{12}K^4}{\omega_{\sf max}} \brac{\frac{\omega_{\sf max}}{\omega_{\sf min}}}^4\brac{\frac{\theta_{\sf max}}{\theta^*_{\sf min}}}\brac{1+\frac{N}{J}+\frac{J}{N}}\log^2(N+J)
\end{align*}
and $NJ\gtrsim \mu_\omega^4\mu^{7/2}_{\bTheta}\kappa^{22}K^3\log(N+J)$. Finally, notice that 
\begin{align*}
	\ab{\sigma_{j,k}^{-1}\frac{1}{\ab{\calC_k}}\sum_{i\in\calC_k}\frac{\delta_i}{\ab{\calC_k}^{1/2}\op{\bU_{i,:}}}R^*_{i,j}}\lesssim \ab{\sigma_{j,k}^{-1}\frac{1}{\ab{\calC_k}}\sum_{i\in\calC_k}\frac{\epsilon_i^2}{\ab{\calC_k}^{1/2}\op{\bU_{i,:}}}R^*_{i,j}}=o\brac{\ab{\sigma_{j,k}^{-1}\frac{1}{\ab{\calC_k}}\sum_{i\in\calC_k}\frac{\epsilon_i}{\ab{\calC_k}^{1/2}\op{\bU_{i,:}}}R^*_{i,j}}},
\end{align*}
due to $\epsilon_i=o(1)$ for all $i\in[N]$. Hence we arrive at 
\begin{align*}
	\ab{\sigma_{j,k}^{-1}\frac{1}{\ab{\calC_k}}\sum_{i\in\calC_k}\brac{\frac{1}{\hat\omega_i}-\frac{1}{\omega_i}}R^*_{i,j}}\lesssim\frac{1}{\sqrt{\log(N+J)}}.
\end{align*}
It suffices to bound the term  $\ab{\frac{1}{\ab{\calC_k}}\sum_{i\in\calC_k}\brac{\frac{1}{\hat\omega_i}-\frac{1}{\omega_i}}E_{i,j}}$ the same as that in the proof of Theorem \ref{thm:Thetaerr-ave}, from which we have
\begin{align*}
	&\ab{\sigma_{j,k}^{-1}\frac{1}{\ab{\calC_k}}\sum_{i\in\calC_k}\brac{\frac{1}{\hat\omega_i}-\frac{1}{\omega_i}}E_{i,j}}\overset{\eqref{eq:omega-crude-bound}}{\lesssim} \brac{\frac{\omega_*}{\omega_{\sf min}}}^2\sqrt{\frac{N\omega_{**}}{K\theta^*_{\sf min}}}\frac{\mu_\omega\mu^{1/2}_{\bTheta}\kappa^4K\xi_{\sf err}}{\sigma^2_K(\bR^*)}\sqrt{\frac{\omega_{\sf max}\theta_{\sf max}\log(N+J)}{N}}\\
	&\lesssim  \mu_\omega^2 \mu^{1/2}_{\bTheta}\kappa^5K\brac{\frac{\omega_{\sf max}}{\omega_{\sf min}}}^3\brac{\frac{\theta_{\sf max}}{\theta^*_{\sf min}}}^{1/2}\frac{\sqrt{\theta_{\sf max}}\log(N+J)}{\sqrt{\omega_{\sf max}}\sigma_\star}\lesssim \frac{1}{\sqrt{\log(N+J)}},
\end{align*}
the last inequality holds provided that 
\begin{align*}
	\frac{\Delta^2}{\theta_{\sf max}}\gtrsim\frac{\mu_\omega^4\mu_{\bTheta}\kappa^{12}K^2}{\omega_{\sf max}} \brac{\frac{\omega_{\sf max}}{\omega_{\sf min}}}^6\brac{\frac{\theta_{\sf max}}{\theta^*_{\sf min}}}\log^3(N+J).
\end{align*}
So far, we establish that under SNR condition \eqref{eq:theta-inf-cond-lem},
\begin{align}\label{eq:theta-res-bound}
	\Prob\brac{\ab{\sigma_{j,k}^{-1}\frac{1}{\ab{\calC_k}}\sum_{i\in\calC_k}\brac{\frac{1}{\hat\omega_i}-\frac{1}{\omega_i}}\brac{R^*_{i,j}+E_{i,j}}}\ge \frac{C}{\sqrt{\log(N+J)}}}=O\brac{\brac{N+J}^{-20}},
\end{align}
for some absolute constant $C>0$. Using \eqref{eq:theta-exact-decomp}, \eqref{eq:theta-main-dist} and \eqref{eq:theta-res-bound}, we thereby have
\begin{align*}
	&\ab{\PP\brac{\sigma_{j,k}^{-1}\brac{\hat \theta_{j,k}-\theta_{j,k}}\II_{\calB_{\sf exact}\cap \wt\calB_{\sf good}}\le  t}-\Phi(t)}\lesssim \frac{1}{\sqrt{\log(N+J)}},
\end{align*}
for any $t\in\RR$. Combined with \eqref{eq:theta-event-small}, we further obtain that 
\begin{align*}
	&\sup_{t\in\RR}\ab{\PP\brac{\sigma_{j,k}^{-1}\brac{\hat \theta_{j,k}-\theta_{j,k}}\le  t}-\Phi(t)}\lesssim \brac{N+J}^{-20}+\frac{1}{\sqrt{\log(N+J)}}\lesssim \frac{1}{\sqrt{\log(N+J)}}
\end{align*}
To complete the proof, it remains to verify $\epsilon_i=o(1)$ for all $i\in[N]$, which boils down to require
\begin{align*}
	\op{\hat\bU\bO^\top -\bU}_{2,\infty}=o\brac{\min_{i\in [N]}\op{\bU_{i,:}}}
\end{align*}
It turns out that on $\wt \calB_{\sf good }$, it suffices for us to assume 
\begin{align*}
\frac{\Delta^2}{\theta_{\sf max}}\ge \frac{\mu_\omega^2\mu_{\bTheta}\kappa^{12}K^2}{\omega_{\sf max}}\brac{\frac{\omega_{\sf max}}{\omega_{\sf min}}}^2\log^2(N+J),
\end{align*}
which is  implied by \eqref{eq:theta-inf-cond-lem}, and this completes the proof.\hfill$\square$

\subsection{Proof of Lemma \ref{lem:consistent-sigma}}
On $\calB_{\sf exact}\cap\wt\calB_{\sf good}$, we obtain that 
\begin{align*}
	\ab{\hat \sigma^2_{j,k}-\sigma_{j,k}^2}&\le \ab{\brac{\hat \theta_{j,k}-\theta_{j,k}}\frac{1}{\ab{ \calC_k}^2}\sum_{i\in\calC_k}\frac{1-\hat\omega_i\hat \theta_{j,k}}{\hat\omega_i}}+ \ab{\theta_{j,k}\frac{1}{\ab{ \calC_k}^2}\sum_{i\in\calC_k}\brac{\frac{1-\hat\omega_i\hat \theta_{j,k}}{\hat\omega_i}-\frac{1-\omega_i\theta_{j,k}}{\omega_i}}}.
\end{align*}
We further consider the event $\calB_{\sf theta}:=\{\eqref{eq:thetaer-event-gen}\text{~holds}\}$. By \eqref{eq:thetaer-event-gen} and \eqref{eq:omega-crude-bound}, we have
\begin{align*}
	&\ab{\theta_{j,k}\frac{1}{\ab{ \calC_k}^2}\sum_{i\in\calC_k}\brac{\frac{1-\hat\omega_i\hat \theta_{j,k}}{\hat\omega_i}-\frac{1-\omega_i\theta_{j,k}}{\omega_i}}}\le \ab{\theta_{j,k}\frac{1}{\ab{ \calC_k}^2}\sum_{i\in\calC_k}\brac{\frac{1}{\hat\omega_i}-\frac{1}{\omega_i}}}+\ab{\theta_{j,k}\frac{1}{\ab{ \calC_k}^2}\sum_{i\in\calC_k}\brac{\hat\theta_{j,k}-\theta_{j,k}}}\\
	&\le \frac{\theta_{j,k}}{\ab{\calC_k}}\brac{\frac{\omega_*}{\omega_{\sf min}}}^2\frac{\mu_\omega \mu^{1/2}_{\bTheta}\kappa^4K^{1/2}\xi_{\sf err}}{\sigma^2_K(\bR^*)}+\frac{\theta_{j,k}}{\ab{\calC_k}}\mu_\omega\mu^{1/2}_{\bTheta}\kappa^6K^{3/2}\brac{\frac{\omega_*}{\omega_{\sf min}}}^2\sqrt{\frac{\omega_{\sf max}\theta_{\sf max}\log(N+J)}{N\wedge J}}\\
	&\overset{\eqref{eq:xi-err-bound}}{\lesssim} \frac{\theta_{j,k}}{\ab{\calC_k}}\mu^2_\omega \mu^{1/2}_{\bTheta}\kappa^5K\brac{\frac{\omega_*}{\omega_{\sf min}}}^2\frac{\sqrt{\theta_{\sf max}\log(N+J)}}{\sqrt{\omega_{\sf max}}\sigma_\star}+\frac{\theta_{j,k}}{\ab{\calC_k}}\mu_\omega\mu^{1/2}_{\bTheta}\kappa^6K^{3/2}\brac{\frac{\omega_*}{\omega_{\sf min}}}^2\sqrt{\frac{\omega_{\sf max}\theta_{\sf max}\log(N+J)}{N\wedge J}}.
\end{align*}
Using $\sigma^{-2}_{j,k}\lesssim \ab{\calC_k}\omega_{**}/{\theta_{j,k}}\le \ab{\calC_k}/{\theta_{j,k}}$ we can proceed as
\begin{align*}
	&\ab{\sigma^{-2}_{j,k}\frac{\theta_{j,k}}{\ab{ \calC_k}^2}\sum_{i\in\calC_k}\brac{\frac{1-\hat\omega_i\hat \theta_{j,k}}{\hat\omega_i}-\frac{1-\omega_i\theta_{j,k}}{\omega_i}}}\\
	&\lesssim \mu^2_\omega \mu^{1/2}_{\bTheta}\kappa^5K\brac{\frac{\omega_*}{\omega_{\sf min}}}^2\frac{\sqrt{\theta_{\sf max}\log(N+J)}}{\sqrt{\omega_{\sf max}}\sigma_\star}+\mu_\omega\mu^{1/2}_{\bTheta}\kappa^6K^{3/2}\brac{\frac{\omega_*}{\omega_{\sf min}}}^2\sqrt{\frac{\omega_{\sf max}\theta_{\sf max}\log(N+J)}{N\wedge J}}\\
	&\lesssim \frac{1}{\sqrt{\log (N+J)}},
\end{align*}
where the last inequality holds under the SNR condition
\begin{align}\label{eq:snr-1}
		\frac{\Delta^2}{\theta_{\sf max}}\gtrsim \frac{\mu_\omega^4\mu_{\bTheta}\kappa^{14}K^3}{\omega_{\sf max}} \brac{\frac{\omega_{\sf max}}{\omega_{\sf min}}}^4\brac{\frac{\theta_{\sf max}}{\theta^*_{\sf min}}}\brac{\frac{J}{N}+1}\log^2(N+J).
\end{align}
Similarly, we note that 
\begin{align*}
	&\ab{\brac{\hat \theta_{j,k}-\theta_{j,k}}\frac{1}{\ab{ \calC_k}^2}\sum_{i\in\calC_k}\frac{1-\hat\omega_i\hat \theta_{j,k}}{\hat\omega_i}}\le\ab{\brac{\hat \theta_{j,k}-\theta_{j,k}}\frac{1}{\ab{ \calC_k}^2}\sum_{i\in\calC_k}\frac{1}{\hat\omega_i}}+\ab{\brac{\hat \theta_{j,k}-\theta_{j,k}}\frac{1}{\ab{ \calC_k}}\hat \theta_{j,k}}\\
	&\lesssim \mu_\omega\mu^{1/2}_{\bTheta}\kappa^6K^{3/2}\brac{\frac{\omega_*}{\omega_{\sf min}}}^2\sqrt{\frac{\omega_{\sf max}\theta_{\sf max}\log(N+J)}{N\wedge J}}\brac{\frac{1}{\ab{\calC_k}\omega_{**}}+\frac{\theta_{j,k}}{\ab{\calC_k}}}.
\end{align*}
Here, the last inequality holds since $\ab{\hat\theta_{j,k}-\theta_{j,k}}<\theta_{j,k}$ under \eqref{eq:snr-1}.
We can then readily obtain that 
\begin{align*}
	&\ab{\sigma^{-2}_{j,k}\brac{\hat \theta_{j,k}-\theta_{j,k}}\frac{1}{\ab{ \calC_k}^2}\sum_{i\in\calC_k}\frac{1-\hat\omega_i\hat \theta_{j,k}}{\hat\omega_i}}\\
	&\lesssim \mu_\omega\mu^{1/2}_{\bTheta}\kappa^6K^{3/2}\brac{\frac{\omega_*}{\omega_{\sf min}}}^2\sqrt{\frac{\omega_{\sf max}\theta_{\sf max}\log(N+J)}{N\wedge J}}\brac{\frac{1}{\omega_{**}\theta^*_{\sf min}}+1}\lesssim\frac{1}{\sqrt{\log (N+J)}},
\end{align*}
provided that 
\begin{align*}
	\frac{\Delta^2}{\theta_{\sf max}}\gtrsim \frac{\mu_\omega^4\mu_{\bTheta}\kappa^{14}K^3}{\omega_{\sf max}} \brac{\frac{\omega_{\sf max}}{\omega_{\sf min}}}^6\brac{\frac{\theta_{\sf max}}{\theta^*_{\sf min}}}\brac{\frac{J}{N}+1}\log^2(N+J).
\end{align*}
Denote $\calB_*:=\wt\calB_{\sf good}\bigcap  \calB_{\sf exact} \bigcap \calB_{\sf theta}$ and note that $\Prob(\calB_*)\ge 1-O(\brac{N+J}^{-20})$. Thus so far, we can establish that 
\begin{align*}
	&\Prob\brac{\ab{\frac{\hat \sigma^2_{j,k}}{\sigma_{j,k}^2}-1}\ge \frac{C}{\sqrt{\log(N+J)}}}\\
	&=\Prob\brac{\left\{\ab{\frac{\hat \sigma^2_{j,k}}{\sigma_{j,k}^2}-1}\ge \frac{C}{\sqrt{\log(N+J)}}\right\}\bigcap \calB_*}+\Prob\brac{\left\{\ab{\frac{\hat \sigma^2_{j,k}}{\sigma_{j,k}^2}-1}\ge \frac{C}{\sqrt{\log(N+J)}}\right\}\bigcap \calB_*^c}\\
	&= O\brac{\brac{N+J}^{-20}}
\end{align*}
for some absolute constant $C>0$ sufficiently large.\hfill$\square$

\subsection{Proof of Lemma \ref{lem:lower-bound}}
Rearranging terms we obtain that the term in bracket is equivalent to 
\begin{align*}
	\sum_{j=1}^J \brac{\bar X_j \log \frac{p_{j,2}(1-p_{j,1})}{p_{j,1}(1-p_{j,2})}-D_{\sf KL}(p_1||p_2)}> 0
\end{align*}
where $\bar X_j=X_j-p_{j,1}$ and $D_{\sf KL}(p_1||p_2):=\sum_{j=1}^J\left[p_{j,1}\log\frac{p_{j,2}}{p_{j,1}}+(1-p_{j,1})\log\frac{1-p_{j,2}}{1-p_{j,1}}\right]$ is the Kullback-Leibler divergence between two Bernoulli random vectors. Denote $Z_j:=\bar X_j \log \frac{p_{j,2}(1-p_{j,1})}{p_{j,1}(1-p_{j,2})}-D_{\sf KL}(p_1||p_2)$,  $\calZ=\{z\in \RR^{J}:0\le \sum_{j=1}^Jz_j\le \tau\}$ for some $\tau>0$ to be specified later, and $h_j(\cdot )$ the probability mass function of $Z_j$. It follows that for any $t\ge 0$
\begin{align*}
	\PP&\brac{\sum_{j=1}^J \brac{\bar X_j \log \frac{p_{j,2}(1-p_{j,1})}{p_{j,1}(1-p_{j,2})}-D_{\sf KL}(p_1||p_2)}> 0}\ge \PP\brac{\tau \ge \sum_{j=1}^JZ_j\ge 0}=\sum_{z\in \calZ}\prod_{j=1}^Jh_j(z_j)\\
	&= \EE\exp\brac{t\sum_{j=1}^JZ_j}\sum_{z\in \calZ}\prod_{j=1}^J\frac{\exp(tz_j)h_j(z_j)}{\exp(th_j)\EE\exp\brac{t\sum_{j=1}^JZ_j}}\\
	&\ge \frac{\EE\exp\brac{t\sum_{j=1}^JZ_j}}{\exp\brac{t\tau }}\sum_{z\in \calZ}\prod_{j=1}^J\frac{\exp(tz_j)h_j(z_j)}{\EE\exp\brac{t\sum_{j=1}^JZ_j}}
\end{align*}
By taking $t=\frac{1}{2}$, we obtain that
\begin{align*}
	\EE\exp\brac{\frac{1}{2}\sum_{j=1}^JZ_j}	&=\prod_{j=1}^J\brac{p_{j,1}e^{t\log\frac{p_{j,2}}{p_{j,1}}}+(1-p_{j,1})e^{-t\log\frac{1-p_{j,1}}{1-p_{j,2}}}}\\
	&=\prod_{j=1}^J\brac{\sqrt{p_{j,1}p_{j,2}}+\sqrt{(1-p_{j,1})(1-p_{j,2})}}\\
	&=\exp\brac{-\frac{1}{2}\sum_{j=1}^JI_{j}}
\end{align*}
where $I_j=-2\log \brac{\sqrt{p_{j,1}p_{j,2}}+\sqrt{(1-p_{j,1})(1-p_{j,2})}}$. Define for any $j\in[J]$,
\begin{align*}
	q_j(y)=\frac{\exp\brac{\frac{y}{2}}h_j(y)}{\EE\exp\brac{\frac{1}{2}\sum_{j=1}^JZ_j}}
\end{align*}
It readily follows that $q_j(y)\ge 0$ and $\sum_{y}q_j(y)=1$, which implies that $q_j(\cdot)$ is a probability mass function. Let $\{Y_j\}_{j=1}^d$ be a sequence of independent random variables such that $Y_j\sim q_j(\cdot)$,  then we have
\begin{align*}
	\PP&\brac{\sum_{j=1}^J \brac{\bar X_j \log \frac{p_{j,2}(1-p_{j,1})}{p_{j,1}(1-p_{j,2})}-D_{\sf KL}(p_1||p_2)}> 0}\ge \exp\brac{-\sum_{j=1}^JI_{j}-\frac{\tau }{2}}\sum_{z\in \calZ}\prod_{j=1}^Jq_j(z_j)\\
	&\ge \exp\brac{-\sum_{j=1}^JI_{j}-\frac{\tau }{2}}\PP\brac{1\ge  \frac{\sum_{j=1}^JY_j}{\tau }\ge 0 }
\end{align*}
Next we will calculate $\textsf{Var}\brac{\sum_{j=1}^JY_j}$. By direct calculation of the moment generating function of $Y_j$, we can obtain that
\begin{align*}
	\textsf{Var}(Y_j)=\frac{\sqrt{p_{j,1}p_{j,2}}(1-p_{j,1})\sqrt{\frac{1-p_{j,1}}{1-p_{j,2}}}\left[\log \frac{p_{j,2}(1-p_{j,1})}{p_{j,1}(1-p_{j,2})}\right]^2}{\left[1-p_{j,1}+\sqrt{p_{j,1}p_{j,2}}\sqrt{\frac{1-p_{j,1}}{1-p_{j,2}}}\right]^2}\asymp \sqrt{p_{j,1}p_{j,2}}\left[\log \frac{p_{j,2}(1-p_{j,1})}{p_{j,1}(1-p_{j,2})}\right]^2
\end{align*}
Without loss of generality, assume $p_{j,1}>p_{j,2}$ and recall that we have $p_{j,1}\asymp p_{j,2}=o(1)$. Now write $p:=p_{j,2}$ and $\epsilon:=p_{j,1}-p_{j,2}$. We consider two cases. First if $\epsilon =o(p)$, then we have
\begin{align*}
	\textsf{Var}(Y_j)\asymp p\left[\log\brac{1+\frac{\epsilon}{p}}+\log \brac{1+\frac{\epsilon}{1-p-\epsilon}}\right]^2\asymp \frac{\epsilon^2}{p}
\end{align*}
On the other hand we have $I_j\asymp \brac{\sqrt{p+\epsilon}-\sqrt{p}}^2\asymp \frac{\epsilon^2}{p}$, which implies that $\textsf{Var}(Y_j)\asymp I_j$. For another case when $\epsilon\asymp p$, we simply have $\textsf{Var}(Y_j)\asymp I_j\asymp p$. Hence we conclude that $\textsf{Var}\brac{\sum_{j=1}^JY_j}\asymp \sum_{j=1}^J I_j\rightarrow \infty$ by assumption. Furthermore, direct computation gives that 
\begin{align*}
	|Y_j-\EE Y_j|\le 1,\quad \EE|Y_j-\EE Y_j|^3\le \textsf{Var}(Y_j)
\end{align*}
We then can take $\tau =\sqrt{\textsf{Var}\brac{\sum_{j=1}^JY_j}}$ and use Berry-Essen theorem to have
\begin{align*}
	\PP&\brac{\sum_{j=1}^J \brac{\bar X_j \log \frac{p_{j,2}(1-p_{j,1})}{p_{j,1}(1-p_{j,2})}-D_{\sf KL}(p_1||p_2)}> 0}\\
	&\ge \exp\brac{-\sum_{j=1}^JI_{j}-\frac{1}{2}\sqrt{\textsf{Var}\brac{\sum_{j=1}^JY_j}}}\PP\brac{1\ge  \frac{\sum_{j=1}^JY_j}{\sqrt{\textsf{Var}\brac{\sum_{j=1}^JY_j}}}\ge 0 }\\
	&\ge \exp\brac{-\sum_{j=1}^JI_j(1+o(1))}\brac{\PP\brac{1\ge N(0,1)\ge 0}-C_0\frac{\sum_{j=1}^J\EE|Y_j-\EE Y_j|^3}{\brac{\sum_{j=1}^J\textsf{Var}(Y_j)}^{{3/2}}}}\\
	&\ge\exp\brac{-\sum_{j=1}^JI_j(1+o(1))}
\end{align*}
as $\sum_{j=1}^J I_j\rightarrow \infty$.  \hfill$\square$

\end{document}